\documentclass[onecolumn,bibtex]{aastex701}
\usepackage{times}
\usepackage{epsfig}
\usepackage{amsmath, amsthm, amssymb}
\usepackage{mathrsfs}
\usepackage{mathtools}
\usepackage{color}
\usepackage{microtype}
\usepackage{float}
\usepackage[caption=false]{subfig}
\usepackage{stfloats}
\usepackage{natbib}
\usepackage{verbatim}
\usepackage{wrapfig}
\hypersetup{backref,breaklinks,colorlinks,citecolor=blue}
\usepackage[all]{hypcap}
\usepackage{my_definitions}
\usepackage{subfiles}
\usepackage{booktabs}
\usepackage[linesnumbered,ruled,vlined]{algorithm2e}
\usepackage{enumerate}
\usepackage{placeins}
\usepackage[inkscapelatex=false]{svg}

\newtheorem{rem}{Remark}

\newcommand{\ye}{Y_{\rm e}}
\newcommand{\mB}{m_{\mbox{\tiny B}}}
\newcommand{\kB}{k_{\mbox{\tiny B}}}
\newcommand{\Rshock}{R_{\mbox{\tiny Sh}}}
\newcommand{\Rpns}{R_{\mbox{\tiny PNS}}}
\newcommand{\Ef}{E_{\rm f}}
\newcommand{\QL}{Q_{\rm L}}
\newcommand{\QM}[1]{{Q_{{\rm M}}}_{#1}}
\newcommand{\QE}{Q_{\rm E}}
\newcommand{\Lo}{\mbox{\tiny L}}
\newcommand{\Hi}{\mbox{\tiny H}}
\newcommand{\Center}{\mbox{\tiny C}}
\newcommand{\Kx}{\vect{K}_{\vect{x}}}
\newcommand{\Kz}{\vect{K}_{\vect{z}}}
\newcommand{\Ax}{\vect{A}_{\vect{x}}}
\newcommand{\Az}{\vect{A}_{\vect{z}}}

\newcommand{\intKz}[1]{\big({#1}\big)_{\Kz^{\eidx}}}
\newcommand{\eidx}{\mathfrak{e}}
\newcommand{\nElementsE}{\mathfrak{N}}
\newcommand{\qidx}{\mathfrak{q}}
\newcommand{\pluseq}{\mathrel{+}=}
\newcommand{\aint}[1]{\langle{#1}\rangle}

\shortauthors{Endeve et al.}
\received{\today}

\shorttitle{\thornado+\flashx\ Neutrino Rad-Hydro}

\begin{document}

\title{\thornado+\flashx: A Hybrid DG-IMEX and Finite-Volume Framework for \\ Neutrino-Radiation Hydrodynamics in Core-Collapse Supernovae
\footnote{This manuscript has been authored in part by UT-Battelle, LLC, under contract DE-AC05-00OR22725 with the US Department of Energy (DOE). 
The US government retains and the publisher, by accepting the article for publication, acknowledges that the US government retains a nonexclusive, paid-up, irrevocable, worldwide license to publish or reproduce the published form of this manuscript, or allow others to do so, for US government purposes. 
DOE will provide public access to these results of federally sponsored research in accordance with the DOE Public Access Plan (http://energy.gov/downloads/doe-public-access-plan).}}

\author[0000-0003-1251-9507]{Eirik Endeve}
\email{endevee@ornl.gov}
\affiliation{Computer Science and Mathematics Division, Oak Ridge National Laboratory, Oak Ridge, TN 37831, USA}
\affiliation{Department of Physics and Astronomy, University of Tennessee Knoxville, Knoxville, TN 37996, USA}

\author[0000-0001-5869-8542]{Vassilios Mewes}
\email{mewesv@ornl.gov}
\affiliation{National Center for Computational Sciences, Oak Ridge National Laboratory, Oak Ridge, TN 37831, USA}

\author[0000-0003-3023-7140]{J. Austin Harris}
\email{harrisja@ornl.gov}
\affiliation{National Center for Computational Sciences, Oak Ridge National Laboratory, Oak Ridge, TN 37831, USA}

\author[0000-0002-2215-6968]{M. Paul Laiu}
\email{laiump@ornl.gov}
\affiliation{Computer Science and Mathematics Division, Oak Ridge National Laboratory, Oak Ridge, TN 37831, USA}

\author[0000-0001-7951-049X]{Ran Chu}
\email{rchu@vols.utk.edu}
\affiliation{Department of Physics and Astronomy, University of Tennessee Knoxville, Knoxville, TN 37996, USA}

\author[0000-0002-3591-123X]{Steven A. Fromm}
\email{frommsa@ornl.gov}
\affiliation{National Center for Computational Sciences, Oak Ridge National Laboratory, Oak Ridge, TN 37831, USA}

\author[0000-0001-9816-9741]{Anthony Mezzacappa}
\email{mezz@utk.edu}
\affiliation{Department of Physics and Astronomy, University of Tennessee Knoxville, Knoxville, TN 37996, USA}

\author[0000-0002-5358-5415]{O. E. Bronson Messer}
\email{bronson@ornl.gov}
\affiliation{National Center for Computational Sciences, Oak Ridge National Laboratory, Oak Ridge, TN 37831, USA}

\author[0000-0002-9481-9126]{W. Raphael Hix}
\email{raph@ornl.gov}
\affiliation{Physics Division, Oak Ridge National Laboratory, Oak Ridge, TN 37831, USA}
\affiliation{Department of Physics and Astronomy, University of Tennessee Knoxville, Knoxville, TN 37996, USA}

\author[0000-0003-0999-5297]{Stephen W. Bruenn}
\email{bruenn@fau.edu}
\affiliation{Department of Physics, Florida Atlantic University, Boca Raton, FL 33431, USA}

\author[0000-0002-5231-0532]{Eric J. Lentz}
\email{elentz@utk.edu}
\affiliation{Department of Physics and Astronomy, University of Tennessee Knoxville, Knoxville, TN 37996, USA}
\affiliation{Physics Division, Oak Ridge National Laboratory, Oak Ridge, TN 37831, USA}

\author[0000-0001-9869-9750]{Klaus Weide}
\email{kweide@uchicago.edu}
\affiliation{Department of Computer Science, University of Chicago, Chicago, IL 60637, USA}

\author[0000-0002-0086-105X]{Christian Y. Cardall}
\email{cardallcy@ornl.gov}
\affiliation{Physics Division, Oak Ridge National Laboratory, Oak Ridge, TN 37831, USA}

\author[0000-0003-2103-312X]{Ann S. Almgren}
\email{asalmgren@lbl.gov}
\affiliation{Center for Computational Science and Engineering, Lawrence Berkeley National Laboratory, CA 94720, USA}

\author[0000-0003-3299-7426]{Anshu Dubey}
\email{adubey@anl.gov}
\affiliation{Mathematics and Computer Science Division, Argonne National Laboratory, Lemont, IL 60439, USA}
\affiliation{Department of Computer Science, University of Chicago, Chicago, IL 60637, USA}

\author[0000-0002-5080-5996]{Sean M. Couch}
\email{scouch@msu.edu}
\affiliation{Department of Physics and Astronomy, Michigan State University, East Lansing, MI 48824, USA}
\affiliation{Department of Computational Mathematics, Science, and Engineering, Michigan State University, East Lansing, MI 48824, USA}
\affiliation{Facility for Rare Isotope Beams, Michigan State University, East Lansing, MI 48824, USA}

\author[0000-0002-9371-1447]{Philipp M{\"o}sta}
\email{p.moesta@uva.nl}
\affiliation{GRAPPA, Anton Pannekoek Institute for Astronomy, Institute of High-Energy Physics, and Institute of Theoretical Physics, University of Amsterdam, Amsterdam, The Netherlands}

\author[0000-0003-2300-5165]{Donald E. Willcox}
\email{eugene.willcox@gmail.com}
\affiliation{Center for Computational Science and Engineering, Lawrence Berkeley National Laboratory, CA 94720, USA}

\correspondingauthor{Eirik Endeve}
\email{endevee@ornl.gov}

\begin{abstract}
	We present neutrino-transport algorithms implemented in the toolkit for high-order neutrino-radiation hydrodynamics (\thornado) and their coupling to self-gravitating hydrodynamics within the adaptive mesh refinement (AMR)-based multiphysics simulation framework \flashx.  
	\thornado, developed primarily for simulations of core-collapse supernovae (CCSNe), employs a spectral, six-species two-moment formulation with algebraic closure and special-relativistic observer corrections accurate to $\mathcal{O}(v/c)$, and uses discontinuous Galerkin (DG) methods for phase-space discretization combined with implicit-explicit time stepping.  
	A key development is a nonlinear neutrino--matter coupling algorithm based on nested fixed-point iteration with Anderson acceleration, enabling fully implicit treatment of collisional processes, including energy-coupling interactions such as neutrino--electron scattering and pair production.
	Coupling to finite-volume (FV) hydrodynamics is achieved through a hybrid DG-FV representation of the fluid variables and operator-split evolution within \flashx.
	The implementation is verified using basic transport tests with idealized opacities and relaxation and deleptonization problems with tabulated microphysics.  
	Spherically symmetric CCSN simulations demonstrate accuracy and robustness of the coupled scheme, including close agreement with the CCSN simulation code \chimera.  
	An axisymmetric CCSN simulation further demonstrates the viability of DG-based neutrino transport for multidimensional supernova modeling within \flashx.  
	\thornado's neutrino-transport solver is GPU-enabled using OpenMP offloading or OpenACC, and all CCSN applications included in this work use the GPU implementation.  
	Together, these results establish a foundation for future enhancements in physics fidelity, numerical algorithms, and computational performance, for increasingly realistic large-scale CCSN simulations.  
\end{abstract}


\section{Introduction}
\label{sec:Intro}

In this paper, we present the formulation, implementation, and initial verification of neutrino transport capabilities in the open-source toolkit for high-order neutrino-radiation hydrodynamics (\thornado).  
The transport model is based on a spectral two-moment formulation with algebraic closure that evolves six distinct neutrino species, includes special-relativistic observer corrections to $\cO(v/c)$, and accounts for spectrally coupled neutrino--matter interactions such as neutrino--electron scattering.  
The numerical method combines a discontinuous Galerkin (DG) phase-space discretization with implicit-explicit (IMEX) time integration, coupled to \flashx's finite-volume methods for Newtonian self-gravitating hydrodynamics, employing adaptive mesh refinement (AMR) to resolve the wide range of spatial scales characteristic of core-collapse supernovae (CCSNe).  
We introduce a novel nonlinear neutrino--matter coupling algorithm based on nested fixed-point iteration with Anderson acceleration, and describe a strategy for coupling DG neutrino transport with finite-volume hydrodynamics.  
Verification is demonstrated on a hierarchy of problems: basic tests with known solutions and idealized opacities, relaxation and transport tests with tabulated microphysics (equation of state and opacities), and full CCSN simulations in spherical and axial symmetry.  

A CCSN is the energetic explosion that marks the death of a massive star --- one with an initial mass greater than about ten solar masses.  
At the end of its life, the star's iron core can no longer support itself against gravity and collapses catastrophically until the central density exceeds that of the atomic nucleus.  
At these densities, the repulsive component of the strong nuclear force halts the collapse and drives a rebound of the inner core, launching the supernova shock wave that propagates outward through the infalling outer layers.  
This shock eventually disrupts the star, leaving behind a compact object in the form of a neutron star or a black hole.  
Neutrinos play a pivotal role in the sequence of events that leads to the explosion of massive stars.  
They provide the primary channel for balancing the gravitational energy released during collapse ($\sim10^{46}$~J), while the kinetic energy of the ejected material is roughly two orders of magnitude smaller ($\sim10^{44}$~J) \citep[e.g.,][]{janka_2001}.  
In the neutrino-heating mechanism of CCSNe, a small fraction of the neutrino energy is deposited behind the shock, driving the explosion dynamics.  

Computational modeling has for about sixty years played, and continues to play, a uniquely important role in the development of CCSN theory.  
Pioneering computational studies include those of \citet{colgateWhite_1966}, who first proposed that supernova explosions are powered by neutrino energy deposition, and \citet{leblancWilson_1970}, who demonstrated that magnetic fields amplified by rotation could, in principle, result in jet-like explosions.  
The modern stalled-shock, delayed-neutrino-reheating paradigm was established with the spherically symmetric simulations of \citet{betheWilson_1985}.  
Since then, computational discoveries, including the identification of neutrino-driven convection \citep{herant_etal_1994,burrows_etal_1995}, the realization that the stalled shock is unstable to nonradial perturbations \citep{blondin_etal_2003}, the establishment that, in general, spherically symmetric models do not produce explosions \citep{ramppJanka_2000,liebendorfer_etal_2001,thompson_etal_2003}, and subsequent successful neutrino-driven explosion simulations in two \citep[e.g.,][]{marekJanka_2009,bruenn_etal_2013} and three \citep[e.g.,][]{lentz_etal_2015,melson_etal_2015a,melson_etal_2015b,burrows_etal_2020} spatial dimensions, have profoundly shaped our current understanding of the explosion mechanism \citep[see, e.g.,][for reviews]{muller_2020,mezzacappa_etal_2020,burrowsVartanyan_2021,yamada_etal_2024,janka_2025,muller_2025,mezzacappaZanolin_2025}.  
Continued modeling will play a crucial role in connecting theory to observations of CCSNe through their gravitational-wave and neutrino signatures.  

Modeling neutrino transport in CCSNe is particularly challenging because it occurs in a multiphysics, multiscale environment characterized by strong coupling between reactive, self-gravitating hydrodynamics, or magnetohydrodynamics, and neutrino transport, in three spatial dimensions.  
The relevant spatial scales connecting explosion dynamics to observable signatures such as nucleosynthetic yields and gravitational wave and neutrino emission range from subkilometers to $\cO(10^{5}~\mathrm{km})$, with corresponding time scales spanning microseconds to seconds.  
Computational and numerical challenges are further amplified by the weakly interacting nature of neutrinos---their mean free path ranges from $\cO(1~\mathrm{m})$ in the proto-neutron star (PNS) to $\cO(10^{3}~\mathrm{km})$ in the neutrino-heating region, which calls for a kinetic description.  
Realistic supernova neutrino kinetics requires the inclusion of a range of neutrino--matter weak interactions, some of which couple globally in momentum space and across neutrino species \citep[e.g.,][]{bruenn_1985,burrows_etal_2006,fischer_etal_2024}, as well as special and general relativistic effects \citep[][]{bruenn_etal_2001}.  
It also demands the evolution of six distinct neutrino species to include muonic effects \citep{bollig_etal_2017,fischer_etal_2020}.  
At the numerical level, simultaneously conserving lepton number and energy \citep[][]{cardallMezzacappa_2003,liebendorfer_etal_2004}, maintaining bounds on the phase-space distribution \citep{endeve_etal_2015}, and constructing asymptotic-preserving schemes \citep{jin_2022} that remain accurate and stable from the diffusive to the free-streaming limits present challenges for practitioners developing numerical methods for sophisticated models.  
	
Boltzmann neutrino-radiation hydrodynamics has been implemented in several CCSN codes evolving in spherical symmetry \citep{liebendorfer_etal_2001,ramppJanka_2002,thompson_etal_2003,sumiyoshi_etal_2005}, where the general-relativistic AGILE-BOLTZTRAN code \citep{liebendorfer_etal_2004} was rigorously compared against the Garching-group code by \citet{liebendorfer_etal_2005}.  
In axial symmetry, early application of multiangle, multigroup neutrino transport were carried out by \cite{ott_etal_2008}, followed by more recent work by \citet{nagakura_etal_2018} and \citet{harada_etal_2019}.  
Simulations with Boltzmann neutrino transport have also been performed in three spatial dimensions \citep{iwakami_etal_2020}, though thus far limited to short (about $10$~ms) post-bounce evolution covering prompt convection.  
In parallel, efforts to incorporate and assess the role of neutrino flavor transformation physics have intensified \citep[e.g.,][]{tamborraShalgar_2021,richersManibrata_2022,volpe_2024,johns_etal_2025}.  
However, because of their enormous computational cost, applications of Boltzmann neutrino transport in multiple spatial dimensions remain limited.  

\thornado\ is an open-source collection of Fortran modules for solving the equations of neutrino-radiation hydrodynamics with application to CCSN models, focusing primarily on computational kernels for updating solution representations organized in logically Cartesian data structures.  
The development of \thornado\ emphasizes specialized solver technologies and node-level performance and portability on heterogeneous computing architectures.  
Its modular design allows \thornado\ to be integrated with relative ease into large-scale simulation frameworks based on block-structured AMR \citep[e.g.,][]{dubey_etal_2014}, as demonstrated here by its coupling with the multiphysics software \flashx\ \citep{dubey_etal_2022}.  
\thornado\ is unique within the CCSN modeling community in its use of DG methods \citep[e.g.,][]{cockburnShu_2001} to discretize the neutrino kinetics equations in phase space (position and momentum space).  
Solver developments within \thornado\ have addressed challenges of maintaining physical bounds and conservation in three classes of models: (i) models incorporating Fermi--Dirac statistics \citep{chu_etal_2019}; (ii) models including special relativistic observer corrections to $\cO(v/c)$ \citep{laiu_etal_2025}; and (iii) models incorporating fully special relativistic observer corrections \citep{hunter_etal_2025}.  
Nonlinear solvers for time-implicit neutrino--matter coupling were introduced by \cite{laiu_etal_2021}, while GPU-targeted implementations using OpenMP offloading or OpenACC have also been assessed \citep{laiu_etal_2020,thavappiragasam_etal_2024,laiu_etal_2025}.  
Flavor transformation physics is not yet part of \thornado's current capabilities but may be incorporated in the future, as physically and numerically robust models continue to mature.  

To balance physical fidelity with computational feasibility, \thornado\ employs a spectral two-moment approximation to neutrino kinetics, in which the zeroth and first angular moments of the neutrino distribution function are evolved, representing spectral energy and momentum densities---or, equivalently the spectral number and number-flux densities.  
The model considered here is derived from the general relativistic moment formulations of \citet{shibata_etal_2011} and \citet{cardall_etal_2013a}, after taking the flat-spacetime and $\cO(v/c)$ limits.  
It therefore retains velocity-dependent (observer) corrections that account for spatial advection, Doppler shift, and angular aberration, while neglecting gravitational redshifts.  
The moment equations are closed using an algebraic closure relation, for which several prescriptions exist in the literature \citep[e.g.,][]{minerbo_1978,levermore_1984,cernohorskyBludman_1994}; in this work, we primarily use the Minerbo closure.  
For comparisons of commonly used two-moment closures in neutrino kinetics, see, for example, \cite{smit_etal_2000}, \citet{murchikova_etal_2017}, and \citet{wangBurrows_2023}.  
The model and its implementation in Cartesian coordinates with idealized opacities are described and analyzed in detail in \citet{laiu_etal_2025}.  
Here, it is extended for more general orthogonal curvilinear coordinates, evolves six neutrino species, incorporates tabulated microphysics and a comprehensive set of neutrino--matter interactions, and is dynamically coupled to Newtonian self-gravitating hydrodynamics.  
	
Spectral moment-based neutrino transport is now the workhorse model in the CCSN modeling community.  
\chimera~\citep{bruenn_etal_2020} employs multigroup flux-limited diffusion, while PROMETHEUS--VERTEX~\citep{ramppJanka_2002} uses a two-moment formulation with a variable Eddington factor closure.  
In multidimensional simulations, both codes apply the ray-by-ray approach \citep{ramppJanka_2002,buras_etal_2006}.  
AENUS--ALCAR~\citep{just_etal_2015}, FLASH~\citep{oconnorCouch_2018}, and \fornax~\citep{skinner_etal_2019} employ fully multidimensional, two-moment transport with algebraic closure, and their transport formulations are most similar to the one adopted here.  
All of the aforementioned codes include velocity-dependent terms accurate to $\cO(v/c)$ and account for gravitational redshift effects, while approximating relativistic gravity by correcting the monopole component of the Newtonian potential \citep{marek_etal_2006}.  
Among fully general relativistic models, NADA--FLD~\citep{rahman_etal_2019} employs multidimensional, multigroup flux-limited diffusion; CoCoNut--VERTEX~\citep{muller_etal_2010} uses the variable Eddington factor method together with the ray-by-ray approach in multidimensional simulations; and GR1D~\citep{oconnor_2015} and \citet{kuroda_etal_2016} employ spectral two-moment transport with algebraic closures in one and multiple spatial dimensions, respectively.  

The present implementation in \thornado\ is distinguished from the aforementioned CCSN simulation frameworks, which are based on finite-volume or finite-difference methods, by employing a DG discretization in phase-space.  
DG methods, which are based on piecewise polynomial representations, combine features of finite-volume and spectral methods \citep[see, e.g.,][for a survey]{shu_2016}.  
They achieve high-order accuracy on compact stencils---data are exchanged only between nearest neighbors, regardless of the order of accuracy, trading higher arithmetic intensity for fewer memory accesses, which promotes parallel scalability on heterogeneous architectures \citep{klockner_etal_2009}.
DG methods can be formulated naturally in curvilinear coordinates, an advantage in numerical relativity and general relativistic astrophysics \citep{teukolsky_2016}.  
The variational formulation and flexibility in the choice of test functions make the DG framework particularly suitable for the design of structure-preserving methods that approximate the continuum model with reduced computational cost.  
This is advantageous for neutrino kinetics, as DG schemes can capture the asymptotic diffusion limit---characterized by frequent scattering with the background---on coarse meshes \citep{larsenMorel_1989,adams_2001,lowrieMorel_2002,guermondKanschat_2010} \emph{without} modifying numerical fluxes \citep[as in, e.g.,][]{audit_etal_2002}.  
The flexibility to more freely specify numerical fluxes while maintaining stability across diffusive and streaming regimes facilitates the construction of conservative \emph{and} bound-preserving methods for two-moment models \citep{chu_etal_2019,laiu_etal_2025,hunter_etal_2025}.  
Finally, the flexibility with respect to test functions makes DG methods attractive for developing schemes that conserve particle number, momentum, and energy \citep[e.g.,][]{ayuso_etal_2011,cheng_etal_2013b}, which can be more difficult to achieve with finite-volume methods and is of particular importance for CCSN applications, where \emph{simultaneous} lepton-number and energy conservation provides increased confidence in the physical fidelity of simulations.  
	
Like many of the aforementioned frameworks, \thornado\ employs a combination of implicit and explicit methods to integrate the semi-discretized neutrino kinetics equations in time, specifically using implicit-explicit (IMEX) Runge--Kutta (RK) methods \citep{ascher_etal_1997,pareschiRusso_2005}.  
The neutrino--matter coupling (collision) terms are integrated implicitly to avoid the need to resolve the small timescales imposed by collisions in the PNS, while the neutrino phase-space advection terms, governed by the speed of light, are advanced explicitly.  
Because, in relativistic systems, other characteristic speeds, such as the sound speed, fluid velocity, and gravitational-wave propagation speed, are comparable to the speed of light, the stability constraint associated with explicit treatment of light-speed transport does not impose a significant additional restriction, unlike in other radiation-hydrodynamics applications \citep[e.g.,][]{jiang_etal_2012,gonzalez_etal_2007}.  
By treating spatially coupled divergence operators explicitly, rather than using a fully implicit method, we avoid solving a large, sparse, and distributed algebraic system at every time step.  
The collision terms couple globally in momentum space but remain local in position space.  
Distributing position space across MPI ranks therefore localizes computations associated with the compute-intensive neutrino--matter coupling, improving parallel scalability.  
	
With diagonally implicit IMEX-RK schemes, the implicit part of each stage can be interpreted as a backward Euler solve.  
Because the nonlinear neutrino--matter coupling is computationally expensive---involving numerous neutrino weak-interaction channels, with the evaluation of tabulated neutrino opacities being particularly costly---it is critical to design an iterative solver that is efficient, robust, extensible, and easily parallelizable.  
To this end, we have extended the \emph{nested fixed-point} algorithm proposed by \citet{laiu_etal_2021} to the $\cO(v/c)$ case with six neutrino species.  
In this nested approach, the matter variables are iterated in an outer layer, while the radiation moments are iterated to convergence for a fixed matter state in an inner layer.  
Opacities are evaluated only in the outer layer, where the matter state is updated.  
Compared to Newton's method, fixed-point methods offer advantages by avoiding Jacobian matrices, which are costly to evaluate and difficult to estimate accurately for tabulated opacities, and the associated dense linear solves.  
Jacobian-free Newton--Krylov methods \citep{knollKeyes_2004} are also not ideal in this context, since opacity evaluations dominate the computational cost.  
Following \citet{laiu_etal_2021}, we employ Anderson acceleration \citep{Anderson-1965,Walker-Ni-2011} in both inner and outer layers, using information from previous iterations to accelerate convergence relative to simple Picard iteration.  
The extension to $\cO(v/c)$ introduces additional equations for the fluid-velocity components in the outer layer.  
The inner layer is significantly more complex than in the zero-velocity case considered by \citet{laiu_etal_2021}, due to the nonlinear relation between the \emph{primitive} moments being iterated and the \emph{conserved} moments evolved with the transport equations.  
The fixed-point map used to update the primitive moments is not unique and requires additional care to maintain robustness of the inner iteration.  
We apply the fixed-point iteration developed for idealized opacities by \citet{laiu_etal_2025}, which incorporates a velocity-dependent step-size parameter to promote realizable moments during inner-layer iterations.  
This algorithm has been implemented in \thornado\ with GPU support using either OpenMP offloading or OpenACC.  

To enable application of the proposed neutrino-radiation transport methods in large-scale CCSN simulations, \thornado\ has been integrated into the multiphysics simulation framework \flashx~\citep{dubey_etal_2022}.  
This integration, distributed through GitHub submodules, combines \thornado's DG neutrino transport with \flashx's mature infrastructure for large-scale scientific computing, which provides distributed memory parallelism with MPI, AMR, parallel I/O, and finite-volume self-gravitating hydrodynamics.  
In \flashx, physics data structures, including radiation moments, are organized as a collection of logically Cartesian mesh blocks managed by a physics-agnostic infrastructure, while \thornado\ kernels are invoked to advance physical fields in time.  
Incorporating \thornado's solvers within \flashx\ presents two main challenges.  
First, matter quantities in \flashx, which are evolved with finite-volume methods, are stored as cell averages, whereas \thornado\ operates on element-local polynomial representations.  
In the proposed algorithm, \thornado\ depends on the fluid three-velocity to evaluate $\cO(v/c)$ observer corrections and updates fluid quantities directly through the neutrino--matter coupling.  
Consequently, matter quantities require a hybrid DG/finite-volume representation.  
Following an approach from the subcell limiting procedure of \citet{dumbser_etal_2014}, the finite-volume grid is viewed as a subcell discretization of the DG mesh, with the number of finite-volume cells within each DG element matching the number of degrees of freedom in the tensor-product polynomial basis used by \thornado.  
Reconstruction and projection operators are used to transform fluid quantities between cell-averages and polynomial representations.  
Second, for advancing the neutrino-radiation hydrodynamics system in time, we employ a Lie--Trotter operator-splitting technique: the self-gravitating hydrodynamics is evolved using explicit RK methods, followed by an IMEX update of the neutrino-transport.  
The coupled algorithm is verified through spherically symmetric simulations of gravitational collapse and early post-bounce evolution via a code-to-code comparison with \chimera, and an axisymmetric simulation is performed to demonstrate the multidimensional transport capability.  

The rest of the paper is organized as follows.  
Section~\ref{sec:model} describes the neutrino-radiation hydrodynamics model, and Section~\ref{sec:method} details the DG method for neutrino transport in \thornado.  
Section~\ref{sec:flashx} presents the coupling to \flashx.  
Sections~\ref{sec:results}, \ref{sec:application_idealized}, and \ref{sec:application_ccsn} describe a multi-fidelity hierarchy of verification cases: basic tests with idealized opacities, tests with tabulated microphysics, and CCSN applications in spherical and axial symmetry, respectively, with particular emphasis on assessing the robustness of the nonlinear neutrino--matter coupling solver in the latter two sections.  

We adopt units in which $c=G=\kB=1$, where $c$, $G$, and $\kB$ are the speed of light, gravitational constant, and Boltzmann's constant, respectively.  
Einstein's summation convention is used, where repeated latin indices imply summation from $1$ to $3$.  

\section{Physical Model}
\label{sec:model}

In this section we provide the equations for nonrelativistic, self-gravitating neutrino-radiation hydrodynamics.  
The equations are formulated in a position space coordinate basis \citep[e.g.,][]{rezzollaZanotti_2013} to accommodate orthogonal curvilinear spatial coordinates, sufficiently general to solve problems in Cartsian, cylindrical, and spherical-polar spatial coordinates.  
The orthogonal coordinate system is encoded in the \emph{diagonal} spatial metric $\gamma_{ij}=\mbox{diag}[\,h_{1}h_{1},h_{2}h_{2},h_{3}h_{3}\,]$, which we take to be \emph{independent of time}.  
Here, $h_{1}$, $h_{2}$, and $h_{3}$ are scale factors (Lam{\'e} coefficients), relating coordinate and proper distances.  
The determinant of the spatial metric is denoted $\gamma=(h_{1}h_{2}h_{3})^{2}$, and the inverse is $\gamma^{ij}$.  
See, e.g., Table~1 in \citet{pochik_etal_2021} for metric quantities in the various coordinate systems.  
The spatial metric and its inverse are, respectively, used to lower and raise indices on vectors and tensors.  
We sometimes use the notation $\bsgamma=(\gamma_{ij})$ to refer to the spatial metric.  

\subsection{Hydrodynamics Equations with Self-Gravity and Neutrino Coupling}

The stellar fluid is modeled with an enhanced version of the nonrelativistic Euler--Poisson system, supplemented with a nuclear equation of state (EoS), which is given by the mass conservation equation
\begin{equation}
  \pd{}{t} \rho + \frac{1}{\sqrt{\gamma}}\,\pd{}{i} \big(\,\sqrt{\gamma}\,\rho\,v^{i}\,\big) = 0,
  \label{eq:massConservation}
\end{equation}
the momentum equation
\begin{equation}
  \pd{}{t}(\rho\,v_{j}) + \frac{1}{\sqrt{\gamma}}\,\pd{}{i}\big(\,\sqrt{\gamma}\,\Pi_{~j}^{i}\,\big)
  = \f{1}{2}\,\Pi^{ik}\,\pd{}{j}\gamma_{ik} - \rho\,\pd{}{j}\Phi - \QM{j},
  \label{eq:fluidMomentumEquation}
\end{equation}
the fluid energy equation
\begin{equation}
  \pd{}{t} \Ef + \frac{1}{\sqrt{\gamma}}\,\pd{}{i}\big(\,\sqrt{\gamma}\,[\,\Ef+p\,]\,v^{i}\,\big)
  = -\rho\,v^{i}\,\pd{}{i}\Phi - \QE,
  \label{eq:fluidEnergyEquation}
\end{equation}
and the electron number equation
\begin{equation}
	\pd{}{t}(\rho\ye) + \frac{1}{\sqrt{\gamma}}\,\pd{}{i}\big(\,\sqrt{\gamma}\,\rho\ye\,v^{i}\,\big) = -\mB\,\QL,
	\label{eq:electronNumberEquation}
\end{equation}
where $\rho$ represents the mass density, $v^{i}$ the components of the fluid three-velocity, $\Pi_{~j}^{i}=\rho\,v^{i}\,v_{j} +p\,\delta_{~j}^{i}$ the fluid stress tensor, $p$ the fluid pressure, $\ye$ the electron fraction, $\Ef=\rho\,(\epsilon +\frac{1}{2}\,v^2)$ the total fluid energy density (internal plus kinetic), and $\epsilon$ the specific internal energy.  
The electron number density is related to the electron fraction by $n_{\rm e}=\rho\ye/\mB$, where $\mB$ is the average baryon mass.  
The fluid equations are closed with a nuclear EoS, where thermodynamic quantities, such as pressure and specific internal energy, are functions of mass density, temperature $T$, and electron fraction; e.g., $p=p(\rho,T,\ye)$.  
The second term on the right-hand side of Equation~\eqref{eq:fluidMomentumEquation} and the first term on the right-hand side of Equation~\eqref{eq:fluidEnergyEquation} are gravitational sources due to the Newtonian gravitational potential $\Phi$, which is obtained by solving the Poisson equation
\begin{equation}
  \f{1}{\sqrt{\gamma}}\pd{}{i}\big(\,\sqrt{\gamma}\,\gamma^{ij}\pd{\Phi}{j}\,\big) = 4\pi\,\rho.  
  \label{eq:poissonEquation}
\end{equation}
On the right-hand sides of Equations~\eqref{eq:fluidMomentumEquation}, \eqref{eq:fluidEnergyEquation}, and \eqref{eq:electronNumberEquation}, the terms $\QM{j}$, $\QE$, and $\QL$ model the exchange of momentum, energy, and lepton number, respectively, between the fluid and the neutrinos.  
These terms depend on the neutrino distribution functions and are detailed in the following section.  

\subsection{Neutrino Kinetics Equations}

We employ a spectral two-moment approximation in the $\mathcal{O}(v/c)$ limit \citep[e.g.,][]{mihalasMihalas_1999} to model neutrino kinetics.  
More specifically, our model is obtained from the general relativistic moment formalism of \cite{shibata_etal_2011,cardall_etal_2013a}, which expresses the equations using laboratory-frame spacetime coordinates ($\vect{x}$ and $t$) and --- to facilitate neutrino--matter coupling --- comoving-frame spherical-polar momentum-space coordinates ($\varepsilon$, $\vartheta$, and $\varphi$).  
Here, $\varepsilon\in\bbR^{+}$ is the neutrino energy measured by an observer comoving with the fluid, and $\vartheta\in[0,\pi]$, $\varphi\in[0,2\pi)$ parametrize the comoving-frame propagation direction $\bsell(\vartheta,\varphi)$ on the unit sphere
\begin{equation}
	\bbS^{2}
	=\{\,\bsell\in\bbR^{3}~\vcentcolon~|\bsell|=1\,\}.  
\end{equation}
We define \emph{primitive} (comoving-frame) angular moments of the distribution function $f_{s}$ for neutrino species $s$ as
\begin{equation}
	\big\{\,\mathcal{D}_{s},\,{\mathcal{I}_{s}}^{i},\,{\mathcal{K}_{s}}^{ij},\,{\mathcal{L}_{s}}^{ijk}\,\big\}(\varepsilon,\vect{x},t)
	= \f{1}{4\pi}\int_{\bbS^{2}}f_{s}(\omega,\varepsilon,\vect{x},t)\big\{\,1,\ell^{i},\ell^{i}\ell^{j},\ell^{i}\ell^{j}\ell^{k}\,\big\}\,d\omega,
	\label{eq:primitiveMoments}
\end{equation}
where $\omega=(\vartheta,\varphi)$, $d\omega=\sin\vartheta d\vartheta d\varphi$, and $\ell^{i}(\omega)$ is a coordinate-basis unit vector ($\ell_{i}\ell^{i}=1$).  
The zeroth and first moments, $\mathcal{D}_{s}$ and ${\mathcal{I}_{s}}^{i}$, correspond to the spectral number density and number flux density, respectively; ${\mathcal{K}_{s}}^{ij}$ is the symmetric rank-two tensor whose product with $\varepsilon$ yields the spectral stress tensor; and ${\mathcal{L}_{s}}^{ijk}$ is the completely symmetric rank-three moment tensor.  
Thus, the primitive moments defined above are functions of laboratory-frame spacetime coordinates $(\vect{x},t)$ and comoving-frame particle energy $\varepsilon$.  

We emphasize that the subscript $s\in\{\nu_{\rm e},\bar{\nu}_{\rm e},\nu_{\mu},\bar{\nu}_{\mu},\nu_{\tau},\bar{\nu}_{\tau}\}$ on radiation moments --- e.g., as in Eq~\eqref{eq:primitiveMoments} --- is reserved for indicating neutrino species.  
It should never be interpreted as a spatial index on tensors.  

To obtain evolution equations for the angular moments, we consider the spectral moment system derived by taking angular moments of the four-momentum conservative formulation of general relativistic particle kinetics from \cite{cardallMezzacappa_2003}, which gives rise to the two-moment system considered in \citet{shibata_etal_2011} and \citet{cardall_etal_2013a} (see their Equation~(3.18) and (40), respectively).  
Then, following a foliation of spacetime into spacelike hypersurfaces, a decomposition of the two-moment system relative to these slices, taking the limit of flat spacetime, and keeping special relativistic corrections to $\mathcal{O}(v/c)$, we arrive at the two-moment system solved by \thornado.  
More specifically, we consider the flat spacetime, $\mathcal{O}(v/c)$ limit of the number conservative, general relativistic two-moment model derived in \cite{mezzacappa_etal_2020}; see their Equations (123) and (125).  
In this limit, the `zeroth moment' equation, governing the spectral neutrino number density, is given by
\begin{equation}
  \pd{}{t}\big(\,\mathcal{D}_{s}+v^{i}\,{\mathcal{I}_{s}}_{i}\,\big)
  +\f{1}{\sqrt{\gamma}}\pd{}{i}\big(\,\sqrt{\gamma}\,\big[\,{\mathcal{I}_{s}}^{i}+\mathcal{D}_{s}\,v^{i}\,\big]\,\big)
  -\f{1}{\varepsilon^{2}}\pd{}{\varepsilon}
  \big(\,\varepsilon^{3}\,\big[\,{\mathcal{I}_{s}}_{i}\,\pd{v^{i}}{t}+{\mathcal{K}_{s}}_{\hspace{2pt}k}^{i}\,\cderiv{v^{k}}{i}\,\big]\,\big)
  =\aint{\mathcal{C}_{s}(\vect{f})},
  \label{eq:spectralNumberDensity}
\end{equation}
where we have introduced the notation $\aint{\ldots}=\f{1}{4\pi}\int_{\bbS^{2}}\ldots d\omega$, and $\mathcal{C}_{s}(\vect{f})$ denotes the collision term describing neutrino--matter (and neutrino--neutrino) interactions.  
We use $\vect{f}=\{f_{s}\}_{s=1}^{N_{\rm s}}$, where $N_{\rm s}$ denotes the total number of neutrino species included in the model (six in this work), in the argument of the collision term for neutrino species $s$ to indicate that interactions can involve other neutrino species $s'\ne s$.  
The covariant derivative associated with the spatial metric $\gamma_{ik}$ is denoted by $\cderiv{}{i}$, and the contraction of a symmetric tensor $\mathcal{T}^{ik}$ with the covariant derivative of the three-velocity can be written explicitly as
\begin{equation}
  \mathcal{T}_{\hspace{4pt}k}^{i}\nabla_{i}v^{k} = \mathcal{T}_{\hspace{4pt}k}^{i}\,\pd{v^{k}}{i} + \f{1}{2}\,\mathcal{T}^{ik}\,v^{l}\,\pd{\gamma_{ik}}{l}.  
\end{equation}
(Note that $\nabla_{i}\gamma_{jk}=0$.  Therefore, $\mathcal{T}_{\hspace{4pt}k}^{i}\nabla_{i}v^{k}=\mathcal{T}^{ik}\nabla_{i}v_{k}$.)  

Equation~\eqref{eq:spectralNumberDensity} evolves the $\mathcal{O}(v/c)$ approximation to the spectral Eulerian number density, 
\begin{equation}
  \mathcal{N}_{s} = \mathcal{D}_{s}+v^{i}\,{\mathcal{I}_{s}}_{i},
  \label{eq:numberDensityEulerian}
\end{equation}
and in the absence of neutrino--matter interactions, when integrated over energy (with weight $4\pi\varepsilon^{2}/h^{3}$), it expresses exact conservation of particles in the Eulerian (laboratory) frame of reference.  
This conservation property, which helps facilitate total lepton number conservation, is an important reason why \thornado\ solves the number conservative two-moment model.  

The `first moment' equation, governing the spectral neutrino number flux density, is given by
\begin{align}
  &\pd{}{t}\big(\,{\mathcal{I}_{s}}_{j}+v^{i}\,{\mathcal{K}_{s}}_{ij}\,\big)
  +\f{1}{\sqrt{\gamma}}\pd{}{i}\big(\,\sqrt{\gamma}\,\big[\,{\mathcal{K}_{s}}^{i}_{\hspace{2pt}j}+{\mathcal{I}_{s}}_{j}\,v^{i}\,\big]\,\big)
  -\f{1}{\varepsilon^{2}}\pd{}{\varepsilon}
  \big(\,\varepsilon^{3}\,\big[\,{\mathcal{K}_{s}}_{jk}\,\pd{v^{k}}{t}+{\mathcal{L}_{s}}_{\hspace{2pt}kj}^{i}\,\cderiv{v^{k}}{i}\,\big]\,\big) \nonumber \\
  &=
  \f{1}{2}\,\big(\,{\mathcal{K}_{s}}^{ik}+{\mathcal{I}_{s}}^{i}\,v^{k}+v^{i}\,{\mathcal{I}_{s}}^{k}\,\big)\,\pd{\gamma_{ik}}{j}
  +\big(\,{\mathcal{K}_{s}}_{jk}\,\pd{v^{k}}{t}+{\mathcal{L}_{s}}_{\hspace{4pt}kj}^{i}\,\cderiv{v^{k}}{i}\,\big)
  -\big(\,\mathcal{D}_{s}\,\pd{v_{j}}{t}+{\mathcal{I}_{s}}^{i}\,\pd{v_{j}}{i}\,\big)
  +\aint{\mathcal{C}_{s}(\vect{f})\ell_{j}}.  
  \label{eq:spectralNumberFluxDensity}
\end{align}

\begin{rem}
	In the absence of collisions, Equations~\eqref{eq:spectralNumberDensity} and \eqref{eq:spectralNumberFluxDensity} describe phase-space advection of neutrinos through a moving fluid.  
	The velocity-dependent terms in the spatial and energy derivatives, and the additional terms on the right-hand side of Equation~\eqref{eq:spectralNumberFluxDensity}, collectively represent spatial advection, Doppler shift, and angular aberration between adjacent comoving observers \cite[e.g., see][]{liebendorfer_etal_2004}.  
	These ``observer corrections'' arise from the choice of comoving-frame momentum coordinates, which simplify the kinematics of neutrino--matter interactions \citep[e.g.,][]{castor_1972,buchler_1979,buchler_1983,mihalasMihalas_1999} and facilitates the construction of closure relations for moment-based models \cite[e.g.,][]{buchler_1983}, but introduces additional complexity in the phase-space discretization --- especially when aiming for simultaneous energy-momentum and lepton number conservation \cite[see][for in-depth discussion]{cardallMezzacappa_2003,cardall_etal_2005,cardall_etal_2013a,cardall_etal_2013b}.  
	\label{rem:observer_corrections}
\end{rem}

The system given by Equations~\eqref{eq:spectralNumberDensity} and \eqref{eq:spectralNumberFluxDensity} contains the higher-order angular moments ${\mathcal{K}_{s}}^{ij}$ and ${\mathcal{L}_{s}}^{ijk}$, and is therefore not closed.  
To close the system of equations, these moments must be related to the lower-order moments through a closure procedure.  
As is commonly done in CCSN codes based on two-moment neutrino kinetics \citep[e.g.,][]{just_etal_2015,oconnor_2015,skinner_etal_2019}, we employ the family of algebraic moment closures in \thornado.  
To this end, the rank-two tensor of moments is related to the number density through ${\mathcal{K}_{s}}^{ij}={\mathsf{k}_{s}}^{ij}\,\mathcal{D}_{s}$, where the approximate Eddington tensor is \citep{levermore_1984}
\begin{equation}
	{\mathsf{k}_{s}}^{ij} 
	= \f{1}{2}\,\big[\,(1-\psi_{s})\,\gamma^{ij}+(3\psi_{s}-1)\,{\hat{n}_{s}}^{~i}\,{\hat{n}_{s}}^{~j}\,\big],
	\label{eq:eddingtonTensor}
\end{equation}
where ${\hat{n}_{s}}^{\hspace{2.5pt}i}={\mathcal{I}_{s}}^{i}/\mathcal{I}_{s}$, with $\mathcal{I}_{s}=\sqrt{\gamma_{ij}{\mathcal{I}_{s}}^{i}{\mathcal{I}_{s}}^{j}}$, is a unit vector parallel to the comoving frame number flux density, and the Eddington factor is defined as
\begin{equation}
	\psi_{s} = {\hat{n}_{si}}\,{\hat{n}_{sj}}\,{\mathsf{k}_{s}}^{ij} = \aint{f_{s} ({\hat{n}_{si}}\ell^{i})^{2}}/\aint{f_{s}}.  
	\label{eq:eddingtonFactor}
\end{equation}
The Eddington tensor in Equation~\eqref{eq:eddingtonTensor} satisfies the trace condition $\gamma_{ij}{\mathsf{k}_{s}}^{ij}=1$.  
In a similar manner, the rank-three tensor of moments is related to the number density through the normalized `heat-flux' tensor by ${\mathcal{L}_{s}}^{ijk}={\mathsf{q}_{s}}^{ijk}\,\mathcal{D}_{s}$, where the normalized heat-flux tensor is approximated as \citep[][]{pennisi_1992,just_etal_2015}
\begin{equation}
	{\mathsf{q}_{s}}^{ijk} 
	= \f{1}{2}\,
	\big[\,
		(h_{s}-\zeta_{s})\,\big(\,{\hat{n}_{s}}^{~i}\,\gamma^{jk}+{\hat{n}_{s}}^{~j}\,\gamma^{ik}+{\hat{n}_{s}}^{~k}\,\gamma^{ij}\,\big)
		+(5\zeta_{s}-3h_{s})\,{\hat{n}_{s}}^{~i}\,{\hat{n}_{s}}^{~j}\,{\hat{n}_{s}}^{~k}
	\,\big],
	\label{eq:heatFluxTensor}
\end{equation}
where $h_{s}=\mathcal{I}_{s}/\mathcal{D}_{s}$ is the flux factor, and the `heat-flux factor' is defined as
\begin{equation}
	\zeta_{s} = {\hat{n}_{si}}\,{\hat{n}_{sj}}\,{\hat{n}_{sk}}\,{\mathsf{q}_{s}}^{ijk} = \aint{f_{s} ({\hat{n}_{si}}\ell^{i})^{3}}/\aint{f_{s}}.  
	\label{eq:heatFluxFactor}
\end{equation}
The rank-three tensor in Equation~\eqref{eq:heatFluxTensor} satisfies the trace condition $\gamma_{ij}\,{\mathsf{q}_{s}}^{ijk}=h_{s}\,{\hat{n}_{s}}^{~k}$.  

The two-moment model is then closed by expressing the Eddington and heat-flux factors in terms of $\mathcal{D}_{s}$ and ${\mathcal{I}_{s}}^{i}$.  
Several closure strategies have been developed, with the most widely used to date falling into two families: maximum entropy closures \citep[e.g.,][]{minerbo_1978,dubrocaFeugeas_1999,cernohorskyBludman_1994,richers_2020} and Kershaw-type closures \citep[e.g.,][]{kershaw_1976,schneider_2016,banachLarecki_2017a}.  
The cited maximum entropy closures differ in the entropy functional employed.  
For neutrino kinetics, the appropriate choice is the Fermi--Dirac entropy, which can yield nontrivial deviations from classical (Maxwell--Boltzmann) or photonic (Bose--Einstein) behavior when the neutrino occupation approaches unity (i.e, when $\mathcal{D}_{s}\lesssim1$).  
In the CCSN environment, such conditions arise for electron neutrinos in the collapsed, degenerate core; however, because the neutrino mean free path in this region is short, the Eddington approximation is generally quite accurate.  
Further out, in the neutrino-heating region, $\mathcal{D}_{s}\ll1$ typically holds for all neutrino species, and the classical maximum entropy closure, i.e., based on Maxwell--Boltzmann statistics, may therefore provide a reasonable approximation for two-moment neutrino transport in CCSN \cite[e.g.,][]{just_etal_2015,murchikova_etal_2017,wangBurrows_2023}, balancing accuracy against model and algorithm complexity.  
Nevertheless, the extent to which transport behavior in the transition from diffusive to free-streaming regimes is sensitive to the choice of entropy functional, particularly whether closures based on Fermi--Dirac statistics are required, warrants further study.  
Developing robust numerical methods in the context of Fermi--Dirac statistics is also considerably more challenging than in the Maxwell--Boltzmann case \citep[e.g.,][]{chu_etal_2019}.  
For these reasons, we primarily employ Minerbo's closure \citep[a maximum entropy closure based on Maxwell--Boltzmann statistics;][]{minerbo_1978}, with the functions in Equations~\eqref{eq:eddingtonFactor} and \eqref{eq:heatFluxFactor} approximated by polynomials in the flux factor as \citep[e.g.,][]{cernohorskyBludman_1994,just_etal_2015}
\begin{align}
	\psi_{s}(h)
	&\vcentcolon=\psi_{s}^{\rm Mi}(h)
	=\f{1}{3}+\f{2}{15}\,\big(\,3-h+3h^{2}\,\big)\,h^{2}, \label{eq:eddingtonFactorMinerbo} \\
	\zeta_{s}(h) 
	&\vcentcolon=\zeta_{s}^{\rm Mi}(h)
	=\f{1}{75}\,\big(\,45+10\,h-12\,h^{2}-12\,h^{3}+38\,h^{4}-12\,h^{5}+18\,h^{6}\,\big)\,h.  \label{eq:heatFluxFactorMinerbo}
\end{align}

In the code-to-code comparison with \chimera\ in Section~\ref{sec:application_ccsn} we include results from models employing the Levermore and Kershaw closures.  
For completeness, we list the corresponding Eddington and heat-flux factors in Appendix~\ref{app:closures}.  

Equations~\eqref{eq:spectralNumberDensity} and \eqref{eq:spectralNumberFluxDensity} are related to the two-moment system solved by AENUS--ALCAR~\citep{just_etal_2015} and \fornax~\citep{skinner_etal_2019}.  
However, they solve for energy and momentum densities associated with a comoving observer.  
(Moment equations for number transport, similar to our Equations~\eqref{eq:spectralNumberDensity} and \eqref{eq:spectralNumberFluxDensity}, are listed in Section~2.2.2 of \citet{just_etal_2015}.)  
Defining the energy moments in terms of the angular moments in Equation~\eqref{eq:primitiveMoments} as 
\begin{equation}
	\big\{\,\mathcal{J}_{s},\,{\mathcal{H}_{s}}^{i},\,{\widehat{\mathcal{K}_{s}}}^{ij},\,{\widehat{\mathcal{L}_{s}}}^{ijk}\,\big\}
	=\varepsilon\,\big\{\,\mathcal{D}_{s},\,{\mathcal{I}_{s}}^{i},\,{\mathcal{K}_{s}}^{ij},\,{\mathcal{L}_{s}}^{ijk}\big\}, 
\end{equation}
the two-moment model for energy--momentum transport is obtained by multiplying Equations~\eqref{eq:spectralNumberDensity} and \eqref{eq:spectralNumberFluxDensity} by the neutrino energy $\varepsilon$ (using the product rule to bring $\varepsilon$ inside the energy derivatives) to give the spectral energy equation
\begin{align}
  &\pd{}{t}\big(\,\mathcal{J}_{s}+v^{i}\,{\mathcal{H}_{s}}_{i}\,\big)
  +\f{1}{\sqrt{\gamma}}\pd{}{i}\big(\,\sqrt{\gamma}\,\big[\,{\mathcal{H}_{s}}^{i}+\mathcal{J}_{s}\,v^{i}\,\big]\,\big)
  -\f{1}{\varepsilon^{2}}\pd{}{\varepsilon}
  \big(\,\varepsilon^{3}\,\big[\,{\mathcal{H}_{s}}_{i}\,\pd{v^{i}}{t}+{\widehat{\mathcal{K}_{s}}}_{\hspace{2pt}k}^{i}\,\cderiv{v^{k}}{i}\,\big]\,\big) \nonumber \\
  &=
  -\big(\,{\mathcal{H}_{s}}_{i}\,\pd{v^{i}}{t}+{\widehat{\mathcal{K}_{s}}}_{\hspace{2pt}k}^{i}\,\cderiv{v^{k}}{i}\,\big)
  +\varepsilon\,\aint{\mathcal{C}_{s}(\vect{f})}
  \label{eq:energyEquationLagrangian}
\end{align}
and the spectral momentum equation
\begin{align}
  &\pd{}{t}\big(\,{\mathcal{H}_{s}}_{j}+v^{i}\,{\widehat{\mathcal{K}_{s}}}_{ij}\,\big)
  +\f{1}{\sqrt{\gamma}}\pd{}{i}\big(\,\sqrt{\gamma}\,\big[\,{\widehat{\mathcal{K}_{s}}}^{i}_{\hspace{2pt}j}+{\mathcal{H}_{s}}_{j}\,v^{i}\,\big]\,\big)
  -\f{1}{\varepsilon^{2}}\pd{}{\varepsilon}
  \big(\,\varepsilon^{3}\,\big[\,{\widehat{\mathcal{K}_{s}}}_{jk}\,\pd{v^{k}}{t}+{\widehat{\mathcal{L}_{s}}}_{\hspace{2pt}kj}^{i}\,\cderiv{v^{k}}{i}\,\big]\,\big)
  \nonumber \\
  &=
  \f{1}{2}\,\big(\,{\widehat{\mathcal{K}_{s}}}^{ik}+{\mathcal{H}_{s}}^{i}\,v^{k}+v^{i}\,{\mathcal{H}_{s}}^{k}\,\big)\,\pd{\gamma_{ik}}{j}
  -\big(\,\mathcal{J}_{s}\,\pd{v_{j}}{t}+{\mathcal{H}_{s}}^{i}\,\pd{v_{j}}{i}\,\big)
  +\varepsilon\,\aint{\mathcal{C}_{s}(\vect{f})\ell_{j}},
  \label{eq:momentumEquationLagrangian}
\end{align}
respectively.  
(\citet{just_etal_2015} and \citet{skinner_etal_2019} define the angular moments in terms of the specific intensity $\propto\varepsilon^{3}f_{s}$.)  
Equations~\eqref{eq:energyEquationLagrangian} and \eqref{eq:momentumEquationLagrangian} differ from those listed in \citet{just_etal_2015} and \citet{skinner_etal_2019}.  
First, \citet{just_etal_2015} and \citet{skinner_etal_2019} ignore the terms proportional to the time derivative of the fluid three-velocity in the energy derivative operator --- the final term on the left-hand sides of Equations~\eqref{eq:energyEquationLagrangian} and \eqref{eq:momentumEquationLagrangian} --- as well as on the right-hand side of Equation~\eqref{eq:momentumEquationLagrangian}.  
This can be justified for nonrelativistic velocities \citep{lowrie_etal_2001}.  
Second, these works do not include the term involving the fluid three-velocity in the time derivative operator --- the first term on the left-hand sides of Equations~\eqref{eq:energyEquationLagrangian} and \eqref{eq:momentumEquationLagrangian} --- so that $\mathcal{J}_{s}$ and ${\mathcal{H}_{s}}_{j}$ are evolved directly by their respective moment equations.  
This is akin to treating the time variation of radiation quantities on the fluid time scale \citep{lowrie_etal_2001}; however, the wave speeds supported by the resulting two-moment system can exceed the speed of light, which is unphysical.  
The wave speeds supported by Equations~\eqref{eq:energyEquationLagrangian} and \eqref{eq:momentumEquationLagrangian} (or equivalently Equations~\eqref{eq:spectralNumberDensity} and \eqref{eq:spectralNumberFluxDensity}) are bounded by the speed of light for all $v<1$ in one spatial dimension, and for $v\lesssim0.25$ in the multidimensional case \citep{laiu_etal_2025}.  
Another consequence of neglecting the velocity-dependent terms in the time derivative is that the correct, to $\mathcal{O}(v/c)$, Eulerian-frame momentum, energy, and number equations cannot be obtained, and this affects momentum, energy, and lepton number conservation properties of the model \citep{lowrie_etal_2001,mezzacappa_etal_2020,laiu_etal_2025}.  
We reemphasize this point below.  
Keeping these terms increases the algorithmic complexity of solving the two-moment model numerically because a nonlinear inversion algorithm is needed to obtain $\mathcal{J}_{s}$ and ${\mathcal{H}_{s}}_{j}$ from the evolved quantities $(\mathcal{J}_{s}+v^{i}\,{\mathcal{H}_{s}}_{i})$ and $({\mathcal{H}_{s}}_{j}+v^{i}\,{\widehat{\mathcal{K}_{s}}}_{ij})$, and the fluid three-velocity $v^{i}$.  
This problem is similar to that encountered in relativistic hydro- and magneto-hydrodynamics \citep[e.g.,][]{noble_etal_2006}.  
We describe our approach to this conserved-to-primitive conversion problem in the context of Equations~\eqref{eq:spectralNumberDensity} and \eqref{eq:spectralNumberFluxDensity} in Section~\ref{sec:method}.  

Equations~\eqref{eq:spectralNumberDensity} and \eqref{eq:spectralNumberFluxDensity} are, to $\mathcal{O}(v/c)$, consistent with the correct equations for the Eulerian neutrino energy and momentum densities.  
To see this, first add $\varepsilon$ times Equation~\eqref{eq:spectralNumberDensity} to $v^{j}$ contracted with $\varepsilon$ times Equation~\eqref{eq:spectralNumberFluxDensity} and drop $\mathcal{O}(v^{2}/c^{2})$ terms to obtain the Eulerian neutrino energy equation
\begin{align}
  \pd{\mathcal{E}_{s}}{t} + \f{1}{\sqrt{\gamma}}\pd{}{i}\big(\,\sqrt{\gamma}\,{\mathcal{F}_{s}}^{i}\,\big)
  -\f{1}{\varepsilon^{2}}\pd{}{\varepsilon}
  \big(\,\varepsilon^{3}\,\big[\,{\mathcal{H}_{s}}_{i}\,\pd{v^{i}}{t}+{\widehat{\mathcal{K}_{s}}}_{\hspace{2pt}k}^{i}\,\cderiv{v^{k}}{i}\,\big]\,\big)
  =\varepsilon\,\big(\aint{\mathcal{C}_{s}(\vect{f})} + v^{j}\,\aint{\mathcal{C}_{s}(\vect{f})\ell_{j}}\big).
  \label{eq:energyEquationEulerian}
\end{align}
Next, add $\varepsilon$ times Equation~\eqref{eq:spectralNumberFluxDensity} to $v_{j}$ multiplied with $\varepsilon$ times Equation~\eqref{eq:spectralNumberDensity} and drop $\mathcal{O}(v^{2}/c^{2})$ terms to obtain the Eulerian neutrino momentum equation
\begin{align}
  \pd{{\mathcal{F}_{s}}_{j}}{t}
  +\f{1}{\sqrt{\gamma}}\pd{}{i}\big(\,\sqrt{\gamma}\,{\mathcal{S}_{s}}_{\hspace{2pt}j}^{i}\,\big)
  -\f{1}{\varepsilon^{2}}\pd{}{\varepsilon}
  \big(\,\varepsilon^{3}\,\big[\,{\widehat{\mathcal{K}_{s}}}_{jk}\,\pd{v^{k}}{t}+{\widehat{\mathcal{L}_{s}}}_{\hspace{2pt}kj}^{i}\,\cderiv{v^{k}}{i}\,\big]\,\big)
  = \f{1}{2}\,{\mathcal{S}_{s}}^{ik}\,\pd{\gamma_{ik}}{j} + \varepsilon\,\big(\aint{\mathcal{C}_{s}(\vect{f})\ell_{j}} + v_{j}\,\aint{\mathcal{C}_{s}(\vect{f})}\big).
  \label{eq:momentumEquationEulerian}
\end{align}
Here, the `Eulerian' moments $\{\cE_{s},{\cF_{s}}^{i},{\cS_{s}}^{ij}\}$ have been expressed in terms of the Lagrangian moments $\{\cJ_{s},{\cH_{s}}^{i},{\widehat{\cK_{s}}}^{ij}\}$ as
\begin{align}
  \mathcal{E}_{s} &= \mathcal{J}_{s} + 2\,v^{i}\,{\mathcal{H}_{s}}_{i}, \label{eq:energyDensityEulerian} \\
  {\mathcal{F}_{s}}^{i} &= {\mathcal{H}_{s}}^{i} + v^{i}\,\mathcal{J}_{s} + v^{j}{\widehat{\mathcal{K}_{s}}}_{\hspace{2pt}j}^{i}, \label{eq:momentumDensityEulerian} \\
  {\mathcal{S}_{s}}^{ij} &= {\widehat{\mathcal{K}_{s}}}^{ij} + {\mathcal{H}_{s}}^{i}\,v^{j} + v^{i}\,{\mathcal{H}_{s}}^{j},  \label{eq:stressTensorEulerian}
\end{align}
where $\mathcal{E}_{s}$, ${\mathcal{F}_{s}}^{i}$, and ${\mathcal{S}_{s}}^{ij}$ represent the spectral Eulerian neutrino energy density, momentum density, and stress, respectively.  
The term `Eulerian' is taken to mean that these moments are the $\mathcal{O}(v/c)$ limit to the components of the Eulerian decomposition of the spectral stress-energy tensor \citep{cardall_etal_2013a}.  
When integrated over the neutrino energy (using the appropriate weighting $4\pi\varepsilon^{2}/h^{3}$), these moments are those measured by an Eulerian observer.  
Moreover, the associated evolution equations, obtained by integrating Equations~\eqref{eq:energyEquationEulerian} and \eqref{eq:momentumEquationEulerian} over energy, become position-space balance equations, which, when combined with Equations~\eqref{eq:fluidEnergyEquation} and \eqref{eq:fluidMomentumEquation}, express conservation of total (matter plus neutrino) energy and momentum, respectively.  

Next, we use conservation principles for momentum, energy, and lepton number to specify the sources $\QM{j}$, $\QE$, and $\QL$ appearing on the right-hand sides of Equations~\eqref{eq:fluidMomentumEquation}, \eqref{eq:fluidEnergyEquation}, and \eqref{eq:electronNumberEquation}, respectively.  
We define the neutrino lepton number density and flux as
\begin{equation}
  \big\{\,N_{\nu},\,{F_{N_{\nu}}}^{i}\,\big\}
  =\sum_{s=1}^{N_{\rm s}}{\mathsf g}_{s}\int_{\bbR^{+}}\{\,\mathcal{N}_{s},\,{\mathcal{I}_{s}}^{i}+\mathcal{D}_{s}\,v^{i}\,\}\,dV_{\varepsilon},
\end{equation}
where $dV_{\varepsilon}=4\pi\varepsilon^{2}d\varepsilon/h^{3}$, and ${\mathsf g}_{s}$ is the lepton number for neutrino species $s$.  
In our model, only electron-flavor neutrinos contribute to lepton exchange with the matter, and we will take ${\mathsf g}_{\nu_{e}}=1$, ${\mathsf g}_{\bar{\nu}_{e}}=-1$, and ${\mathsf g}_{s}=0$ for $s\notin\{\nu_{e},\bar{\nu}_{e}\}$.  
Then, integrating the neutrino number equation in Equation~\eqref{eq:spectralNumberDensity} over energy, and combining the result with Equation~\eqref{eq:electronNumberEquation}, we obtain the conservation law for the total lepton number density
\begin{equation}
  \pd{}{t}\big(\,n_{\rm e} + N_{\nu}\,\big)+\f{1}{\sqrt{\gamma}}\pd{}{i}\big(\,\sqrt{\gamma}\,[\,n_{\rm e}\,v^{i}+{F_{N_{\nu}}}^{i}\,]\,\big) = 0,
  \label{eq:totalLeptonNumber}
\end{equation}
provided the lepton number exchange rate on the right-hand side of Equation~\eqref{eq:electronNumberEquation} is defined as
\begin{equation}
  \QL = \sum_{s=1}^{N_{\rm s}}{\mathsf g}_{s}\int_{\bbR^{+}}\aint{\mathcal{C}_{s}(\vect{f})}\,dV_{\varepsilon}.  
  \label{eq:leptonNumberExchangeRate}
\end{equation}
Similarly, we define the Eulerian neutrino energy, momentum, and stress as
\begin{equation}
  \big\{\,E_{\nu},\,{F_{\nu}}^{i},\,{S_{\nu}}^{ij}\,\big\}
  =\sum_{s=1}^{N_{\rm s}}\int_{\bbR^{+}}\,\{\,\mathcal{E}_{s},\,{\mathcal{F}_{s}}^{i},\,{\mathcal{S}_{s}}^{ij}\,\}\,dV_{\varepsilon}.  
  \label{eq:eulerianGrayMoments}
\end{equation}
Then, integrating Equation~\eqref{eq:momentumEquationEulerian} over energy, and combining with Equation~\eqref{eq:fluidMomentumEquation}, we obtain the total (matter plus neutrino) momentum equation
\begin{equation}
  \pd{}{t}\big(\,\rho v_{j}+{F_{\nu}}_{j}\,\big)
  +\f{1}{\sqrt{\gamma}}\pd{}{i}\big(\,\sqrt{\gamma}\,[\,\Pi_{\hspace{2pt}j}^{i}+{S_{\nu}}_{\hspace{2pt}j}^{i}\,]\,\big)
  =\f{1}{2}\,\big(\,\Pi^{ik}+{S_{\nu}}^{ik}\,\big)\,\pd{\gamma_{ik}}{j} - \rho\pd{\Phi}{j},
  \label{eq:totalMomentum}
\end{equation}
while  integrating Equation~\eqref{eq:energyEquationEulerian} over energy, and combining with Equation~\eqref{eq:fluidEnergyEquation}, we obtain the total (matter plus neutrino) energy equation
\begin{equation}
  \pd{}{t}\big(\,\Ef+E_{\nu}\,\big) + \f{1}{\sqrt{\gamma}}\pd{}{i}\big(\,\sqrt{\gamma}\,[\,(\Ef+p)\,v^{i}+{F_{\nu}}^{i}\,]\,\big) = - \rho v^{i}\pd{\Phi}{i},
  \label{eq:totalEnergy}
\end{equation}
provided the momentum and energy exchange rates due to neutrino--matter interactions in Equations~\eqref{eq:fluidMomentumEquation} and \eqref{eq:fluidEnergyEquation} are defined, respectively, as
\begin{align}
  \QM{j}
  &=\sum_{s=1}^{N_{\rm s}}\int_{\bbR^{+}}\big(\,\aint{\mathcal{C}_{s}(\vect{f})\ell_{j}} + v_{j}\,\aint{\mathcal{C}_{s}(\vect{f})\,}\,\big)\,\varepsilon\,dV_{\varepsilon},  
  \label{eq:momentumExchangeRate} \\
  \QE
  &=\sum_{s=1}^{N_{\rm s}}\int_{\bbR^{+}}\big(\,\aint{\mathcal{C}_{s}(\vect{f})} + v^{j}\,\aint{\mathcal{C}_{s}(\vect{f})\ell_{j}}\,\big)\,\varepsilon\,dV_{\varepsilon}.  
  \label{eq:energyExchangeRate}
\end{align}
The exchange rates in Equations~\eqref{eq:leptonNumberExchangeRate}, \eqref{eq:momentumExchangeRate}, and \eqref{eq:energyExchangeRate} are, to $\mathcal{O}(v/c)$, consistent with conservation of lepton number, momentum, and energy due to neutrino--matter interactions.  
We enforce these conservation laws in the algorithm for modeling neutrino--matter coupling described in Section~\ref{sec:method}.  
We mention that Equation~\eqref{eq:totalEnergy} can, with the aid of Equation~\eqref{eq:poissonEquation}, be brought into an exact conservation law for matter plus neutrino plus gravitational potential energy, but we do not include these details here.  

\subsection{Weak Interactions and Collision Terms}
\label{sec:weakInteractionsAndCollisionTerms}

Next, we describe the opacities included in the model and incorporated in the neutrino--matter coupling algorithm presented in Section~\ref{sec:method}.  
Table~\ref{tab:opacities} summarizes the full opacity set, providing relevant references and pointing to the expressions for the corresponding collision terms that enter on the right-hand sides of Equations~\eqref{eq:spectralNumberDensity} and \eqref{eq:spectralNumberFluxDensity}.  
For numerical evaluation, the opacity and EoS tables---as well as the infrastructure needed for on-the-fly interpolation---are supplied by \weaklib.

\begin{table}[H]
  \caption{Opacities incorporated in \thornado's neutrino--matter coupling algorithm include neutrino ($\nu\in\{\nu_{\rm e},\nu_{\mu},\nu_{\tau}\}$) and/or antineutrino ($\bar{\nu}\in\{\bar{\nu}_{\rm e},\bar{\nu}_{\mu},\bar{\nu}_{\tau}\}$) interactions with electrons ($e^{-}$), positrons ($e^{+}$), neutrons ($n$), protons ($p$), heavy nuclei ($A$), and nucleons ($N$).\label{tab:opacities}}
  \begin{center}
    \begin{tabular}{cccc}
      \midrule\midrule
      & Process & Citation(s) & Equation(s) \\
      \midrule
        I & $e^{-}+p\rightleftharpoons n+\nu_{\rm e}$ 
        & \cite{Reddy_etal1998,Horowitz_weak_magnetism2002} 
        & Eq.~\eqref{eq:momentsCollisionTerm_EmAb} \\

        II & $e^{+}+n\rightleftharpoons p+\bar{\nu}_{\rm e}$ 
        & \cite{,Reddy_etal1998,Horowitz_weak_magnetism2002} 
        & Eq.~\eqref{eq:momentsCollisionTerm_EmAb} \\

        III & $\nu_{\rm e} + A' \rightleftharpoons e^{-}+A$
        & \cite{Langanke_etal2003,Hix_etal2003} 
        & Eq.~\eqref{eq:momentsCollisionTerm_EmAb} \\

	IV & $\nu+\{n,p\}\rightleftharpoons\nu+\{n,p\}$ 
        & \cite{bruenn_1985,Horowitz_weak_magnetism2002} 
        & Eq.~\eqref{eq:momentsCollisionTerm_Iso} \\

	V & $\nu+A\rightleftharpoons\nu+A$ 
        & \cite{bruenn_1985,Bruenn_Mezzacappa1997,Horowitz_IonIon1997} 
        & Eq.~\eqref{eq:momentsCollisionTerm_Iso} \\

	VI & $\nu+e^{\pm}\rightleftharpoons\nu'+e^{\pm}\,'$ 
        & \cite{bruenn_1985,mezzacappaBruenn_1993c} 
        & Eqs.~\eqref{eq:zerothMomentCollisionTerm_NES}, \eqref{eq:firstMomentCollisionTerm_NES} \\
 
	VII & $e^{-}+e^{+}\rightleftharpoons\nu+\bar{\nu}$ 
        & \cite{bruenn_1985} 
        & Eqs.~\eqref{eq:zerothMomentCollisionTerm_Pair}, \eqref{eq:firstMomentCollisionTerm_Pair} \\

	VIII & $\nu_{\rm e}+\bar{\nu}_{\rm e}\Rightarrow\nu_{\mu,\tau}+\bar{\nu}_{\mu,\tau}$ 
        & \cite{buras_etal_2003} 
        & Eqs.~\eqref{eq:zerothMomentCollisionTerm_Pair}, \eqref{eq:firstMomentCollisionTerm_Pair} \\

	IX & $N+N\rightleftharpoons N'+N'+\nu+\bar{\nu}$ 
        & \cite{hannestadRaffelt_1998} 
        & Eqs.~\eqref{eq:zerothMomentCollisionTerm_Pair}, \eqref{eq:firstMomentCollisionTerm_Pair} \\
      \midrule\midrule
    \end{tabular}
  \end{center}
\end{table}

The opacity set listed in Table~\ref{tab:opacities} includes the dominant weak interactions, such as emission and absorption (I-III), scattering (IV-VI), and pair processes (VII-IX).  
This set does not represent the current state of the art in CCSN theory \citep[cf. reviews by, e.g.,][]{burrows_etal_2006,janka_etal_2007,janka_2012,mezzacappa_etal_2020,fischer_etal_2023}, but captures the main opacities used in contemporary models, and the formalism we employ to include this set is general enough to accommodate more modern sets by updating the cross-sections and/or improving the treatment of a specific process; e.g., treating neutrino--nucleon scattering as inelastic \citep[e.g.,][]{muller_etal_2012}, as opposed to isoenergetic (as we do here).  
Also, we do not include muons and their associated weak interactions \citep{bollig_etal_2017,fischer_etal_2020}; nor do we include the many-body corrections to IV, derived by \citet{Horowitz2017}.  
For emission and absorption on nucleons, (I and II), the cross sections from~\cite{Reddy_etal1998} contain corrections due to nucelon 
recoil, we therefore follow~\cite{buras_etal_2006} to separate the weak magnetism and recoil corrections from~\cite{Horowitz_weak_magnetism2002} 
(see also~\cite{bruenn_etal_2020}).
For electron capture on heavy nuclei (III), we store the full nuclear statistical equilibrium (NSE)-folded tabular data from~\cite{Langanke_etal2003,Hix_etal2003} and renormalize the spectrum of emitted neutrinos to its original tabulated value after interpolation onto the DG spectral grid in \thornado.  
We do include strange-quark contributions to neutral-current neutrino--nucleon interactions~\citep{Horowitz_weak_magnetism2002} with a value of the strangeness contribution to the axial-vector coupling constant of $g_{\rm A}^{\rm s} = -0.1$~\citep{Hobbs_2016_ga_strange}.  
The annihilation rates for $\nu_e+\bar{\nu}_e\Rightarrow\nu_{\mu,\tau}+\bar{\nu}_{\mu,\tau}$ (VIII) are approximated using the same opacity kernels of VII, but using $\eta_{\nu_{\rm e}}=\mu_{\nu_{\rm e}}/T$, as opposed to $\eta_{\rm e}=\mu_{\rm e}/T$, as input for table interpolations, where $\mu_{\nu_{\rm e}}$ and $\mu_{\rm e}$ are the electron neutrino and electron chemical potential, respectively, and using weak coupling constants $C_{\rm V}=C_{\rm A}=\frac{1}{2}$. 
As in \citet{buras_etal_2003}, we only use VIII as a production mechanism for $\nu_{\mu,\tau}$ and $\bar{\nu}_{\mu,\tau}$, and only in regions where $\rho \geq 10^{12}$~g~cm$^{-3}$ to ensure that the assumption that $\nu_{\rm e}$ and $\bar{\nu}_{\rm e}$ are in local thermal equilibrium holds.  
We will report on improvements to the opacity set and their numerical treatment in \thornado\ in future studies.  

To accommodate the opacity set in Table~\ref{tab:opacities}, we write the collision term as
\begin{align}
	\mathcal{C}_{s}(\vect{f}) = \mathcal{C}_{s}^{\EmAb}(f_{s}) + \mathcal{C}_{s}^{\Iso}(f_{s}) + \mathcal{C}_{s}^{\NES}(f_{s}) + \mathcal{C}_{s}^{\Pair}(f_{s},\bar{f}_{s}),
	\label{eq:collisionTerm_AllProcesses}
\end{align}
where $\mathcal{C}_{s}^{\EmAb}$, $\mathcal{C}_{s}^{\Iso}$, $\mathcal{C}_{s}^{\NES}$, and $\mathcal{C}_{s}^{\Pair}$ represent contributions from emission and absorption (I-III), isoenergetic scattering on nucleons (IV) and nuclei (V), neutrino--electron scattering (NES; VI), and pair processes (VII-IX), respectively.  
Pair processes include neutrino pair production from electron--positron annihilation (VII), electron-neutrino pair annihilation (VIII), and nucleon-nucleon bremsstrahlung (IX).  
Here, $\bar{f}_{s}$ is the distribution of the antiparticle to neutrino species $s$, with distribution function $f_{s}$; e.g., for electron neutrinos, $s=\nu_{\rm e}$, $\bar{f}_{\nu_{\rm e}}=f_{\bar{\nu}_{\rm e}}$.  

Expressions for the terms on the right-hand side of Equation~\eqref{eq:collisionTerm_AllProcesses} are given in Appendix~\ref{app:collisionTerms}.  
For the two-moment model, angular moments of the collision term in Equation~\eqref{eq:collisionTerm_AllProcesses} are needed.  
Specifically, for the collision terms on the right-hand sides of Equations~\eqref{eq:spectralNumberDensity} and \eqref{eq:spectralNumberFluxDensity} we write
\begin{equation}
	\aint{\mathcal{C}_{s}(\vect{f})\{1,\ell_{j}\}}
	= \aint{\mathcal{C}_{s}^{\EmAb}(f_{s})\{1,\ell_{j}\}} 
	+ \aint{\mathcal{C}_{s}^{\Iso}(f_{s})\{1,\ell_{j}\}} 
	+ \aint{\mathcal{C}_{s}^{\NES}(f_{s})\{1,\ell_{j}\}} 
	+ \aint{\mathcal{C}_{s}^{\Pair}(f_{s},\bar{f}_{s})\{1,\ell_{j}\}},
	\label{eq:momentCollisionTerm_AllProcesses}
\end{equation}
where the contributions from emission and absorption are
\begin{align}
  \aint{\mathcal{C}_{s}^{\EmAb}(f_{s})}
  =\eta_{s}^{\EmAb}-\chi_{s}^{\EmAb}\,\mathcal{D}_{s} \quad\text{and}\quad
  \aint{\mathcal{C}_{s}^{\EmAb}(f_{s})\,\ell_{j}}
  =-\chi_{s}^{\EmAb}\,{\mathcal{I}_{s}}_{j},
  \label{eq:momentsCollisionTerm_EmAb}
\end{align}
where $\eta_{s}^{\EmAb}$ is the emissivity and $\chi_{s}^{\EmAb}$ is the absorption opacity, corrected for ``stimulated absorption'' \citep[e.g.,][]{mezzacappaMatzner_1989}.  
Contributions from isoenergetic scattering are
\begin{align}
  \aint{\mathcal{C}_{s}^{\Iso}(f_{s})} = 0 
  \quad\text{and}\quad
  \aint{\mathcal{C}_{s}^{\Iso}(f_{s})\,\ell_{j}} = - \sigma_{s}^{\Iso}\,{\mathcal{I}_{s}}_{j},
  \label{eq:momentsCollisionTerm_Iso}
\end{align}
where we have defined the total isoenergetic scattering opacity as
\begin{equation}
  \sigma_{s}^{\Iso} = \f{4\pi\varepsilon^{2}}{h^{3}}\,\big[\,\Phi_{s,0}^{\Iso}-\f{1}{3}\,\Phi_{s,1}^{\Iso}\,\big],
\end{equation}
and where $\Phi_{s,0}^{\Iso}$ and $\Phi_{s,1}^{\Iso}$ are expansion coefficients resulting from a Legendre expansion of the scattering kernel to linear order in the scattering angle (see Equation~\eqref{eq:kernelExpansion_Iso}).  

The contributions from NES can be written as
\begin{align}
  \aint{\mathcal{C}_{s}^{\NES}(f_{s})}
  &=\eta_{s}^{\NES} - \chi_{s}^{\NES}\,\mathcal{D}_{s}
  -\big[\,
	\langle\Phi_{s,1}^{\In}\,{\mathcal{I}_{s}}^{i}\rangle_{\bbR^{+}}
	-\langle\Phi_{s,1}^{\Out}\,{\mathcal{I}_{s}}^{i}\rangle_{\bbR^{+}}
  \,\big]\,{\mathcal{I}_{s}}_{i}, \label{eq:zerothMomentCollisionTerm_NES} \\
  \aint{\mathcal{C}_{s}^{\NES}(f_{s})\,\ell_{j}}
  &=-\chi_{s}^{\NES}\,{\mathcal{I}_{s}}_{j}
  +\f{1}{3}\big(\,\gamma_{ij}-3{\mathcal{K}_{s}}_{ij}\,\big)\,\langle\Phi_{s,1}^{\In}\,{\mathcal{I}_{s}}^{i}\rangle_{\bbR^{+}}
  +{\mathcal{K}_{s}}_{ij}\,\langle\Phi_{s,1}^{\Out}\,{\mathcal{I}_{s}}^{i}\rangle_{\bbR^{+}},
  \label{eq:firstMomentCollisionTerm_NES}
\end{align}
where the emissivity and opacity are defined in terms of the zeroth-order expansion coefficients (see Equation~\eqref{eq:collisionTermKernelExpanded_NES}) as
\begin{equation}
	\eta_{s}^{\NES}
	=\int_{\bbR^{+}}\Phi_{s,0}^{\In}(\varepsilon,\varepsilon')\,\mathcal{D}_{s}(\varepsilon')\,dV_{\varepsilon'}
	\quad\text{and}\quad
	\chi_{s}^{\NES}
	=\eta_{s}^{\NES}+\int_{\bbR^{+}}\Phi_{s,0}^{\Out}(\varepsilon,\varepsilon')\,(1-\mathcal{D}(\varepsilon'))\,dV_{\varepsilon'}, 
	\label{eq:etaChiNES}
\end{equation}
respectively, and the linear-order corrections are given by
\begin{equation}
	\langle\Phi_{s,1}^{\In/\Out}\,{\mathcal{I}_{s}}^{i}\rangle_{\bbR^{+}}
	=\int_{\bbR^{+}}\Phi_{s,1}^{\In/\Out}(\varepsilon,\varepsilon')\,{\mathcal{I}_{s}}^{i}(\varepsilon')\,dV_{\varepsilon'}.  
	\label{eq:linearCorrectionsNES}
\end{equation}
Contributions from pair processes can similarly be written as
\begin{align}
  \aint{\mathcal{C}_{s}^{\Pair}(f_{s},\bar{f}_{s})}
  &=\eta_{s}^{\Pair}-\chi_{s}^{\Pair}\,\mathcal{D}_{s}
  +\big[\,\langle\Phi_{s,1}^{\Pro}\,{\bar{\mathcal{I}_{s}}}^{i}\rangle_{\bbR^{+}}-\langle\Phi_{s,1}^{\Ann}\,{\bar{\mathcal{I}_{s}}}^{i}\rangle_{\bbR^{+}}\,\big]\,{\mathcal{I}_{s}}_{i},
  \label{eq:zerothMomentCollisionTerm_Pair} \\
  \aint{\mathcal{C}_{s}^{\Pair}(f_{s},\bar{f}_{s})\,\ell_{j}}
  &=
  -\chi_{s}^{\Pair}\,{\mathcal{I}_{s}}_{j}
  -\f{1}{3}\big(\,\gamma_{ij}-3{\mathcal{K}_{s}}_{ij}\,\big)\,\langle\Phi_{s,1}^{\Pro}\,{\bar{\mathcal{I}_{s}}}^{i}\rangle_{\bbR^{+}}
  -{\mathcal{K}_{s}}_{ij}\,\langle\Phi_{s,1}^{\Ann}\,{\bar{\mathcal{I}_{s}}}^{i}\rangle_{\bbR^{+}}, \label{eq:firstMomentCollisionTerm_Pair}
\end{align}
where pair emissivities and opacities are defined in terms of the zeroth-order expansion coefficients
\begin{equation}
	\eta_{s}^{\Pair}
  	=\int_{\bbR^{+}}\Phi_{s,0}^{\Pro}(\varepsilon,\varepsilon')\,(1-\bar{\mathcal{D}}_{s}(\varepsilon'))\,dV_{\varepsilon'}
	\quad\text{and}\quad
  	\chi_{s}^{\Pair}
 	=\eta_{s}^{\Pair} + \int_{\bbR^{+}}\Phi_{s,0}^{\Ann}(\varepsilon,\varepsilon')\,\bar{\mathcal{D}}_{s}(\varepsilon')\,dV_{\varepsilon'},
	\label{eq:etaChiPair}
\end{equation}
respectively, and the linear-order corrections are
\begin{equation}
	\langle\Phi_{s,1}^{\Pro/\Ann}\,{\bar{\mathcal{I}_{s}}}^{i}\rangle_{\bbR^{+}}
	=\int_{\bbR^{+}}\Phi_{s,1}^{\Pro/\Ann}(\varepsilon,\varepsilon')\,{\bar{\mathcal{I}_{s}}}^{i}(\varepsilon')\,dV_{\varepsilon'}.
	\label{eq:linearCorrectionsPair}
\end{equation}

For the interactions in Table~\ref{tab:opacities}, {\weaklib} provides tabulated opacities and kernels that are evaluated via interpolation.  
For the charged-current reactions I and II, $\chi_{s}^{\EmAb}(\varepsilon,\rho,T,Y_e)$ is tabulated and evaluated via tetralinear interpolation.  
The emissivity from electron capture on heavy nuclei (III) is obtained from the full NSE-folded tabular data.  
For isoenergetic scattering on nucleons (IV) and nuclei (V), the kernels $\Phi_{s,0}^{\Iso}(\varepsilon,\rho,T,Y_e)$ and $\Phi_{s,1}^{\Iso}(\varepsilon,\rho,T,Y_e)$ are tabulated and evaluated via tetralinear interpolation.
Kernels for VI, VII, and VIII are constructed from structure functions tabulated as functions of $(\varepsilon,\varepsilon',\eta,T)$ \citep[see][Appendix~C; in particular Equation~(C50) for VI and Equation~(C63) for VII and VIII]{bruenn_1985}.  
These structure functions are first interpolated bilinearly in $(\varepsilon,\varepsilon')$ onto \thornado's energy grid at the beginning of a simulation, and subsequently interpolated bilinearly in degeneracy parameter and temperature $(\eta,T)$ during runtime to obtain the kernels $\Phi_{s,\ell}^{\In/\Out}$ and $\Phi_{s,\ell}^{\Pro/\Ann}$, with $\ell\in\{0,1\}$.  
For process IX, we tabulate the kernel function $S_{\sigma}(\varepsilon,\varepsilon', \rho, T)$ from \citet{hannestadRaffelt_1998} and, as for VI-VIII, we initially interpolate it onto \thornado's energy grid. 
During the simulation, three bilinear interpolations of $S_{\sigma}$ are performed and combined as $S_{\sigma} = S_{\sigma}(\rho \, X_n, T) + S_{\sigma}(\rho \, X_p, T)+ (28/3) \, S_{\sigma}\left( \rho \, \sqrt{X_n \, X_p}, T \right)$~\citep{Brinkmann_Turner1988,thompson_etal_2000,ramppJanka_2002}, where $X_n$ and $X_p$ are the neutron and proton mass fractions, respectively. 
$S_{\sigma}$ is then used to obtain the kernels $\Phi_{s,\ell}^{\Pro/\Ann}$.  

\begin{rem}
	The collision terms involve several features that make neutrino--matter coupling the most computationally demanding component of CCSN neutrino transport.  
	NES and pair processes couple moments across energies and, in the case of pair processes, across neutrino species, through integral operators; e.g., as in Equations~\eqref{eq:etaChiNES} and \eqref{eq:etaChiPair}.
	Fermi blocking factors, $(1-\cD_{s})$, introduce nonlinearity, in addition to the nonlinear dependencies of opacities and kernels on the local thermodynamic state of the fluid.  
	\label{rem:collision_term_complexities}
\end{rem}

\subsection{Neutrino-Radiation Hydrodynamics Equations}
\label{sec:neutrinoRadiationHydrodynamicsEquations}

Here we summarize the neutrino-radiation hydrodynamics equations, and introduce compact notation that will be used in the description of the numerical method in Section~\ref{sec:method}.  
We write the hydrodynamics equations in Equations~\eqref{eq:massConservation}-\eqref{eq:electronNumberEquation} in compact form as
\begin{equation}
	\pd{\vect{u}}{t} + \f{1}{\sqrt{\gamma}}\pd{}{i}(\,\sqrt{\gamma}\,\vect{F}^{i}(\vect{u})) = \vect{S}(\vect{u},\Phi) + \vect{C}(\vect{u},\vect{\mathcal{U}}),
	\label{eq:hydroCompact}
\end{equation}
where the solution vector, flux vectors, and the sources due to curvilinear coordinates and gravity, and neutrino--matter interactions are
\begin{equation}
	\vect{u}
	=
	\left[\begin{array}{c}
		\rho \\
		\rho v_{j} \\
		E_{\rm f} \\
		\rho\ye
	\end{array}\right],
	\quad
	\vect{F}^{i}
	=
	\left[\begin{array}{c}
		\rho v^{i} \\
		\Pi_{~j}^{i} \\
		(E_{\rm f}+p)v^{i} \\
		\rho\ye v^{i}
	\end{array}\right],
	\quad
	\vect{S}
	=
	\left[\begin{array}{c}
		0 \\
		\f{1}{2}\,\Pi^{ik}\,\pd{}{j}\gamma_{ik} - \rho\,\pd{}{j}\Phi \\
		-\rho\,v^{i}\,\pd{}{i}\Phi \\
		0
	\end{array}\right], 
	\quad\text{and}\quad
	\vect{C}
	=-
	\left[\begin{array}{c}
		0 \\
		\QM{j} \\
		\QE \\
		\mB\,\QL
	\end{array}\right],
	\label{eq:solutionFluxAndSourceVectors}
\end{equation}
respectively.  
($\bcU$, representing the neutrino moments, is defined below.)  
We write the Poisson equation in Equation~\eqref{eq:poissonEquation} in compact form as the constraint equation
\begin{equation}
	\mathsf{F}(\Phi,\vect{u}) = 0.  
	\label{eq:poissonCompact}
\end{equation}

The neutrino kinetic equations, given by the two-moment model in Equations~\eqref{eq:spectralNumberDensity} and \eqref{eq:spectralNumberFluxDensity}, are written in compact form as
\begin{equation}
	\pd{\vect{\mathcal{U}}}{t} + \f{1}{\sqrt{\gamma}}\pd{}{i}(\,\sqrt{\gamma}\,\vect{\mathcal{F}}^{i}(\vect{\mathcal{U}},\vect{v}))
	+\f{1}{\varepsilon^{2}}\pd{}{\varepsilon}(\,\varepsilon^{3}\,\vect{\mathcal{F}}^{\varepsilon}(\vect{\mathcal{U}},\vect{v}))
	=\vect{\mathcal{S}}(\vect{\mathcal{U}},\vect{v}) + \vect{\mathcal{C}}(\vect{\mathcal{U}},\vect{u}),
	\label{eq:kineticCompact}
\end{equation}
where the solution vector and the position- and energy-space fluxes include all the neutrino species; i.e., we let $\vect{\mathcal{U}}=\{ \vect{\mathcal{U}}_{s} \}_{s=1}^{N_{\rm s}}$, $\vect{\mathcal{F}}^{i}=\{ {\vect{\mathcal{F}}_{s}}^{i} \}_{s=1}^{N_{\rm s}}$, and $\vect{\mathcal{F}}^{\varepsilon}=\{ {\vect{\mathcal{F}}_{s}}^{\varepsilon} \}_{s=1}^{N_{\rm s}}$, where
\begin{equation}
	\vect{\mathcal{U}}_{s}
	=
	\left[\begin{array}{c}
		\mathcal{D}_{s}+v^{i}\,{\mathcal{I}_{s}}_{i} \\
		{\mathcal{I}_{s}}_{j}+v^{i}\,{\mathcal{K}_{s}}_{ij}
	\end{array}\right], 
	\quad
	{\vect{\mathcal{F}}_{s}}^{i}
	=
	\left[\begin{array}{c}
		{\mathcal{I}_{s}}^{i}+\mathcal{D}_{s}\,v^{i} \\
		{\mathcal{K}_{s}}^{i}_{\hspace{2pt}j}+{\mathcal{I}_{s}}_{j}\,v^{i}
	\end{array}\right], 
	\quad\text{and}\quad
	{\vect{\mathcal{F}}_{s}}^{\varepsilon}
	=-
	\left[\begin{array}{c}
		{\mathcal{K}_{s}}_{\hspace{2pt}k}^{i} \\
		{\mathcal{L}_{s}}_{\hspace{2pt}kj}^{i}
	\end{array}\right]\,\cderiv{v^{k}}{i}.
	\label{eq:evolutionAndFluxVectors}
\end{equation}
Similarly, the sources on the right-hand side of Equation~\eqref{eq:kineticCompact} are $\vect{\mathcal{S}}=\{ \vect{\mathcal{S}}_{s} \}_{s=1}^{N_{\rm s}}$ and $\vect{\mathcal{C}}=\{ \vect{\mathcal{C}}_{s} \}_{s=1}^{N_{\rm s}}$, where
\begin{equation}
	\vect{\mathcal{S}}_{s}
	=
	\left[\begin{array}{c}
		0 \\
		\f{1}{2}\,\big(\,{\mathcal{K}_{s}}^{ik}+{\mathcal{I}_{s}}^{i}\,v^{k}+v^{i}\,{\mathcal{I}_{s}}^{k}\,\big)\,\pd{\gamma_{ik}}{j}
  		+\big(\,{\mathcal{L}_{s}}_{\hspace{4pt}kj}^{i}\,\cderiv{v^{k}}{i} - {\mathcal{I}_{s}}^{i}\,\pd{v_{j}}{i}\,\big)
	\end{array}\right]
	\quad\text{and}\quad
	\vect{\mathcal{C}}_{s}
	=
	\left[\begin{array}{c}
		\aint{\mathcal{C}_{s}(\vect{f})} \\
		\aint{\mathcal{C}_{s}(\vect{f})\ell_{j}}
	\end{array}\right].  
\end{equation}
In Equation~\eqref{eq:kineticCompact}, when compared with Equations~\eqref{eq:spectralNumberDensity} and \eqref{eq:spectralNumberFluxDensity}, we have dropped terms containing the time-derivative of the fluid three-velocity, as was also done by \citet{just_etal_2015} and \citet{skinner_etal_2019}.  
The dependence of $\vect{\mathcal{F}}^{i}$, $\vect{\mathcal{F}}^{\varepsilon}$, and $\vect{\mathcal{S}}$ on the three-velocity $\vect{v}$ is made explicit in Equation~\eqref{eq:kineticCompact}, and it is understood that the components of $\vect{v}$ can be obtained from components of $\vect{u}$ (i.e., $v^{i}=\gamma^{ij}(\rho v_{j})/\rho$).  

Finally, we introduce the vector of primitive moments $\vect{\mathcal{M}}_{s}=\big[\,\mathcal{D}_{s},\,{\mathcal{I}_{s}}^{i}\,\big]^{\intercal}$.  
Then, with a moment closure giving ${\mathcal{K}_{s}}^{ij}={\mathcal{K}_{s}}^{ij}(\vect{\mathcal{M}}_{s})$ , the solution vector can be expressed as a function of the primitive moments and the fluid three-velocity, $\vect{v}$, as $\vect{\mathcal{U}}_{s}=\vect{\mathcal{U}}_{s}(\vect{\mathcal{M}}_{s},\vect{v})$, and their explicit relationship can be written as
\begin{equation}
	\vect{\mathcal{U}}_{s} = \mathcal{A}_{s}(\vect{\mathcal{M}}_{s},\vect{v})\,\vect{\mathcal{M}}_{s},
	\quad\text{where}\quad
	\mathcal{A}_{s}(\vect{\mathcal{M}}_{s},\vect{v})
	=\left[\begin{array}{cc}
		1 & v^{k}\gamma_{kj} \\
		v^{k}{\mathsf{k}_{s}}_{ki}(\vect{\mathcal{M}}_{s}) & \gamma_{ij}
	\end{array}\right].
	\label{eq:conservedToPrimitive}
\end{equation}
For given $\vect{\mathcal{U}}_{s}$ and $\vect{v}$, Equation~\eqref{eq:conservedToPrimitive} defines a nonlinear system that can be solved for $\vect{\mathcal{M}}_{s}$.  
This problem must be solved repeatedly during the numerical integration of the two-moment model; e.g., to evaluate the fluxes in Equation~\eqref{eq:evolutionAndFluxVectors}.  

\section{Numerical Method}
\label{sec:method}

The neutrino-radiation hydrodynamics equations summarized in Section~\ref{sec:neutrinoRadiationHydrodynamicsEquations} support dynamics spanning a wide range of spatial and temporal scales.  
To address this multiscale structure, we use a combination of explicit and implicit time integration, where fluid flows and neutrino phase-space advection are integrated with explicit methods, and neutrino--matter coupling is integrated with implicit methods.  
In the context of Lie--Trotter splitting, combining explicit RK methods for the Euler--Poisson system and IMEX-RK methods for neutrino transport, it is useful to split the coupled problem given by Equations~\eqref{eq:hydroCompact}, \eqref{eq:poissonCompact}, and \eqref{eq:kineticCompact} into the following three subproblems:
\begin{itemize}
	\item {\bf Subproblem~1:} Hydrodynamics with self-gravity (Euler--Poisson)
		\begin{subequations}\label{eq:subproblem1}
			\begin{align}
				&\pd{\vect{u}}{t} + \f{1}{\sqrt{\gamma}}\pd{}{i}(\,\sqrt{\gamma}\,\vect{F}^{i}(\vect{u})) = \vect{S}(\vect{u},\Phi), \label{eq:subproblem1a} \\
				&\mathsf{F}(\Phi,\vect{u}) = 0. \label{eq:subproblem1b}
			\end{align}
		\end{subequations}
	\item {\bf Subproblem~2:} Neutrino phase-space advection
		\begin{equation}
			\pd{\vect{\mathcal{U}}}{t} + \f{1}{\sqrt{\gamma}}\pd{}{i}(\,\sqrt{\gamma}\,\bcF^{i}(\vect{\mathcal{U}},\vect{v}))
			+\f{1}{\varepsilon^{2}}\pd{}{\varepsilon}(\,\varepsilon^{3}\,\vect{\mathcal{F}}^{\varepsilon}(\vect{\mathcal{U}},\vect{v}))
			=\vect{\mathcal{S}}(\vect{\mathcal{U}},\vect{v}).
			\label{eq:subproblem2}
		\end{equation}
	\item {\bf Subproblem~3:} Neutrino--matter coupling
		\begin{subequations}\label{eq:subproblem3}
			\begin{align}
				&\pd{\vect{\mathcal{U}}}{t} = \vect{\mathcal{C}}(\vect{\mathcal{U}},\vect{u}), \label{eq:subproblem3a} \\
				&\pd{\vect{\mathsf{U}}}{t} = 0, \label{eq:subproblem3b}
			\end{align}
		\end{subequations}
\end{itemize}
where, in Equation~\eqref{eq:subproblem3b}, we have introduced the vector of `collision invariants' 
\begin{equation}
	\vect{\mathsf{U}}(\vect{\mathcal{U}},\vect{u})=(\,\rho,\,\mathsf{M}_{j},\,\mathsf{E},\,\mathsf{N}\,)^{\intercal}, 
	\label{eq:collision_invariants}
\end{equation}
where $\mathsf{M}_{j}=\rho v_{j}+{F_{\nu}}_{j}$ is the total momentum density, $\mathsf{E}=E_{\rm f}+E_{\nu}$ the total energy density, and $\mathsf{N}=n_{\rm e}+N_{\nu}$ the total lepton number density; see Equations~\eqref{eq:totalMomentum}, \eqref{eq:totalEnergy}, and \eqref{eq:totalLeptonNumber}, respectively.  

In this paper, we will mainly discuss the methods to solve the phase-space advection problem in Equation~\eqref{eq:subproblem2} (subproblem~2) and the neutrino--matter coupling problem in Equation~\eqref{eq:subproblem3} (subproblem~3).  
To solve Equation~\eqref{eq:subproblem1} (subproblem~1), we use the hydrodynamics and gravity solvers provided by \flashx\ \citep{dubey_etal_2022}; specifically, we use \spark\ \citep{couch_etal_2021} with the multipole gravity solver of \cite{couch_etal_2013}.  
We describe how the full system is integrated in time in Section~\ref{sec:timeIntegration}.  
Subproblem~2 models the propagation of neutrinos in phase-space, and couples nearest neighbors through the divergence operator on the left-hand side.  
The solution method employed for this explicit part is an extension of the method proposed in \citet{laiu_etal_2025} to curvilinear spatial coordinates and multiple neutrino species.  
In subproblem~3, we solve the space-homogeneous kinetic equations together with corresponding equations for the matter, which is updated by enforcing preservation of the collision invariants in Equation~\eqref{eq:collision_invariants}.  
This part is local in position space, but couples across neutrino energies and species.  
The implicit solution method employed for this nonlinear problem is an extension of the nested fixed-point iteration scheme proposed by \citet[][Section~4.3]{laiu_etal_2021}.  

\subsection{Phase-Space Discretization}

\subsubsection{Discontinuous Galerkin Method}

In \thornado, the DG method \citep[e.g.,][]{cockburnShu_2001} is used to discretize the neutrino moment equations in phase-space \citep[see also][]{chu_etal_2019,laiu_etal_2021,laiu_etal_2025}.  
The method described here is an extension of that proposed by \citet{laiu_etal_2025} to orthogonal curvilinear coordinates and more general collision operators, and we largely adopt their notation \citep[see also][]{pochik_etal_2021}.  
The $d_{\vect{x}}$-dimensional spatial domain $D_{\vect{x}}\subset\bbR^{d_{\vect{x}}}$ is subdivided into the collection $\mathcal{T}_{\vect{x}}$ of nonoverlapping spatial elements $\Kx$ so that $D_{\vect{x}}=\cup_{\Kx\in\mathcal{T}_{\vect{x}}}\Kx$.  
Each element is a logically Cartesian box
\begin{equation}
	\Kx = \big\{\,\vect{x} : x^{i}\in K_{\bx}^{i} \vcentcolon= (x_{\Lo}^{i},x_{\Hi}^{i}), i=1,\ldots,d_{\vect{x}}\,\big\},
\end{equation}
where $x_{\Lo}^{i}$ and $x_{\Hi}^{i}$ are the low and high boundaries of the element in the $i$th spatial dimension, respectively.  
Similarly, the neutrino energy domain $D_{\varepsilon}=[\varepsilon_{\min},\varepsilon_{\max}]\subset\bbR^{+}$ is subdivided into the collection $\mathcal{T}_{\varepsilon}$ of $\mathfrak{N}$ energy elements (or bins) $K_{\varepsilon}^{\eidx}=(\varepsilon_{\eidx-\f{1}{2}},\varepsilon_{\eidx+\f{1}{2}})$, where
\begin{equation}
	\varepsilon_{\min} = \varepsilon_{\f{1}{2}} < \ldots < \varepsilon_{\eidx-\f{1}{2}} < \varepsilon_{\eidx+\f{1}{2}} < \ldots < \varepsilon_{\mathfrak{N}+\f{1}{2}} = \varepsilon_{\max}.  
\end{equation}
Therefore, $D_{\varepsilon}=\cup_{\eidx=1}^{\nElementsE}K_{\varepsilon}^{\eidx}$.  
(We reserve the Fraktur index $\eidx$ to exclusively refer to the $\eidx$th energy element.)  
We emphasize that the index $i$, e.g., used in $K_{\bx}^{i}$, refers to spatial dimension, while the index $\eidx$ refers to a specific energy element.  
The index $\eidx$ on the energy elements is needed later when we describe the discretization of collision operators, which couple all $K_{\varepsilon}^{\eidx}\in D_{\varepsilon}$, while a corresponding index to refer to a specific spatial element is not needed.  
We define the phase-space coordinates as $\vect{z}=(\vect{x},\varepsilon)$ and denote the phase-space element by 
\begin{equation}
	\Kz^{\eidx}=\Kx\times K_{\varepsilon}^{\eidx}, \quad\eidx\in\{1,\ldots,\nElementsE\}.  
\end{equation}
Let $D_{\bz}= D_{\bx}\times D_{\varepsilon}$ denote the $d_{\bz}$-dimensional phase-space domain, where $d_{\bz}=1+d_{\bx}$.  
We further define the surface elements $\Ax^{i}=\times_{j\ne i}K_{\bx}^{j}$ and $\Az^{i,\eidx}=\Ax^{i}\times K_{\varepsilon}^{\eidx}$, and note that $\Kx=\Ax^{i}\times K_{\bx}^{i}$ and $\Kz^{\eidx}=\Az^{i,\eidx}\times K_{\bx}^{i}$, for any $i\in\{1,\ldots,d_{\vect{x}}\}$.  

We also let $\vect{x}=\{x^{i},\tilde{\vect{x}}^{i}\}$ and $\vect{z}=\{x^{i},\tilde{\vect{z}}^{i}\}$, to distinguish spatial and phase-space coordinates along ($x^{i}$) and transverse ($\tilde{\bx}^{i}$ and $\tilde{\bz}^{i}$) to the $i$th spatial dimension; e.g., for $i=1$, $\tilde{\bx}^{1}=\{x^{2},x^{3}\}$ and $\tilde{\bz}^{1}=\{x^{2},x^{3},\varepsilon\}$.  
We denote the spatial and phase-space volume elements, respectively, by
\begin{equation}
	V_{\Kx} = \int_{\Kx}dV_{h}^{\vect{x}},
	\quad\text{where}\quad
	dV_{h}^{\vect{x}}=\sqrt{\gamma_{h}}d\vect{x},
\end{equation}
and
\begin{equation}
	V_{\Kz^{\eidx}} = \int_{\Kz^{\eidx}}dV_{h}^{\bz},
	\quad\text{where}\quad
	dV_{h}^{\vect{z}}=\sqrt{\gamma_{h}}\varepsilon^{2}d\vect{z}=dV_{h}^{\vect{x}}\varepsilon^{2}d\varepsilon.
\end{equation}
Here, as in \citet{pochik_etal_2021}, $\gamma_{h}$ is an approximation (as indicated by the subscript $h$) to the determinant of the spatial metric $\gamma$.  
Specifically, we employ polynomial approximations to the scale factors, denoted $h_{1,h}$, $h_{2,h}$, and $h_{3,h}$, from which we obtain the components of the spatial metric $(\gamma_{h})_{ij}$ and $\gamma_{h}={\rm det}[(\gamma_{h})_{ij}]$.  
Spatial derivatives of components of the spatial metric---e.g., as appearing in the sources of Equations~\eqref{eq:fluidMomentumEquation} and \eqref{eq:spectralNumberFluxDensity}---are computed as in \citet[][Section~3.1.1]{pochik_etal_2021}.  

Next we define the approximation spaces for the DG method, consisting of square integrable functions $\psi_{h}$.  
For approximating matter quantities, depending on position space coordinates $\vect{x}$, we set
\begin{equation}
	\bbV_{\vect{x}}^{k}
	=\{\,\psi_{h}\in L^{2}(D_{\vect{x}})~:~\psi_{h}|_{\Kx}\in\bbQ^{k}(\Kx),\forall\Kx\in\mathcal{T}_{\vect{x}}\,\},
	\label{eq:approximationSpacePositionSpace}
\end{equation}
where, on each element $\Kx$, $\bbQ^{k}(\Kx)$ is the tensor product space of one-dimensional polynomials of maximum degree $k$.  
Similarly, for approximating neutrino quantities, depending on phase-space coordinates $\vect{z}$, we set
\begin{equation}
	\bbV_{\vect{z}}^{k}
	=\{\,\psi_{h}\in L^{2}(D_{\vect{z}})~:~\psi_{h}|_{\Kz^{\eidx}}\in\bbQ^{k}(\Kz^{\eidx}),\forall\Kz^{\eidx}\in\mathcal{T}_{\vect{z}}\,\}.  
	\label{eq:approximationSpacePhaseSpace}
\end{equation}

We can now state the semi-discrete DG formulation for the neutrino moment equations.  
To this end, we seek $\vect{\mathcal{U}}_{h}\in\bbV_{\vect{z}}^{k}$, the approximation to $\vect{\mathcal{U}}$ in Equation~\eqref{eq:kineticCompact}, such that
\begin{equation}
	\intKz{\,\pd{}{t}\vect{\mathcal{U}}_{h},\psi_{h}\,}
	=\vect{\mathcal{B}}_{h}\big(\,\vect{\mathcal{U}}_{h},\vect{u}_{h},\psi_{h}\,\big)_{\Kz^{\eidx}}
	+\intKz{\,\vect{\mathcal{C}}_{h}(\vect{\mathcal{U}}_{h},\vect{u}_{h}),\psi_{h}\,}
	\label{eq:kineticCompactWeak}
\end{equation}
holds for all $\psi_{h}\in\bbV_{\vect{z}}^{k}$ and all $\Kz^{\eidx}\in\mathcal{T}_{\vect{z}}$, and where $\vect{u}_{h}\in\bbV_{\vect{x}}^{k}$ is the approximation to the hydrodynamics variables $\vect{u}$ in Equation~\eqref{eq:hydroCompact}.  
It is here understood that the semi-discrete DG problem in Equation~\eqref{eq:kineticCompactWeak} holds for each (moment and species) component of $\vect{\mathcal{U}}_{h}$.  
Equation~\eqref{eq:kineticCompactWeak} must be supplied with initial and boundary conditions, which we leave unspecified for now.  

In Equation~\eqref{eq:kineticCompactWeak}, $\intKz{\ldots}$ denotes the $\varepsilon^{2}\sqrt{\gamma_{h}}$-weighted integral over the phase-space element $\Kz^{\eidx}$; e.g.,
\begin{equation}
	\intKz{\,\pd{}{t}\vect{\mathcal{U}}_{h},\psi_{h}\,} = \int_{\Kz^{\eidx}}\pd{}{t}\vect{\mathcal{U}}_{h}\,\psi_{h}\,dV_{h}^{\vect{z}},
\end{equation}
and we have defined the discretization of the phase-space advection terms on $\Kz^{\eidx}$ by
\begin{align}
	&\vect{\mathcal{B}}_{h}\big(\vect{\mathcal{U}}_{h},\vect{u}_{h},\psi_{h}\big)_{\Kz^{\eidx}}
	=-\sum_{i=1}^{d_{\bx}}\int_{\Az^{i,\eidx}}
	\big(
		\sqrt{\gamma_{h}}\,\widehat{\vect{\mathcal{F}}^{i}}(\vect{\mathcal{U}}_{h},\vect{u}_{h})\,\psi_{h}\big|_{x_{\Hi}^{i}}
		-\sqrt{\gamma_{h}}\,\widehat{\vect{\mathcal{F}}^{i}}(\vect{\mathcal{U}}_{h},\vect{u}_{h})\,\psi_{h}\big|_{x_{\Lo}^{i}}
	\big)\,\varepsilon^{2}d\tilde{\bz}^{i}
	+\sum_{i=1}^{d_{\bx}}\intKz{\vect{\mathcal{F}}^{i}(\vect{\mathcal{U}}_{h},\vect{u}_{h}),\pd{\psi_{h}}{i}} \nonumber \\
	&\hspace{0pt}
	-\int_{\Kx}
	\big(
		\varepsilon^{3}\,\widehat{\vect{\mathcal{F}}^{\varepsilon}}(\vect{\mathcal{U}}_{h},\vect{u}_{h})\,\psi_{h}\big|_{\varepsilon_{\eidx+\f{1}{2}}}
		-\varepsilon^{3}\,\widehat{\vect{\mathcal{F}}^{\varepsilon}}(\vect{\mathcal{U}}_{h},\vect{u}_{h})\,\psi_{h}\big|_{\varepsilon_{\eidx-\f{1}{2}}}
	\big)\,dV_{h}^{\bx}
	+\intKz{\varepsilon\,\vect{\mathcal{F}}^{\varepsilon}(\vect{\mathcal{U}}_{h},\vect{u}_{h}),\pd{\psi_{h}}{\varepsilon}}
	+\intKz{\vect{\mathcal{S}}(\vect{\mathcal{U}}_{h},\vect{u}_{h}),\psi_{h}}.  
	\label{eq:kineticCompactBLF}
\end{align}
Here, $\widehat{\vect{\mathcal{F}}^{i}}$ and $\widehat{\vect{\mathcal{F}}^{\varepsilon}}$ are numerical fluxes, approximating $\vect{\mathcal{F}}^{i}$ and $\vect{\mathcal{F}}^{\varepsilon}$, on the surface elements $\Az^{i,\eidx}$ and $\Kx$, respectively.  
For the spatial divergence in Equation~\eqref{eq:kineticCompactBLF}, letting $x^{i,\pm}=\lim_{\delta\to0^{+}}x^{i}\pm\delta$, the numerical flux, 
\begin{equation}
	\widehat{\vect{\mathcal{F}}^{i}}(\vect{\mathcal{U}}_{h},\vect{u}_{h};x^{i},\tilde{\bz}^{i})
	\vcentcolon= \mathscr{F}_{\rm LF}^{i}(\vect{\mathcal{U}}_{h}(x^{i,-},\tilde{\bz}^{i}),\vect{\mathcal{U}}_{h}(x^{i,+},\tilde{\bz}^{i}),\hat{\vect{v}}(x^{i},\tilde{\bx}^{i})),
\end{equation}
is evaluated using a global Lax--Friedrichs (LF) flux
\begin{equation}
	\mathscr{F}_{\rm LF}^{i}(\vect{\mathcal{U}}_{a},\vect{\mathcal{U}}_{b},\hat{\vect{v}})
	=\f{1}{2}
	\Big\{\,
		\vect{\mathcal{F}}^{i}(\vect{\mathcal{U}}_{a},\hat{\vect{v}})+\vect{\mathcal{F}}^{i}(\vect{\mathcal{U}}_{b},\hat{\vect{v}})
		-\alpha^{i}\,\big(\,\vect{\mathcal{U}}(\vect{\mathcal{M}}_{b},\hat{\vect{v}}^{i})-\vect{\mathcal{U}}(\vect{\mathcal{M}}_{a},\hat{\vect{v}}^{i})\,\big)
	\,\Big\},
	\label{eq:numericalFluxSpatial}
\end{equation}
where $\alpha^{i}$ is the largest absolute eigenvalue of the flux Jacobian $\partial\vect{\mathcal{F}}^{i}/\partial\vect{\mathcal{U}}$.  
(For simplicity, since we consider neutrinos propagating at the speed of light, we set $\alpha^{i}=1$.)  
The components of the fluid three-velocity at element interfaces, $\hat{\vect{v}}$, are set to the average of the left and right states; i.e., 
\begin{equation}
	\hat{\vect{v}}(x^{i},\tilde{\bx}^{i}) \vcentcolon= \f{1}{2}\big(\,\vect{v}(x^{i,-},\tilde{\bx}^{i})+\vect{v}(x^{i,+},\tilde{\bx}^{i})\,\big).  
	\label{eq:velocityFace}
\end{equation}
For the dissipation term (proportional to $\alpha^{i}$) in the numerical flux in Equation~\eqref{eq:numericalFluxSpatial}, the primitive moments $\vect{\mathcal{M}}_{a/b}$ are first obtained by solving $\vect{\mathcal{U}}_{a/b}=\vect{\mathcal{U}}(\vect{\mathcal{M}}_{a/b},\hat{\vect{v}})$.  
Then, $\vect{\mathcal{U}}(\vect{\mathcal{M}}_{a/b},\hat{\vect{v}}^{i})$ is computed using $\hat{\vect{v}}^{i}=(\delta^{i1}\hat{v}^{1},\delta^{i2}\hat{v}^{2},\delta^{i3}\hat{v}^{3})^{\intercal}$.  
When compared to the standard LF flux, which uses $\vect{\mathcal{U}}_{a/b}$ in the dissipation term, we have found the numerical flux in Equation~\eqref{eq:numericalFluxSpatial} to result in a more robust scheme in terms of maintaining physically realizable moments \citep[][Remark~3]{laiu_etal_2025}.  

Similarly, for the energy divergence in Equation~\eqref{eq:kineticCompactBLF}, the numerical flux on energy element interfaces is evaluated as
\begin{equation}
	\widehat{\vect{\mathcal{F}}^{\varepsilon}}(\vect{\mathcal{U}}_{h},\vect{u}_{h};\varepsilon,\bx)
	\vcentcolon=\mathscr{F}_{\rm LF}^{\varepsilon}(\vect{\mathcal{U}}_{h}(\varepsilon^{-},\bx),\vect{\mathcal{U}}_{h}(\varepsilon^{+},\bx),\vect{v}_{h}(\bx)),
\end{equation}
where the flux function is given by the LF-type expression
\begin{equation}
	\mathscr{F}_{\rm LF}^{\varepsilon}(\vect{\mathcal{U}}_{a},\vect{\mathcal{U}}_{b},\vect{v})
	=\f{1}{2}
	\Big\{\,
		\vect{\mathcal{F}}^{\varepsilon}(\vect{\mathcal{U}}_{a},\vect{v})+\vect{\mathcal{F}}^{\varepsilon}(\vect{\mathcal{U}}_{b},\vect{v})
		-\alpha^{\varepsilon}\,\big(\,\vect{\mathcal{M}}_{b}-\vect{\mathcal{M}}_{a}\,\big)
	\,\Big\},
	\label{eq:numericalFluxEnergy}
\end{equation}
where $\vect{\mathcal{M}}_{a/b}$ are obtained by solving $\vect{\mathcal{U}}_{a/b}=\vect{\mathcal{U}}(\vect{\mathcal{M}}_{a/b},\vect{v})$.  
In Equation~\eqref{eq:numericalFluxEnergy}, $\alpha^{\varepsilon}$ is an estimate on the largest eigenvalue of the flux Jacobian $\partial\vect{\mathcal{F}}^{\varepsilon}/\partial\vect{\mathcal{U}}$.  
We set $\alpha^{\epsilon}=\lambda_{A}$, where $\lambda_{A}$ is the largest absolute eigenvalue of the symmetric matrix $A_{ij}=\f{1}{2}\,(\nabla_{i}v_{j}+\nabla_{j}v_{i})$.  
The use of the primitive moments, rather than conserved moments, in the dissipative term of the numerical flux in Equation~\eqref{eq:numericalFluxEnergy} is motivated by robustness considerations \citep[][Remark~4]{laiu_etal_2025}.  

In the semi-discrete scheme in Equation~\eqref{eq:kineticCompactWeak}, the dependence on the fluid field --- due to the velocity-dependent terms in the phase-space divergence and the weak interaction rates' dependence on the thermodynamic state of the fluid --- is made explicit through $\vect{u}_{h}$.  
The fluid is partially updated in Subproblem~1 by \flashx, using a FV representation, and partially updated in Subproblem~3, using a DG representation.  
Here, for the purpose of the phase-space advection update (Subproblem~2), we assume that $\bu_{h}\in\bbV_{\bx}^{k}$ is known, and we postpone the detailed DG formulation of Subproblem~3 to Section~\ref{sec:neutrinoMatterCoupling}.  
We describe the reconstruction and projection steps to map between the FV and DG representations in Section~\ref{sec:flashx}.  

The fluxes in energy space, $\bcF^{\varepsilon}$, depend on spatial derivatives of the three-velocity.  
To this end, for given $v_{j,h}\in\bbV_{\bx}^{k}$, we approximate the spatial derivatives in each element $\Kx$ in a weak sense by finding $(\pd{v_{j}}{i})_{h}\in\bbV_{\bx}^{k}$ such that
\begin{equation}
	\int_{\Kx}(\pd{v_{j}}{i})_{h}\psi_{h}d\bx
	=\int_{\Ax^{i}}\big[\,\hat{v}_{j}(x_{\Hi}^{i},\tilde{\bx}^{i})\,\psi_{h}(x_{\Hi}^{i,-},\tilde{\bx}^{i})-\hat{v}_{j}(x_{\Lo}^{i},\tilde{\bx}^{i})\,\psi_{h}(x_{\Lo}^{i,+},\tilde{\bx}^{i})\,\big]\,d\tilde{\bx}^{i}
	-\int_{\Kx}v_{j,h}\,\pd{\psi_{h}}{i}\,d\bx
	\label{eq:velocityWeakDerivative}
\end{equation}
holds for all $\psi_{h}\in\bbV_{\bx}^{k}$.  
In Equation~\eqref{eq:velocityWeakDerivative}, the velocity on the faces of the spatial element are computed using the average in Equation~\eqref{eq:velocityFace}.  

\subsubsection{Nodal Collocation DG Scheme}

We employ a spectral-type nodal collocation DG method \citep[e.g.,][]{bassi_etal_2013} to solve the moment equations in \thornado.  
For convenience, we introduce multi-index notation.  
We define the $d_{\bz}$-tuples 
\begin{equation}
	\bi=\{i^{\varepsilon},\bi_{\bx}\}
	\quad\text{and}\quad 
	\bN=\{N,\bN_{\bx}\}, 
\end{equation}
where $\bi_{\bx}=\{i^{1},\ldots,i^{d_{\bx}}\}$ and $\bN_{\bx}=\{N,\ldots,N\}$ are $d_{\bx}$-tuples, and $N=k+1$ ($k$ is the polynomial degree).  
The multi-index notation is used generically: in addition to $\bi$ (and $\bi_{\bx})$, we will also make use of $\bj$, $\bk$, $\bp$, and $\bq$, together with their corresponding spatial components $\bj_{\bx}$, $\bk_{\bx}$, $\bp_{\bx}$, and $\bq_{\bx}$.  
Further, we define $\tilde{\vect{\imath}}_{\bx}^{j}=\bi_{\bx}-\{i^{j}\}$ to write $\bi_{\bx}=\{\tilde{\vect{\imath}}_{\bx}^{j},i^{j}\}$ $\forall j\in\{1,\ldots,d_{\bx}\}$; for example, with $d_{\bx}=3$ and $j=1$, $\tilde{\vect{\imath}}_{\bx}^{1}=\{i^{2},i^{3}\}$.

\begin{figure}[h]
	\begin{center}
		\includegraphics[width=0.85\linewidth]{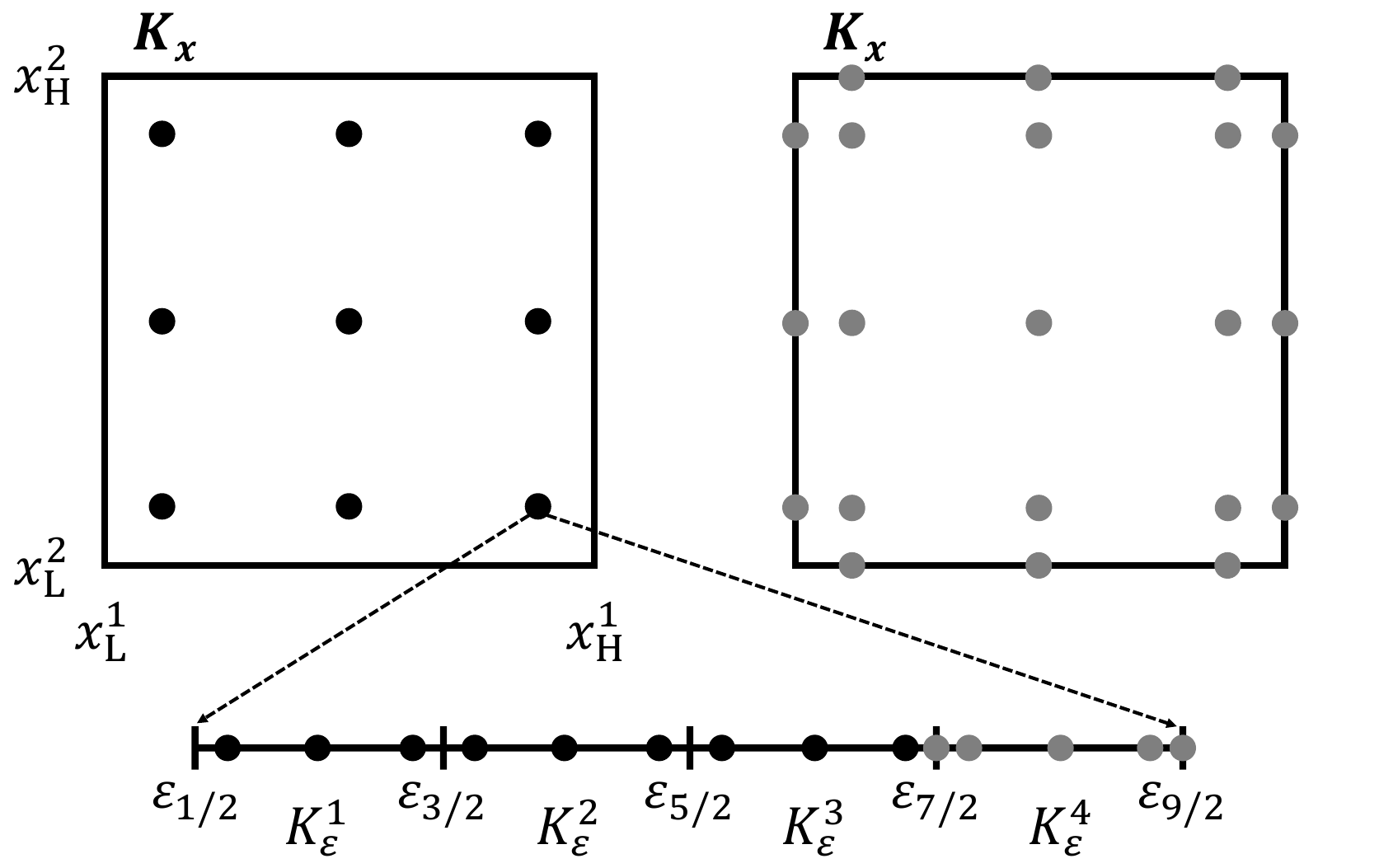}
		\caption{Illustration of key point sets of the nodal collocation DG scheme for the case with $d_{\bx}=2$ for quadratic elements: $N=k+1=3$.  
		The left spatial element shows the LG spatial node set $S_{\bx}$ (black dots).  
		The right spatial element shows the spatial node set $\brmS_{\bx}$ (grey dots).  
		An energy grid $\cT_{\varepsilon}$ is attached to each spatial node $\bx\in S_{\bx}$, as illustrated below the spatial elements for the case with $\nElementsE=4$.  
		For the first three elements, the LG node set $S_{\varepsilon}^{\eidx}$ ($\eidx\in\{1,2,3\}$) is shown (black dots), while the node set $\brmS_{\varepsilon}^{\eidx}$ ($\eidx=4$) is shown for the last element (grey dots).  
		The point set $S_{\bz}^{\eidx}=S_{\bx}\otimes S_{\varepsilon}^{\eidx}$ is used to build the polynomial representation in Equation~\eqref{eq:nodalExpansion}, while the polynomial representation is evaluated in the point set $\brmS_{\bz}^{\eidx}=\brmS_{\varepsilon}^{\eidx}\otimes\brmS_{\varepsilon}^{\eidx}$ for each element $\Kz^{\eidx}$ in order to update the expansion coefficients in Equation~\eqref{eq:nodalExpansion} by the nodal collocation DG scheme.}
        		\label{fig:SpectralElements}
        \end{center}
\end{figure}

On each phase-space element, the moments are approximated with the polynomial expansion
\begin{equation}
	\bcU_{h}(\bz,t)|_{\Kz^{\eidx}} = \sum_{\bi=\vect{1}}^{\bN}\bcU_{\bi}^{\eidx}(t)\,\phi_{\bi}(\bz),
	\label{eq:nodalExpansion}
\end{equation}
where the \emph{nodal} basis functions,
\begin{equation}
	\phi_{\bi}(\bz(\vect{\xi})) = \ell_{i^{\varepsilon}}(\varepsilon(\xi^{\varepsilon}))\times\ell_{i^{1}}(x^{1}(\xi^{1}))\times\ldots\times\ell_{i^{d_{\bx}}}(x^{d_{\bx}}(\xi^{d_{\bx}})),
\end{equation}
are given by Lagrange polynomials of degree $k$, constructed from the set of interpolation points (nodes) $S=\{\xi_{1},\ldots,\xi_{N}\}$,
\begin{equation}
	\ell_{p}(\xi) = \prod_{\substack{q = 1 \\ q \ne p}}^{N}
	\f{\xi-\xi_{q}}{\xi_{p}-\xi_{q}},
	\label{eq:lagrangePolynomial}
\end{equation}
defined on the unit reference interval $I=\{\xi:\xi\in(-0.5,0.5)\}$.  
We take the interpolation points to build the Lagrange polynomials on the reference interval in each phase-space dimension to be the Legendre--Gauss (LG) quadrature points.  
In a phase-space element $\Kz^{\eidx}$, we denote the sets of $N$-point LG quadrature points on $K_{\varepsilon}^{\eidx}$ and $K_{\bx}^{i}$ by $S_{\varepsilon}^{\eidx}=\{\varepsilon_{q}^{\eidx}\}_{q=1}^{N}$ and $S_{\bx}^{i}=\{x_{q}^{i}\}_{q=1}^{N}$, respectively.  
The set of LG interpolation points on a spatial element $\Kx$ is denoted $S_{\bx}=\otimes_{i=1}^{d_{\bx}}S_{\bx}^{i}$, while the set of LG interpolation points on a phase-space element $\Kz^{\eidx}$ is denoted $S_{\bz}^{\eidx}=S_{\varepsilon}^{\eidx}\otimes S_{\bx}$.  
In addition, we denote LG points in the surface elements $\Ax^{i}$ and $\Az^{i,\eidx}$ by $\tilde{S}_{\bx}^{i}=\otimes_{j\ne i}S_{\bx}^{j}$ and $\tilde{S}_{\bz}^{i,\eidx}=\tilde{S}_{\bx}^{i}\otimes S_{\varepsilon}^{\eidx}$, respectively.  
Specifically, $\tilde{S}_{\bx}^{i,\Lo/\Hi}$ denote the quadrature points in the surface elements at $x_{\Lo/\Hi}^{i}$, $\tilde{S}_{\bx}=\{\tilde{S}_{\bx}^{i,\Lo},\tilde{S}_{\bx}^{i,\Hi}\}_{i=1}^{d_{\bx}}$, and $\brmS_{\bx}=\{S_{\bx},\tilde{S}_{\bx}\}$.  
Similarly, $\tilde{S}_{\varepsilon}^{\eidx}=\{\varepsilon_{\eidx-\f{1}{2}},\varepsilon_{\eidx+\f{1}{2}}\}$ and $\brmS_{\varepsilon}^{\eidx}=\{S_{\varepsilon}^{\eidx},\tilde{S}_{\varepsilon}^{\eidx}\}$, and $\brmS_{\bz}^{\eidx}=\brmS_{\bx}\otimes\brmS_{\varepsilon}^{\eidx}$.  
Figure~\ref{fig:SpectralElements} illustrates key point sets of the nodal collocation DG scheme implemented in \thornado.  

Since Lagrange polynomials satisfy $\ell_{p}(\xi_{q})=\delta_{pq}$ for any $\xi_{q}\in S$, the basis functions in the expansion in Equation~\eqref{eq:nodalExpansion} satisfy $\phi_{\bi}(\bz_{\bj})=\delta_{\bi\bj}$ for any $\bz_{\bj}\in S_{\bz}^{\eidx}\subset \Kz^{\eidx}$, and $\bcU_{h}(\bz_{\bj})=\bcU_{\bj}^{\eidx}$.  
That is, the expansion coefficients in Equation~\eqref{eq:nodalExpansion} represent the moments evaluated in the LG points $S_{\bz}^{\eidx}$.  
Neutrino energy coordinates on $K_{\varepsilon}^{\eidx}$ are related to local coordinates by $\varepsilon(\xi^{\varepsilon})=\varepsilon_{\eidx}+\Delta\varepsilon_{\eidx}\,\xi^{\varepsilon}$, where $\varepsilon_{\eidx}=0.5\times(\varepsilon_{\eidx-\f{1}{2}}+\varepsilon_{\eidx+\f{1}{2}})$ and $\Delta \varepsilon_{\eidx}=\varepsilon_{\eidx+\f{1}{2}}-\varepsilon_{\eidx-\f{1}{2}}$.  
Specifically, we denote $\varepsilon_{q}^{\eidx}=\varepsilon(\xi_{q}^{\varepsilon})=\varepsilon_{\eidx}+\Delta\varepsilon_{\eidx}\,\xi_{q}^{\varepsilon}$.  
Similarly, on $K_{\bx}^{i}$, $x^{i}(\xi^{i})=x_{\Center}^{i}+\Delta x^{i}\,\xi^{i}$, where $x_{\Center}^{i}=0.5\times(x_{\Lo}^{i}+x_{\Hi}^{i})$ and $\Delta x^{i}=x_{\Hi}^{i}-x_{\Lo}^{i}$.  

The integrals in Equation~\eqref{eq:kineticCompactWeak} are evaluated with quadrature rules.  
To this end, we denote the weights associated with the $N$-point Gaussian quadrature on the unit interval $I$ by $\{w_{1},\ldots,w_{N}\}$, normalized such that $\sum_{i=1}^{N}w_{i}=1$.  
Multidimensional integrals are evaluated by tensorization of one-dimensional quadratures; e.g., with weights $w_{\bi}=w_{i^{\varepsilon}}\times w_{i^{1}}\times\ldots\times w_{i^{d_{\bx}}}$ for phase-space volume integrals.  
In the spectral-type nodal collocation DG method employed by \thornado, the interpolation points $S_{\bz}^{\eidx}$ are also used as quadrature points for evaluating integrals \citep{pochik_etal_2021}.  
Then, the delta-function property, $\phi_{\bi}(\bz_{\bj})=\delta_{\bi\bj}$, can be exploited to improve the computational efficiency of the DG method.  
For example, when using curvilinear coordinates, the weak form of the DG scheme in Equation~\eqref{eq:kineticCompactWeak} is formulated with the $\varepsilon^{2}\sqrt{\gamma_{h}}$ weight.  
Exact integration for all test functions $\psi_{h}$ will then typically require the number of quadrature points in each dimension to exceed $N$, and the mass matrix, which must be inverted, becomes non-diagonal and spatially dependent.  
With the collocation of interpolation and integration points, at the expense of some integration errors, the mass matrix becomes diagonal and can be trivially inverted.  
This `decoupling' of quadrature points is also beneficial for the neutrino--matter collision solve, because spatial nodes become decoupled and more parallelism is exposed.  
We use the LG quadrature points to define the expansion in Equation~\eqref{eq:nodalExpansion} because the $N$-point Gaussian quadrature is exact for polynomials of degree up to $2N-1$, as opposed to $2N-3$ for the Legendre--Gauss--Lobatto rule, and thus provides better accuracy \citep[see, e.g.,][]{bassi_etal_2013,chan_etal_2019,ruedaRamirez_etal_2023}.  

Using the $N$-point LG quadrature to evaluate integrals, the semi-discrete DG scheme in Equation~\eqref{eq:kineticCompactWeak} for the expansion coefficients $\bcU_{\bp}^{\eidx}$ can be written as
\begin{equation}
	\pd{}{t}\bcU_{\bp}^{\eidx} 
	=\bcB(\bcU_{h},\bu_{h})_{\bp}^{\eidx} + \bcC(\bcU_{h},\bu_{h})_{\bp}^{\eidx},
\end{equation}
where the phase-space advection term is
\begin{equation}
	\bcB(\bcU_{h},\bu_{h})_{\bp}^{\eidx}
	=\bcB_{\bx}(\bcU_{h},\bu_{h})_{\bp}^{\eidx}
	+\bcB_{\varepsilon}(\bcU_{h},\bu_{h})_{\bp}^{\eidx}
	+\bcS(\bcU_{h},\bu_{h};\varepsilon_{p_{\varepsilon}}^{\eidx},\bx_{\bp_{\bx}}),
\end{equation}
with position-space and energy-space divergences given by
\begin{align}
	\bcB_{\bx}(\bcU_{h},\bu_{h})_{\bp}^{\eidx}
	&=-\sum_{i=1}^{d_{\bx}}\f{1}{\sqrt{\gamma_{h}(\bx_{\bp_{\bx}})}\,\Delta x^{i}\,w_{p_{i}}} \nonumber \\
	&\hspace{12pt}\times
	\Big\{\,
		\Big(\,
			\sqrt{\gamma_{h}(x_{\Hi}^{i},\tilde{\bx}_{\tilde{\bp}_{\bx}^{i}}^{i})}\,\widehat{\bcF^{i}}(\bcU_{h},\bu_{h};\varepsilon_{p_{\varepsilon}}^{\eidx},x_{\Hi}^{i},\tilde{\bx}_{\tilde{\bp}_{\bx}^{i}}^{i})\,\ell_{p_{i}}(x_{\Hi}^{i,-}) \nonumber \\
			&\hspace{42pt}
			-\sqrt{\gamma_{h}(x_{\Lo}^{i},\tilde{\bx}_{\tilde{\bp}_{\bx}^{i}}^{i})}\,\widehat{\bcF^{i}}(\bcU_{h},\bu_{h};\varepsilon_{p_{\varepsilon}}^{\eidx},x_{\Lo}^{i},\tilde{\bx}_{\tilde{\bp}_{\bx}^{i}}^{i})\,\ell_{p_{i}}(x_{\Lo}^{i,+})
		\,\Big) \nonumber \\
		&\hspace{32pt}
		-\sum_{q_{i}=1}^{N}w_{q_{i}}\,\sqrt{\gamma_{h}(x_{q_{i}}^{i},\tilde{\bx}_{\tilde{\bp}_{\bx}^{i}}^{i})}\,\bcF^{i}(\varepsilon_{p_{\varepsilon}}^{\eidx},x_{q_{i}}^{i},\tilde{\bx}_{\tilde{\bp}_{\bx}^{i}}^{i})\,\pderiv{\ell_{p_{i}}}{\xi^{i}}(\xi_{q_{i}}^{i})
	\,\Big\}
\end{align}
and
\begin{align}
	\bcB_{\varepsilon}(\bcU_{h},\bu_{h})_{\bp}^{\eidx}
	&=-\f{1}{(\varepsilon_{p_{\varepsilon}}^{\eidx})^{2}\,\Delta\varepsilon_{\eidx}\,w_{p_{\varepsilon}}} \nonumber \\
	&\hspace{12pt}\times
	\Big\{\,
		\Big(\,
			\varepsilon_{\eidx+\f{1}{2}}^{3}\,\widehat{\bcF^{\varepsilon}}(\bcU_{h},\bu_{h};\varepsilon_{\eidx+\f{1}{2}},\bx_{\bp_{\bx}})\,\ell_{p_{\varepsilon}}(\varepsilon_{\eidx+\f{1}{2}}^{-}) \nonumber \\
			&\hspace{42pt}
			-\varepsilon_{\eidx-\f{1}{2}}^{3}\,\widehat{\bcF^{\varepsilon}}(\bcU_{h},\bu_{h};\varepsilon_{\eidx-\f{1}{2}},\bx_{\bp_{\bx}})\,\ell_{p_{\varepsilon}}(\varepsilon_{\eidx-\f{1}{2}}^{+})
		\,\Big) \nonumber \\
		&\hspace{32pt}
		-\sum_{q_{\varepsilon}=1}^{N}
		w_{q_{\varepsilon}}\,(\varepsilon_{q_{\varepsilon}}^{\eidx})^{3}\,\bcF^{\varepsilon}(\bcU_{h},\bu_{h};\varepsilon_{q_{\varepsilon}}^{\eidx},\bx_{\bp_{\bx}})\,\pderiv{\ell_{p_{\varepsilon}}}{\xi^{\varepsilon}}(\xi_{q_{\varepsilon}}^{\varepsilon})
	\,\Big\},
\end{align}
respectively.  

The solution to Subproblem~2 is then approximated by solving
\begin{equation}
	\pd{}{t}\bcU_{\bp}^{\eidx} 
	=\bcB(\bcU_{h},\bu_{h})_{\bp}^{\eidx}
	\label{eq:subproblem2Nodal}
\end{equation}
for all $\bz_{\bp}\in S_{\bz}^{\eidx}\subset\Kz^{\eidx}$ and all $\Kz^{\eidx}\in\cT_{\bz}=\cT_{\bx}\otimes\cT_{\varepsilon}$.  

We postpone the details on the nodal collision term, $\bcC(\bcU_{h},\bu_{h})_{\bp}^{\eidx}$, until Section~\ref{sec:neutrinoMatterCoupling}.  
However, we note that when coupling the fluid and neutrino moments, the collision invariants are enforced point-wise for each $\bx_{\bp_{\bx}}\in\Kx$; i.e., the solution to Subproblem~3 is approximated by solving
\begin{subequations}\label{eq:subproblem3Nodal}
	\begin{align}
		\pd{}{t}\bcU_{\bp}^{\eidx} 
		&=\bcC(\bcU_{h},\bu_{h})_{\bp}^{\eidx}, \label{eq:subproblem3aNodal} \\
		\pd{\vect{\mathsf{U}}(\bcU_{h},\bu_{h})_{\bp_{\bx}}}{t} 
		&= 0,
		\label{eq:subproblem3bNodal}
	\end{align}
\end{subequations}
for all $\bz_{\bp}=\{\varepsilon_{p^{\varepsilon}},\bx_{\bp_{\bx}}\}\in S_{\bz}^{\eidx}$ and all $\Kz^{\eidx}\in\cT_{\bz}$.  

When evaluating collision operators in the nodal DG scheme, it is useful to consider a single spatial point in isolation.  
Evaluating Equation~\eqref{eq:nodalExpansion} in $\bx_{\bp_{\bx}}\in S_{\bx}$ gives
\begin{equation}
	\bcU_{h}(\varepsilon,\bx_{\bp_{\bx}})|_{\Kz^{\eidx}}
	= \sum_{q^{\varepsilon}=1}^{N}\sum_{\bq_{\bx}=\vect{1}}^{\bN_{\bx}}\bcU_{\{q^{\varepsilon},\bq_{\bx}\}}^{\eidx}\phi_{\{q^{\varepsilon},\bq_{\bx}\}}(\varepsilon,\bx_{\bp_{\bx}})
	= \sum_{q^{\varepsilon}=1}^{N}\bcU_{\{q^{\varepsilon},\bp_{\bx}\}}^{\eidx}\ell_{q^{\varepsilon}}(\varepsilon).  
	\label{eq:nodalExpansionEnergy}
\end{equation}
That is, in each spatial interpolation point, the expansion in Equation~\eqref{eq:nodalExpansion} reduces to a polynomial solely in the energy dimension.  
Integration over the entire energy domain $D_{\varepsilon}$ results in
\begin{equation}
	\int_{D_{\varepsilon}}\bcU_{h}(\varepsilon,\bx_{\bp_{\bx}})\,dV_{\varepsilon}
	=\f{4\pi}{h^{3}}\sum_{\eidx=1}^{\nElementsE}\sum_{q^{\varepsilon}=1}^{N}w_{q^{\varepsilon}}\,(\varepsilon_{q^{\varepsilon}}^{\eidx})^{2}\,\Delta\varepsilon_{\eidx}\,\bcU_{\{q^{\varepsilon},\bp_{\bx}\}}^{\eidx}
	=\sum_{\qidx=1}^{N^{\varepsilon}}W_{\qidx}^{(2)}\,\bcU_{\{\qidx,\bp_{\bx}\}},
	\label{eq:energyIntegral}
\end{equation}
where we have defined the global energy integration weights
\begin{equation}
	W_{\qidx}^{(2)}=4\pi\,w_{q^{\varepsilon}}\,(\varepsilon_{q^{\varepsilon}}^{\eidx})^{2}\,\Delta\varepsilon_{\eidx}/h^{3}, 
	\label{eq:energy_integration_weights}
\end{equation}
the expansion coefficients $\bcU_{\{\qidx,\bp_{\bx}\}}\vcentcolon=\bcU_{\{q^{\varepsilon},\bp_{\bx}\}}^{\eidx}$, and where the global energy index is given by $\qidx=(\eidx-1)\times N+q^{\varepsilon}$, which runs from $1$ to $N^{\varepsilon}=\nElementsE\times N$ for $\eidx\in\{1,\ldots,\nElementsE\}$ and $q^{\varepsilon}\in\{1,\ldots,N\}$.  
This global energy index notation will be useful when discretizing collision operators.  
With $N=(k+1)$-point LG quadrature, the integral in Equation~\eqref{eq:energyIntegral}, a polynomial of degree $k+2$, is exact for $k\ge1$.  
We will also integrate functions against higher powers in neutrino energy with the $N$-point LG quadrature.  
To this end, we define the integration weights $W_{\qidx}^{(2+p)}=W_{\qidx}^{(2)}\times(\varepsilon_{\qidx})^{p}$.  
When using these weights, integration against polynomials of degree $k$ may no longer be exact.  

\subsection{Time Integration}
\label{sec:timeIntegration}

We use IMEX methods \citep[e.g.,][]{ascher_etal_1997,pareschiRusso_2005} to integrate Subproblem~2 and Subproblem~3 in a coupled fashion, using the semi-discrete approximations given by Equation~\eqref{eq:subproblem2Nodal} and Equation~\eqref{eq:subproblem3Nodal}, respectively.  
Then, in one time step $\dt=t^{n+1}-t^{n}$, the solution is advanced from $\{\bcU_{\bp}^{\eidx,n},\bu_{\bp_{\bx}}^{n}\}$ to $\{\bcU_{\bp}^{\eidx,n+1},\bu_{\bp_{\bx}}^{n+1}\}$, with an $s$-stage IMEX-RK method as follows: 
For all $\bz_{\bp}=\{\varepsilon_{p^{\varepsilon}},\bx_{\bp_{\bx}}\}\in S_{\bz}^{\eidx}\subset\Kz^{\eidx}$ and all $\Kz^{\eidx}\in\cT_{\bz}$:
\begin{itemize}
	\item[1.] For $i=1,\ldots,s$ solve the coupled nonlinear system for $\widetilde{\bcU}_{h}^{(i)}$ and $\tilde{\bu}_{h}^{(i)}$
	\begin{subequations}\label{eq:imexImplicit}
		\begin{align}
			\widetilde{\bcU}_{\bp}^{\eidx,(i)} 
			&= \bcU_{\bp}^{\eidx,(i\star)} + \alpha_{ii}\,\dt\,\bcC(\widetilde{\bcU}_{h}^{(i)},\widetilde{\bu}_{h}^{(i)})_{\bp}^{\eidx}, \label{eq:imexImplicitMoments} \\
			\vect{\mathsf{U}}(\widetilde{\bcU}_{h}^{(i)},\widetilde{\bu}_{h}^{(i)})_{\bp_{\bx}} 
			&= \vect{\mathsf{U}}(\bcU_{h}^{(i\star)},\bu_{h}^{(i\star)})_{\bp_{\bx}}, \label{eq:imexImplicitFluid}
		\end{align}
	\end{subequations}
	set $\big[\,\delta\bu/\delta t\,\big]_{\bp_{\bx}}^{(i)}\vcentcolon=(\widetilde{\bu}_{\bp_{\bx}}^{(i)}-\bu_{\bp_{\bx}}^{(i\star)})/(\alpha_{ii}\dt)$, and apply limiters\footnote{Tildes indicate pre-limiter states in the IMEX scheme.}
	\begin{equation}
		\bcU_{h}^{(i)}\vcentcolon=\Lambda^{\bcU}\big\{\widetilde{\bcU}_{h}^{(i)}\big\}
		\quad\text{and}\quad
		\bu_{h}^{(i)}\vcentcolon=\Lambda^{\bu}\big\{\widetilde{\bu}_{h}^{(i)}\big\}.  
	\end{equation}
	In Equation~\eqref{eq:imexImplicit}, we have defined $\bcU_{h}^{(i\star)}\vcentcolon=\Lambda^{\bcU}\big\{\widetilde{\bcU}_{h}^{(i\star)}\big\}$ and $\bu_{h}^{(i\star)}\vcentcolon=\Lambda^{\bu}\big\{\widetilde{\bu}_{h}^{(i\star)}\big\}$, where
	\begin{subequations}
		\begin{align}
			\widetilde{\bcU}_{\bp}^{\eidx,(i\star)} 
			&= \bcU_{\bp}^{\eidx,n} + \dt\sum_{j=1}^{i-1}\Big\{\,\tilde{\alpha}_{ij}\,\bcB(\bcU_{h}^{(j)},\bu_{h}^{(j)})_{\bp}^{\eidx}+\alpha_{ij}\,\bcC(\widetilde{\bcU}_{h}^{(j)},\widetilde{\bu}_{h}^{(j)})_{\bp}^{\eidx}\,\Big\}, \\
			\widetilde{\bu}_{\bp_{\bx}}^{(i\star)}
			&=\bu_{\bp_{\bx}}^{n}+\dt\sum_{j=1}^{i-1}\alpha_{ij}\,\big[\,\delta\bu/\delta t\,\big]_{\bp_{\bx}}^{(j)}.
		\end{align}
	\end{subequations}
	\item[2.] Assemble
	\begin{subequations}\label{eq:imexAssembly}
		\begin{align}
			\widetilde{\bcU}_{\bp}^{\eidx,n+1}
			&=\bcU_{\bp}^{\eidx,n} + \dt\sum_{i=1}^{s}\Big\{\,\tilde{\beta}_{i}\,\bcB(\bcU_{h}^{(i)},\bu_{h}^{(i)})_{\bp}^{\eidx}+\beta_{i}\,\bcC(\widetilde{\bcU}_{h}^{(i)},\widetilde{\bu}_{h}^{(i)})_{\bp}^{\eidx}\,\Big\}, \label{eq:imexAssemblyMoments} \\
			\widetilde{\bu}_{\bp_{\bx}}^{n+1}
			&=\bu_{\bp_{\bx}}^{n}+\dt\sum_{i=1}^{s}\beta_{i}\,\big[\,\delta\bu/\delta t\,\big]_{\bp_{\bx}}^{(i)}, \label{eq:imexAssemblyFluid}
		\end{align}
	\end{subequations}
	and apply limiters $\bcU_{h}^{n+1}\vcentcolon=\Lambda^{\bcU}\big\{\widetilde{\bcU}_{h}^{n+1}\big\}$ and $\bu_{h}^{n+1}\vcentcolon=\Lambda^{\bu}\big\{\widetilde{\bu}_{h}^{n+1}\big\}$.  
\end{itemize}
Here $\tilde{\alpha}_{ij}$ and $\alpha_{ij}$ are coefficients of $s\times s$ matrices $\tilde{A}=(\tilde{\alpha}_{ij})$ and $A=(\alpha_{ij})$, respectively, and $\tilde{\vect{b}}=(\tilde{\beta}_{1},\ldots,\tilde{\beta}_{s})^{\intercal}$ and $\vect{b}=(\beta_{1},\ldots,\beta_{s})^{\intercal}$ are vectors.  
The coefficient matrix $\tilde{A}$ is strictly lower diagonal, $\tilde{\alpha}_{ij}=0$ for $i\ge j$; therefore, the phase-space advection term is treated explicitly.  
The matrix $A$ is lower diagonal, $\alpha_{ij}=0$ for $i>j$; therefore, the collision operator is treated implicitly.  
The lower diagonal structure of $A$ gives rise to the family of diagonally implicit RK schemes and allows the implicit solves to be written as modified backward Euler updates, as in Equation~\eqref{eq:imexImplicitMoments}, with initial state $\bcU_{\bp}^{\eidx,(i\star)}$ and time step $\alpha_{ii}\dt$.  

The operators $\Lambda^{\bcU}\big\{\ldots\big\}$ and $\Lambda^{\bu}\big\{\ldots\big\}$ denote limiters applied to the radiation moments and fluid fields, respectively.  
For the fluid fields, one can apply the limiters described in \citet{pochik_etal_2021} to prevent oscillations and non-physical states.  
For the radiation moments, similar limiters can be applied, which we summarize in more detail in the next subsection.  

In the IMEX scheme, to avoid application of limiters during the iterative solution procedure, the implicit operator, $\bcC()_{\bp}^{\eidx}$, is evaluated with unlimited stage values, while the explicit operator, $\bcB()_{\bp}^{\eidx}$, is evaluated with limited stage values.  

The time step for the IMEX scheme is set by stability conditions on the explicit part \citep[e.g.,][]{cockburnShu_2001}
\begin{equation}
	\dt = \f{1}{d_{\bx}(2k+1)}\min_{i\in\{1,\ldots,d_{\bx}\}}(\Delta x^{i}\,h_{i}) \quad(\mbox{no summation over $i$}).
	\label{eq:cflConditionTransport}
\end{equation}
($c=1$ in our units; $h_{i}$ are the spatial scale factors.)

The coefficients that make up the matrices $\tilde{A}/A$ and vectors $\tilde{\vect{b}}/\vect{b}$ must satisfy certain order conditions for the desired order of accuracy, and there is a large body of literature devoted to IMEX methods for kinetic equations \citep[see, e.g.,][and references therein]{dimarcoPareschi_2014}.  
Here, for reasons delineated in \citet{chu_etal_2019}, we use the two-stage IMEX scheme dubbed PD-ARS, characterized by the Butcher tableau
\begin{equation}
  \begin{array}{c | c}
    \tilde{\vect{c}} & \tilde{A} \\
    \hline
    & \tilde{\vect{b}}^{\intercal}
  \end{array}
  =
  \begin{array}{c | ccc}
    0 & 0 & 0 & 0 \\
    1 & 1 & 0 & 0 \\
    1 & 1/2 & 1/2 & 0 \\
    \hline
    & 1/2 & 1/2 & 0 
  \end{array}
  \quad\text{and}\quad
  \begin{array}{c | c}
    \vect{c} & A \\
    \hline
    & \vect{b}^{\intercal}
  \end{array}
  =
  \begin{array}{c | ccc}
    0 & 0 & 0 & 0 \\
    1 & 0 & 1 & 0 \\
    1 & 0 & 1/2 & 1/2 \\
    \hline
    & 0 & 1/2 & 1/2 
  \end{array},
  \label{eq:butcherPDARS}
\end{equation}
where the coefficients $\tilde{c}_{i}=\sum_{j=1}^{i-1}\tilde{\alpha}_{ij}$ and $c_{i}=\sum_{j=1}^{i}\alpha_{ij}$ are needed only for non-autonomous systems.  
This IMEX scheme has two stages (two implicit solves per time step), and is globally stiffly accurate, $\tilde{\beta}_{i}=\tilde{\alpha}_{si}$ and $\beta_{i}=\alpha_{si}$; therefore, the assembly step in Equation~\eqref{eq:imexAssembly} is not needed.  

For some tests in Section~\ref{sec:results}, we use the optimal second-order or third-order strong stability-preserving (SSP) RK method of \citet{shuOsher_1988}.  
In this case, all the components of the implicit Butcher tableau ($A$, $\vect{c}$, and $\vect{b}$) are set to zero, while the explicit Butcher tableau are set to
\begin{equation}
	\begin{array}{c | c}
    \tilde{\vect{c}} & \tilde{A} \\
    \hline
    & \tilde{\vect{b}}^{\intercal}
  \end{array}
  =
  \begin{array}{c | cc}
    0 & 0 & 0 \\
    1 & 1 & 0 \\
    \hline
    & 1/2 & 1/2
  \end{array}
  \quad\text{and}\quad
  \begin{array}{c | c}
    \tilde{\vect{c}} & \tilde{A} \\
    \hline
    & \tilde{\vect{b}}^{\intercal}
  \end{array}
  =
  \begin{array}{c | ccc}
    0 & 0 & 0 & 0 \\
    1 & 1 & 0 & 0 \\
    1/2 & 1/4 & 1/4 & 0 \\
    \hline
    & 1/6 & 1/6 & 2/3 
  \end{array},
\end{equation}
for SSP-RK2 and SSP-RK3, respectively.  

\subsection{Limiters}
\label{sec:limiters}

Limiters can be invoked to modify the polynomial representation of the radiation moments to reduce oscillations near discontinuities and to prevent nonphysical states.  
In \thornado, we have implemented the simple minmod-type limiter \citep[][]{cockburnShu_2001} to reduce oscillations, and the realizability-enforcing limiter \citep{chu_etal_2019} to ensure that the moments satisfy certain constraints in a discrete set of points on each element.  
We briefly describe these procedures next.  

\subsubsection{Slope Limiter}

In \thornado, slope limiting of radiation moments is applied only in the spatial dimensions.  
Evaluating Equation~\eqref{eq:nodalExpansion} in $\varepsilon_{i^{\varepsilon}}^{\eidx}\in S_{\varepsilon}^{\eidx}$ gives the polynomial representation for $\cU_{h}$ (a component of $\bcU_{h}$) on the spatial element $\Kx$
\begin{equation}
	\cU_{h}(\varepsilon_{i^{\varepsilon}},\bx,t)|_{\Kx} = \sum_{\bi_{\bx}=\bOne}^{\bN_{\bx}}\cU_{\{i^{\varepsilon},\bi_{\bx}\}}^{\eidx}(t)\,\phi_{\bi_{\bx}}(\bx).
	\label{eq:nodalExpansionSlopeLimiter}
\end{equation}
Since polynomial representations associated with different energy nodes are limited independently, we drop the energy index for the remainder of this subsection; i.e., $\cU_{h}(\varepsilon_{i^{\varepsilon}},\bx,t)|_{\Kx}\to\cU_{h}(\bx,t)|_{\Kx}$ and $\cU_{\{i^{\varepsilon},\bi_{\bx}\}}^{\eidx}\to\cU_{\bi_{\bx}}$.  

We employ a standard minmod-type limiter \citep[e.g.,][]{cockburnShu_1998} on the evolved moments.  
To this end, we define the linear approximation $\cU_{h}^{1}\in\bbP^{1}(\Kx)$, 
\begin{equation}
	\cU_{h}^{1}(\bx,t)|_{\Kx} \vcentcolon= \sum_{\bi_{\bx}=\bZero}^{|\bi_{\bx}|\le 1}\cC_{\bi_{\bx}}(t)\,P_{\bi_{\bx}}(\bx),
	\label{eq:modalExpansionP1}
\end{equation}
where $P_{\bi_{\bx}}(\bx(\vect{\xi}))=P_{i^{1}}(\xi^{1})\times\ldots\times P_{i^{d_{\bx}}}(\xi^{d_{\bx}})$, and where $P_{0}(\xi)=1$ and $P_{1}(\xi)=\xi$ are defined on the unit reference interval $I$, defined below Equation~\eqref{eq:lagrangePolynomial}.  
Here, $\bbP^{1}(\Kx)$ is the space of polynomials of total degree less than or equal to $1$.  
The expansion coefficients in Equation~\eqref{eq:modalExpansionP1} are obtained by a projection of $\cU_{h}$ onto $\bbP^{1}$,
\begin{equation}
	\cC_{\bi_{\bx}} = \f{\int_{\Kx}\cU_{h}(\bx)\,P_{\bi_{\bx}}(\bx)\,d\bx}{\int_{\Kx}P_{\bi_{\bx}}(\bx)\,P_{\bi_{\bx}}(\bx)\,d\bx}.  
\end{equation}
Note that $\cC_{\bZero}$ is the cell average of $\cU_{h}$, while $\cC_{\{1,0,0\}}$, $\cC_{\{0,1,0\}}$, and $\cC_{\{0,0,1\}}$ are cell-averaged first derivatives (slopes) of $\cU_{h}^{1}$, with respect to $x^{1}$, $x^{2}$, and $x^{3}$, respectively.  
Limited slopes are computed with the minmod function; e.g., in the $x^{1}$-dimension
\begin{equation}
	\widetilde{\cC}_{\{1,0,0\}}
	=\mbox{minmod}\big(\,\cC_{\{1,0,0\}},\,\beta_{\mbox{\tiny TVD}}\,(\cC_{\bZero}^{+}-\cC_{\bZero}),\,\beta_{\mbox{\tiny TVD}}\,(\cC_{\bZero}-\cC_{\bZero}^{-})\,\big),
	\label{eq:slopeLimiting}
\end{equation}
where 
\begin{equation}
	\mbox{minmod}(a,b,c) = 
	\left\{\begin{array}{cl}
		s\times\min(|a|,|b|,|c|), & \text{ if } s=\mbox{sign}(a)=\mbox{sign}(b)=\mbox{sign}(c) \\
		0 & \text{otherwise}
	\end{array}\right.
\end{equation}
and $\beta_{\mbox{\tiny TVD}}\in[1,2]$, and where $\cC_{\bZero}^{+}$ and $\cC_{\bZero}^{-}$ are cell averages in the next and previous element relative to $\Kx$ in the $x^{1}$-dimension, respectively.  
Then, if $\max_{|\bi_{\bx}|=1}(|\widetilde{\cC}_{\bi_{\bx}}-\cC_{\bi_{\bx}}|)>\texttt{tol}\,|\cC_{\bZero}|$\footnote{The small tolerance is introduced to avoid unnecessary re-evaluation of expansion coefficients when the limited and unlimited slopes are nearly identical.}, the expansion coefficients in Equation~\eqref{eq:nodalExpansionSlopeLimiter} are replaced with
\begin{equation}
	\widetilde{\cU}_{\bi_{\bx}}^{1}=\sum_{\bk_{\bx}=\bZero}^{|\bk_{\bx}|\le1}\widetilde{\cC}_{\bk_{\bx}}\,P_{\bk_{\bx}}(\bx_{\bi_{\bx}}),
\end{equation}
where $\widetilde{\cC}_{\bZero}=\cC_{\bZero}$.  
Finally, if limiting is done, the polynomial is scaled by a factor $\alpha=\cU_{\Kx}/\widetilde{\cU}_{\Kx}^{1}$, where $\cU_{\Kx}$ and $\widetilde{\cU}_{\Kx}^{1}$ are, respectively, cell averages of the original and limited polynomials with respect to the weight $\sqrt{\gamma_{h}}$.  
The scaling with factor $\alpha$ is included to maintain conservation during limiting when using curvilinear coordinates \citep{pochik_etal_2021}.  
By default, we set the tolerance to trigger limiting to $\texttt{tol}=10^{-6}$.  

For simplicity, limiting in Equation~\eqref{eq:slopeLimiting} is applied separately to each component of $\bcU_{h}$ (componentwise), rather than to the preferred characteristic fields \citep[e.g.][]{cockburn_etal_1989}.  
It is well known that the minmod limiter reduces the accuracy to first order at smooth extrema.  
To avoid excessive limiting, we use the troubled-cell indicator (TCI) of \citet{fuShu_2017}, which flags cells in the vicinity of steep gradients in selected variables.  
\citep[See][for the use of this TCI in \thornado's hydrodynamics.]{pochik_etal_2021}
For the two-moment solver, only the Eulerian-frame number density is used to detect troubled cells.  
We mention that $\beta_{\mbox{\tiny TVD}}=2$ in Equation~\eqref{eq:slopeLimiting} results in the \emph{double minmod} limiter, which was analyzed by \citet{mcclarrenLowrie_2008} in a one-dimensional setting and shown to preserve the asymptotic diffusion limit.  
For the multidimensional case, \citet{adams_2001} showed that the local approximation must (at least) belong to $\bbQ^{1}(\Kx)$ (and \emph{not} $\bbP^{1}(\Kx)$) to preserve the diffusion limit.  
For this reason, the TCI, which in our experience is effective at preventing limiting in the diffusion limit, is essential in order to preserve accuracy in this regime.  
Most of the test cases in Sections~\ref{sec:results} and \ref{sec:application_idealized}, and the CCSN simulations in Section~\ref{sec:application_ccsn}, were run without slope limiting, using only the realizability-enforcing limiter presented in the next section.  

\subsubsection{Realizability-Enforcing Limiter}

The realizability-enforcing limiter described here \citep[see also][]{chu_etal_2019,laiu_etal_2025} is essential to maintaining stability during time integration of the two-moment model.  
When using higher-order elements (i.e., $k\ge1$), the polynomial representation in Equation~\eqref{eq:nodalExpansion} can evaluate to non-realizable (i.e., non-physical) moments within certain elements $\Kz^{\eidx}$; specifically for $\bz\in\brmS_{\bz}^{\eidx}\subset\Kz^{\eidx}$ (Figure~\ref{fig:SpectralElements}).  
When this happens, the two-moment closure problem becomes ill posed, and the overall solution algorithm will eventually fail.  
The realizability-enforcing limiter aims to prevent this.  
Since we consider the Minerbo closure, based on Maxwell--Boltzmann statistics, we take the set of admissible kinetic distributions to be\footnote{$\mathfrak{R}$ is also the appropriate admissible set for the Levermore and Kershaw closures listed in Appendix~\ref{app:closures}}
\begin{equation}
	\mathfrak{R} = \big\{\, f ~ | ~ f\ge 0 ~ \text{and} ~ \aint{f} > 0 \,\big\}.
	\label{eq:admissibleDistributions}
\end{equation}
Then, the moments $\bcU$, defined in Equation~\eqref{eq:conservedToPrimitive}, are realizable if they can be obtained from a kinetic distribution $f\in\mathfrak{R}$.  
The set of all realizable moments is defined by the following convex domain
\begin{equation}
	\cR = \big\{\,\bcU=(\cN,\bcG)^{\intercal} ~|~ \cN > 0 ~ \text{and} ~ \Gamma(\bcU,\bsgamma) \ge 0 \,\big\},
	\label{eq:realizableSet}
\end{equation}
where $\Gamma(\bcU,\bsgamma)=\cN-|\bcG|$, $\cN=\cD+v^{i}\cI_{i}$, and $\cG_{j}=\cI_{j}+v^{i}\cK_{ij}$, and where the norm evaluated using the spatial metric $\bsgamma$, i.e., $|\bcG|=\sqrt{\gamma_{ij}\cG^{i}\cG^{j}}$.  
(To simplify notation we omit the species subscript $s$ in this section.)  
We emphasize that for particles obeying Bose--Einstein statistics, the admissible set of distributions in Equation~\eqref{eq:admissibleDistributions} and the realizable set in Equation~\eqref{eq:realizableSet} remain unchanged, and the realizability-enforcing limiter remains applicable.  
For particles obeying Fermi--Dirac statistics, both $\mathfrak{R}$ and $\cR$ change \citep[e.g.,][]{lareckiBanach_2011}, and modifications along the lines proposed by \citet{chu_etal_2019} for the static case must be considered.  
For reasons delineated in Section~\ref{sec:model}, we only consider distributions in $\mathfrak{R}$ as defined in Equation~\eqref{eq:admissibleDistributions}.  

Since the realizability-enforcing limiter is completely local to each phase-space element, we provide the description for a specific element $\Kz^{\eidx}$.  
The aim is to ensure that the polynomial representation $\bcU_{h}|_{\Kz^{\eidx}}$ is realizable at each point $\bz_{\bq}\in\brmS_{\bz}^{\eidx}$.  
Let the cell average of $\bcU_{h}=(\cN_{h},\bcG_{h})^{\intercal}$ on $\Kz^{\eidx}$ be denoted $\bcU_{\Kz^{\eidx}}=(\bcU_{h})_{\Kz^{\eidx}}/V_{\Kz^{\eidx}}=(\cN_{\Kz^{\eidx}},\bcG_{\Kz^{\eidx}})^{\intercal}$.  
Assume that $\bcU_{\Kz^{\eidx}}\in\cR$; i.e., $\cN_{\Kz^{\eidx}}>0$ and
\begin{equation}
	\Gamma\big(\bcU_{\Kz^{\eidx}},\bsgamma(\bx)\big) \ge 0
	\label{eq:realizabilityCondition2}
\end{equation}
holds for all $\bx\in\brmS_{\bx}$.  
Then the limiter consists of the following two steps:
\begin{itemize}
	\item {\bf Step~1:} Ensure positive spectral number density by
	\begin{equation}
		\widetilde{\cN}_{h} \vcentcolon= (1-\vartheta_{1})\,\cN_{\Kz^{\eidx}} + \vartheta_{1}\,\cN_{h},
		\label{eq:RealizabilityLimiterStep1}
	\end{equation}
	where
	\begin{equation}
		\vartheta_{1} = \min\Big\{\,1,\,\f{\cN_{\Kz^{\eidx}}}{\cN_{\Kz^{\eidx}}-\min_{\bz\in\brmS_{\bz}^{\eidx}}\cN_{h}(\bz)}\,\Big\}.  
		\label{eq:theta1}
	\end{equation}
	Set $\widetilde{\bcU}_{h}\vcentcolon=(\widetilde{\cN}_{h},\bcG_{h})^{\intercal}$.  
	\item {\bf Step~2:} Ensure bounded number flux density by
	\begin{equation}
		\widehat{\bcU}_{h} \vcentcolon= (1-\vartheta_{2})\,\bcU_{\Kz^{\eidx}} + \vartheta_{2}\,\widetilde{\bcU}_{h},
		\label{eq:RealizabilityLimiterStep2}
	\end{equation}
	where
	\begin{equation}
		\vartheta_{2}
		={\text{arg}\,\min}_{\vartheta}
		\Big\{\,
			\vartheta\in[0,1] ~ : ~ \Gamma\big((1-\vartheta)\,\bcU_{\Kz^{\eidx}} + \vartheta\,\widetilde{\bcU}_{h}(\bz),\bsgamma(\bx)\big) \ge 0, \, \forall \bz=(\varepsilon,\bx)\in \brmS_{\bz}^{\eidx}
		\,\Big\}.
		\label{eq:theta2}
	\end{equation}
	Set $\bcU_{h}\vcentcolon=\widehat{\bcU}_{h}$.  
\end{itemize}

In Equations~\eqref{eq:RealizabilityLimiterStep1} and \eqref{eq:RealizabilityLimiterStep2}, the limiter parameters $\vartheta_{1},\vartheta_{2}\in[0,1]$.  
Then if the original polynomial is realizable in all points $\bz\in\brmS_{\bz}^{\eidx}$, $\vartheta_{1}=\vartheta_{2}=1$, the limiter steps do nothing, and $\widehat{\bcU}_{h}=\bcU_{h}$.  
In the worst case, $\vartheta_{1}=\vartheta_{2}=0$, the polynomial representation in an element $\Kz^{\eidx}$ is replaced by the cell averages; i.e., $\widehat{\bcU}_{h}=\bcU_{\Kz^{\eidx}}$.  
Note that both limiter steps preserve the cell average.  

The realizability-enforcing limiter assumes that the cell average is realizable.  
However, this is in general not guaranteed for the nodal collocation DG scheme implemented in \thornado.  
\citet{laiu_etal_2025} designed a realizability-preserving method for the case with Cartesian coordinates and idealized collision terms.  
The construction of a corresponding provably realizability-preserving method for the more general setting with curvilinear spatial coordinates and full neutrino--matter coupling is beyond the scope of this work.  
Nevertheless, components such as the numerical fluxes in Equations~\eqref{eq:numericalFluxSpatial} and \eqref{eq:numericalFluxEnergy}, motivated by the analysis in \citet{laiu_etal_2025}, have been found empirically to improve the robustness of the scheme when applied to this more general case.  
In the event that $\cN_{\Kz^{\eidx}}\le0$, we set $\cN_{h}|_{\Kz^{\eidx}}=\delta_{\cN}$ (small positive number) and $\bcG_{h}|_{\Kz^{\eidx}}=\bZero$.  
In the event that $\cN_{\Kz^{\eidx}}>0$, but Equation~\eqref{eq:realizabilityCondition2} is not satisfied for all $\bx\in\brmS_{\bx}$, we set $\cN_{h}|_{\Kz^{\eidx}}=\cN_{\Kz^{\eidx}}$ and $\bcG_{h}|_{\Kz^{\eidx}}=(1-\delta_{\bcG})\,\cN_{\Kz^{\eidx}}\,(\bcG_{\Kz^{\eidx}}/\max_{\bz\in\brmS_{\bz}^{\eidx}}|\bcG_{h}(\bz)|)$, where $\delta_{\bcG}>0$ is a small, user-specified number.  
In both of these scenarios, the polynomial representation is replaced by constants within the element.  
In the first case, particles are added to the system, while in the second case, the magnitude of the number flux is reduced relative to the number density.  
Importantly, we did not observe the need to invoke these ad hoc procedures in the numerical examples presented in Sections~\ref{sec:results}, \ref{sec:application_idealized}, and \ref{sec:application_ccsn}.  

Provided the first of the aforementioned ad hoc procedures is not invoked, the Eulerian-frame number density is preserved by the realizability-enforcing limiter.  
However, the Eulerian-frame energy, $\cE=\varepsilon\,(\cN+v^{i}\cG_{i})+\cO(v^{2})$, is changed.  
To promote both Eulerian-frame number and energy conservation of the realizability-enforcing limiter, we invoke the spectral redistribution algorithm proposed by \citet{laiu_etal_2025} after completion of the second step.  

\subsection{Primitive Moments Recovery}
\label{sec:conToPrim}

The number conservative two-moment model evolves the \emph{conserved} moments $\bcU=(\cN,\cG_{j})^{\intercal}$, which are related to the \emph{primitive} moments $\bcM=(\cD,\cI_{j})^{\intercal}$ as stated by Equation~\eqref{eq:conservedToPrimitive}.  
(To simplify notation we omit the species subscript $s$ in this section.)  
For given $\bcU$, three-velocity and spatial metric components, $v^{i}$ and $\gamma_{ij}$, respectively, the recovery of $\bcM$ requires the solution of a nonlinear system.  
To this end, we use the approach proposed and analyzed by \citet{laiu_etal_2025} to write Equation~\eqref{eq:conservedToPrimitive} as
\begin{equation}
	\left[\begin{array}{c}
		\cD \\
		\cI_{j}
	\end{array}\right]
	=
	\left[\begin{array}{c}
		\cN - v^{i}\cI_{i} \\
		\cG_{j} - v^{i}\cK_{ij}(\bcM)
	\end{array}\right].
\end{equation}
Multiplying by $\lambda\in(0,1]$ and rearranging we can write this as
\begin{equation}
	\bcM = (1-\lambda)\,\bcM
	+\lambda\,
	\left[\begin{array}{c}
		\cN - v^{i}\cI_{i} \\
		\cG_{j} - v^{i}\cK_{ij}(\bcM)
	\end{array}\right]
	\equiv
	\bsfg(\bcM,\vect{v}).
	\label{eq:c2p_FP}
\end{equation}
We use this mapping to solve for the primitive moments with the fixed-point iteration sequence
\begin{equation}
	\bcM^{[k+1]} = \bsfg(\bcM^{[k]},\vect{v}), \quad k=0,\ldots,
	\label{eq:c2p_sequence}
\end{equation}
where the initial guess is taken as $\bcM^{[0]}=\bcU$.  
The sequence in Equation~\eqref{eq:c2p_sequence} is iterated until the difference between successive iterates satisfies
\begin{equation}
	\|\bcM^{[k+1]}-\bcM^{[k]}\|\le\texttt{tol}\,\|\bcU\|,
\end{equation}
where $\|\cdot\|$ is the Euclidean norm and $\texttt{tol}$ is a relative tolerance.  
(By default, we set $\texttt{tol}=10^{-8}$).  

The sequence in Equation~\eqref{eq:c2p_sequence} can be viewed as a pseudo time-stepping approach with step size $\lambda$.  
Typically, a smaller value for $\lambda$ requires more iterations to reach convergence but provides a means to improve stability and robustness.  
Indeed, the analysis by \citet{laiu_etal_2025} revealed that realizability of successive iterates is preserved with the choice $\lambda=1/(1+|\vect{v}|)$, which is the value we use.  
Moreover, when $|\vect{v}|<\sqrt{2}-1\approx0.41$, $\bsfg$ is a contraction operator, which guarantees the convergence of Equation~\eqref{eq:c2p_sequence}.  
This restriction is acceptable when considering the applicability of the $\cO(v)$ two-moment model.  
We employ Anderson acceleration to Equation~\eqref{eq:c2p_sequence} to improve the rate of convergence \citep[see][for further details]{laiu_etal_2025}.  

\subsection{Neutrino--matter Coupling}
\label{sec:neutrinoMatterCoupling}

In this section we describe the solution procedure of the IMEX-RK scheme for the nonlinear coupling problem given by Equation~\eqref{eq:imexImplicit}.  
Since the nodal collocation DG scheme results in complete decoupling between spatial nodes within an element, it is sufficient to describe the algorithm for a single spatial node.  
To simplify the notation from Equation~\eqref{eq:imexImplicit}, we drop the tilde and reference to spatial nodes and IMEX stage, and use the global energy indexing introduced in Equation~\eqref{eq:energyIntegral}; i.e., $\{\widetilde{\bcU}_{\bp}^{\eidx,(i)},\widetilde{\bcU}_{\bp}^{\eidx,(i\star)}\}\to\{\bcU_{\qidx},\bcU_{\qidx}^{(\star)}\}$ and $\{\widetilde{\bu}_{\bp_{\bx}}^{(i)},\widetilde{\bu}_{\bp_{\bx}}^{(i\star)}\}\to\{\bu,\bu^{(\star)}\}$.  
We also replace $\alpha_{ii}\dt$ with the generic time step $\dtau$.  

For clarity of presentation, we describe the algorithm with only the isotropic part of the NES and pair kernels included (i.e., $\Phi_{s,0}^{\In/\Out}$ and $\Phi_{s,0}^{\Pro/\Ann}$).  
The inclusion of linear corrections to these processes is a straightforward extension.  

\subsubsection{Neutrino Equations}

The implicit system for $\{\bcU_{\qidx}\}_{\qidx=1}^{N^{\varepsilon}}=\{\,\{[\cN_{s}]_{\qidx},[{\cG_{s}}_{j}]_{\qidx}\}_{s=1}^{N_{s}}\,\}_{\qidx=1}^{N^{\varepsilon}}$ in Equation~\eqref{eq:imexImplicitMoments}, for the number and number flux equations, respectively, can be written as
\begin{equation}
	\big([\cD_{s}]_{\qidx}+v^{i}\,[{\cI_{s}}_{i}]_{\qidx}\big)
	=[\cN]_{\qidx}^{(\star)} + \dtau\,\big(\,[\eta_{s}]_{\qidx}-[\chi_{s}]_{\qidx}\,[\cD_{s}]_{\qidx}\,\big)
	\label{eq:implicitNumberNodal}
\end{equation}
and
\begin{equation}
	\big([{\cI_{s}}_{j}]_{\qidx}+v^{i}\,[{\cK_{s}}_{ij}]_{\qidx}\big)
	=[{\cG_{s}}_{j}]_{\qidx}^{(\star)} - \dtau\,[\kappa_{s}]_{\qidx}\,[{\cI_{s}}_{j}]_{\qidx}.  
	\label{eq:implicitNumberFluxNodal}
\end{equation}
Here, we have written the number and number flux densities at the new time step, $[\cN_{s}]_{\qidx}$ and $[{\cG_{s}}_{j}]_{\qidx}$, in terms of the primitive moments, $[\cD_{s}]_{\qidx}$ and $[{\cI_{s}}_{j}]_{\qidx}$, which will be treated as the unknowns in the solution procedure to satisfy Equations~\eqref{eq:implicitNumberNodal} and \eqref{eq:implicitNumberFluxNodal}.  
(We use square brackets on moment variables to separate the species and tensor component indices from the global energy-node index, $\qidx$, and a fixed-point iteration counter, which we introduce later.)  

In Equations~\eqref{eq:implicitNumberNodal} and \eqref{eq:implicitNumberFluxNodal}, we have defined the generalized emissivity and opacity evaluated in the global energy-node coordinate, $\varepsilon_{\qidx}$,
\begin{equation}
	[\eta_{s}]_{\qidx} = [\eta_{s}^{\EmAb}]_{\qidx} + [\eta_{s}^{\NES}]_{\qidx} + [\eta_{s}^{\Pair}]_{\qidx}
	\label{eq:emissivityNodal}
\end{equation}
and
\begin{equation}
	[\chi_{s}]_{\qidx} = [\chi_{s}^{\EmAb}]_{\qidx} + [\chi_{s}^{\NES}]_{\qidx} + [\chi_{s}^{\Pair}]_{\qidx},
	\label{eq:opacityNodal}
\end{equation}
respectively, as the sum of contributions from emission and absorption, NES, and pair processes.  
The transport opacity,
\begin{equation}
	[\kappa_{s}]_{\qidx} = [\chi_{s}]_{\qidx} + [\sigma_{s}^{\Iso}]_{\qidx},
	\label{eq:transportOpacityNodal}
\end{equation}
used in Equation~\eqref{eq:implicitNumberFluxNodal}, contains the additional contribution from isoenergetic scattering.  

For emission and absorption, the emissivity is related to the opacity by the detailed balance relation
\begin{equation}
	[\eta_{s}^{\EmAb}]_{\qidx} 
	\vcentcolon= [\chi_{s}^{\EmAb}]_{\qidx}\,[\cD_{s}^{\Eq}]_{\qidx},
	\label{eq:emissivityEmAbNodal}
\end{equation}
where $[\cD_{s}^{\Eq}]_{\qidx}$ is an approximation to the isotropic equilibrium Fermi--Dirac distribution in Equation~\eqref{eq:FermiDiracDistribution} at neutrino energy $\varepsilon_{\qidx}$.  
The emissivities and opacities due to NES and pair processes involve integrals over all energies, which are approximated with LG quadrature, using the weights defined in Equation~\eqref{eq:energy_integration_weights}; i.e.,
\begin{equation}
	[\eta_{s}^{\NES}]_{\qidx}
	\vcentcolon=\sum_{\qidx'=1}^{N^{\varepsilon}}W_{\qidx'}^{(2)}\,[\Phi_{s,0}^{\In}]_{\qidx\qidx'}\,[\cD_{s}]_{\qidx'}
	\quad\text{and}\quad
	[\chi_{s}^{\NES}]_{\qidx}
	\vcentcolon=[\eta_{s}^{\NES}]_{\qidx} + \sum_{\qidx'=1}^{N^{\varepsilon}}W_{\qidx'}^{(2)}\,[\Phi_{s,0}^{\Out}]_{\qidx\qidx'}\,(1-[\cD_{s}]_{\qidx'}),
	\label{eq:nesRatesNodal}
\end{equation}
and
\begin{equation}
	[\eta_{s}^{\Pair}]_{\qidx}
	\vcentcolon=\sum_{\qidx'=1}^{N^{\varepsilon}}W_{\qidx'}^{(2)}\,[\Phi_{s,0}^{\Pro}]_{\qidx\qidx'}\,(1-[\bar{\cD}_{s}]_{\qidx'})
	\quad\text{and}\quad
	[\chi_{s}^{\Pair}]_{\qidx}
	\vcentcolon=[\eta_{s}^{\Pair}]_{\qidx}+\sum_{\qidx'=1}^{N^{\varepsilon}}W_{\qidx'}^{(2)}\,[\Phi_{s,0}^{\Ann}]_{\qidx\qidx'}\,[\bar{\cD}_{s}]_{\qidx'}.  
	\label{eq:pairRatesNodal}
\end{equation}

As discussed in Section~\ref{sec:weakInteractionsAndCollisionTerms}, the emissivities and opacities depend on the thermodynamic state of the fluid and are obtained by interpolation of tabulated data provided by \weaklib\ \citep[see also][]{pochik_etal_2021,laiu_etal_2021}.

The approximate equilibrium distribution, $\cD_{s,h}^{\Eq}|_{K_{\varepsilon}^{\eidx}}$, is obtained by a weighted $L^{2}$ projection.  
That is, for each energy element $K_{\varepsilon}^{\eidx}$ and each basis function $\ell_{q^{\varepsilon}}$, we first find $\widetilde{\cD}_{s,h}^{\Eq}$ such that
\begin{equation}
	\int_{K_{\varepsilon}^{\eidx}}\widetilde{\cD}_{s,h}^{\Eq}(\varepsilon)\ell_{q^{\varepsilon}}(\varepsilon)\varepsilon^{2}d\varepsilon
	=\int_{K_{\varepsilon}^{\eidx}}f_{s}^{\Eq}(\varepsilon)\ell_{q^{\varepsilon}}(\varepsilon)\varepsilon^{2}d\varepsilon.  
	\label{eq:fermiDiracProjection}
\end{equation}
Then, $\cD_{s,h}^{\Eq}(\varepsilon)\vcentcolon=\Lambda^{\cD}\big\{\widetilde{\cD}_{s,h}^{\Eq}(\varepsilon)\big\}$, where $\Lambda^{\cD}\big\{\ldots\}$ represents a bound-enforcing limiter \cite[see, e.g.,][]{liuOsher_1996,zhangShu_2010a} that keeps $\cD_{s,h}^{\Eq}(\varepsilon)\in[\cD_{\min}^{\Eq},\cD_{\max}^{\Eq}]$ for all $\varepsilon\in \brmS_{\varepsilon}^{\eidx}$ ($\eidx=1,\ldots,\nElementsE$), without changing the cell average; i.e., $\int_{K_{\varepsilon}^{\eidx}}\cD_{s,h}^{\Eq}\varepsilon^{2}d\varepsilon=\int_{K_{\varepsilon}^{\eidx}}f_{s}^{\Eq}\varepsilon^{2}d\varepsilon$.  
Here, $\cD_{\min}^{\Eq}>0$ and $\cD_{\max}^{\Eq}<1$ are specified numerical bounds.  
Finally, the nodal values of the Fermi--Dirac distribution are approximated as $[\cD_{s}^{\Eq}]_{\qidx}\vcentcolon=\cD_{s,h}^{\Eq}(\varepsilon_{q^{\varepsilon}})|_{K_{\varepsilon}^{\eidx}}$ ($\qidx=(\eidx-1)\times N+q^{\varepsilon}$).  

We have found this limiting procedure of the Fermi--Dirac distribution to be crucial when neutrinos are strongly coupled with the fluid.  
The reason is that the phase-space advection update requires the moments to be realizable in an extended set of quadrature points, including the points on the element interfaces in the energy dimension; i.e., $\{\varepsilon_{\eidx-\f{1}{2}},\varepsilon_{\eidx+\f{1}{2}}\}$.  
When neutrino--matter coupling is strong, the collision term will drive the comoving-frame number density towards the equilibrium $\cD_{s,h}^{\Eq}$.  
Then, in the steep exponential tail of the Fermi--Dirac distribution, with finite resolution in energy space, the projection in Equation~\eqref{eq:fermiDiracProjection} can result in approximations that violate the bounds of the Fermi--Dirac distribution, including the points examined by the realizability-enforcing limiter in Section~\ref{sec:limiters}.  
Without the bound-enforcing limiter on $\widetilde{\cD}_{s,h}^{\Eq}$, the implicit collision solve can relax the distribution towards an equilibrium that is locally non-realizable.  
When this occurs, the realizability-enforcing limiter must be applied, which may in turn push the solution away from equilibrium.  
By enforcing bounds on the equilibrium distribution, we ensure that the collision solver relaxes the solution towards a realizable equilibrium, avoiding this undesirable interaction resulting in artificial energy and lepton exchange between neutrinos and the fluid.  
The bound-enforcing limiter helps keep $\cD_{s,h}^{\Eq}\in[\cD_{\min}^{\Eq},\cD_{\max}^{\Eq}]$ by the collision solve, and ensures that the system can relax to steady state equilibrium.  
The bound-enforcing limiter does not destroy the high-order accuracy of $\cD_{s,h}^{\Eq}$ \citep[see, e.g., Lemma~2.3 in][]{zhangShu_2011}.  

\begin{figure}[h]
        \captionsetup[subfigure]{justification=centering}
        \subfloat[High collisionality]
        {\begin{minipage}{0.3\textwidth}
                        \includegraphics[width=\linewidth]{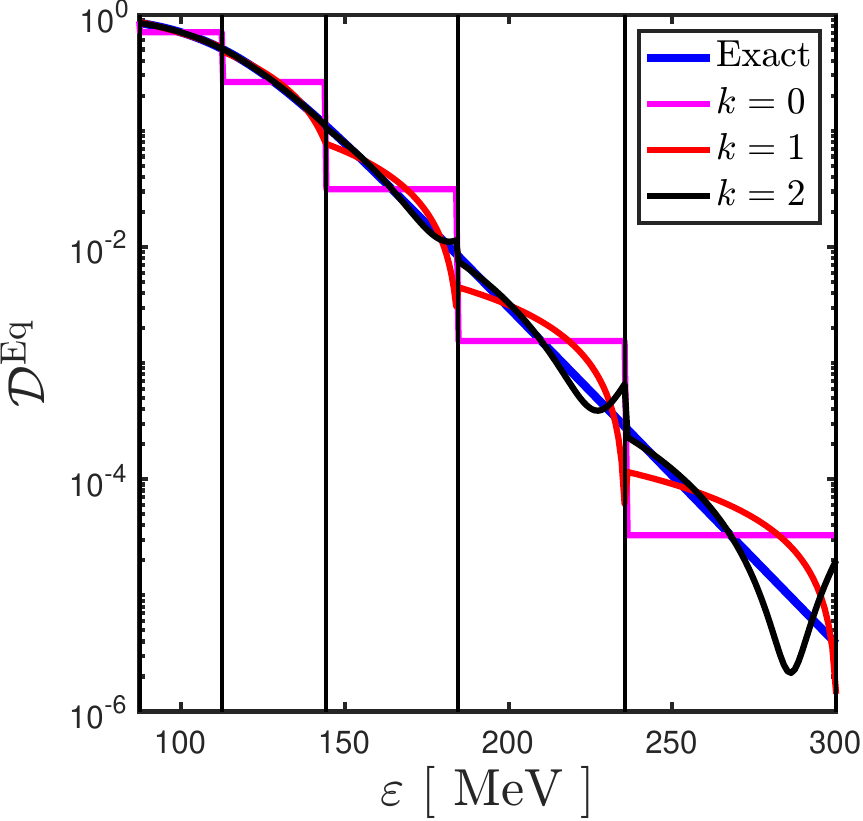}
                        \label{fig:}
         \end{minipage}
        }~~~~~
        \subfloat[Low Collisionality]
        {\begin{minipage}{0.3\textwidth}
                        \includegraphics[width=\linewidth]{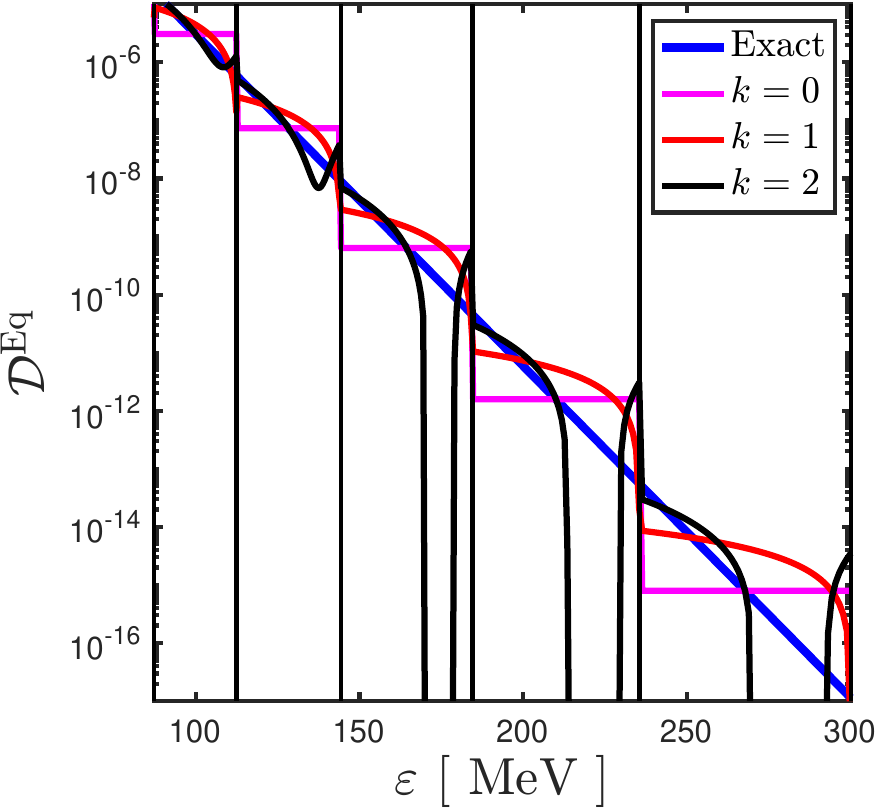}
                        \label{fig:}
         \end{minipage}
        }~~~~~
        \subfloat[Error]
        {\begin{minipage}{0.3\textwidth}
                        \includegraphics[width=\linewidth]{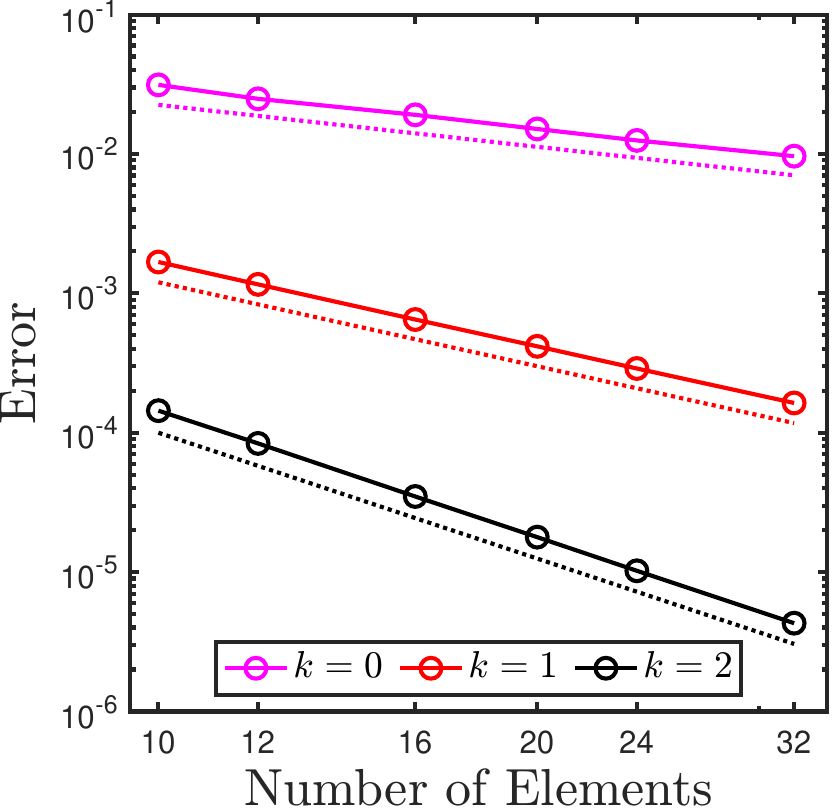}
                        \label{fig:}
         \end{minipage}
        }
        \caption{DG representations of the equilibrium Fermi--Dirac distribution with the bound-enforcing limiter applied for electron neutrinos in conditions from a collapsed stellar core around 100~ms after core bounce.  
        The left panel shows results for $T=15$~MeV and $\mu_{\nu_{\rm e}}=113$~MeV, characteristic of conditions for electron neutrinos at high density ($\rho\sim10^{14}$~g~cm$^{-3}$ and $Y_{\rm e}\sim0.27$).  
	The middle panel shows results for $T=7.6$~MeV and $\mu_{\nu_{\rm e}}=3.5$~MeV, characteristic of conditions at lower density ($\rho\sim10^{12}$~g~cm$^{-3}$ and $Y_{\rm e}\sim0.13$).  
	The right panel shows the root-mean-square error of the approximation against the exact Fermi--Dirac distribution for the low-density case in the middle panel versus the number of energy elements $\nElementsE$ for various polynomial degrees $k$, using the energy grids specified in Table~\ref{tab:energyGrids}.}
        \label{fig:FermiDiracDG}
\end{figure}

Figure~\ref{fig:FermiDiracDG} shows the approximate Fermi--Dirac distribution, $\cD_{\nu_{\rm e},h}^{\Eq}$ (i.e., with the bound-enforcing limiter applied), for electron neutrinos in the exponential tail ($\varepsilon\in[87,300]$~MeV) for various polynomial degrees, compared with the exact distribution $f_{\nu_{\rm e}}^{\Eq}$.  
Here, the energy domain is discretized with a geometrically progressing grid, $\Delta\varepsilon_{\eidx+1}=\alpha\,\Delta\varepsilon_{\eidx}$, where $\alpha$ is a constant `zoom factor' (see Table~\ref{tab:energyGrids} for a collection of energy grids used in this paper).  
The chosen thermodynamic conditions, taken from \citet{liebendorfer_etal_2005}, are characteristic of conditions in a collapsed stellar core about 100~ms after core bounce: one for high density ($\rho\sim10^{14}$~g~cm$^{-3}$; left panel) and one for lower density ($\rho\sim10^{12}$~g~cm$^{-3}$; middle panel).  
For $k=0$ (constant elements) and $k=1$ (linear elements), the approximate equilibrium distribution is nonnegative everywhere inside each element.  
For $k=2$ (quadratic elements), the approximate equilibrium is negative in portions of the last few energy bins, but not in the LG quadrature points or on the element interfaces (see solid black lines in the middle panel).  
Grid~C from Table~\ref{tab:energyGrids} was used in the left and middle panels.  
In the right panel, the root mean square error relative to the exact equilibrium distribution is plotted versus number of energy elements, using all energy grids listed in Table~\ref{tab:energyGrids}.  
In spite of using nonuniform energy grids, we observe that first-, second-, and third-order accuracy of the approximation is maintained for $k=0$, $1$, and $2$, respectively, and that higher-order polynomials provide improved overall accuracy.  

\begin{table}[h]
  \caption{Grid name, number of elements, zoom factors, inner and outer energy bin widths for geometrically progressing energy grids.\label{tab:energyGrids}}
  \small
  \vspace{-6pt}
  \begin{center}
  \begin{tabular}{ccccc}
    \midrule
    Grid & $\nElementsE$ & Zoom Factor & $\Delta\varepsilon_{1}$ [ MeV ] & $\Delta\varepsilon_{\nElementsE}$ [ MeV ] \\
    \midrule
    \midrule
	A  & $10$	& $1.473936813452146$  & $3.0$ 		& $98.5$ \\
	B  & $12$	& $1.386077933697496$  & $2.35$ 		& $85.3$ \\
	C  & $16$	& $1.266038160710160$  & $1.875$		& $64.5$ \\
	D  & $20$	& $1.205682366195175$  & $1.5$ 		& $52.4$ \\
	E  & $24$	& $1.167614387258235$  & $1.25$ 		& $44.1$ \\
	F  & $32$	& $1.119237083677839$  & $1.0$		& $32.9$ \\
    \midrule
    \midrule
  \end{tabular}
  \end{center}
\end{table}

For particle number conservation to hold by NES, the usual in-out invariance is imposed \citep{cernohorsky_1994}; i.e.,
\begin{equation}
	[\Phi_{s,0}^{\In}]_{\qidx\qidx'} = [\Phi_{s,0}^{\Out}]_{\qidx'\qidx}.
	\label{eq:inOutInvarianceNES}
\end{equation}
In addition, the NES collision operator must vanish when the neutrino distribution equals the equilibrium Fermi--Dirac distribution; i.e., when $[\cD_{s}]_{\qidx}=[\cD_{s}^{\Eq}]_{\qidx}$.  
This holds if \citep[cf. Equation~(2.5) in][]{cernohorsky_1994}
\begin{equation}
	[\Phi_{s,0}^{\In}]_{\qidx\qidx'}
	=\f{[\cD_{s}^{\Eq}]_{\qidx}\,(1-[\cD_{s}^{\Eq}]_{\qidx'})}{[\cD_{s}^{\Eq}]_{\qidx'}\,(1-[\cD_{s}^{\Eq}]_{\qidx})}\,[\Phi_{s,0}^{\Out}]_{\qidx\qidx'}.  
	\label{eq:detailedBalanceNES}
\end{equation}
When $[\cD_{s}^{\Eq}]_{\qidx}\vcentcolon=f_{s}^{\Eq}(\varepsilon_{\qidx})$, the factor in front of $[\Phi_{s,0}^{\Out}]_{\qidx\qidx'}$ reduces to $e^{-(\varepsilon_{\qidx}-\varepsilon_{\qidx'})/T}$.  
However, with the projection in Equation~\eqref{eq:fermiDiracProjection}, followed by the bound-enforcing limiter, we generally have $[\cD_{s}^{\Eq}]_{\qidx}\ne f_{s}^{\Eq}(\varepsilon_{\qidx})$ (equality is enforced weakly, rather than pointwise), and the detailed balance relation in Equation~\eqref{eq:detailedBalanceNES} must be enforced instead.  

Similarly, for lepton number conservation by pair processes, we must have
\begin{equation}
	[\Phi_{s,0}^{\Pro/\Ann}]_{\qidx\qidx'}
	= [\bar{\Phi}_{s,0}^{\Pro/\Ann}]_{\qidx'\qidx}, 
	\label{eq:inOutInvariancePair}
\end{equation}
where $\bar{\Phi}_{s,0}^{\Pro/\Ann}$ are the pair rates for antineutrinos of neutrino species $s$ (e.g., $\bar{\Phi}_{\nu_{\rm e},0}^{\Pro/\Ann}=\Phi_{\bar{\nu}_{\rm e},0}^{\Pro/\Ann}$).  
For the pair-process collision term to vanish in equilibrium, the following detailed balance relation must hold
\begin{equation}
	[\Phi_{s,0}^{\Pro}]_{\qidx\qidx'}
	=\f{[\cD_{s}^{\Eq}]_{\qidx}\,[\bar{\cD}_{s}^{\Eq}]_{\qidx'}}{(1-[\cD_{s}^{\Eq}]_{\qidx})\,(1-[\bar{\cD}_{s}^{\Eq}]_{\qidx'})}\,[\Phi_{0,s}^{\Ann}]_{\qidx\qidx'}.  
	\label{eq:detailedBalancePair}
\end{equation}
Here, when $[\cD_{s}^{\Eq}]_{\qidx}\vcentcolon=f_{s}^{\Eq}(\varepsilon_{\qidx})$ and $[\bar{\cD}_{s}^{\Eq}]_{\qidx}\vcentcolon=\bar{f}_{s}^{\Eq}(\varepsilon_{\qidx})$, the factor in front of $[\Phi_{0,s}^{\Ann}]_{\qidx\qidx'}$ reduces to $e^{-(\varepsilon_{\qidx}+\varepsilon_{\qidx'})/T}$.  

Next, we seek to rewrite Equations~\eqref{eq:implicitNumberNodal} and \eqref{eq:implicitNumberFluxNodal} in a form suitable for fixed-point iteration.  
To this end, following \citet{laiu_etal_2025}, for some $\lambda>0$, Equations~\eqref{eq:implicitNumberNodal} and \eqref{eq:implicitNumberFluxNodal} can be written as
\begin{equation}
	[\bcM_{s}]_{\qidx} 
	= 
	\left(\begin{array}{c} 
		{[}\cD_{s}{]}_{\qidx} \\
		{[}{\cI_{s}{}}_{j}]_{\qidx}
	\end{array}\right)
	=\Lambda
	\left(\begin{array}{c}
		(1-\lambda)\,[\cD_{s}]_{\qidx}+\lambda\,\big(\,[\cN]_{\qidx}^{(\star)}-v^{i}[{\cI_{s}}_{i}]_{\qidx}+\dtau\,[\eta_{s}]_{\qidx}\,\big) \\
		(1-\lambda)\,[{\cI_{s}}_{j}]_{\qidx}+\lambda\,\big(\,[{\cG_{s}}_{j}]_{\qidx}^{(\star)}-v^{i}[{\cK_{s}}_{ij}]_{\qidx}\,\big)
	\end{array}\right)
	\equiv \bcQ(\bcM,\bsfy)_{s,\qidx},
	\label{eq:implicitMomentsNodal}
\end{equation}
where $\Lambda=\mbox{diag}\big((1+\lambda\dtau [\chi_{s}]_{\qidx})^{-1},(1+\lambda\dtau[\kappa_{s}]_{\qidx})^{-1}\big)$.  
In the arguments of $\bcQ$, defined as the right-hand side of Equation~\eqref{eq:implicitMomentsNodal}, we have defined $\bcM=\{\{[\bcM_{s}]_{\qidx}\}_{s=1}^{N_{s}}\}_{\qidx=1}^{N^{\varepsilon}}$, indicating that the collision term can couple moments across all energies and all species.  
We have also explicitly included the dependence on the primitive fluid variables $\bsfy=(v_{j},\epsilon_{\rm f},Y_{\rm e})^{\intercal}$.  
(The map $\bcQ$ also depends on the mass density $\rho$, which we exclude from the fluid state vector $\bsfy$ because $\rho$ remains unchanged during the collision update.)

By defining $\bcQ(\bcM,\bsfy)\vcentcolon=\{\{\bcQ(\bcM,\bsfy)_{s,\qidx}\}_{s=1}^{N_{s}}\}_{\qidx=1}^{N^{\varepsilon}}$, we write Equation~\eqref{eq:implicitMomentsNodal} for all species and energies simply as
\begin{equation}
	\bcM = \bcQ(\bcM,\bsfy).  
	\label{eq:implicitMomentsNodalFull}
\end{equation}

\begin{rem}
	In Equation~\eqref{eq:implicitMomentsNodal}, we set the parameter $\lambda$ equal to $1/(1+|v|)$, where $|v|=\sqrt{\gamma_{ij}v^{i}v^{j}}$.  
	\citet{laiu_etal_2025}, using the map in Equation~\eqref{eq:implicitMomentsNodal} for the implicit solve in the context of simplified opacities, proved that this choice for $\lambda$ preserves moment realizability for Picard iteration.  
	Moreover, for $|v|<\sqrt{2}-1\approx0.41$, $\bcQ$ is a contraction operator, which is a sufficient condition to guarantee convergence.  
	We use the same choice for $\lambda$ here.  
	However, no analogous convergence proof is available for the full nonlinear solver with realistic opacities, and we do not pursue such an analysis here.  
	Instead, we investigate the convergence properties of the proposed algorithm in the CCSN context with numerical examples in Sections~\ref{sec:application_idealized} and \ref{sec:application_ccsn}.
        \label{rem:stepSizeParameter}
\end{rem}

\subsubsection{Matter Equations}

The update of matter quantities follows from enforcing conservation laws for mass (trivially conserved during the collision update; i.e., $\rho = \rho^{(\star)}$), momentum, energy, and lepton number; i.e.,
\begin{equation}
	\bsfy
	= \left(\begin{array}{c}
		v_{j} \\ \epsilon_{\rm f} \\ Y_{\rm e}
	\end{array}\right)
	= \left(\begin{array}{c}
		v_{j}^{(\star)} - \big(\,{F_{\nu}}_{j} - {F_{\nu}}_{j}^{(\star)}\,\big)/\rho^{(\star)} \\
		\epsilon_{\rm f}^{(\star)}-\big(\,E_{\nu} - E_{\nu}^{(\star)}\,\big)/\rho^{(\star)} \\
		Y_{\rm e}^{(\star)} -\mB\,\big(\,N_{\nu} - N_{\nu}^{(\star)}\,\big)/\rho^{(\star)}
	\end{array}\right)
	 \equiv \bsfq(\bcM,\bsfy),
	 \label{eq:implicitFluidNodal}
\end{equation}
where the specific internal energy is obtained from $\epsilon=\epsilon_{\rm f}-\f{1}{2}v_{i}v^{i}$, and the Eulerian neutrino momentum-, energy-, and lepton number-densities are computed with LG quadrature rules as
\begin{align}
	{F_{\nu}}_{j} 
	&\vcentcolon= \sum_{s=1}^{N_{s}}\sum_{\qidx=1}^{N^{\varepsilon}}W_{\qidx}^{(3)}\big(\,[{\cI_{s}}_{j}]_{\qidx}+v_{j}[\cD_{s}]_{\qidx}+v^{k}[{\cK_{s}}_{jk}]_{\qidx}\,\big), 
	\label{eq:eulerianMomentumLG} \\
	E_{\nu} 
	&\vcentcolon=\sum_{s=1}^{N_{s}}\sum_{\qidx=1}^{N^{\varepsilon}}W_{\qidx}^{(3)}\big(\,[\cD_{s}]_{\qidx}+2v^{j}[{\cI_{s}}_{j}]_{\qidx}\,\big), 
	\label{eq:eulerianEnergyLG} \\
	N_{\nu}
	&\vcentcolon=\sum_{s=1}^{N_{s}}\sum_{\qidx=1}^{N^{\varepsilon}}W_{\qidx}^{(2)}\sfg_{s}\big(\,[\cD_{s}]_{\qidx}+v^{i}[{\cI_{s}}_{j}]_{\qidx}\,\big),
	\label{eq:eulerianNumberLG}
\end{align}
where $W_{\qidx}^{(2)}$ is defined in Equation~\eqref{eq:energy_integration_weights} and $W_{\qidx}^{(3)}=W_{\qidx}^{(2)}\times(\varepsilon_{\qidx})$.  

Thermodynamic quantities needed to evaluate neutrino opacities---e.g., chemical potentials and compositions---are tabulated by \weaklib\ in terms of $\rho$, $T$, and $\ye$, and are obtained by table interpolation.  
Once $\epsilon$ and $\ye$ are updated, the corresponding temperature is obtained by root finding, using the EoS table.  

\subsubsection{Nested Fixed-point Iteration Algorithm}

Following our earlier work \citep{laiu_etal_2021}, we formulate the coupled nonlinear system given by Equations~\eqref{eq:implicitMomentsNodalFull} and \eqref{eq:implicitFluidNodal} as a nested fixed-point problem with two levels, as illustrated with pseudocode in Algorithm~\ref{alg:neutrinoMatterCoupling}.  
Given a time step $\dtau$ and old states $\bsfy^{(\star)}$ and $\bcM^{(\star)}$, which are set as initial guesses in Line~4, the matter quantities are iterated in an outer loop (beginning on Line~8), using converged moments from an inner loop (beginning on Line~15) in which the matter states are held fixed.  
The nested structure is motivated by the desire to reduce the number of computationally expensive opacity evaluations, which involve multidimensional table lookups and interpolations, and occur after each update of the matter states (Line~12) in a fully implicit solve.  
In our experience, albeit using a simpler model with electron flavor neutrinos and no velocity-dependent terms, fully converging the moments in the inner loop with fixed matter states results in fewer overall calls to evaluate opacities than is required when solving Equations~\eqref{eq:implicitMomentsNodalFull} and \eqref{eq:implicitFluidNodal} on a single level \citep[as in Section~4.1 in][]{laiu_etal_2021}.  
By bypassing the opacity evaluations in Line~12, the algorithm can easily be modified to become semi-implicit, where opacities are evaluated once using the old state in Line~5.  

Additional benefits of solving the neutrino--matter coupling problem with fixed-point iteration, as opposed to Newton's method, are that the Jacobian matrix and the associated inversion of a dense linear system are not needed.  
Forming the Jacobian matrix consisting of derivatives of tabulated opacity kernels with respect to thermodynamic states is a tedious and error-prone process.  
Because of this, it is easier to incorporate additional neutrino weak interaction channels with the fixed-point iteration method.  

Beyond the evaluation of opacities, additional compute intensive parts of Algorithm~\ref{alg:neutrinoMatterCoupling} include the computation of the rates in Line~18, the evaluation of the moment right-hand sides in Line~19, and, to a lesser extent, the evaluation of the matter right-hand sides in Line~25.  
The rate evaluations equate to dense matrix-vector multiplications; e.g., due to Equations~\eqref{eq:nesRatesNodal} and \eqref{eq:pairRatesNodal}.  
The majority of the cost associated with evaluating the matter right-hand sides is due to the moment integrals (scalar products) in Equations~\eqref{eq:eulerianEnergyLG}-\eqref{eq:eulerianNumberLG}.  
Because fixed-point iteration exhibits slower convergence relative to Newton's method, both the inner and outer loops are equipped with Anderson acceleration \citep{Anderson-1965,Walker-Ni-2011}, which uses residuals and right-hand side evaluations from prior iterates, as determined by the memory parameters $\texttt{M{\_}outer}$ and $\texttt{M{\_}inner}$, to improve the rate of convergence.  
As can be seen in Lines~20 and 26, least squares problems that minimize a combination of residuals are solved for coefficients $\{\alpha_{j}\}$ and $\{\beta_{j}\}$, which are subsequently used in the updates in Lines~21 and 27, respectively.  
Setting the memory parameter to unity results in Picard iteration.  
In our experience, setting the memory parameters larger than one reduces the iteration count significantly, but we do not observe much benefit from increasing the memory parameters beyond a few.  
We investigate the effect of varying both the inner and the outer memory parameters in Section~\ref{sec:application_idealized}.  
The inner and outer loops are iterated until convergence, as established by Algorithms~\ref{alg:innerConverged} and \ref{alg:outerConverged}, respectively.  

\begin{algorithm}[h]
	\SetNoFillComment
	\normalsize
	\medskip
	\caption{$[\,\bcM,\,\bsfy\,]=\texttt{SolveNeutrinoMatterCoupling}(\,\dtau,\bcM^{(\star)},\,\bsfy^{(\star)}\,)$}
	\label{alg:neutrinoMatterCoupling}
	{\bf Inputs:} time step $\dtau$, old states $\bcM^{(\star)}$ and $\bsfy^{(\star)}$,  \\
	{\bf Outputs:} new states $\bcM$ and $\bsfy$ \\
	{\bf Parameters:} $\texttt{tol{\_}outer}$, $\texttt{tol{\_}inner}$, $\texttt{M{\_}outer}$, $\texttt{M{\_}inner}$ \\
	$\bsfy^{[1]}\leftarrow\bsfy^{(\star)}$; $\bcM^{[1,1]}\leftarrow\bcM^{(\star)}$ \\
	\texttt{ComputeOpacities}(\,$\bsfy^{[1]}$\,) \\
	$\texttt{OuterConverged}\vcentcolon=\texttt{.false.}$; $\texttt{InnerConverged}\vcentcolon=\texttt{.false.}$ \\
	$k=0$ \\
	\While{( .NOT. \texttt{OuterConverged} )}
	{
		$k\pluseq1$
		
		$M_{k}=\min(\,k,\,\texttt{M{\_}outer}\,)$
		
		\If{( $k>1$ )}
		{
			\texttt{ComputeOpacities}(\,$\bsfy^{[k]}$\,)
			
			$\bcM^{[k,1]}\leftarrow\bcM^{[k-1,*]}$
		}
		$\ell=0$ \\
		\While{( .NOT. \texttt{InnerConverged} )}
		{
			$\ell\pluseq1$
			
			$M_{\ell}=\min(\,\ell,\,\texttt{M{\_}inner}\,)$
			
			\texttt{ComputeRates}(\,$\bcM^{[k,\ell]}$\,)
			
			\tcc{Evaluate moment right-hand sides and residuals}
			
			$\bcQ^{[\ell]}\vcentcolon=\bcQ(\bcM^{[k,\ell]},\bsfy^{[k]})$; $\bcF^{[\ell]}\vcentcolon=\bcQ^{[\ell]}-\bcM^{[k,\ell]}$
			
			\tcc{Solve inner least squares problem}
			\begin{equation*}
				\min_{\alpha_{j}}\f{1}{2}\| \sum_{j=1}^{M_{\ell}} \alpha_{j}\,\bcF^{[\ell-M_{\ell}+j]} \|^{2}
				\quad
				\text{subject to}
				\quad
				\sum_{j=1}^{M_{\ell}}\alpha_{j}=1
			\end{equation*}
			
			\tcc{Update moments}
			
			$\bcM^{[k,\ell+1]} = \sum_{j=1}^{M_{\ell}}\alpha_{j}\bcQ^{[\ell-M_{\ell}+j]}$
			
			\tcc{Check convergence of inner loop}
			
			\texttt{InnerConverged} = \texttt{CheckConvergence{\_}Inner}(\,$\bcM^{[k,\ell+1]} - \bcM^{[k,\ell]}$,\, $\|\bcD^{[k,1]}\|_{W^{(2)}}$,\,$\texttt{tol{\_}inner}$\,)
			
			\If{( \texttt{InnerConverged} )}{$\bcM^{[k,*]}\leftarrow\bcM^{[k,\ell+1]}$}
		}
		
		\tcc{Evaluate matter right-hand sides and residuals}
		
		$\bsfq^{[k]}\vcentcolon=\bsfq(\bcM^{[k,*]},\bsfy^{[k]})$; $\bsff^{[k]}\vcentcolon=\bsfq^{[k]}-\bsfy^{[k]}$
		
		\tcc{Solve outer least squares problem}
		\begin{equation*}
			\min_{\beta_{j}}\f{1}{2}\| \sum_{j=1}^{M_{k}}\beta_{j}\,\bsff^{[k-M_{k}+j]}\|^{2}
			\quad
			\text{subject to}
			\quad
			\sum_{j=1}^{M_{k}}\beta_{j} = 1
		\end{equation*}
		
		\tcc{Update matter quantities}
		
		$\bsfy^{[k+1]} = \sum_{j=1}^{M_{k}}\beta_{j}\,\bsfq^{[k-M_{k}+j]}$
		
		\tcc{Check convergence of outer loop}
		
		\texttt{OuterConverged} = \texttt{CheckConvergence{\_}Outer}(\,$\bsfy^{[k+1]}-\bsfy^{[k]}$,\,$\bsfy^{(\star)}$,\,$\texttt{tol{\_}outer}$\,)
		
		\If{( \texttt{OuterConverged} )}
		{
			$\bcM = \bcM^{[k,*]}$ and $\bsfy = \bsfy^{[k+1]}$
		}
		
	}
\end{algorithm}

\begin{algorithm}[h]
	\SetNoFillComment
	\normalsize
	\medskip
	\caption{$[\,\texttt{Converged}\,]=\texttt{CheckConvergence{\_}Inner}(\,\delta\bcM,\,\|\bcD\|,\,\texttt{tol}\,)$}
	\label{alg:innerConverged}
	{\bf Inputs:} Change $\delta\bcM=\{\{[\delta\bcM_{s}]_{\qidx}\}_{s=1}^{N^{s}}\}_{\qidx=1}^{N^{\varepsilon}}$, norm of number densities $\|\bcD\|=\{\|\cD_{s}\|\}_{s=1}^{N_{s}}$, and relative tolerance $\texttt{tol}$ \\
	{\bf Outputs:} Query result $\texttt{Converged}$ \\
	$\texttt{Converged}\vcentcolon=\texttt{.true.}$\\
	\For{( $s = 1, N_{s}$ )}
	{
		\tcc{Compute weighted norm of all moments for species $s$}
		
		$\|\delta\bcM\|=0$		
		
		\For{( $\qidx=1, N^{\varepsilon}$ )}
		{
			$\|\delta\bcM\| \pluseq W_{\qidx}^{(2)}\,([\delta\bcM_{s}]_{\qidx})^{2}$
		}
		
		\tcc{Select largest across all moments}
		
		$\|\delta\bcM\| \vcentcolon= \sqrt{\max(\|\delta\bcM\|)}$
		
		$\texttt{Converged} = \texttt{Converged} \texttt{.and.}\|\delta\bcM\|\le\texttt{tol}\,\|\cD_{s}\|$
		
	}
\end{algorithm}

\begin{algorithm}[h]
	\SetNoFillComment
	\normalsize
	\medskip
	\caption{$[\,\texttt{Converged}\,]=\texttt{CheckConvergence{\_}Outer}(\,\delta\bsfy,\,\bsfy,\,\texttt{tol}\,)$}
	\label{alg:outerConverged}
	{\bf Inputs:} Change $\delta\bsfy$, reference matter state $\bsfy$, and relative tolerance $\texttt{tol}$ \\
	{\bf Outputs:} Query result $\texttt{Converged}$
	
	$\texttt{Converged} \vcentcolon= \texttt{all}( \, \texttt{abs}(\delta\bsfy) \le \texttt{tol} * \texttt{abs}(\bsfy) \, )$
	
\end{algorithm}

\section{Incorporation in \flashx\ and Coupling to Finite-Volume Hydrodynamics}
\label{sec:flashx}

Here we describe how the algorithms for two-moment spectral neutrino transport in \thornado\ have been incorporated into the multiphysics software system \flashx\ \citep{dubey_etal_2022} to enable neutrino-radiation hydrodynamics simulations.  
(\thornado\ can be imported in \flashx\ as a Git submodule, and the \flashx\ code hosted is configured to facilitate this by default.)
\flashx\ is a performance portable framework that provides a mature infrastructure for large-scale scientific computations, which includes distributed parallelism with MPI, a portability layer to orchestrate data movement between devices, compiler directives for GPU computing, scalable I/O, and adaptive mesh refinement (AMR) capabilities, all of which are either necessary or desirable for three-dimensional simulations of CCSNe.  
Moreover, \flashx\ provides solvers for compressible hydrodynamics and magnetohydrodynamics, which we have coupled with the two-moment solver in \thornado.  
The approach to coupling is agnostic with respect to any of the finite-volume based hydrodynamics solvers available in \flashx, but we exclusively use \spark\ \citep{couch_etal_2021} in this paper.  
We have extended \spark\ to use the 5th order variant of the custom 9th order WENO-Z9 reconstruction described by~\cite{Ji_FFT_filter2023}, which uses building blocks from~\citet{Balsara_WENO_2016,Castro_WENOZ_2011,Tchekhovskoy_WHAM_2007}.  
We use the HLLC Riemann solver~\citep{HLLC_1994}.  
In the vicinity of shocks we employ the HLLE Riemann solver~\citep{EinfeldtHLLE1988,HartenHLLE1983} and use a more diffusive and robust linear reconstruction with a minmod slope limiter \citep[see, e.g.,][]{vanLeer1979,Harten1983,cockburnShu_1998,Kurganov2000}.  
For the purpose of consistent coupling, we have incorporated \weaklib\ as an importable module through Git submodules and created wrappers in \flashx\ to enable simulations to use the same tabulated nuclear equation of state as \thornado.  

We use Lie--Trotter operator splitting to couple the Euler--Poisson and spectral two-moment models.  
Given initial states for the fluid, neutrinos, and the gravitational potential at $t^{n}$, $\{\bu^{n},\bcU^{n},\Phi^{n}\}$, and a time step $\dt$, the updated quantities at $t^{n+1}=t^{n}+\dt$, $\{\bu^{n+1},\bcU^{n+1},\Phi^{n+1}\}$, are obtained in the following schematic way
\begin{itemize}
	\item[1.] Solve Subproblem~1 (Euler--Poisson) with an explicit RK method: 
	\begin{equation}
		\{\bu^{n},\Phi^{n}\}\to\{\bu^{\star},\Phi^{n+1}\}.
		\label{eq:lieTrotterEulerPoisson}
	\end{equation}
	\item[2.] Solve Subproblems~2 and 3 (phase-space advection and neutrino--matter coupling) with an IMEX method:
	\begin{equation}
		\{\,\bu^{\star},\,\bcU^{n}\,\}\to\{\,\bu^{n+1},\,\bcU^{n+1}\,\}.
		\label{eq:lieTrotterTwoMoment}
	\end{equation}
\end{itemize}
The time step is determined by the explicit part of the moments update, as given by Equation~\eqref{eq:cflConditionTransport}.  
When running with AMR, the time step is given by the smallest time step across the entire domain, across all levels; i.e., we do not employ subcycling in time, where different levels of the mesh hierarchy are evolved with different time steps \citep[as in, e.g.,][]{almgren_etal_2010}.  

In Equation~\eqref{eq:lieTrotterEulerPoisson}, the Euler--Poisson system is evolved by \flashx, using either SSP-RK2 or SSP-RK3.  
To avoid solving for the gravitational potential at intermediate stages of the RK integrator, potentials at $t^{n-1}$ and $t^{n}$ are used to extrapolate to intermediate times to evaluate the gravitational source terms.  
In Equation~\eqref{eq:lieTrotterTwoMoment}, the fluid and neutrino moments are evolved by \thornado\ with the IMEX scheme in Section~\ref{sec:timeIntegration}, using the Butcher tableau in Equation~\eqref{eq:butcherPDARS}.  
Since the Newtonian gravitational potential only depends on the mass density, which remains unchanged during the update in Equation~\eqref{eq:lieTrotterTwoMoment}, the potential is fully updated by Equation~\eqref{eq:lieTrotterEulerPoisson}.  
The Lie--Trotter operator splitting method used here introduces a first-order temporal splitting error, which is separate from the temporal accuracy of the individual time stepping methods used in Equations~\eqref{eq:lieTrotterEulerPoisson} and \eqref{eq:lieTrotterTwoMoment}, respectively.  
Because the IMEX scheme used in Equation~\eqref{eq:lieTrotterTwoMoment} is formally only first-order accurate, we accept this splitting error here, but intend to investigate the utility of more advanced integration strategies---e.g., multirate methods \citep{Sandu_2019,rujekoReynolds_2021}---in the future.  

Two remaining challenges associated with coupling the two-moment solver in \thornado\ with the Euler--Poisson solver in \flashx\ are: (1) the need for a hybrid representation --- on FV and finite-element grids --- of fluid variables, which are updated by both \flashx\ and \thornado, and (2) prolongation and restriction operators needed to evolve the neutrino radiation field on adaptively refined grids.  
We briefly describe our approach to these next.  

\subsection{Finite-Volume and Discontinuous Galerkin Representation of Fluid Variables}

\begin{figure}[h]
	\begin{center}
		\includegraphics[width=0.9\linewidth]{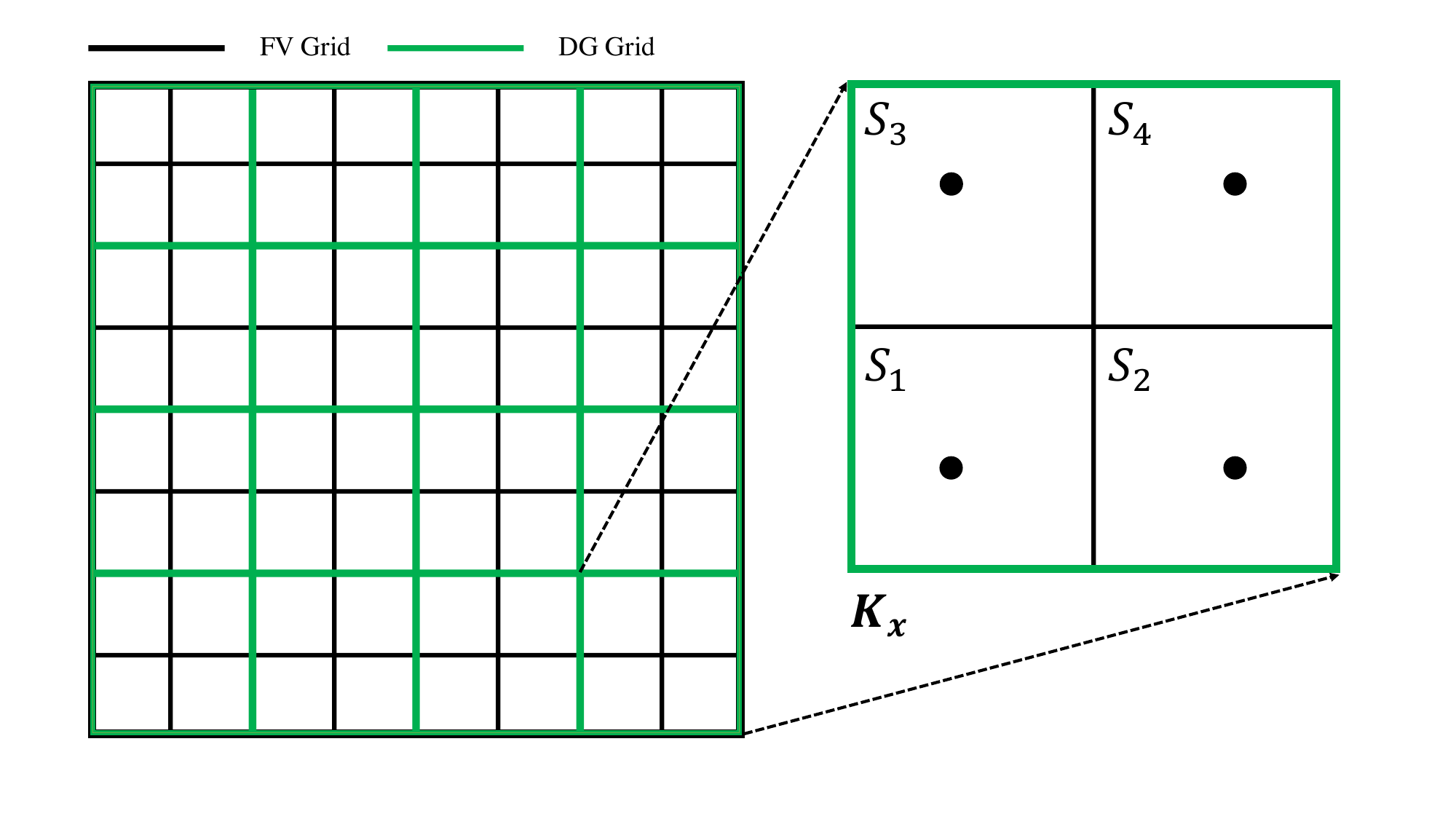}
		\caption{Illustration of a 2D ($d_{\bx}=2$) \flashx\ FV grid block for $\texttt{nxb}=\texttt{nyb}=8$ (black) combined with a DG grid (green) of linear elements ($k=1$).  
		In this case, a spatial element $\Kx$ with GL spatial nodes ($\bx\in\bS_{\bx}$; black dots) covers four subcells $S=\cup_{j=1}^{4} S_{j}$.}
        		\label{fig:dgFvGrids}
        \end{center}
\end{figure}

In \flashx, the three-dimensional computational domain is discretized by a mesh composed of a collection of logically Cartesian blocks of size $\texttt{nxb}\times\texttt{nyb}\times\texttt{nzb}$, where $\texttt{nxb}$ is the number of cells per block in the $x$-dimension, etc.  
Similarly, in \thornado, the computational spatial domain is discretized by a logically Cartesian mesh consisting of $\texttt{nX}(1)\times\texttt{nX}(2)\times\texttt{nX}(3)$ elements.  
Natively, \thornado\ can only evolve on a single-level grid, but all computational kernels are written to update logically Cartesian blocks, which eases the incorporation of its solvers in frameworks for block-structured AMR.  
The solvers for compressible hydrodynamics in \flashx\ are based on the finite-volume method, where fluid variables are represented by cell averages.  
In \thornado, fluid variables are represented by a polynomial representation in each spatial element $\Kx$ (similar to Equation~\eqref{eq:nodalExpansion} for the radiation moments)
\begin{equation}
	\bu_{h}(\bx,t)|_{\Kx} = \sum_{i=1}^{|\bN_{\bx}|}\bu_{i}(t)\phi_{i}(\bx),
	\label{eq:nodalExpansionX}
\end{equation}
where we use the tensor product basis; i.e., $\phi_{i}|_{\Kx}\in\bbQ^{k}(\Kx)$ and $|\bN_{\bx}|=(k+1)^{d_{\bx}}$.  
(In Equation~\eqref{eq:nodalExpansionX}, the basis functions and expansion coefficients are ordered such that the mapping from multi-index $\bi_{\bx}=\{i^{1},i^{2},i^{3}\}$ to scalar index $i$ is of the form $i(\bi_{\bx})=(i^{3}-1)\times N^{2}+(i^{2}-1)\times N+i^{1}$.) 
When $k>0$, the DG representation comprises more degrees of freedom per element than the FV representation per cell.  
For this reason, we let each DG element cover multiple FV cells such that the total number of degrees of freedom per block remains the same; i.e., we enforce the relation $\texttt{nX}(1)\times(k+1) = \texttt{nxb}$ (and similarly for the other spatial dimensions).  
This approach is related to that of the finite volume subcell limiter technique for DG methods proposed by \citet{dumbser_etal_2014}.  
Figure~\ref{fig:dgFvGrids} illustrates the FV and DG grids comprising a \flashx\ grid block in two-spatial dimensions ($d_{\bx}=2$) with linear elements ($k=1$).  

Our approach for Cartesian coordinates was described in \citet{harris_etal_2022}.  
Here we provide the essential elements of the approach, generalized to curvilinear coordinates.  
We let the DG element $\Kx\subset\bbR^{d_{\bx}}$ be subdivided into a subgrid $S$ consisting of $|\bN_{\bx}|$ nonoverlapping FV cells $S_{j}$, $S=\cup_{j=1}^{|\bN_{\bx}|}S_{j}=\Kx$ (i.e., $\Kx\setminus S=\emptyset$).  
On each subcell $S_{j}$, the solution is given by the FV representation (cell average) $\bar{\bu}_{j}$.  
We then find the expansion coefficients in Equation~\eqref{eq:nodalExpansionX} by demanding that
\begin{equation}
	\int_{S_{j}}\bu_{h}(\bx)\sqrt{\gamma_{h}}(\bx)\,d\bx = |S_{j}|\,\bar{\bu}_{j}
	\label{eq:reconstructionAnsatz}
\end{equation}
holds for all subcells $S_{j}$.  
Here, $|S_{j}|$ is the proper volume of subcell $S_{j}$.  
In \thornado, the square root of the metric determinant is computed from the scale factors ($h_{1,h}$, $h_{2,h}$, and $h_{3,h}$), which are approximated by polynomial expansions of the form in Equation~\eqref{eq:nodalExpansionX}.  
Inserting these approximations in Equation~\eqref{eq:reconstructionAnsatz} results in
\begin{equation}
	|S|^{-1}\,P\,\bu = \bar{\bu},
	\label{eq:projection}
\end{equation}
where $\bu=(\bu_{1},\ldots,\bu_{|\bN_{\bx}|})^{\intercal}$, $\bar{\bu}=(\bar{\bu}_{1},\ldots,\bar{\bu}_{|\bN_{\bx}|})^{\intercal}$, $|S|=\mbox{diag}(|S|_{1},\ldots,|S|_{|\bN_{\bx}|})$, and $P=(P_{ij})$ is the $|\bN_{\bx}|\times|\bN_{\bx}|$ \emph{projection matrix} with components
\begin{equation}
	P_{ij} = \int_{S_{i}}\phi_{j}(\bx)\,\sqrt{\gamma_{h}}(\bx)\,d\bx.
	\label{eq:projectionMatrix}
\end{equation}
Defining the \emph{reconstruction matrix} $R=P^{-1}$, the expansion coefficients in Equation~\eqref{eq:nodalExpansionX} are obtained from cell averages by the reconstruction
\begin{equation}
	\bu = R\,|S|\,\bar{\bu}.  
	\label{eq:reconstruction}
\end{equation}

To summarize, after the fluid has been evolved with the Euler--Poisson system using the FV-based solver in \flashx\ in Equation~\eqref{eq:lieTrotterEulerPoisson}, the DG representation is reconstructed using Equation~\eqref{eq:reconstruction} before the coupled two-moment update in Equation~\eqref{eq:lieTrotterTwoMoment}.  
After this step, the FV representation of fluid variables is recovered by the projection step in Equation~\eqref{eq:projection}.  
The construction of the projection matrix and its inverse (one of each per element $\Kx$) is done each time the \thornado\ two-moment solver is called.  
This on-the-fly computation is done to avoid storing $\texttt{nX}(1)\times\texttt{nX}(2)\times\texttt{nX}(3)$ $R$ and $P$ matrices per \flashx\ grid block, due to their dependence on the spatially-dependent metric determinant.  
Because the fluid variables constitute a relatively small fraction of the total number of degrees of freedom evolved by \thornado, combined with the relatively high cost of the two-moment update, the added computations incurred in the mapping of fluid variables between FV and DG representations do not add significantly to overall run times.  
Two-moment variables are always maintained in their DG representation in \flashx\ data structures.  

Unit conversions are needed when transferring variables between \flashx, which uses CGS units by default, and \thornado, which uses natural units ($c=G=\kB=1$).
Also, vectors and tensors in \flashx\ are represented in a local orthonormal basis, while in \thornado\ they are represented in the coordinate basis.  
For an arbitrary vector $\ba$, its components in the coordinate basis, $a^{i}$, are related to the components in the local orthonormal basis, $a^{\bar{\imath}}$, through the transformation $a^{i}=e^{i}_{\hspace{2pt}\bar{\imath}}\,a^{\bar{\imath}}$, where $e^{i}_{\hspace{2pt}\bar{\imath}}$ transforms the spatial metric $\gamma_{ij}$ into the Euclidean metric: $\delta_{\bar{\imath}\bar{\jmath}}=e^{i}_{\hspace{2pt}\bar{\imath}}\,e^{j}_{\hspace{2pt}\bar{\jmath}}\,\gamma_{ij}$.  
The inverse transformation, providing $a^{\bar{\imath}}=e^{\bar{\imath}}_{\hspace{2pt}i}\,a^{i}$, satisfies $e^{i}_{\hspace{2pt}\bar{\imath}}\,e^{\bar{\imath}}_{\hspace{2pt}j}=\delta^{i}_{\hspace{2pt}j}$.  
With the diagonal spatial metric introduced in Section~\ref{sec:model}, we simply have $e^{i}_{\hspace{2pt}\bar{\imath}}=\mbox{diag}[h_{1}^{-1},h_{2}^{-1},h_{3}^{-1}]$ and $e^{\bar{\imath}}_{\hspace{2pt}i}=\mbox{diag}[h_{1},h_{2},h_{3}]$.  

\subsection{Prolongation and Restriction of Radiation variables}

\begin{figure}[h]
	\begin{center}
		\includegraphics[width=0.8\linewidth]{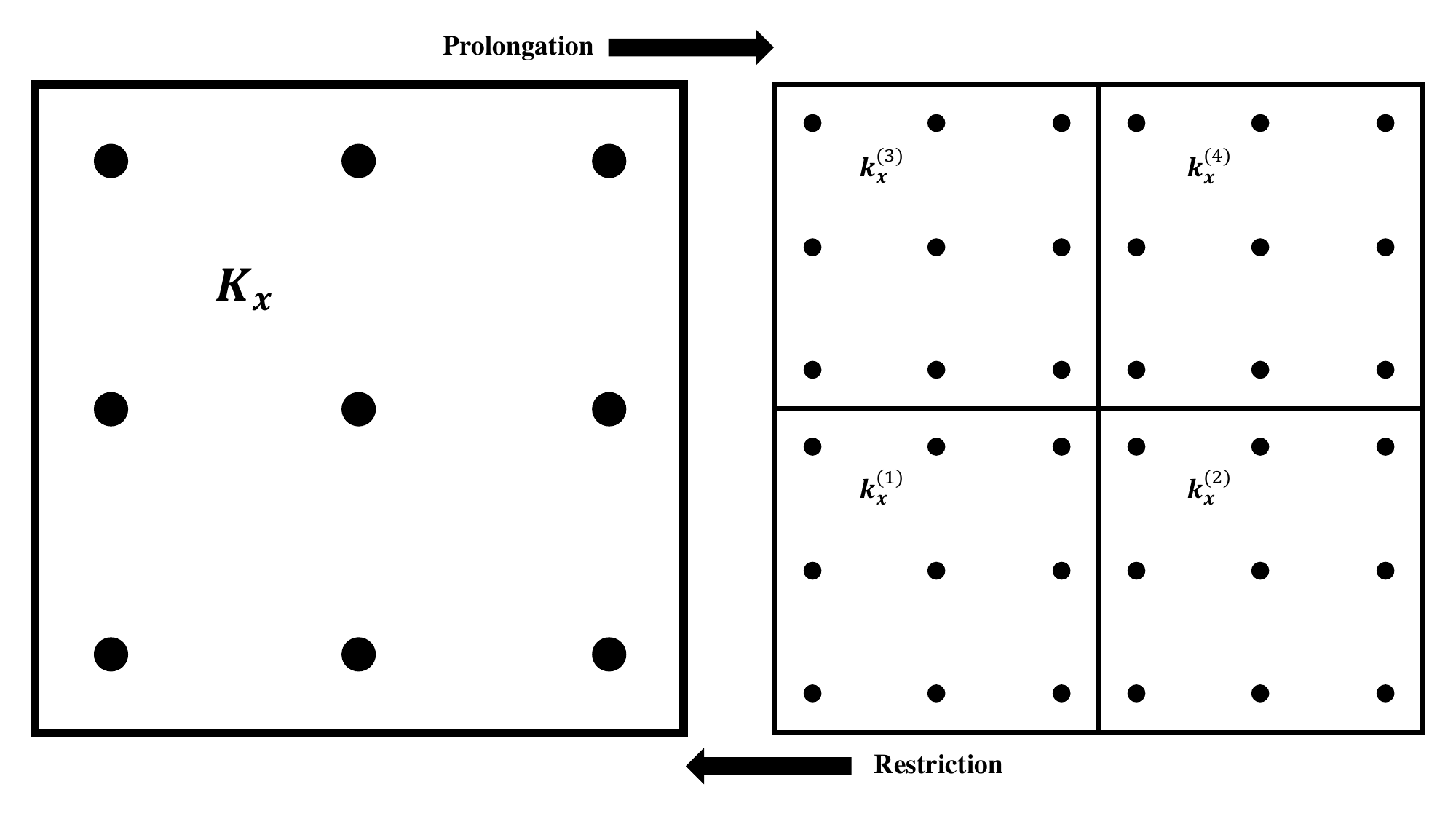}
		\caption{Illustration of refinement of an element $\Kx$ into multiple finer elements $\{\bk_{\bx}^{(j)}\}$ (prolongation; left to right) and coarsening of multiple elements into a single element (restriction; right to left) for the case with polynomial degree $k=2$ and spatial dimensionality $d_{\bx}=2$.}
        		\label{fig:refineCoarsen}
        \end{center}
\end{figure}

\flashx\ provides AMR capabilities in position space ($h$-refinement) to furnish spatial resolution where needed; e.g., to capture the matter distribution during gravitational collapse and to provide desired resolution around the surface of the PNS during the neutrino reheating phase after bounce.  
When a region is refined, variables need to be interpolated to the refined region using data from the overlaying coarser mesh (prolongation).  
Conversely, when a region is coarsened, variables from the finer mesh are averaged to provide data for the coarser mesh (restriction).  
Prolongation and restriction is also used to provide ghost cell data on coarse--fine mesh boundaries.  
\flashx\ already provides routines to handle these operations for variables with FV representation.  
Here we briefly describe how neutrino moment variables with DG representation are refined and coarsened in position space, which is similar to that of \citet{schaal_etal_2015} for solving the Euler equations, but describes our generalization to orthogonal curvilinear spatial coordinates.  
We only consider the case where the resolution changes by a factor of two between adjacent refinement levels.  
Moreover, when refinement (coarsening) occurs, each spatial dimension is refined (coarsened) by a factor of two.  
It is sufficient to consider the refinement of a single element, and the coarsening of $2^{d_{\bx}}$ elements to a single element.  
The prolongation/restriction process for the two-dimensional case with quadratic elements is illustrated in Figure~\ref{fig:refineCoarsen}.  

Let $\Kx$ denote a coarse element, and let $\{\bk_{\bx}^{(j)}\}$ denote a set of non-overlapping refined elements so that $\Kx=\cup_{j=1}^{2^{d_{\bx}}}\bk_{\bx}^{(j)}$.  
On the coarse element, we recall the expansion for radiation moments $\cU$ in Equation~\eqref{eq:nodalExpansionX}; i.e.,
\begin{equation}
	\cU_{h}(\bx)|_{\Kx} = \sum_{\bi_{\bx}=\bOne}^{\bN_{\bx}}\cU_{\bi_{\bx}}\,\phi_{\bi_{\bx}}(\bx),
	\label{eq:nodalExpansionCoarse}
\end{equation}
where $\phi_{\bi_{\bx}}|_{\Kx}\in\bbQ^{k}(\Kx)$.  
Similarly, on the fine elements we let the corresponding neutrino moments be represented by
\begin{equation}
	u_{h}(\bx)|_{\bk_{\bx}^{(j)}} = \sum_{\bi_{\bx}=\bOne}^{\bN_{\bx}}u_{\bi_{\bx}}^{(j)}\,\phi_{\bi_{\bx}}^{(j)}(\bx), 
	\label{eq:nodalExpansionFine}
\end{equation}
where $\phi_{\bi_{\bx}}^{(j)}|_{\bk_{\bx}^{(j)}}\in\bbQ^{k}(\bk_{\bx}^{(j)})$.  
(In this subsection, the symbol $u$ should not be confused with the conserved fluid variables; here it denotes the radiation moments $\cU$ restricted to a refined element.)  
Since refinement and coarsening is only applied in space, we omit energy and species indices for notational convenience, and note that identical operations are performed for all neutrino moments, species, and energies.  
On the coarse element, the refined data provides the piecewise polynomial representation
\begin{equation}
	u_{h}(\bx)|_{\Kx} = \sum_{j=1}^{2^{d_{\bx}}}u_{h}(\bx)|_{\bk_{\bx}^{(j)}}.  
	\label{eq:fineOnCoarse}
\end{equation}

The radiation moments on the refined element $\bk_{\bx}^{(j)}$ are obtained by weighted $L^{2}$ projection of the polynomial representation on the coarse element $\Kx$ onto the basis defined on the refined element; i.e., we find $u_{h}\in\bbQ^{k}(\bk_{\bx}^{(j)})$ such that
\begin{equation}
	\int_{\bk_{\bx}^{(j)}} u_{h}(\bx)\,\phi_{\bi_{\bx}}^{(j)}(\bx)\,\sqrt{\gamma_{h}^{(j)}}(\bx)\,d\bx
	=\int_{\bk_{\bx}^{(j)}}\cU_{h}(\bx)\,\phi_{\bi_{\bx}}^{(j)}(\bx)\,\sqrt{\gamma_{h}^{(j)}}\,d\bx
	\label{eq:projectionCoarseToFine}
\end{equation}
holds for all $\phi_{\bi_{\bx}}^{(j)}(\bx)\in\bbQ^{k}(\bk_{\bx}^{(j)})$, where $\sqrt{\gamma_{h}^{(j)}}$ is evaluated using polynomial approximations of the scale factors on $\bk_{\bx}^{(j)}$.  
Similarly, when coarsening, the polynomial on a coarse element is obtained by weighted $L^{2}$ projection of the piecewise polynomial in Equation~\eqref{eq:fineOnCoarse} onto the basis defined on the coarse element; i.e., we find $\cU_{h}\in\bbQ^{k}(\Kx)$ such that
\begin{equation}
	\int_{\Kx}\cU_{h}(\bx)\,\phi_{\bi_{\bx}}(\bx)\,\sqrt{\gamma_{h}}(\bx)\,d\bx
	=\sum_{j=1}^{2^{d_{\bx}}}\int_{\bk_{\bx}^{(j)}}u_{h}(\bx)\,\phi_{\bi_{\bx}}(\bx)\,\sqrt{\gamma_{h}^{(j)}}\,d\bx
	\label{eq:projectionFineToCoarse}
\end{equation}
holds for all $\phi_{\bi_{\bx}}\in\bbQ^{k}(\Kx)$.  

The prolongation and restriction operations defined by Equations~\eqref{eq:projectionCoarseToFine} and \eqref{eq:projectionFineToCoarse} can be formulated in terms of basic linear algebra operations.  
To this end, define the vectors $\bu^{(j)}=(u_{\bOne}^{(j)},\ldots,u_{\bN_{\bx}}^{(j)})^{\intercal}$ and $\bcU=(\cU_{\bOne},\ldots,\cU_{\bN_{\bx}})^{\intercal}$.  
Then, inserting Equations~\eqref{eq:nodalExpansionCoarse} and \eqref{eq:nodalExpansionFine} into Equation~\eqref{eq:projectionCoarseToFine} one obtains
\begin{equation}
	\bu^{(j)}=(m^{(j)})^{-1}\,\cP^{(j)}\,\bcU,
	\label{eq:projectionCoarseToFineLA}
\end{equation}
where $m^{(j)}=(m_{\bi_{\bx}\bj_{\bx}}^{(j)})$ is the mass matrix and $\cP^{(j)}=(\cP_{\bi_{\bx}\bj_{\bx}}^{(j)})$ are the prolongation matrices, with components
\begin{equation}
	m^{(j)}_{\bi_{\bx}\bj_{\bx}}=\int_{\bk_{\bx}^{(j)}}\phi_{\bi_{\bx}}^{(j)}(\bx)\,\phi_{\bj_{\bx}}^{(j)}(\bx)\,\sqrt{\gamma_{h}^{(j)}}(\bx)\,d\bx
	\quad\text{and}\quad
	\cP_{\bi_{\bx}\bj_{\bx}}^{(j)}=\int_{\bk_{\bx}^{(j)}}\phi_{\bi_{\bx}}^{(j)}(\bx)\,\phi_{\bj_{\bx}}(\bx)\,\sqrt{\gamma_{h}^{(j)}}(\bx)\,d\bx,
\end{equation}
respectively.  
The mass matrix $m^{(j)}$ is in general dense because of the weight $\sqrt{\gamma_{h}^{(j)}}$, and some additional care is required to compute the (nonsymmetric) prolongation matrices $\cP^{(j)}$ because $\phi_{\bj_{\bx}}$ has support on $\Kx$, which extends outside the refined element $\bk_{\bx}^{(j)}$.  
Similarly, by inserting Equations~\eqref{eq:nodalExpansionCoarse} and \eqref{eq:nodalExpansionFine} into Equation~\eqref{eq:projectionFineToCoarse} one obtains
\begin{equation}
	\bcU = \cM^{-1}\,\sum_{j=1}^{2^{d_{\bx}}}\cR^{(j)}\,\bu^{(j)},
	\label{eq:projectionFineToCoarseLA}
\end{equation}
where $\cM=(\cM_{\bi_{\bx}\bj_{\bx}})$ is the mass matrix with components
\begin{equation}
	\cM_{\bi_{\bx}\bj_{\bx}} = \int_{\Kx}\phi_{\bi_{\bx}}(\bx)\,\phi_{\bj_{\bx}}(\bx)\,\sqrt{\gamma_{h}}(\bx)\,d\bx,
\end{equation}
and the restriction matrices are simply $\cR^{(j)}=(\cP^{(j)})^{\intercal}$.  

\section{Results from Basic Verification Tests}
\label{sec:results}

In this section we present results to demonstrate the performance of the DG method for the two-moment model implemented in \thornado\ on some basic test problems with well-known solution characteristics.  
To this end, we consider a single species, and use simplified, analytic opacities due to emission and absorption, and isotropic and isoenergetic scattering; i.e., we set
\begin{equation}
	\aint{\mathcal{C}(\vect{f})}
	:=\chi\,\big(\,\cD^{\Eq}-\cD\,\big)
	\quad
	\text{and}
	\quad
	\aint{\mathcal{C}(\vect{f})\,\ell_{j}}
	:=-\big(\,\chi+\sigma\,\big)\,\cI_{j},
\end{equation}
where $\chi$ is the absorption opacity, $\cD^{\Eq}$ an equilibrium distribution, and $\sigma$ the scattering opacity.  
Results from some tests of this type have already been presented for the case of Cartesian spatial coordinates by \citet{laiu_etal_2025}.  
Because of this, we focus on relevant tests employing spherical-polar and Cylindrical spatial coordinates; except for the test in Section~\ref{sec:results.sine_wave_streaming_amr}, which uses Cartesian coordinates and AMR with \flashx.  
For all tests involving opacities, we use the implicit collision solver based on fixed-point iteration proposed in \citet{laiu_etal_2025}.  
Except for the test in Section~\ref{sec:results.sine_wave_streaming_amr}, all numerical tests use dimensionless units.  

\subsection{Diffusion limit}

Here, we consider a diffusion problem in spherical symmetry, involving a static background with constant scattering opacity~$\sigma$.  
(The absorption opacity is set to zero.)  
This problem is adapted from \citet{pons_etal_2000} \citep[see also][]{oconnor_2015}.  
For sufficiently high scattering opacity, the moment equations limit to a diffusion equation for the number density.  
With a Gaussian initial distribution $\cD_{0}(r)=\exp(-3\,\sigma\,r^{2}/(4\,t_{0}))$, the analytical solution to the limiting diffusion equation is given by
\begin{equation}
	\cD(r,t)=\Big(\f{\sigma}{t_{0}+t}\Big)^{3/2}\,\exp\Big\{\,-\f{3\,\sigma\,r^{2}}{4\,(t_{0}+t)}\,\Big\},
\end{equation}
while the radial component of the number flux density is obtained from $\cI^{r}=-\pd{\cD}{r}/(3\,\sigma)$.  

We consider two values for the scattering opacity: one with $\sigma=10^{2}$ (thin test) and one with $\sigma=10^{5}$ (thick test).  
In the thin test, the computational domain extends over $r\in[0,5/3]$, $t_{0}=1$, and is run until $t+t_{0}=5$.  
In the thick test, the computational domain extends over $r\in[0,5/6]$ and $t_{0}=200$, and is run until $t+t_{0}=400$.  
In both tests, we use 50 quadratic elements ($k=2$) and the PD-ARS IMEX scheme.  
With the mean-free path defined as $\lambda=1/\sigma$, the ratio of the mean-free path to the cell width $\Delta r$ (the Knudsen number, ${\rm Kn}$) is $3\times10^{-1}$ for the thin test, and $6\times10^{-4}$ for the thick test.  

\begin{figure}[h]
	\begin{center}
        \captionsetup[subfigure]{justification=centering}
        \subfloat[$\sigma=10^{2}$]
        {\begin{minipage}{0.45\textwidth}
                        \includegraphics[width=\linewidth]{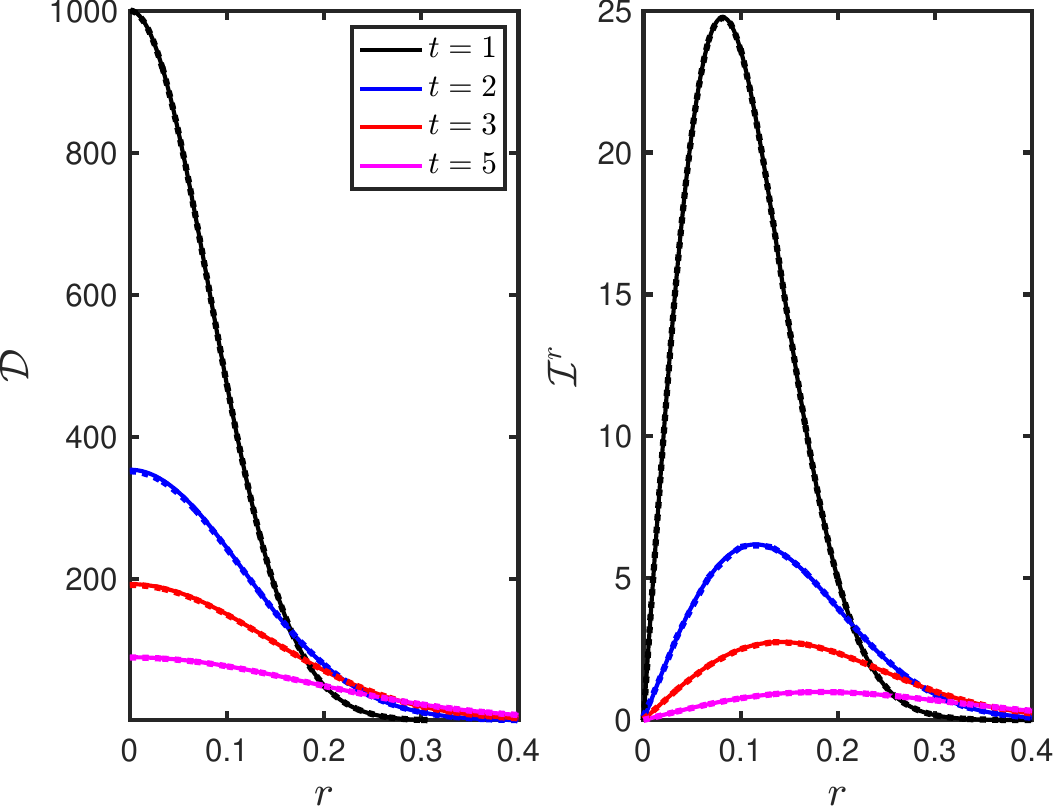}
                        \label{fig:}
                \end{minipage}
        }~~~~~~~
        \subfloat[$\sigma=10^{5}$]
        {\begin{minipage}{0.45\textwidth}
                        \includegraphics[width=\linewidth]{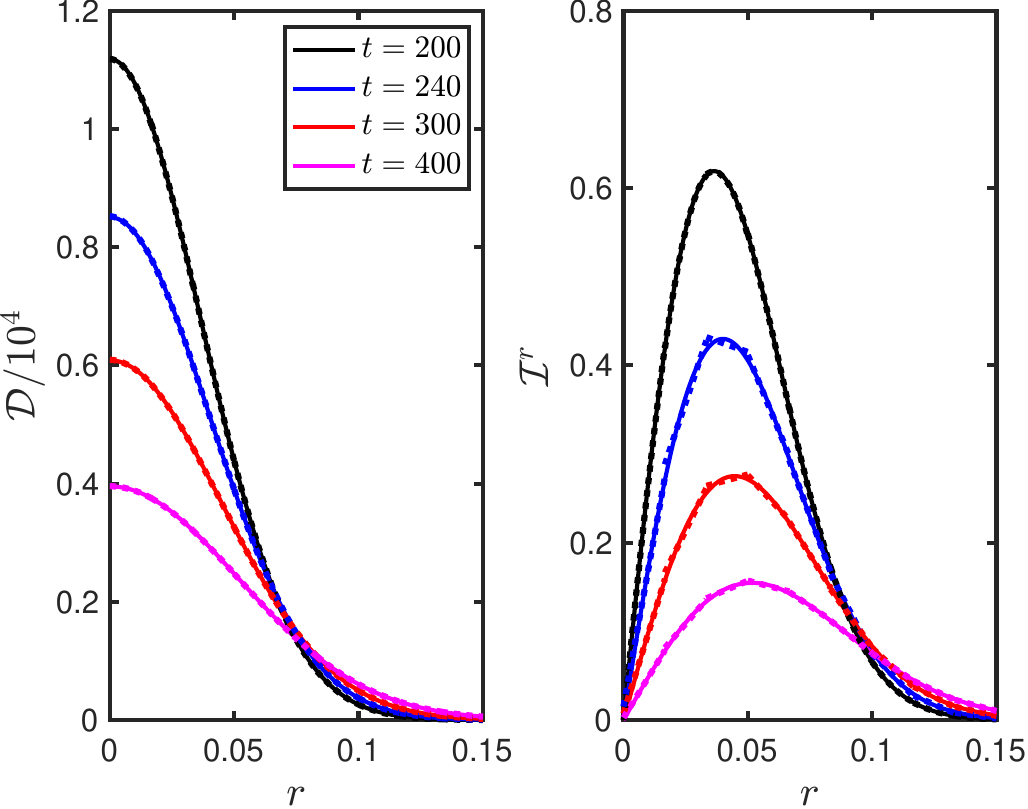}
                        \label{fig:}
                \end{minipage}
        }
        \caption{Results for the diffusion limit problem for the thin ($\sigma=10^{2}$; two leftmost panels) and thick ($\sigma=10^{2}$; two rightmost panels) configurations.  For each configuration, the number density and number flux density are plotted versus radius for select times.  The numerical solution is plotted with dotted lines, while the analytical solution is plotted with solid lines.}
        \label{fig:sphericalDiffusion}
        \end{center}
\end{figure}

Results are plotted in Figure~\ref{fig:sphericalDiffusion}.  
There is good agreement between the numerical and analytical solutions, for both the thin (two leftmost panels) and thick (two rightmost panels) cases.  
We emphasize that the use of finite-volume methods in diffusive regimes (${\rm Kn}\ll1$) requires modifications to the numerical fluxes to reduce numerical dissipation to perform well on this test.  
For $k\ge1$, DG methods are able to capture the diffusion limit without modified numerical fluxes \citep[e.g.,][]{larsenMorel_1989,adams_2001}.  

\subsection{Radiating sphere}

This test, adapted from \citet{muller_etal_2010}, considers free-streaming radiation propagating through a spatially varying velocity field.  
The absorption and scattering opacities are both set to zero.  
We consider a one-dimensional (spherically symmetric) spatial domain $D_{\bx}=[10^{1},10^{4}]$, and let the velocity profile be given by
\begin{equation}
	v^{r} = 
	\left\{
	\begin{array}{ll}
		0, & r \le 135 \\
		-v_{\max}^{r}\times(r-135)/15, & r\in(135,150] \\
		-v_{\max}^{r}\times(150/r)^{2}, & r>150
	\end{array}
	\right.,
\end{equation}
where we set $v_{\max}^{r}=0.2$.  
We set the energy domain to $D_{\varepsilon}=[0,300]$.
In this test, we discretize the spatial and energy domains into $200$ and $32$ elements, respectively, and use linear elements ($k=1$) and SSPRK2 time stepping.
We use geometrically increasing elements in both phase-space dimensions.  
For the spatial grid, the width of the first radial element is $(\Delta r)_{1}=1$, which increases with a constant factor of about $1.0287$, so that the width of the last element is $(\Delta r)_{200}=280$.  
Similarly, for the energy grid, the width of the first element is $\Delta\varepsilon_{1}=1$, which increases with a constant factor of $1.1192$, resulting in $\Delta\varepsilon_{32}=32.9$.  
In the computational domain, the moments are initially set to $\cD=1\times10^{-40}$ and $\cI^{r}=0$ for all $(r,\varepsilon)\in D_{\bx}\times D_{\varepsilon}$.
At the inner spatial boundary, we impose an incoming, forward-peaked radiation field with a Fermi-Dirac spectrum; i.e., we set $\cD(\varepsilon,r=10)=1/[\exp(\varepsilon/3-3)+1]$ and $\cI^{r}(\varepsilon,r=10)=0.999\times\cD(\varepsilon,r=10)$, so that the flux factor $h\approx1$.
(We impose outflow boundary conditions at the outer spatial boundary.)
Then, for $t>0$, a radiation front propagates through the computational domain, and a steady state is established for $t\gtrsim10^{4}$, where the spectrum is Doppler-shifted according to the velocity field.
From special relativistic considerations, assuming forward-peaked and free-streaming radiation, the analytical spectral number density in the steady state can be written as \citep[cf.][]{just_etal_2015}
\begin{equation}
        \cD_{\rm A}(\varepsilon,r)=\Big(\f{10}{r}\Big)^{2}\times\f{s^{2}}{\exp(s\varepsilon/3-3)+1},
        \label{eq:dopplerSpectraSR}
\end{equation}
where $s=\sqrt{(1+v^{r})/(1-v^{r})}$.  
We run all simulations until $t=2\times10^{4}$, when an approximate steady state has been reached, and we use the expression in Equation~\eqref{eq:dopplerSpectraSR} to compare with our numerical solutions.  

\begin{figure}[h]
	\begin{center}
        \captionsetup[subfigure]{justification=centering}
       {\begin{minipage}{0.45\textwidth}
                        \includegraphics[width=\linewidth]{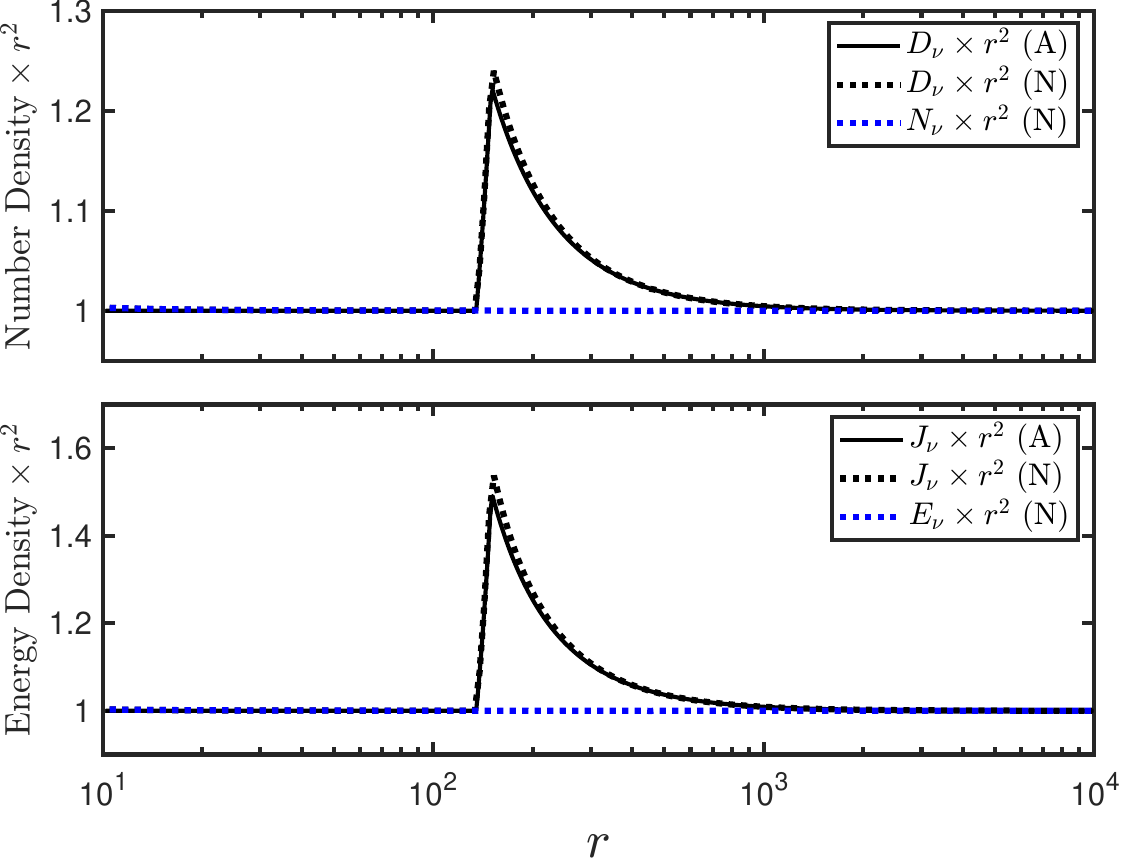}
                        \label{fig:}
                \end{minipage}
        }~~~~~~~
        {\begin{minipage}{0.425\textwidth}
                        \includegraphics[width=\linewidth]{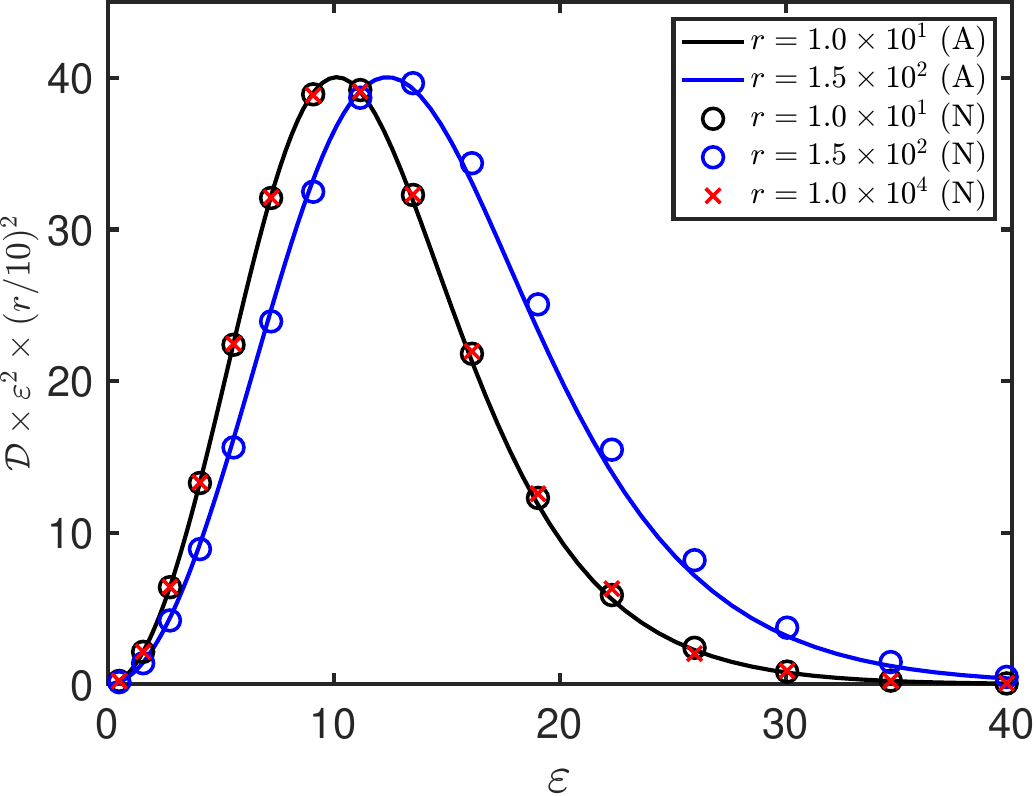}
                        \label{fig:}
                \end{minipage}
        }
        \caption{Results for the radiating sphere test at $t=10^{4}$.  Left panels show energy-integrated number densities (top panel) and energy densities (bottom panel) multiplied by $r^{2}$, versus radius.  
        In each of these panels we show the comoving- and Eulerian-frame number and energy densities, and results have been scaled to values at the outer radial boundary.  
        In the top left panel, we plot the comoving-frame number density $[D_{\nu}(r)/D_{\nu}(r=10^{4})]\times(r/10^{4})^{2}$ (dotted black) and the Eulerian-frame number density $[N_{\nu}(r)/N_{\nu}(r=10^{4})]\times(r/10^{4})^{2}$ (dotted blue).  
        In the bottom left panel, we plot the comoving-frame energy density $[J_{\nu}(r)/J_{\nu}(r=10^{4})]\times(r/10^{4})^{2}$ (dotted black) and the Eulerian-frame energy density $[E_{\nu}(r)/E_{\nu}(r=10^{4})]\times(r/10^{4})^{2}$ (dotted blue).  
        In both of these panels, the analytical solution, obtained using Equation~\eqref{eq:dopplerSpectraSR}, is plotted for the comoving-frame quantity (solid black).
        In the right panel, spectral particle distributions are plotted for various radii.  Analytic spectra are plotted for $r=10^{1}$ (solid black) and $r=1.5\times10^{2}$ (solid blue).  Numerically obtained spectra are plotted for $r=10^{1}$ (black circles), $r=1.5\times10^{2}$ (blue circles), and $r=10^{4}$ (red crosses).}
        \label{fig:radiatingSphereCharacteristics}
        \end{center}
\end{figure}

Figure~\ref{fig:radiatingSphereCharacteristics} shows the obtained solution characteristics for the radiating sphere test.  
The left panels plot the energy-integrated comoving-frame number density (top) and energy density (bottom) multiplied by $r^{2}$ and scaled to values at the outer spatial boundary.  
Because the radiation field is forward-peaked and free-streaming ($I_{\nu}^{r}\approx D_{\nu}$), these quantities also correspond to number and energy luminosities, respectively.
The numerically obtained values (dotted black lines) compare well with the analytic solutions (solid black lines) obtained by integrating Equation~\eqref{eq:dopplerSpectraSR} over energy with appropriate weights.  
We also plot the corresponding Eulerian-frame number and energy densities multiplied by $r^{2}$ (dotted blue line, respectively, in each panel).  
These quantities should be unaffected by the velocity field, and remain constant with radius.  
As can be seen in the figure, this property is maintained well in the numerical solutions.  

The right panel in Figure~\ref{fig:radiatingSphereCharacteristics} shows computed comoving-frame spectral particle distributions at various radii (symbols), compared with analytic spectra obtained from Equation~\eqref{eq:dopplerSpectraSR} (solid lines).  
The black circles show the spectra of the incoming radiation field at $r=10^{1}$, where each circle represents the cell average in an element.  
As expected, there is good agreement with the analytic spectrum (solid black line).  
At $r=1.5\times10^{2}$, at the peak velocity magnitude ($v^{r}=0.2$), the comoving-frame spectrum is significantly blue-shifted (see blue solid line and circles).  
There is still good agreement between the numerical and analytic spectra at this location, with the numerical solution slightly above the analytic solution past the peak of the spectrum.  
At the outer boundary, where the velocity is almost back to zero, the numerical spectrum has shifted back (red crosses) and agrees well with the incoming spectrum.  

\subsection{Differentially expanding isothermal atmosphere}

This problem was studied by \citet{mihalas_1980}, using a special relativistic radiative transfer model, and later considered by \citet{ramppJanka_2002} and \citet{just_etal_2015}, using $\cO(v)$ models.  
It considers the transport of neutral particles (photons) through an expanding atmosphere, where the photons interact with the background through emission and absorption.  
We consider a one-dimensional (spherically symmetric) spatial domain, and let $D_{\bx}\times D_{\varepsilon}=[0,15]\times[0,12]$.  
We discretize the spatial and energy domains into $200$ and $50$ linear elements ($k=1$), respectively, and use the PD-ARS IMEX time stepping scheme.
For $r\in[0,R)$, the absorption opacity is specified as
\begin{equation}
	\chi(\varepsilon,r) 
	=
	\left\{\begin{array}{ll}
		\alpha\times\big\{\,10\times\exp\big[-(\varepsilon-\varepsilon_{0})^{2}/\Delta^{2}\big]+\big(\,1-\exp\big[-(\varepsilon-\varepsilon_{0})^{2}/\Delta^{2}\big]\,\big)\,\big\} / r^{2} & \varepsilon\le\varepsilon_{0} \\
		10\times\alpha/r^{2} & \varepsilon>\varepsilon_{0}
	\end{array}\right.,
	\label{eq:expandingAtmosphereOpacity}
\end{equation}
while $\chi=0$ for $r\ge R$.  
In the interaction region ($r<R$), the opacity falls off as $r^{-2}$.  
At a given radius, the opacity is energy-independent except within a transition region of width $\sim\Delta$ centered at $\varepsilon_{0}$, over which it increases by approximately one order of magnitude.  
The equilibrium distribution is given by that of a Bose--Einstein distribution with unit temperature; i.e., $\cD^{\Eq}=1/(e^{\varepsilon}-1)$.  
In Equation~\eqref{eq:expandingAtmosphereOpacity}, $\alpha=10.9989$, $\Delta=0.2$, $\varepsilon_{0}=3$, and $R=11$.  
The additional vacuum layer outside $r=R$ is included to ease the treatment of the outer boundary, where we impose outflow boundary conditions \citep{just_etal_2015}.  
Reflecting boundary conditions are imposed at $r=0$.  
The radial velocity profile is given by
\begin{equation}
	v^{r} = 
	\left\{
	\begin{array}{ll}
		0, & r < r_{\min} \\
		v_{\max}^{r}\times\f{r-r_{\min}}{(r_{\max}-r_{\min})}, & r\in[r_{\min},r_{\max}] \\
		v_{\max}^{r}, & r>r_{\max}
	\end{array}
	\right.,
	\label{eq:expandingAtmosphereVelocity}
\end{equation}
where $r_{\min}=1$ and $r_{\max}=R$.  
That is, the radial velocity increases linearly from zero to $v_{\max}^{r}$ for $r\in[r_{\min},r_{\max}]$ and is kept constant at $v_{\max}^{r}$ for $r>r_{\max}$.  
Initially, the moments are set to $\cD=1\times10^{-10}$ and $\cI^{r}=0$ for all $(r,\varepsilon)\in D_{\bx}\times D_{\varepsilon}$, and we evolve to $t=15$, when an approximate steady state is obtained.  
We compute solutions for $v_{\max}^{r}\in\{0.0,0.1,0.3\}$.  

\begin{figure}[h]
	\begin{center}
        \captionsetup[subfigure]{justification=centering}
       {\begin{minipage}{0.425\textwidth}
                        \includegraphics[width=\linewidth]{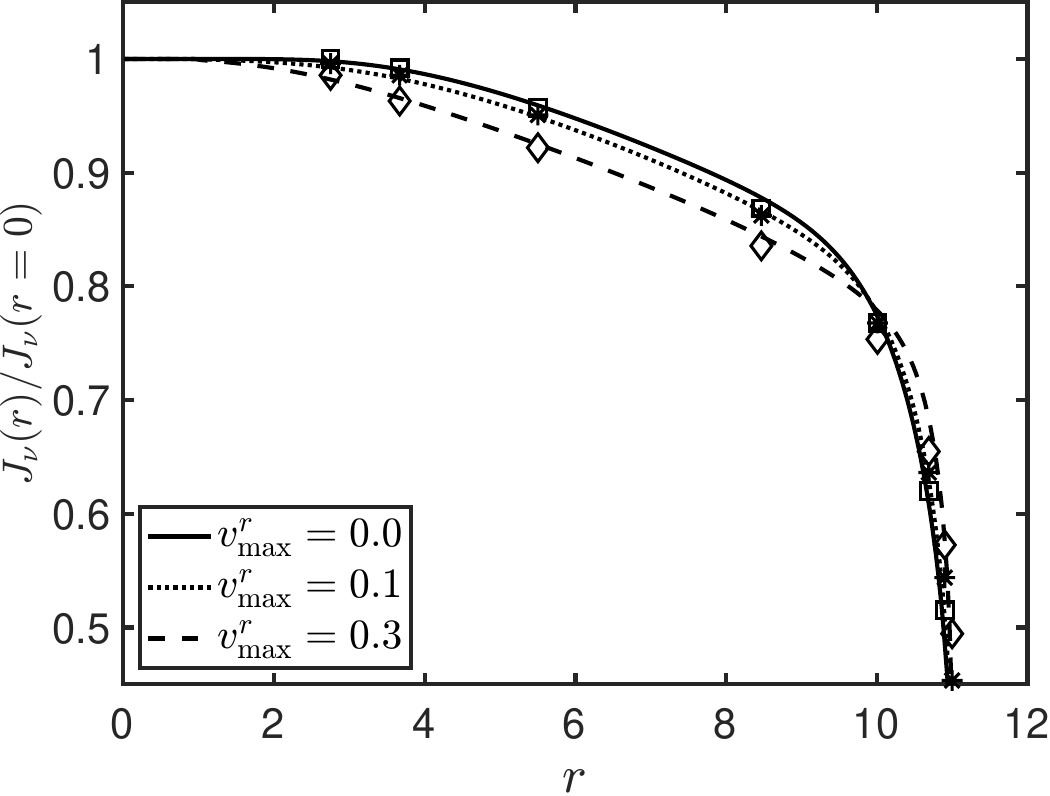}
                        \label{fig:}
                \end{minipage}
        }~~~~~~~
        {\begin{minipage}{0.425\textwidth}
                        \includegraphics[width=\linewidth]{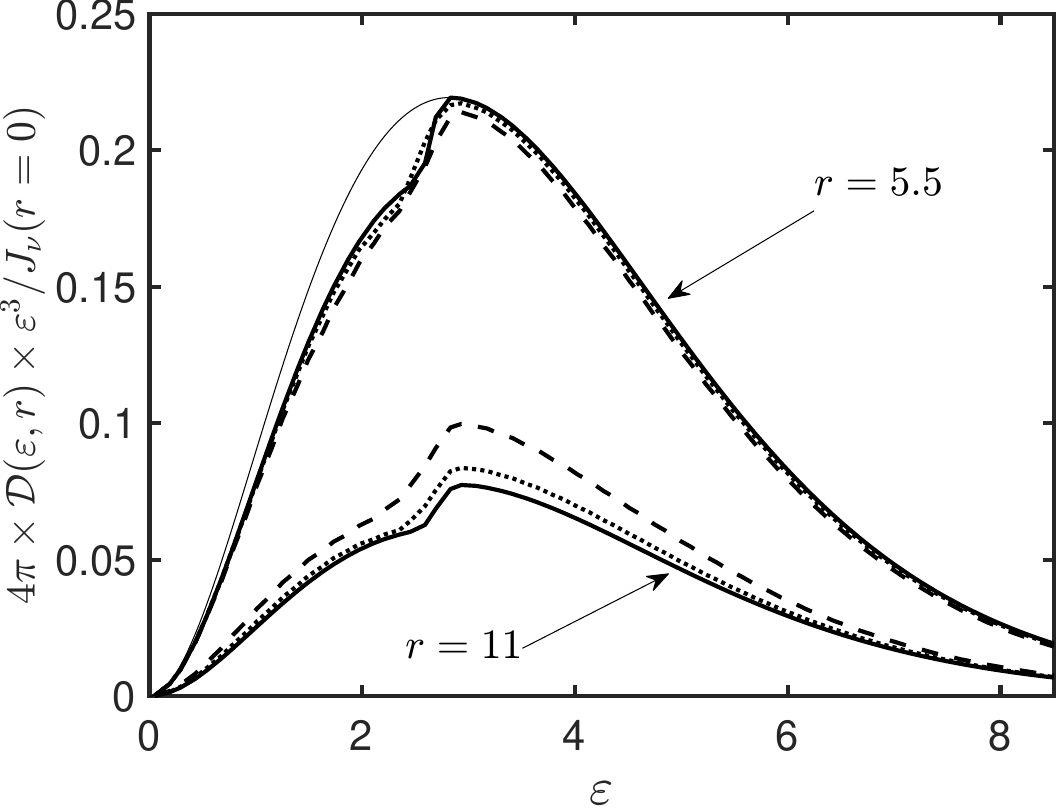}
                        \label{fig:}
                \end{minipage}
        }
        \caption{Results for the differentially-expanding isothermal atmosphere test.  The left panel shows the comoving-frame energy density versus radius, scaled to the value at $r=0$, for three values of $v_{\max}^{r}$ in Equation~\eqref{eq:expandingAtmosphereVelocity}: $0.0$ (solid), $0.1$ (dotted), and $0.3$ (dashed).  The symbols --- boxes ($v_{\max}^{r}=0.0$), stars ($v_{\max}^{r}=0.1$), and diamonds ($v_{\max}^{r}=0.3$) --- correspond to special relativistic solutions read off from \cite{mihalas_1980}, provided by \citet[][private communication]{just_etal_2015}.  The right panel shows scaled energy spectra for the models plotted in the left panel (with mathching line styles) at two select radii: $r=5.5$ and $r=11$.  In this panel, the thin solid black line represents the equilibrium spectrum.}
        \label{fig:expandingAtmosphere}
        \end{center}
\end{figure}

Figure~\ref{fig:expandingAtmosphere} plots the comoving-frame energy density versus radius (left panel) and spectral energy distributions at $r=5.5$ and $r=11$ (right panel).  
The figure shows good qualitative agreement with Figure~2 in \citet{just_etal_2015}, exhibiting the same qualitative solution features for both panels.  
The symbols (boxes, stars, and diamonds) in the left panel, extracted from Figure~5 of \citet{mihalas_1980} and kindly provided by Oliver Just (private communication), show good agreement with solutions of the special-relativistic radiative transfer equation.  
Inside $r\approx 5$, where the background medium is relatively opaque and velocities are non-relativistic for all models, the agreement with the transport solutions is very good.  
For larger radii both the two-moment and $\cO(v)$ approximations contribute to deviations from the relativistic transport solutions, but the general trends are still well captured.  

\subsection{Shadow Casting -- Cylindrical Coordinates}

Next, we consider a shadow casting test, which is often used to highlight the advantage of two-moment models, over flux-limited diffusion models, in their ability to capture shadow regions \citep[e.g.,][]{hayesNorman_2003,skinnerOstriker_2013,just_etal_2015,kuroda_etal_2016,oconnorCouch_2018}.  
\citet{kuroda_etal_2016} also used a version of this test to emphasize challenges of transitioning to free-streaming radiation, where the particle flux approaches the density (i.e., the boundary of the realizable domain).  
To exercise the use of cylindrical $(R,z)$ coordinates in the implementation of the two-moment model in \thornado, we adopt the setup from \citet{oconnorCouch_2018}.  

\begin{figure}[h]
	\begin{center}
		\includegraphics[width=0.6\linewidth]{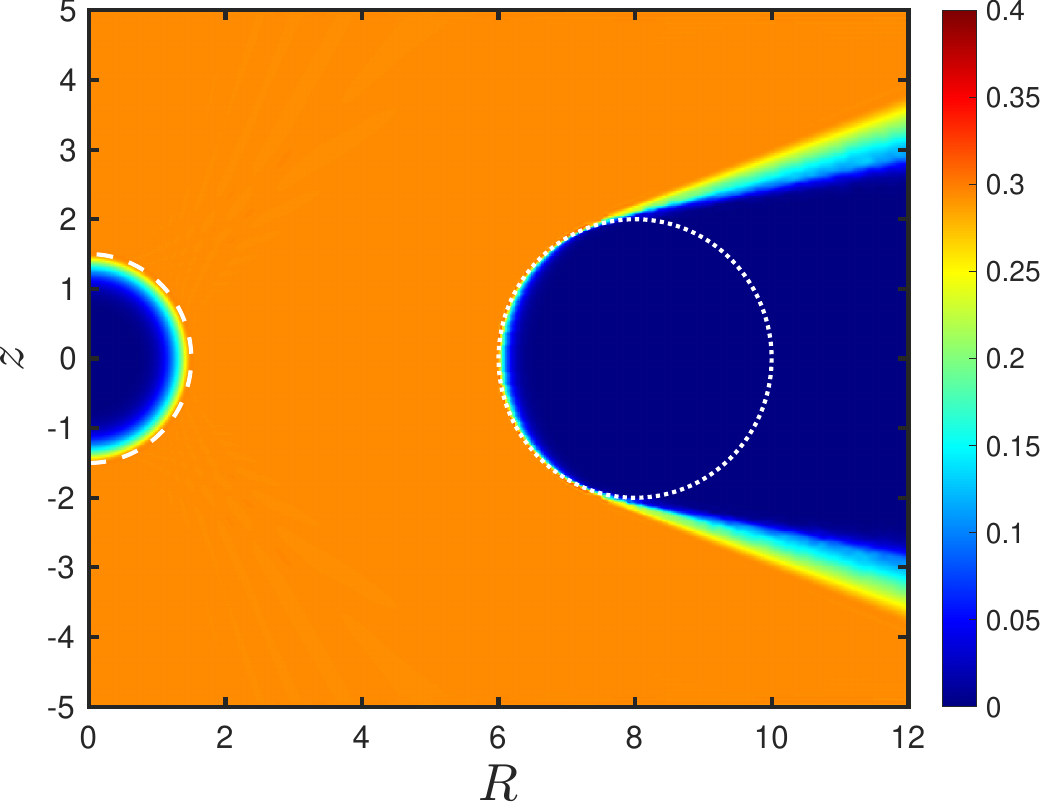}
        		\caption{Results for the shadow casting test in cylindrical coordinates.  
		The colormap displays $2\pi\,r^{2}\,|\bI|$, where $|\bI|$ is the magnitude of the particle flux density and $r^{2}=R^{2}+z^{2}$.  
		The white dashed line indicates the radius of the source regions, while the white dotted line traces out the boundary of the absorbing region.}
        		\label{fig:shadowCasting}
        \end{center}
\end{figure}

We consider a two-dimensional spatial domain, $D_{\bx}=(R,z)\in[0,12]\times[-5,5]$, which is discretized into $168\times140$ quadratic elements ($k=2$).  
Because the background is at rest (i.e., no relativistic observer corrections contribute), we perform this test for a single particle energy.  
The numerical setup consists of a source region centered on the origin, $\cS=\{(R,z)~|~r^{2}=R^{2}+z^{2}\le R_{\cS}^{2}\}$, and an absorbing region centered on $(R_{\cA,0},0)$, $\cA=\{(R,z)~|~(R-R_{\cA,0})^{2}+z^{2}\le R_{\cA}^{2}\}$, where we set $R_{\cS}=1.5$, $R_{\cA,0}=8$, and $R_{\cA}=2$.  
Inside the source region, the absorption opacity is set to $\chi=10\times\exp[\,-2\,(r/\cR_{\cS})^{2}\,]$ and $\cD^{\Eq}=0.1$.  
Inside the absorbing region, $\chi=10$ and $\cD^{\Eq}=0$.  
The scattering opacity is set to zero everywhere.  
To avoid numerical artifacts associated with discontinuous opacities across the source radius, we apply a small amount of smoothing (on the order of two elements) of the source region.  
Initially, we set $\cD=10^{-10}$ and $\cI^{R}=\cI^{z}=0$ in the computational domain.  
We use reflecting boundaries along the symmetry axis ($R=0$) and outflow conditions elsewhere.  
We use the PD-ARS IMEX time stepping scheme, and integrate until an approximate steady state has been reached ($t=15$).  

Figure~\ref{fig:shadowCasting} shows a color map of $2\pi r^{2}|\bI|$ \citep[cf.][]{oconnorCouch_2018} at $t=15$.  
In a steady state, away from the source and absorbing regions, this quantity should be constant, and our numerical results agree with this expectation.  
A shadow is clearly seen behind the absorbing region, opposite of the source region.  
There is a gradual transition between the shadow and the free-streaming region.  
This is partially due to numerical diffusion, but also because the source is extended (not a point source).  

\subsection{Transparent Vortex -- Spherical Polar Coordinates}

This test, adapted from \citet[][]{just_etal_2015}, considers the propagation of free-streaming radiation through a velocity vortex.  
See also \citet{laiu_etal_2025} for results obtained with \thornado\ using Cartesian coordinates.  
Here, we employ spherical polar $(r,\theta)$ coordinates in two spatial dimensions, and consider a spatial domain $D_{\bx}=(r,\theta)\in[2,10]\times[0,\pi/2]$, which is discretized into $48\times48$ quadratic elements ($k=2$).  
We set the energy domain to $D_{\varepsilon}=[0,50]$, which is discretized into $32$ quadratic elements.  
In the computational domain, the moments are initially set to $\cD=1\times10^{-8}$ and $\cI^{r}=\cI^{\theta}=0$ for all $(r,\theta,\varepsilon)\in D_{\bx}\times D_{\varepsilon}$.
As in the radiating sphere test, at the inner spatial boundary we impose an incoming, forward-peaked radiation field with a Fermi-Dirac spectrum; i.e., we set $\cD(\varepsilon,r=2,\theta)=1/[\exp(\varepsilon/3-3)+1]$, $\cI^{r}(\varepsilon,r=2,\theta)=0.999\times\cD(\varepsilon,r=2,\theta)$, and $\cI^{\theta}(\varepsilon,r=2,\theta)=0$.  
We impose outflow boundary conditions at the outer radial boundary, and reflecting boundary conditions at $\theta=0$ and $\theta=\pi/2$.  

\begin{figure}[h]
	\begin{center}
        \captionsetup[subfigure]{justification=centering}
       {\begin{minipage}{0.32\textwidth}
                        \includegraphics[width=\linewidth]{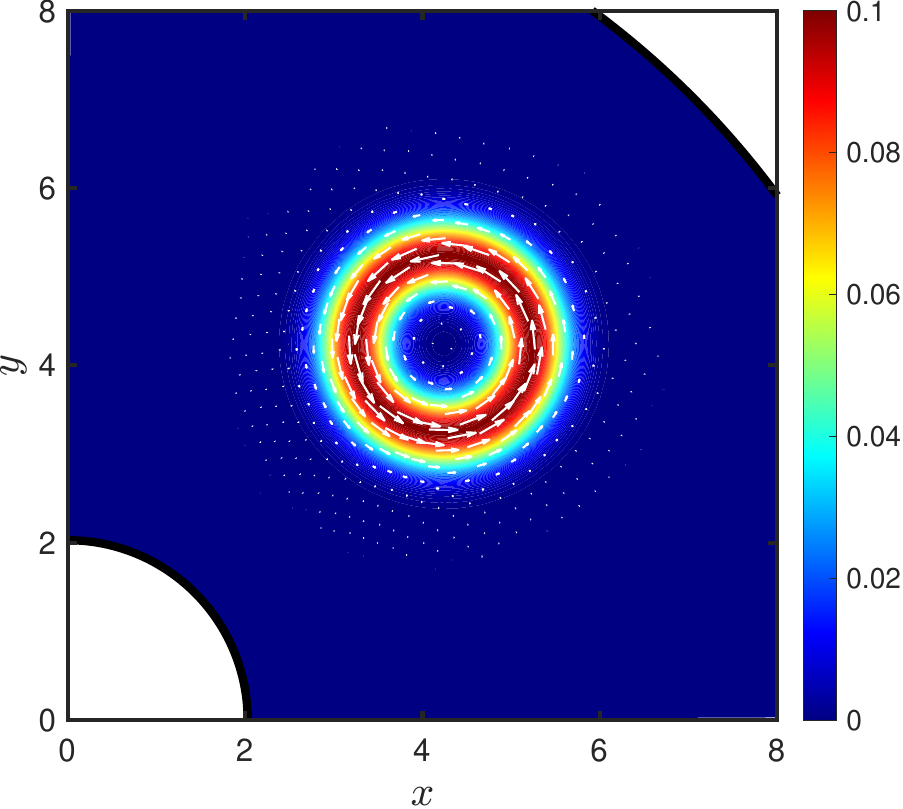}
                        \label{fig:}
                \end{minipage}
        }~~~~
        {\begin{minipage}{0.32\textwidth}
                        \includegraphics[width=\linewidth]{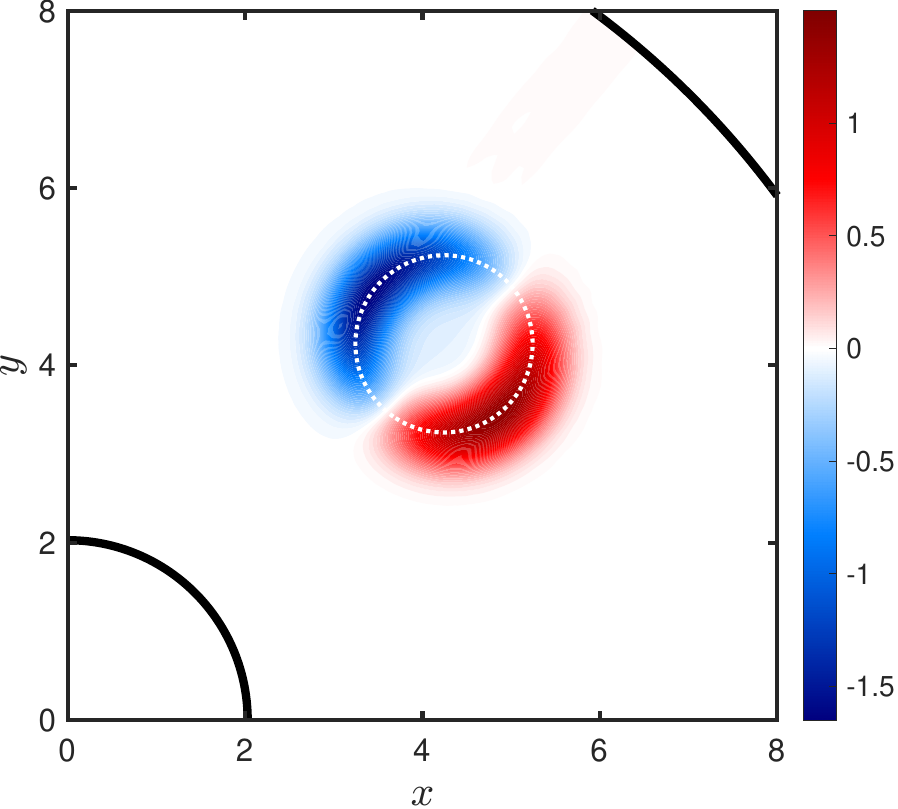}
                        \label{fig:}
                \end{minipage}
        }~~~~
        {\begin{minipage}{0.30\textwidth}
                        \includegraphics[width=\linewidth]{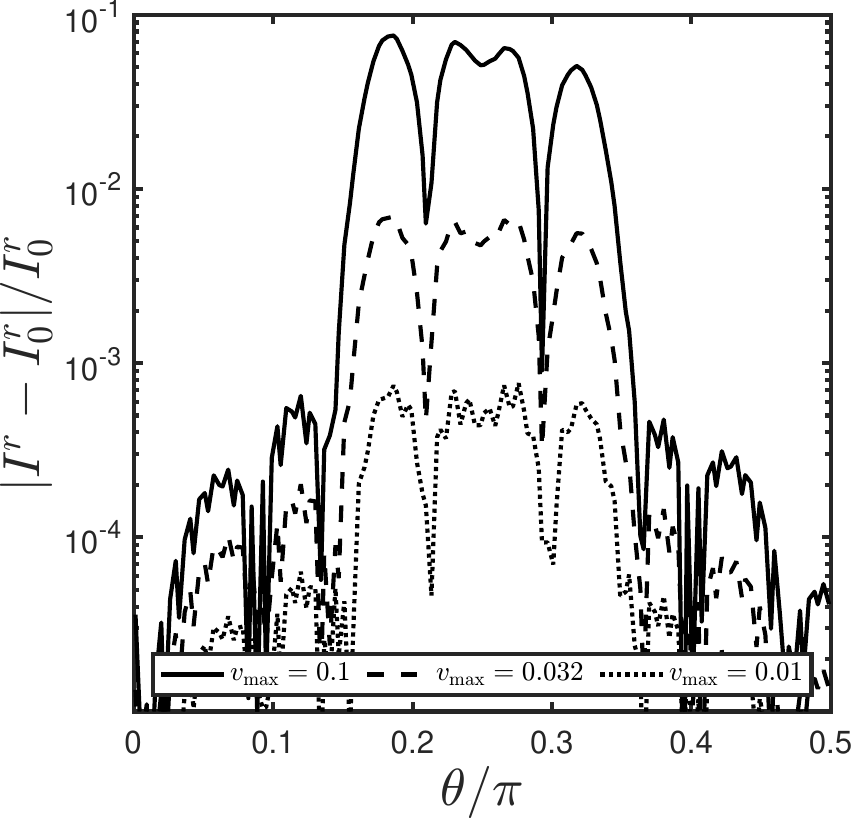}
                        \label{fig:}
                \end{minipage}
        }
        \caption{Results for the transparent vortex problem.
        The colormap (left panel) displays the magnitude of the velocity field for the model with $v_{\max}=0.1$, with velocity vectors overlaid (in white).
        The colormap in the middle panel displays the root--mean--squared energy, $\varepsilon_{\rm RMS,0}-\varepsilon_{\rm RMS}$, where $\varepsilon_{\rm RMS,0}\approx15.6$ is the value at the inner boundary ($r=2$).  
        In this panel, the dotted white line traces out the circle where $v=v_{\max}=0.1$.  
        The right panel plots the relative difference in the radial component of the energy-integrated number flux (relative to a model with $v_{\max}=0$) versus polar angle for models with various values of $v_{\max}$: $0.1$ (solid), $0.032$ (dashed), and $0.01$ (dotted).}
        \label{fig:transparentVortex}
        \end{center}
\end{figure}

For $t>0$, some of the radiation emanating from the inner boundary propagates towards, and then through, a velocity vortex centered on $(x_{0},y_{0})^{\intercal}=(6,6)^{\intercal}/\sqrt{2}$.  
The Cartesian components of the velocity field are given by
\begin{equation}
	v^{x} = -\f{(y-y_{0})}{R_{0}}\times V
	\quad\text{and}\quad
	v^{y} = \f{(x-x_{0})}{R_{0}}\times V,
	\label{eq:vortexVelocityField}
\end{equation}
respectively, where
\begin{equation}
	V=v_{\max}\times\exp\Big[-\Big(\f{R_{0}-1}{0.4}\Big)^{2}\Big]
\end{equation}
and $R_{0}=\sqrt{(x-x_{0})^{2}+(y-y_{0})^{2}}$.  
Given the velocity components in Equation~\eqref{eq:vortexVelocityField}, the spherical-polar components are $v^{r}=(v^{x}\,\sin\theta+v^{y}\,\cos\theta)$ and $r v^{\theta}=(v^{x}\,\cos\theta-v^{y}\,\sin\theta)$.  
Figure~\ref{fig:transparentVortex} (left panel) shows the velocity field for a model with $v_{\max}=0.1$.  
For $t\gtrsim8$, the system establishes a steady state, where inflow at the inner boundary is balanced by outflow at the outer boundary.  

\begin{figure}[h]
	\begin{center}
		\captionsetup[subfigure]{justification=centering}
       		{\begin{minipage}{0.425\textwidth}
                        \includegraphics[width=\linewidth]{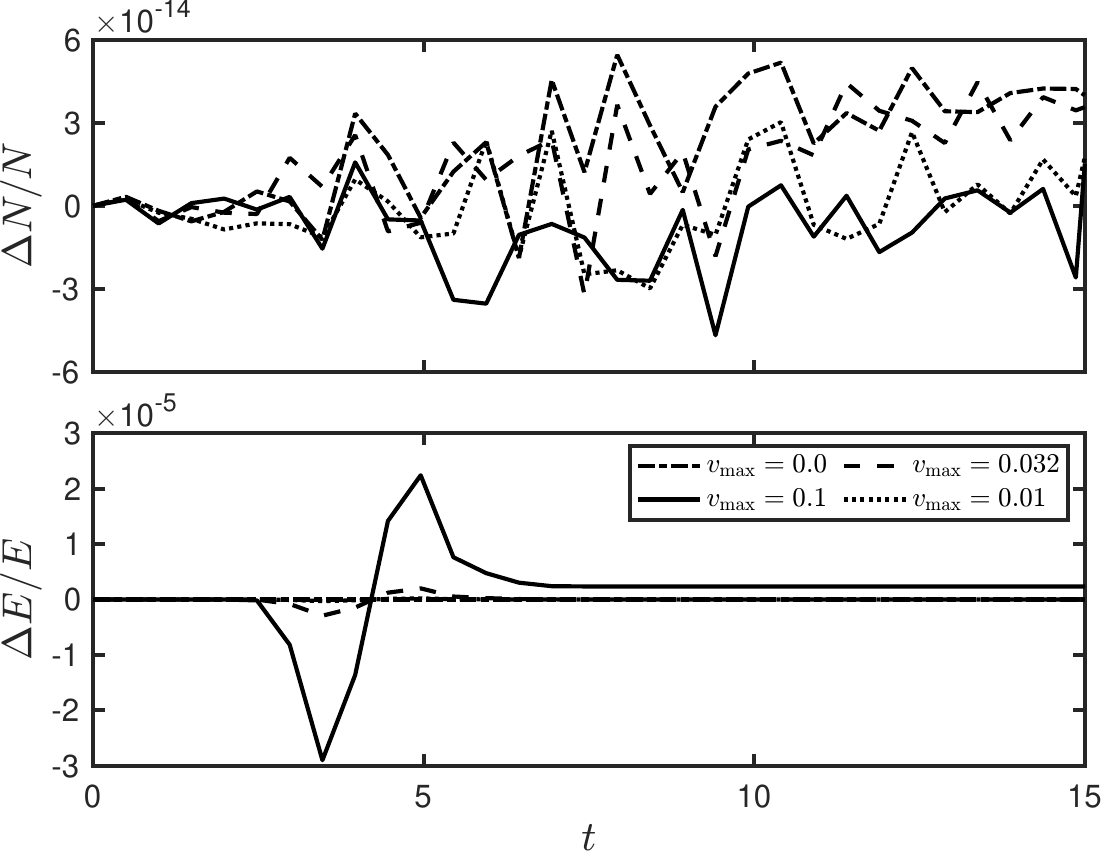}
                        \label{fig:}
                	\end{minipage}
        		}~~~~~~~
        		{\begin{minipage}{0.4\textwidth}
                        \includegraphics[width=\linewidth]{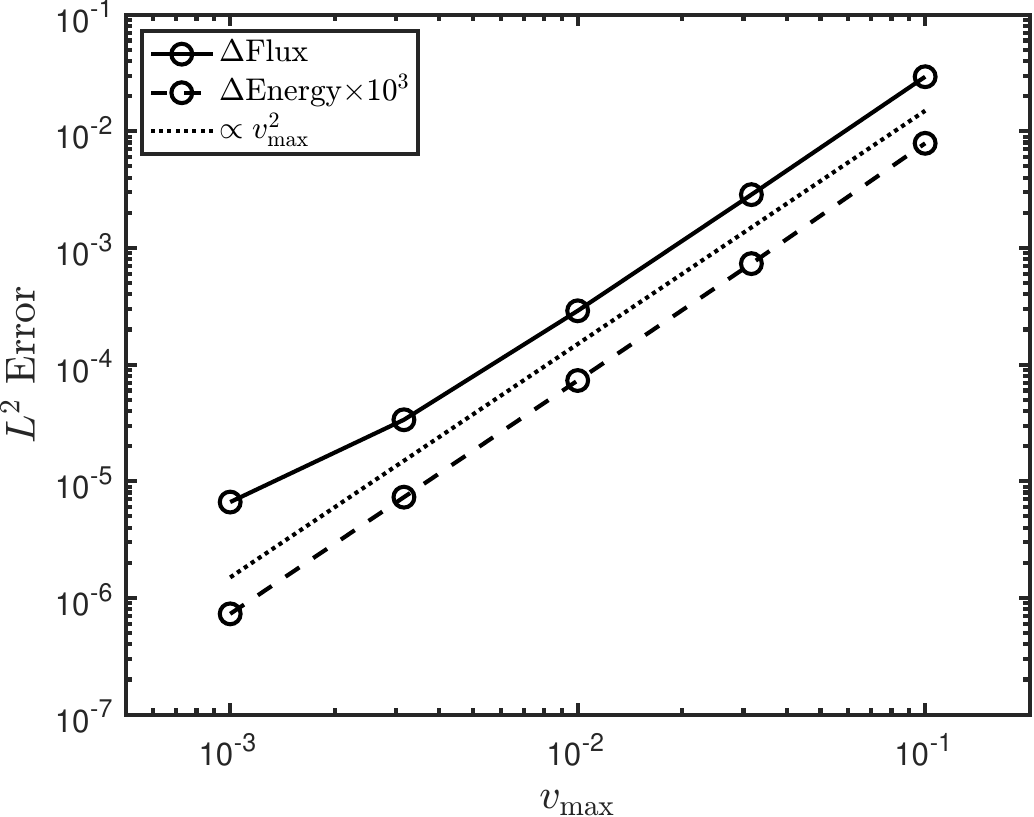}
                        \label{fig:}
                	\end{minipage}
        		}
        		\caption{Left panels show Eulerian-frame number and energy conservation for the transparent vortex test.  
		The upper left panel shows the relative change in the Eulerian-frame particle number versus time, while the lower left panel shows the relative change in the Eulerian-frame particle energy versus time.
		In each of these panels, the change is relative to the respective value at $t=15$ (see text for details).  
		In both left panels, results are plotted for models with different values of $v_{\max}$: $0.0$ (dash dot), $0.1$ (solid), $0.032$ (dashed), and $0.01$ (dotted).  
		In the right panel, we plot (versus $v_{\max}$) the $L^{2}$ error norm of the particle flux at the outer boundary (solid line with circles) and the relative change in the Eulerian-frame energy (dashed line with symbols).  (For plotting purposes, the energy change has been multiplied by a factor $10^{3}$ to make the numerical values comparable to those for the solid line.)  
		The dotted reference line is proportional to $v_{\max}^{2}$.}
        		\label{fig:transparentVortexConservation}
        \end{center}
\end{figure}

For $v_{\max}\ne0$, the comoving-frame radiation moments are modified by the spatially varying velocity field.  
For example, a comoving observer instantaneously moving towards the inner boundary will measure a blue-shifted spectrum due to the relativistic Doppler effect.  
The middle panel of Figure~\ref{fig:transparentVortex} shows a colormap of the root mean square (RMS) energy, 
\begin{equation}
	\varepsilon_{\rm RMS} = \sqrt{\int_{D_{\varepsilon}}\cD(\varepsilon)\varepsilon^{5}d\varepsilon/\int_{D_{\varepsilon}}\cD(\varepsilon)\varepsilon^{3}d\varepsilon},
	\label{eq:rmsEnergy}
\end{equation}
subtracted from the RMS energy at the inner boundary $\varepsilon_{{\rm RMS},0}$ for a model with $v_{\max}=0.1$ (i.e., $\varepsilon_{{\rm RMS},0}-\varepsilon_{{\rm RMS}}$ is plotted).  
The spectral shift (i.e., blueshift versus redshift) in the vortex agrees qualitatively with what can be expected given the velocity field in the left panel.  
In the wake of the vortex ($r\gtrsim7.5$), where the velocity is again close to zero, the RMS energy is in close agreement with that of the incident spectrum.  
Similar to \citet{just_etal_2015}, in the right panel in Figure~\ref{fig:transparentVortex} we plot the relative difference in the energy-integrated number-flux density at $r=10$ versus polar angle $\theta$ for models with various values of $v_{\max}$.  
The relative difference is computed with respect to the model with $v_{\max}=0$.  
While this difference should be identically zero, several factors can contribute to non-zero differences.  
\citet{just_etal_2015} point out both discretization and model errors contribute, where model errors are caused by neglecting velocity-dependent terms of $\cO(v^{2})$ and higher.  
\citet{laiu_etal_2025} propose that the two-moment model's inability to adequately capture angular aberration effects also contributes.  
In the right panel in Figure~\ref{fig:transparentVortexConservation} (solid black line with circles), we plot the $L^{2}$ norm of the quantity shown in the right panel in Figure~\ref{fig:transparentVortex} versus $v_{\max}$.  
The error increases as $v_{\max}^{2}$ (cf. dotted black line), from which we infer that we have accounted for all relevant $\cO(v)$ terms.  

The relative change in the Eulerian-frame particle number and energy --- Equations~\eqref{eq:numberDensityEulerian} and \eqref{eq:energyDensityEulerian}, respectively, integrated over the phase-space domain --- are plotted versus time in Figure~\ref{fig:transparentVortexConservation} (top and bottom left panels, respectively) for models with various values of $v_{\max}$.  
When computing the number and energy change, the fluxes through the domain boundaries have been taken into account \citep[as in][]{laiu_etal_2025}.  
Because the initial number density is essentially zero, the relative change is computed with respect to values at $t=15$.  
Since \thornado\ solves the equation for the Eulerian-frame number density in conservative form, the associated particle number is conserved to machine accuracy for all values of $v_{\max}$.  
However, as discussed in Section~\ref{sec:model}, and more extensively in \citet{laiu_etal_2025}, conservation of the Eulerian-frame energy can only be achieved approximately, even at the continuum level, because we neglect terms of $\cO(v^{2})$ and higher.  
As can be seen in the bottom left panel in Figure~\ref{fig:transparentVortexConservation}, the relative change in the Eulerian-frame energy increases with $v_{\max}$, and, for a given value of $v_{\max}$, peaks around the time when the initial radiation pulse, driven by the inner boundary condition, propagates through the velocity vortex.  
We plot the $L^{2}$ error of the relative change in the Eulerian-frame energy (integrated from $t=0$ to $t=15$) versus $v_{\max}$ in the right panel in Figure~\ref{fig:transparentVortexConservation} (dashed black line with circles).  
The $L^{2}$ error also grows as $v_{\max}^{2}$, which provides additional evidence that all the relevant $\cO(v)$ terms have been included \emph{and} that these terms have been discretized in a consistent manner to achieve Eulerian-frame energy conservation properties at $\cO(v)$, given that we solve a number-conservative $\cO(v)$ two-moment model.  

\subsection{Streaming Sine Wave with AMR}
\label{sec:results.sine_wave_streaming_amr}

This test, adapted from \cite{laiu_etal_2025}, models free-streaming radiation propagating along the $x^{1}$-direction through a background with a spatially and temporally constant velocity field.
The scattering and absorption opacities are set to zero, while the background velocity is prescribed as $\vect{v}=[v,0,0]^{\intercal}$ with $v=0.1c$.  
We consider a periodic spatial domain $D_{\bx}=[0,L]^{3}$, with $L=100$~cm, in Cartesian coordinates.  
(Distance and time for this problem are provided in centimeters and seconds, respectively, for compatibility with the CGS units used in \flashx.)  
The initial number density and flux are given by
\begin{equation}
	\cD(\bx,0)=\cD_{0}(\bx)=0.5+0.49\times\sin(2\pi x^{1}), \quad \cI^{1}(\bx,0)=\cD_{0}(\bx), \quad \cI^{2}(\bx,0)=\cI^{3}(\bx,0)=0.
\end{equation}
With this initialization, the flux factor is unity, and the nontrivial components of the analytic solution are
\begin{equation}
	\cD(\bx,t)=\cI^{1}(\bx,t)=\cD_{0}(x^{1}-t);
\end{equation}
i.e., the initial profile propagates in the $x^{1}$-direction at the (unit) speed of light.
To investigate the effect of AMR on the evolution of the radiation moment, we initialize the background mass density with a top hat profile in the $x^{1}$-direction.  
The mesh is configured to track the evolution of the density discontinuities, independent of the radiation field.  
The simulations are run until $t=L/(0.1~c) \approx 3.34\times 10^{-8}$~s, corresponding to one full crossing of the density profile through the computational domain and ten crossings of the radiation profile.

\begin{figure}[h]
    \centering
    \includegraphics[width=1.0\linewidth]{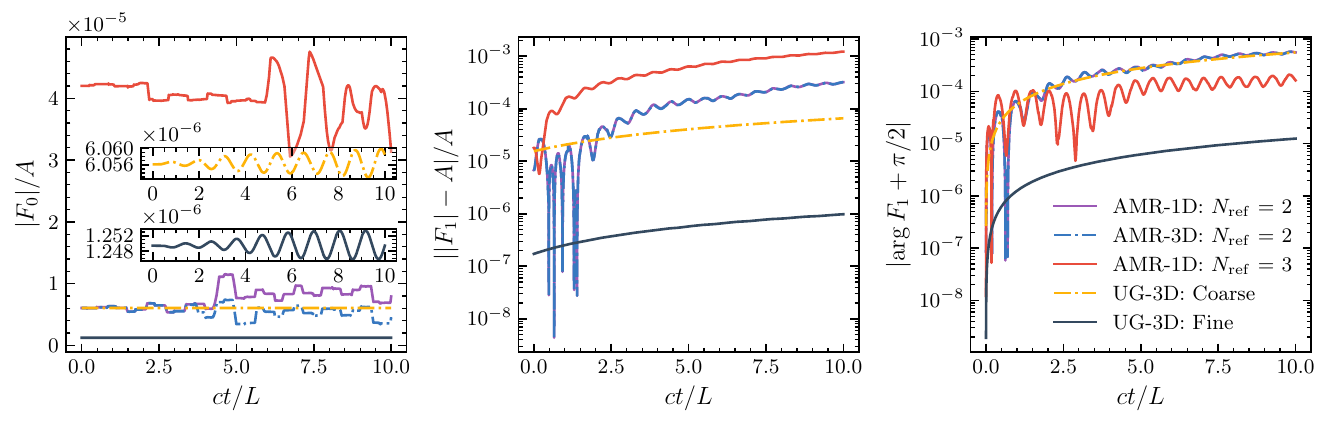}
    \caption{Results for the streaming sine wave test with AMR.  Each panel shows the drift (left), change in amplitude (middle), and change in phase (right) of the wave relative to the expected solution.  Results show the magnitudes and argument of the first few components of the spatial Fourier transform of the wave [see Equation~\eqref{eq:fourierTransform}] and are scaled relative to the initial amplitude ($A$) and phase.  Three AMR simulations, two-level refinement ($N_\text{ref} = 2$) in 1D (solid purple) and 3D (dash-dotted blue), and a three-level refinement ($N_\text{ref} = 3$) in 1D (solid red), are compared against UG grids matching the coarsest (dash-dotted yellow) and second (solid dark gray) refinement levels (the maximum refinement level for the $N_\text{ref} = 2$ case).  Insets in the left plot zoom in on the behavior of the drift for the two UG simulations.}
    \label{fig:StreamingSineWaveErrors}
\end{figure}

Figure~\ref{fig:StreamingSineWaveErrors} provides a quantitative analysis of the evolving waveform for different AMR and uniform grid (UG) configurations.  
With UG, the coarse and fine grid simulations were run with FV resolutions of $\Delta x^{1}=L/256\approx3.91\times10^{-1}$~cm and $\Delta x^{1}=L/512\approx1.95\times10^{-1}$~cm, respectively.
For AMR configurations, we set the coarsest refinement level to match the coarse UG configuration, and each higher refinement level increases resolution by a factor of two.
We use linear polynomials for the DG method, so that the element width is twice that of the FV cell width.  
We determine the relative drift, change in amplitude, and change in phase of the waveform compared to the expected solution by examining the first few components of the spatial Fourier transform of the Eulerian number density at each time step:
\begin{equation}
    F_k(t) = \int_{D_{\bx^1}} \left[\mathcal{N}(\bx, t) - 0.5\right]\exp\left[-2\pi i k(x^1 - t)\right] dx^{1}.
    \label{eq:fourierTransform}
\end{equation}
The magnitude of the zeroth component ($k=0$) quantifies the drift of the solution away from its initial offset, while the magnitude and argument of the first component ($k=1$) are used to quantify the relative change in amplitude and phase, respectively, compared to the advected initial profile.  
In all cases we find that the code is capable of advecting the radiation profile through the jumps in grid refinement.  
Both the 1D and 3D simulations using two refinement levels show comparable results, but we note that the accuracy of the solutions are bounded by the UG grid sharing the resolution of the coarsest AMR level.  
Increasing the number of refinement levels slightly degrades the overall accuracy of the solution, which we attribute to an increased number of interpolations across coarse-fine mesh boundaries and possibly the lack of flux corrections applied to spatial fluxes of the evolved radiation moments at coarse-fine mesh boundaries \citep[as described in, e.g.,][Section~3]{bergerColella_1989}.  

\section{Results from Idealized Tests with Tabulated Microphysics}
\label{sec:application_idealized}

In this section we present results from applications of the neutrino transport algorithms implemented in \thornado\ that make use of \weaklib-tabulated microphysics.  

\subsection{Relaxation Problem}
\label{sec:relaxation}

We begin with a version of the relaxation problem from \citet{laiu_etal_2021}, modified for the $\cO(v)$ model and including all six neutrino species (electron, muon, and tau neutrinos and antineutrinos).  
This test solves the space-homogeneous problem (Subproblem~3) given by Equation~\eqref{eq:subproblem3}, and considers neutrinos initially out of equilibrium with matter.  
For the initial matter configurations, we set $\{\rho,T,\ye\}=\{\,10^{12}~\text{g cm}^{-3},\,7.6~\text{MeV},\,0.135\,\}$, corresponding to conditions in a collapsed stellar core at an approximate radius of $40$~km about $100$~ms after bounce \citep{liebendorfer_etal_2005}, and $\vect{v}=(0.1,0,0)^{\intercal}$.  
All neutrino species are given the same initial distribution,
\begin{equation}
	f_{0}(\varepsilon,\mu) = 0.99\times\exp\Big[\,-\f{1}{2}\,\Big(\f{\varepsilon-\varepsilon_{0}}{\sigma_{\varepsilon}}\Big)^{2}\,\Big]\times H(\mu-\mu_{0}),
\end{equation}
where $\varepsilon_{0}=2\,T$, $\sigma_{\varepsilon}=10$~MeV, $H$ is the Heaviside step function, and $\mu_{0}=0$.  
We then set the initial comoving-frame moments as 
\begin{equation}
	\{\cD,\cI^{1}\}(\varepsilon)
	=\f{1}{2}\int_{-1}^{1}f_{0}(\varepsilon,\mu)\{1,\mu\}d\mu
\end{equation}
and $\cI^{2}=\cI^{3}=0$.  
This initial distribution, Gaussian in energy and anisotropic in angle, is far away from the equilibrium Fermi-Dirac distribution.  
The energy domain $D_{\varepsilon}=[0,300]$~MeV is discretized into 16 linear elements, and we use a geometrically progressing grid, $\Delta\varepsilon_{\eidx+1}/\Delta\varepsilon_{\eidx}=1.266$, with $\Delta\varepsilon_{1}=1.875$~MeV (Grid~C in Table~\ref{tab:energyGrids}).  
The system of equations is integrated with the backward Euler method, and we set $\texttt{tol{\_}outer}=\texttt{tol{\_}inner}=10^{-8}$ for all the tests in this section.  

\begin{figure}[h]
	\begin{center}
        \captionsetup[subfigure]{justification=centering}
        {\begin{minipage}{0.45\textwidth}
                        \includegraphics[width=\linewidth]{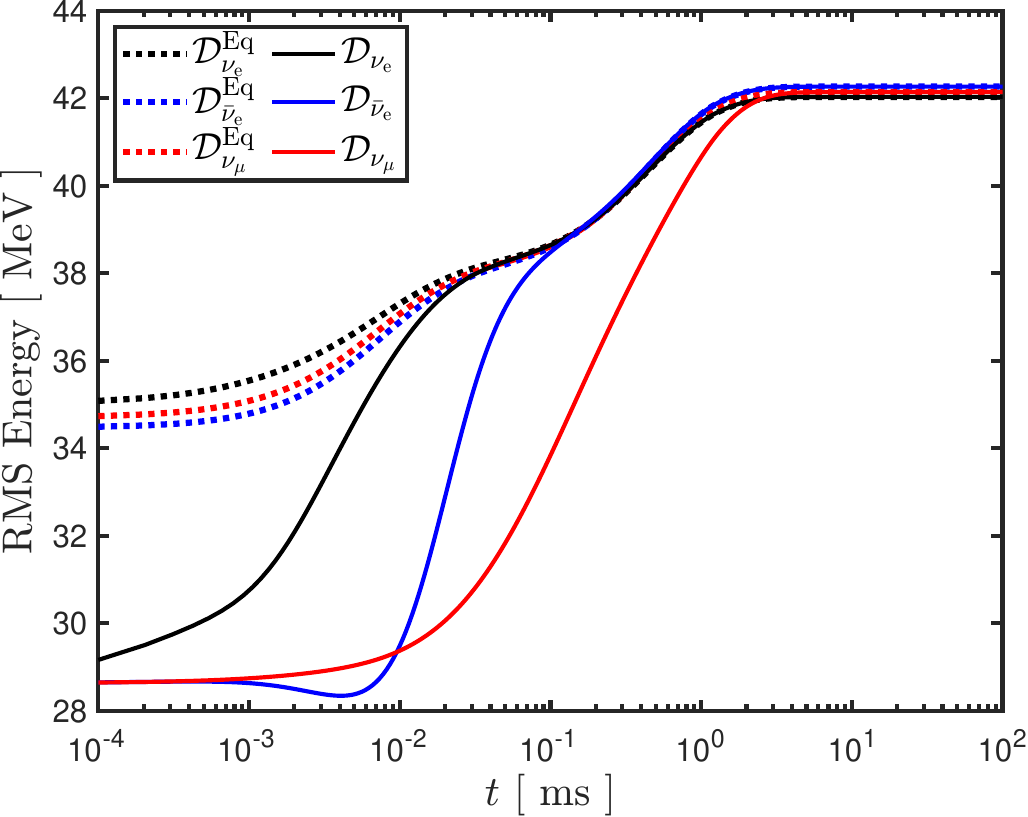}
                        \label{fig:}
                \end{minipage}
        }~~~~~~~
        {\begin{minipage}{0.45\textwidth}
                        \includegraphics[width=\linewidth]{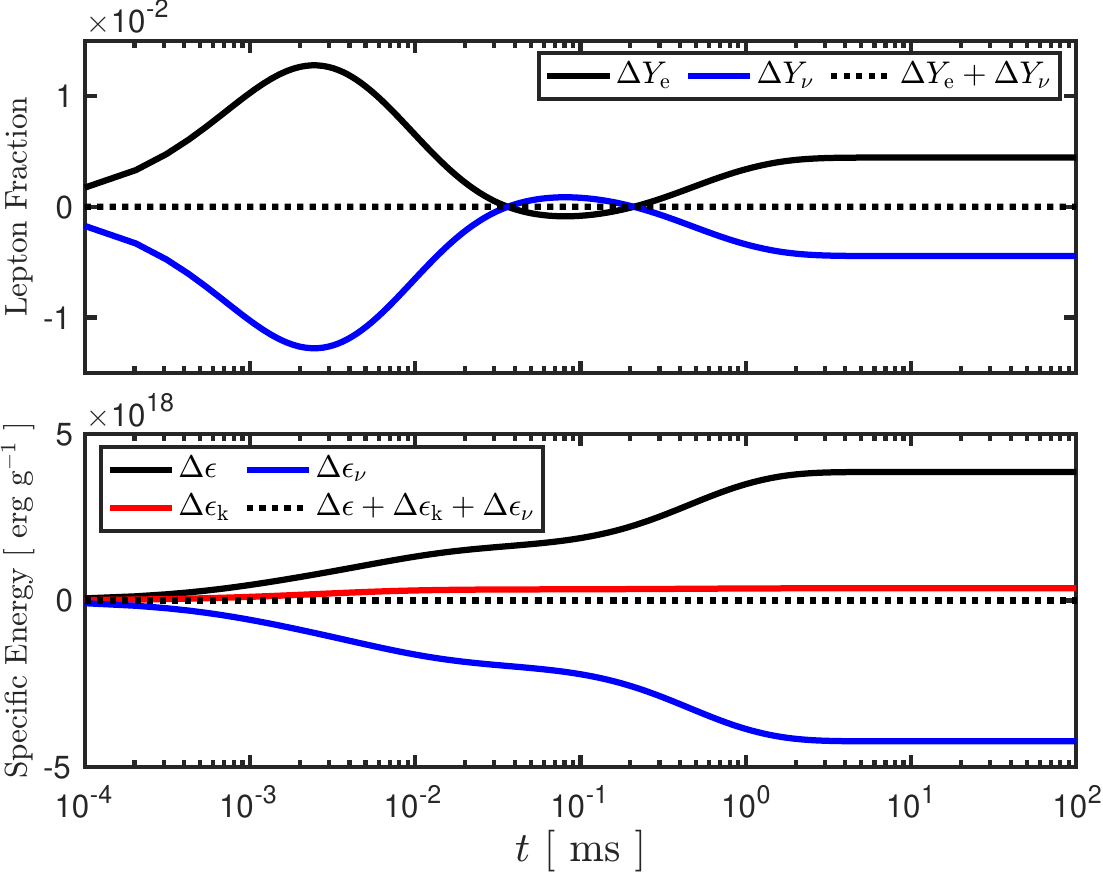}
                        \label{fig:}
                \end{minipage}
        }
        \caption{Results for the relaxation problem.  
        		     The left panel plots the RMS energy, defined in Equation~\eqref{eq:rmsEnergy}, versus time for electron neutrinos (solid black), electron antineutrinos (solid blue), and muon neutrinos (solid red).  
		     The dotted black, blue, and red lines represent the corresponding RMS energies computed with the respective equilibrium distributions.  
		     The top right panel shows the change in the electron fraction (solid black), neutrino lepton fraction (solid blue), and total lepton fraction (dotted black) versus time.  
		     The bottom right panel shows the change in the specific internal energy (solid black), specific kinetic energy (solid red), specific Eulerian neutrino energy (solid blue), and total specific energy (dotted black) versus time.}
        \label{fig:relaxationOverview}
        \end{center}
\end{figure}

Figure~\ref{fig:relaxationOverview} shows results obtained with the nested fixed-point algorithm in Listing~\ref{alg:neutrinoMatterCoupling}, with $\texttt{M{\_}outer}=\texttt{M{\_}inner}=2$.  
The simulation was run until $t=100$~ms, with an initial time step $\dt=10^{-4}$~ms, increasing by a factor of $1.04$ per time step, until a maximum of $\dt=10$~ms.  
The left panel illustrates the approach to equilibrium for electron neutrinos and antineutrinos, as well as muon neutrinos.  
The evolution of muon antineutrinos and tau neutrinos and antineutrinos is omitted from the plot, but we mention that the evolution of muon and tau neutrinos is identical, and the evolution of muon and tau antineutrinos is identical, and follow very closely that of the muon neutrinos.  
All species eventually reach a steady state, where the distribution is given by the local Fermi--Dirac distribution.  
Electron neutrinos equilibrate first, around $t=0.1$~ms, followed by electron antineutrinos, and finally muon neutrinos around $t=4$~ms.  
This ordering is expected because of the relative strength of coupling to matter for the different neutrino species.  
The top right panel in Figure~\ref{fig:relaxationOverview} shows the change in the electron fraction $\Delta \ye=\ye(t)-\ye(t=0)$, the neutrino lepton fraction $\Delta Y_{\nu}=Y_{\nu}(t)-Y_{\nu}(t=0)$, where $Y_{\nu}=\mB N_{\nu}/\rho$, and the total lepton fraction $\Delta \ye+\Delta Y_{\nu}$ versus time.  
Similarly, the bottom right panel in Figure~\ref{fig:relaxationOverview} shows the change in specific internal, kinetic ($\f{1}{2}|\vect{v}|^{2}$), Eulerian neutrino ($E_{\nu}/\rho$), and total energy versus time.  
The plots illustrate the lepton and energy exchange between the fluid and neutrinos during the relaxation process, and show that the total remains relatively constant.  
For this simulation, the relative change in total lepton number and energy are both on the order of $10^{-6}$.  
Since the matter update is formulated by enforcing total lepton, momentum, and energy conservation, their relative change is sensitive to solver tolerance.  

\begin{figure}[h]
	\begin{center}
		\includegraphics[width=0.5\linewidth]{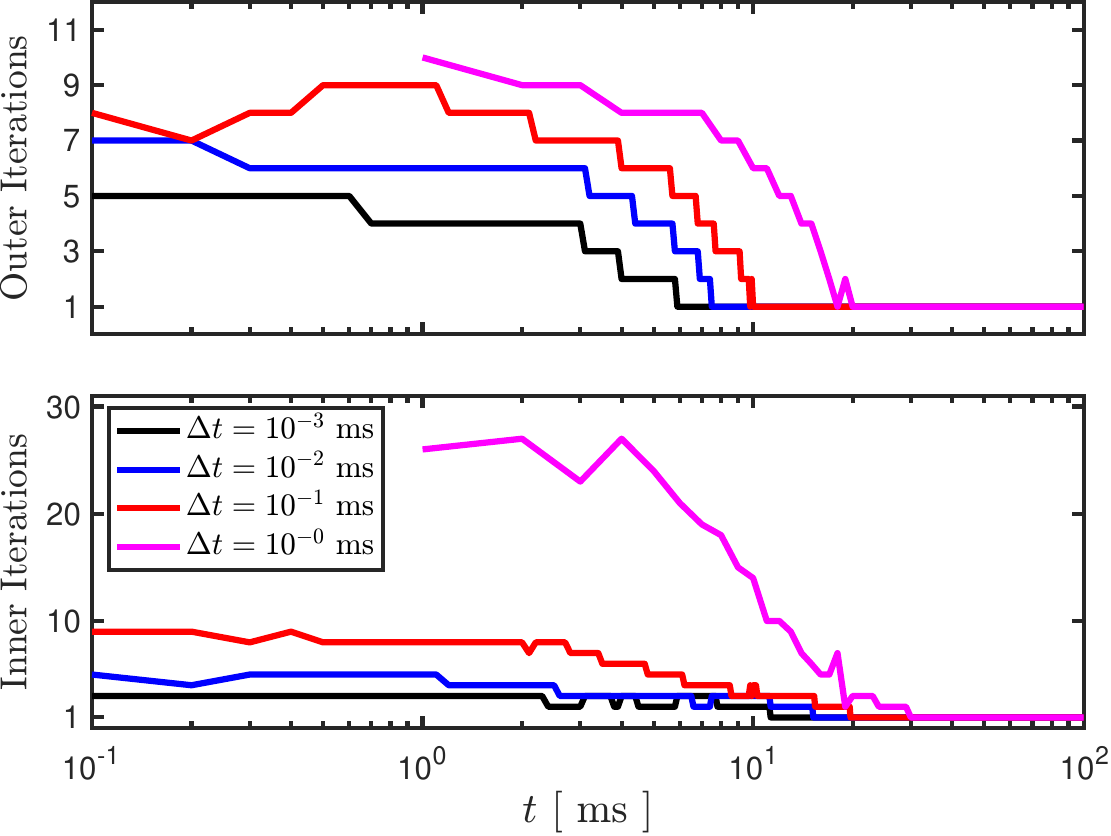}
        		\caption{Results from running the relaxation problem using different time steps $\dt$.  
			     The upper panel shows the number of outer iterations per time step versus time.  
			     The lower panel shows the average number of inner iterations per outer iteration per time step versus time.  
			     Results are plotted for $\dt=10^{-3}$~ms (black), $10^{-2}$~ms (blue), $0.1$~ms (red), and $1$~ms (magenta).}
        		\label{fig:relaxation_dt}
        \end{center}
\end{figure}

Next, we repeat the simulation discussed above, but keep the time step fixed throughout, and present results from simulations with different time steps: $\dt\in\{10^{-3},10^{-2},0.1,1\}$~ms.  
Again, we let $\texttt{M{\_}outer}=\texttt{M{\_}inner}=2$.  
We investigate the effect of varying the time step on the convergence properties of the algorithm; i.e., iteration counts per time step and the final steady state solution.  
Figure~\ref{fig:relaxation_dt} plots the number of outer iterations (top panel) and the average number of inner iterations per outer iteration per time step (bottom panel) versus time.  
We observe that the number of outer iterations increases somewhat with increasing time step size: for $\dt=10^{-3}$~ms, the maximum number of outer iterations is $5$, while for $\dt=1$~ms, it is $10$.  
The average number of inner iterations stays below $10$ for $\dt\le0.1$~ms, while it is at most $27$ for $\dt=1$~ms.  
We find that all simulations reach the same steady state (within $\sim10^{-6}$ in $\ye$, $T$, and RMS energy), although through different (but similar) trajectories because of different time steps.  
We conclude from this that the algorithm is robust with respect to the step size.  

\begin{figure}[h]
	\begin{center}
		\includegraphics[width=0.5\linewidth]{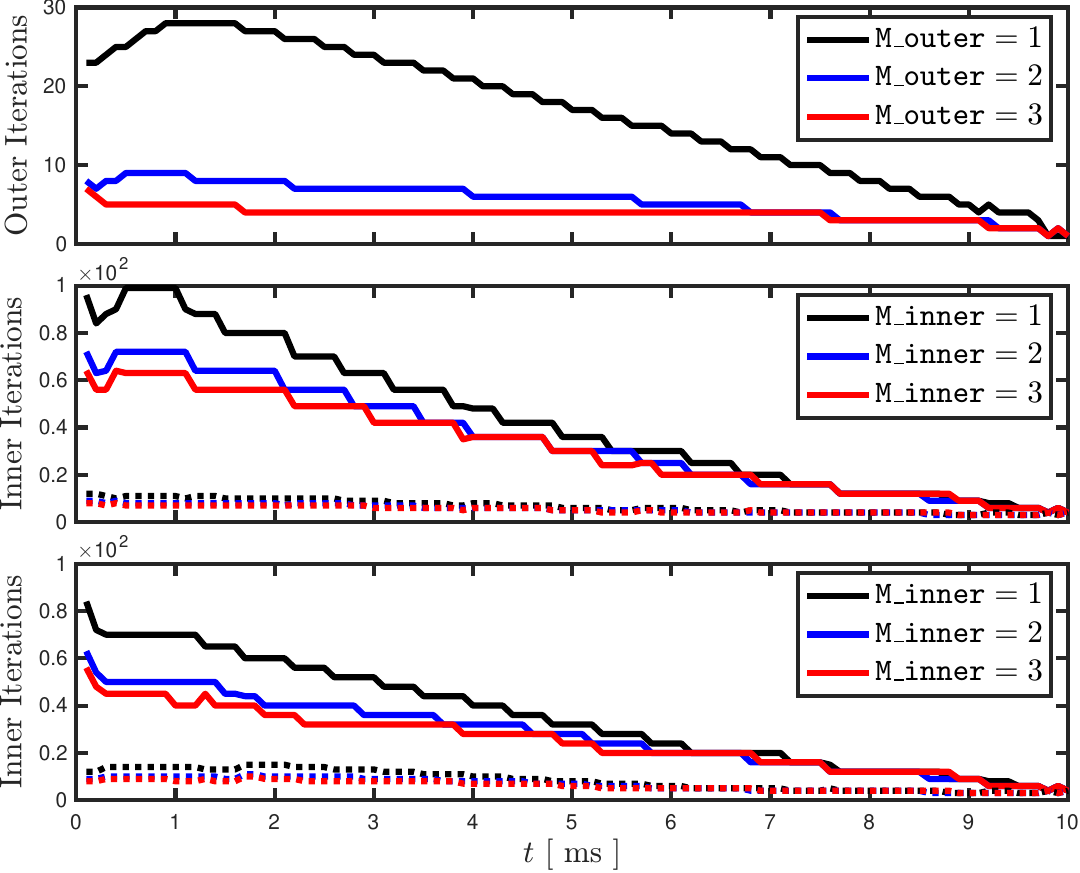}
        		\caption{Results from running the relaxation problem for various values of $\texttt{M{\_}outer}$ and $\texttt{M{\_}inner}$ for $\dt=0.1$~ms.
		The top panel shows the number of outer iterations per time step versus time for $\texttt{M{\_}inner}=5$ and $\texttt{M{\_}outer}=1$ (black), $2$ (blue), and $3$ (red).
		The middle panel shows the total (solid) and average (dotted) number of inner iterations per time step versus time for $\texttt{M{\_}outer}=2$ and $\texttt{M{\_}inner}=1$ (black), $2$ (blue), and $3$ (red).  
		The bottom panel shows the same quantities as the middle panel but for results obtained with $\texttt{M{\_}outer}=3$.}
        		\label{fig:relaxation_Memory}
        \end{center}
\end{figure}

Finally, we investigate the effect of varying the Anderson acceleration memory parameters, $\texttt{M{\_}outer}$ and $\texttt{M{\_}inner}$, on the convergence properties of the nested fixed-point iteration algorithm.  
We let $\dt=0.1$~ms, run to $t=10$~ms, and vary the outer and inner memory parameters separately.  
The top panel in Figure~\ref{fig:relaxation_Memory} shows the number of outer iterations per time step versus time for various $\texttt{M{\_}outer}$ and $\texttt{M{\_}inner}=5$.  
Setting $\texttt{M{\_}outer}=2$ or $3$ has a significant impact on the early iteration counts when compared to Picard iteration (\texttt{M{\_}outer}=1).  
Increasing $\texttt{M{\_}outer}$ from $2$ to $3$ reduces the iteration count further, but the largest gain is seen when going from $1$ to $2$.  
The total number of outer iterations for each simulation (from $t=0$ to $t=10$~ms) is $1662$, $556$, and $384$, for $\texttt{M{\_}outer}=1$, $2$, and $3$, respectively.  
Increasing $\texttt{M{\_}outer}$ further does not reduce the number of outer iterations, one reason being that the fluid system solved in the outer loop consists of only three unknowns for this problem.  
The middle panel shows the total number of inner iterations per time step versus time for the case with $\texttt{M{\_}outer}=2$.  
Again, the largest reduction in iteration count is seen when going from Picard iteration ($\texttt{M{\_}inner}=1$) to $\texttt{M{\_}inner}=2$.  
For this case the total number of inner iterations for each simulation is $4314$, $3459$, and $3183$ for $\texttt{M{\_}inner}=1$, $2$, and $3$, respectively.  
Increasing $\texttt{M{\_}outer}$ to $3$ does reduce the total number of inner iterations.  
In this case the total number of inner iterations is $3519$, $2739$, and $2485$ for $\texttt{M{\_}inner}=1$, $2$, and $3$, respectively.  
Further increasing $\texttt{M{\_}inner}$ does not reduce the total number of inner iterations for $\texttt{M{\_}outer}=2$ or $3$.  
These observations are consistent with those made by \citet{laiu_etal_2021}.  

\subsection{Deleptonization Problem}
\label{sec:deleptonization}

Next, we consider the deleptonization problem from \citet{laiu_etal_2021}, extended to include six neutrino species.  
We consider a spatial domain $D_{\bx}=[0,512]$~km, discretized with $256$ uniform elements, and an energy domain $D_{\varepsilon}=[0,300]$~MeV, discretized in the same manner as for the relaxation problem, using Grid~C in Table~\ref{tab:energyGrids}.  
To initialize matter quantities, we adopt profiles for mass density, temperature, and electron fraction from \citet{liebendorfer_etal_2005}, as obtained with the \textsc{VERTEX} code at $100$~ms after core bounce using a $15~M_{\odot}$ progenitor from \citet{woosleyWeaver_1995}.  
We set the initial velocity to zero, and keep it fixed during the solution procedure by setting the first three components of the right-hand side of Equation~\eqref{eq:implicitFluidNodal} to zero, so that only the specific internal energy and electron fraction are evolved.  
The electron neutrinos and antineutrinos are initialized as described in \citet{laiu_etal_2021}, using elements of the homogeneous sphere solution \citep[e.g.,][]{smit_etal_1997}.  
Muon and tau neutrinos and antineutrinos are initialized with zeroth and first moments set as $\cD=1.49\times10^{-154}$ and $\bcI=0$.  
We use linear elements ($k=1$), and integrate the system of equations using the PD-ARS IMEX scheme with Butcher tableau given in Equation~\eqref{eq:butcherPDARS}.  
We let $\texttt{M{\_}outer}=\texttt{M{\_}inner}=2$, set the solver tolerances to $\texttt{tol{\_}outer}=\texttt{tol{\_}inner}=10^{-8}$, and evolve until $t=100$~ms with $\dt=10^{-3}$~ms.  

\begin{figure}[h]
	\begin{center}
        \captionsetup[subfigure]{justification=centering}
        {\begin{minipage}{0.31\textwidth}
                        \includegraphics[width=\linewidth]{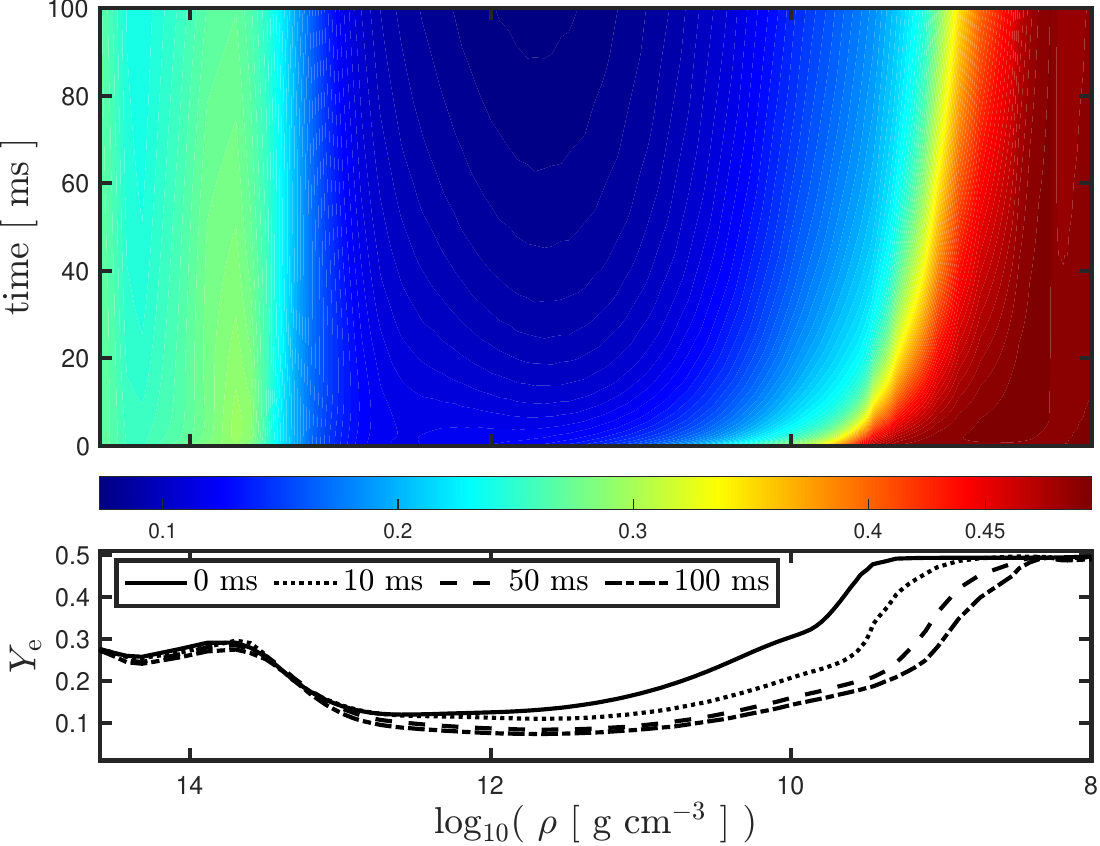}
                        \label{fig:}
                \end{minipage}
        }~~~
        {\begin{minipage}{0.31\textwidth}
                        \includegraphics[width=\linewidth]{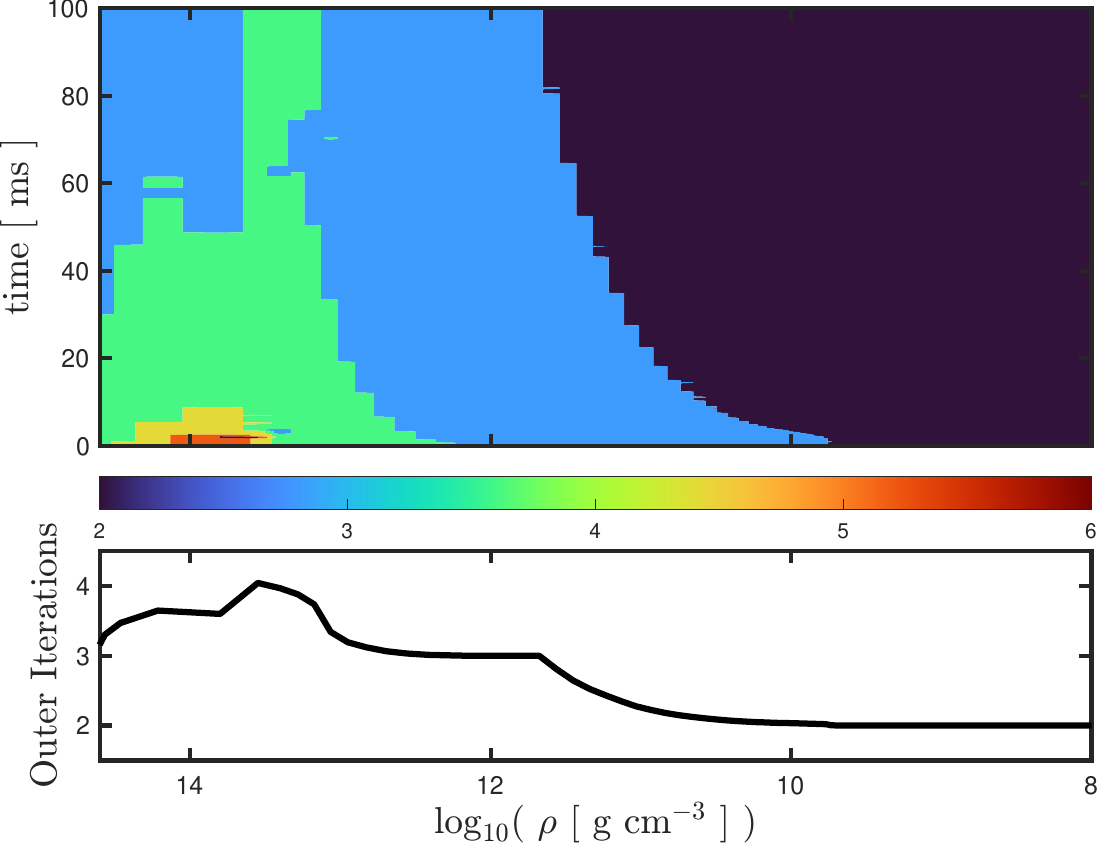}
                        \label{fig:}
                \end{minipage}
        }~~~
        {\begin{minipage}{0.31\textwidth}
                        \includegraphics[width=\linewidth]{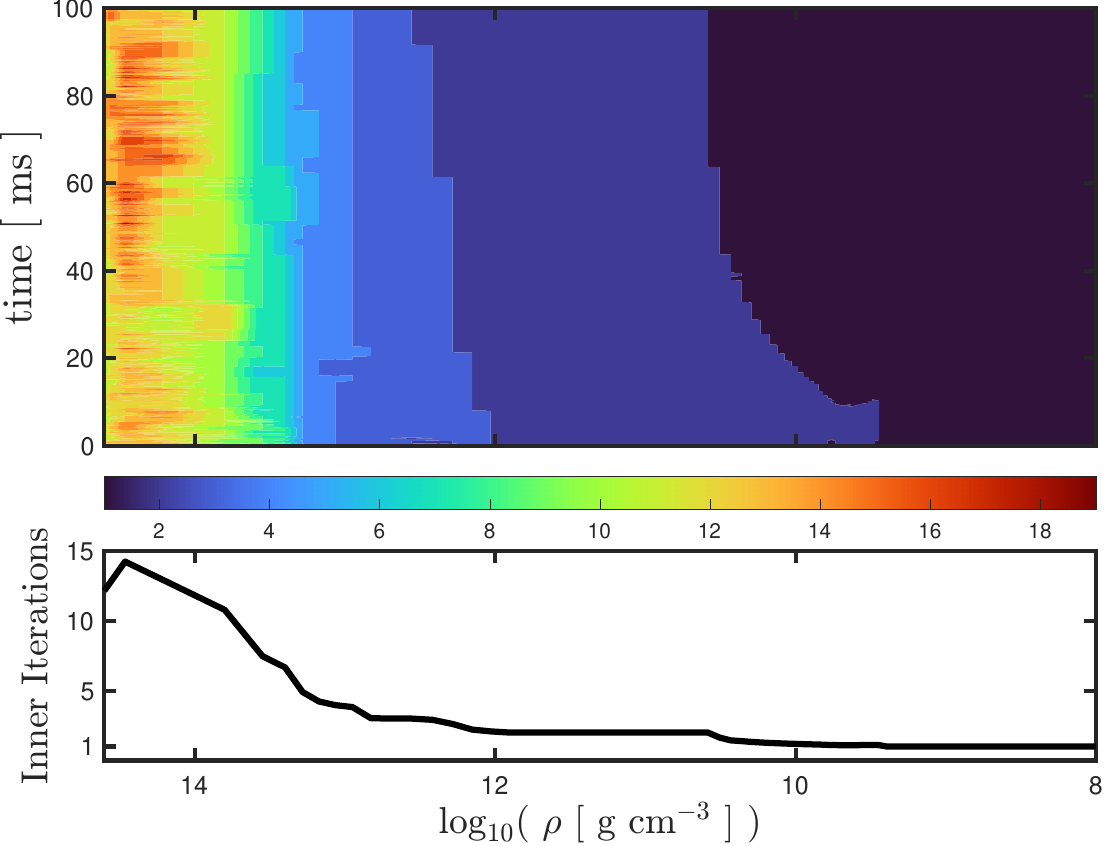}
                        \label{fig:}
                \end{minipage}
        }
        \caption{Results from the deleptonizarion problem.  The color maps in the top row show the electron fraction (left), number of outer iterations (middle), and the average number of inner iterations per outer iteration (right) in the $\log_{10}(\rho)$-$t$ plane.  The line plots in the bottom row show the electron fraction at select times (left), the time-averaged number of outer iterations (middle), and the time-averaged average number of inner iterations per outer iteration (right) versus $\log_{10}(\rho)$.}
        \label{fig:deleptonization}
        \end{center}
\end{figure}

Figure~\ref{fig:deleptonization} provides an evolutionary overview of the deleptonization problem.  
Electron-type neutrinos produced at higher mass density are transported out of the core, resulting in a gradual reduction in the electron fraction; i.e., deleptonization (left panels).  
The top middle panel shows the number of outer iterations as a function of mass density and time.  
There is an early hotspot of outer iterations around $\rho=10^{14}$~g~cm$^{-3}$, where six outer iterations are needed for convergence.  
Otherwise, between two and four iterations are needed, with more iterations generally needed for higher mass densities, as can be seen in the bottom middle panel, where the time-averaged number of outer iterations (averaged from $t=0$ to $t=100$~ms) is plotted versus mass density.  
Similar to the middle panels, the right panels show the average number of inner iterations per outer iteration versus mass density and time.  
Because the neutrino mean free path decreases and the system becomes increasingly stiff at higher densities, the inner iteration count increases steadily with mass density (aside from a slight dip in the center), averaging around one iteration for $\rho\in[10^{8},10^{10}]$~g~cm$^{-3}$, remaining below two iterations up to $\rho=10^{12}$~g~cm$^{-3}$, and exceeding ten iterations for $\rho\ge10^{14}$~g~cm$^{-3}$.  
There are some fluctuations ($\pm$ a few) in the number of inner iterations at high density, while the iteration count stays constant, or decreases slightly with time, at lower densities.  
The iteration counts observed here are in qualitative agreement with those reported for the nested~AA algorithm in \citet[][cf. their Figure~15]{laiu_etal_2021}, even though we use an expanded opacity set, six neutrino species (versus two), and about a factor of two larger time step, $\dt=10^{-3}$~ms.  

\begin{figure}[h]
	\begin{center}
		\includegraphics[width=0.5\linewidth]{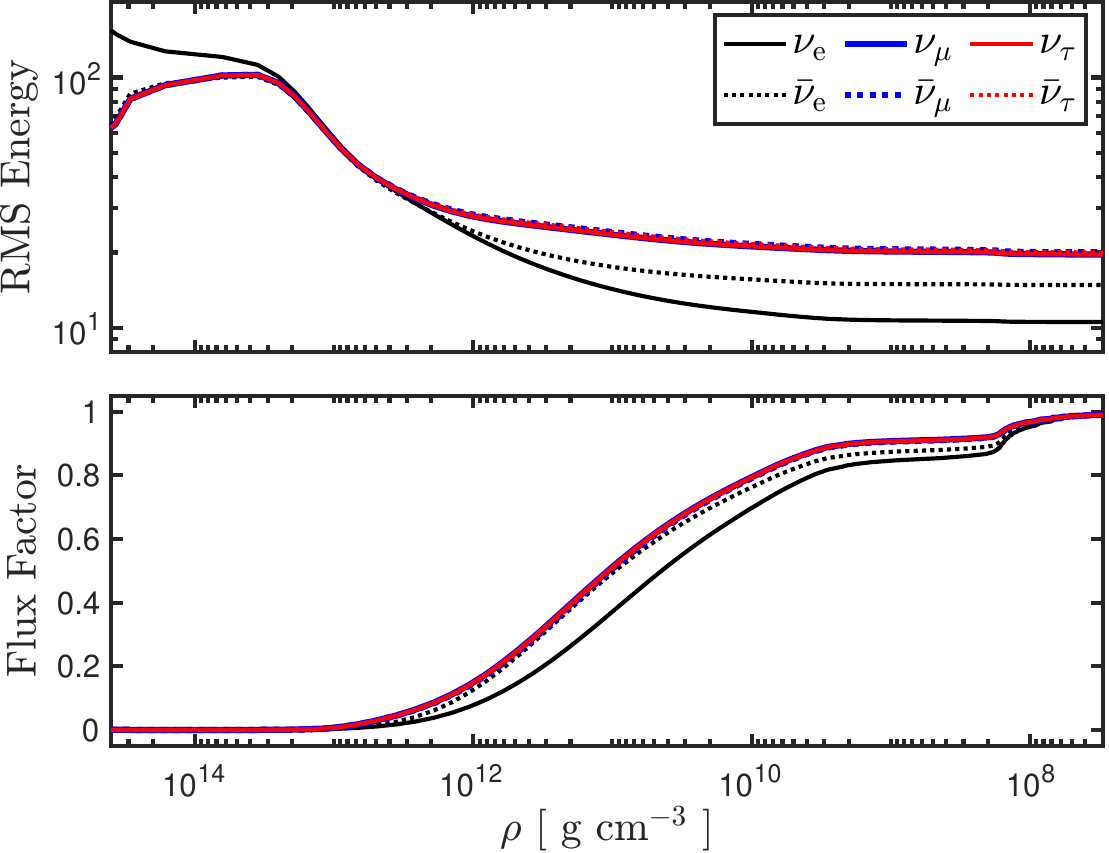}
        		\caption{Plot of neutrino RMS energy (top panel), as defined in Equation~\eqref{eq:rmsEnergy}, and energy-averaged flux factor (bottom panel) versus mass density for the deleptonization problem after $100$~ms of evolution.  
		In each panel, values are plotted for all six neutrino species: $\nu_{\rm e}$ (solid black), $\bar{\nu}_{\rm e}$ (dotted black), $\nu_{\mu}$ (solid blue), $\bar{\nu}_{\mu}$ (dotted blue), $\nu_{\tau}$ (solid red), and $\bar{\nu}_{\tau}$ (dotted red).}
        		\label{fig:deleptonization_neutrino}
        \end{center}
\end{figure}

Figure~\ref{fig:deleptonization_neutrino} provides an overview of the neutrino RMS energy and flux factor versus mass density after $100$~ms of evolution, when the system has reached a quasi--steady state and neutrino quantities evolve on a long time scale due to continued deleptonization of the core.  
Electron neutrinos are degenerate at high densities, where their RMS energy exceeds $100$~MeV.  
Electron antineutrinos and heavy-flavor (i.e., muon and tau) neutrinos have lower and similar RMS energies in the core.  
At lower densities, the RMS energies decrease with mass density for all species, and reach asymptotic values as neutrinos decouple from matter.  
Asymptotic RMS energies are about $20$~MeV for muon and tau neutrinos and antineutrinos, $14.8$~MeV for electron antineutrinos, and $10.5$~MeV for electron neutrinos.  
This ordering arises because muon and tau neutrinos decouple in deeper layers, at higher density and temperature, while electron antineutrinos, followed by electron neutrinos, decouple further out.  
The decoupling of neutrinos and matter is also evident in the flux factor, which is essentially zero at the highest densities, where neutrinos are in equilibrium with matter and their distribution functions (nearly) isotropic.  
As neutrinos propagate radially, decouple from matter, and become free-streaming, their angular distribution become forward-peaked, causing the flux factor to approach unity.  
This transition happens deeper (at higher mass density) for muon and tau neutrinos, followed by electron antineutrinos and electron neutrinos.  
Because we do not include any weak interaction channels that distinguish between muon and tau neutrinos \citep[cf.][]{bollig_etal_2017}, we observe that the RMS energy and flux factor for muon and tau neutrinos are identical, and the same is true for muon and tau antineutrinos, which is a good consistency check when evolving six distinct neutrino species.  

\begin{figure}[h]
	\begin{center}
		\includegraphics[width=0.5\linewidth]{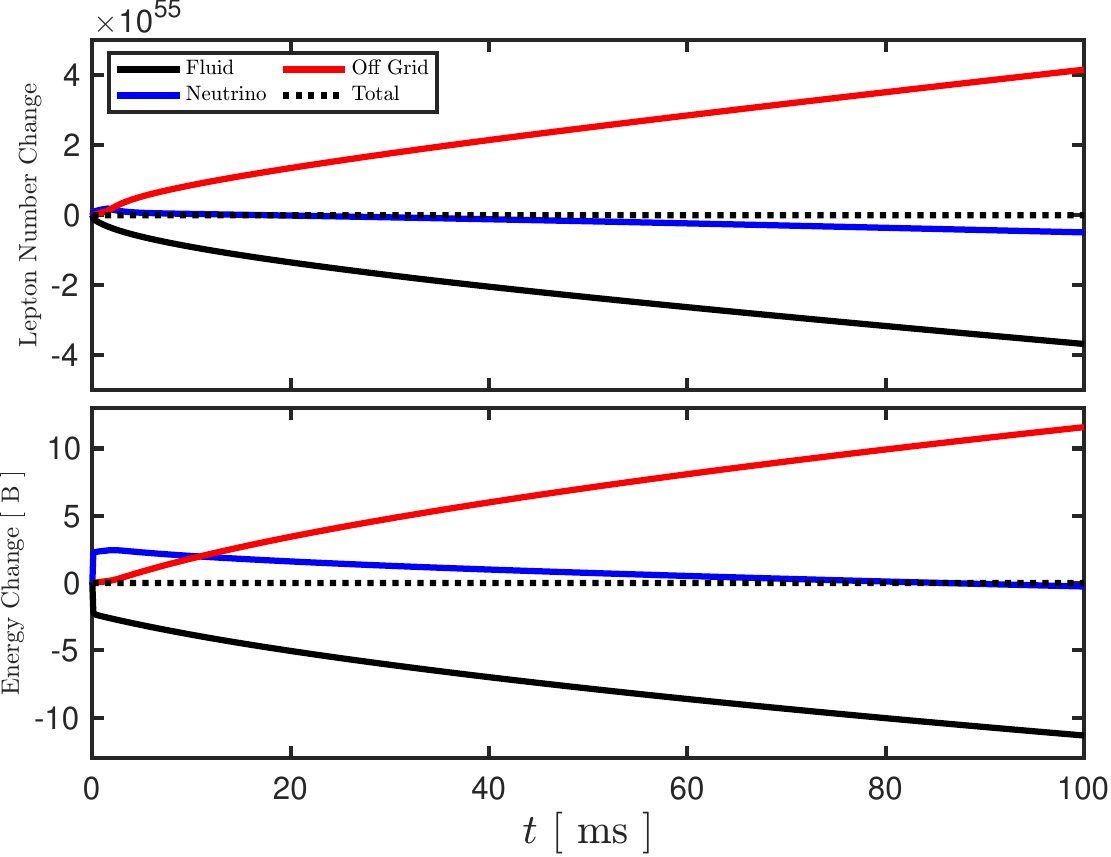}
        		\caption{Total lepton number and energy balance for the deleptonization problem.  
		The top panel plots the change in the fluid lepton number (solid black), change in neutrino lepton number (blue), cumulative neutrino lepton number that flows off the computational grid through the outer boundary (red), and the change in the total lepton number (including off-grid contributions; dotted black).
		Similarly, the bottom panel plots the change in the fluid energy (solid black), change in neutrino energy (blue), cumulative neutrino energy through the outer boundary (red), and the change in the total energy (including off-grid contributions; dotted black), as a function of time.}
        		\label{fig:deleptonization_conservation}
        \end{center}
\end{figure}

In Figure~\ref{fig:deleptonization_conservation}, we plot the total lepton number and energy conservation properties of the algorithms implemented in \thornado\ for the deleptonization problem.  
By Equations~\eqref{eq:totalLeptonNumber} and \eqref{eq:totalEnergy}, for a static background, the total lepton number and energy are exactly conserved by the model.  
However, conservation of these quantities is sensitive to tolerances of the neutrino--matter solver algorithm, and energy conservation is affected by limiters applied to the neutrino moments, for which the spectral redistribution algorithm mentioned in Section~\ref{sec:limiters} aims to correct.  
The top panel shows that the fluid lepton number decreases steadily throughout the simulation (black line).  
The neutrino lepton number increases slightly early on, and then decreases slowly with time (blue line).  
The decrease in the fluid lepton number is mainly balanced by the neutrino lepton number flux through the outer boundary (red line).  
The evolution of the energy components, shown in the bottom panel, follow trajectories that are very similar to the corresponding lepton number components, except for an initial sharp rise of about 2.3~B in the neutrino energy (blue line), which is balanced by a corresponding drop in the fluid energy (black line).  
The muon and tau neutrinos and antineutrinos are initially set to zero, and are immediately populated through pair processes (e.g., electron-positron annihilations and nucleon-nucleon Bremsstrahlung) after the onset of the simulation.  
This dynamics is not seen in the top panel because pair production does not contribute to the total lepton number.  
Both the total lepton number and energy (dotted black) remain relatively constant throughout, with relative changes on the order of $10^{-5}$.  

\section{Application to Core-Collapse Supernova Simulations}
\label{sec:application_ccsn}

In this section, we apply the neutrino transport algorithms implemented in \thornado\ to model CCSNe in spherical and axial symmetry.  
These models involve dynamic coupling to self-gravitating fluid flows, evolved with \flashx\ using AMR, as outlined in the beginning of Section~\ref{sec:method}.  
Our main objective is to carry out a code-to-code comparison against the established CCSN simulation code \chimera\ \citep{bruenn_etal_2020}.  
For this purpose, we consider the 15~$M_{\odot}$ progenitor of \citet{woosleyHeger_2007} as the pre-collapse initial condition, and evolve with the SFHo EoS \citep{steiner_etal_2013} and the neutrino opacities listed in Table~\ref{tab:opacities}.  
We evolve all six neutrino species; i.e., $s\in\{\nu_{\rm e},\bar{\nu}_{\rm e},\nu_{\mu},\bar{\nu}_{\mu},\nu_{\tau},\bar{\nu}_{\tau}\}$.  

The spatial domain, $r\in[0,8192]$~km, is discretized using a block-based adaptive mesh with up to 8 refinement levels ($\ell=0,\ldots,7$).
The fixed block size is $64$. 
The coarsest level ($\ell=0$) is composed of two base blocks for a total of 128 FV cells, each with width $(\Delta r)^{\ell=0} = 64$~km.  
With a refinement ratio of two between successive levels, the resolution on the finest level ($\ell=7$) is $(\Delta r)^{\ell=7}=2^{-\ell}\times(\Delta r)^{0}=0.5$~km.  
For the DG discretization, we employ linear basis functions (polynomial degree $k=1$), such that each DG element spans two FV cells.  
Consequently, DG spatial elements are $1$~km wide on the finest refinement level.  
During collapse, the refinement strategy aims to capture the change in spatial scale of the central mass distribution as the central density increases.  
In this phase, the maximum refinement level increases progressively from $\ell_{\max}=3$ to $\ell_{\max}=7$, with additional levels activated when the central density satisfies $\log_{10}(\rho_{\rm c}/[{\rm g}~{\rm cm}^{-3}])=11$, $11.5$, $12$, and $12.5$.  
The innermost $600$~km is always maximally refined (at $\ell_{\max}$).  
This ensures that the FV resolution in the inner core is $0.5$~km during core bounce and the subsequent shock propagation.  
The spectral domain, $\varepsilon\in[0,300]$~MeV, is discretized with 16 elements, using a geometrically progressing grid with parameters corresponding to Grid~C in Table~\ref{tab:energyGrids}.  
The neutrino evolution, including its coupling to matter (Subproblems~2 and 3), is advanced using the PDARS IMEX scheme, while hydrodynamics with self-gravity (Subproblem~1) is evolved with the SSP-RK2 scheme.  
The time step is determined by the explicit part of the neutrino evolution and constrained by the speed of light; i.e., 
\begin{equation}
	\dt=2^{-\ell_{\max}}\,\dt^{0}, \qquad\text{where}\qquad \dt^{0}=\f{1}{2k+1}\f{(\Delta r)^{0}}{c}.
	\label{eq:cflConditionCCSN}
\end{equation}
With the adaptive mesh described above, the time step decreases from about $8.9\times10^{-6}$~s at the onset of collapse to about $5.6\times10^{-7}$~s after the maximum refinement level is reached.  
Unless stated otherwise, the nonlinear neutrino--matter coupling solver uses $\texttt{M{\_}outer}=\texttt{M{\_}inner}=2$ and tolerances $\texttt{tol{\_}outer}=\texttt{tol{\_}inner}\equiv\texttt{tol}=10^{-8}$.  
All spherically symmetric simulations were performed on Perlmutter at NERSC, using one compute node per run with 4~GPUs and 40~MPI ranks (10 ranks per GPU).  

\begin{figure}[h]
    \begin{center}
    	\includegraphics[width=\textwidth]{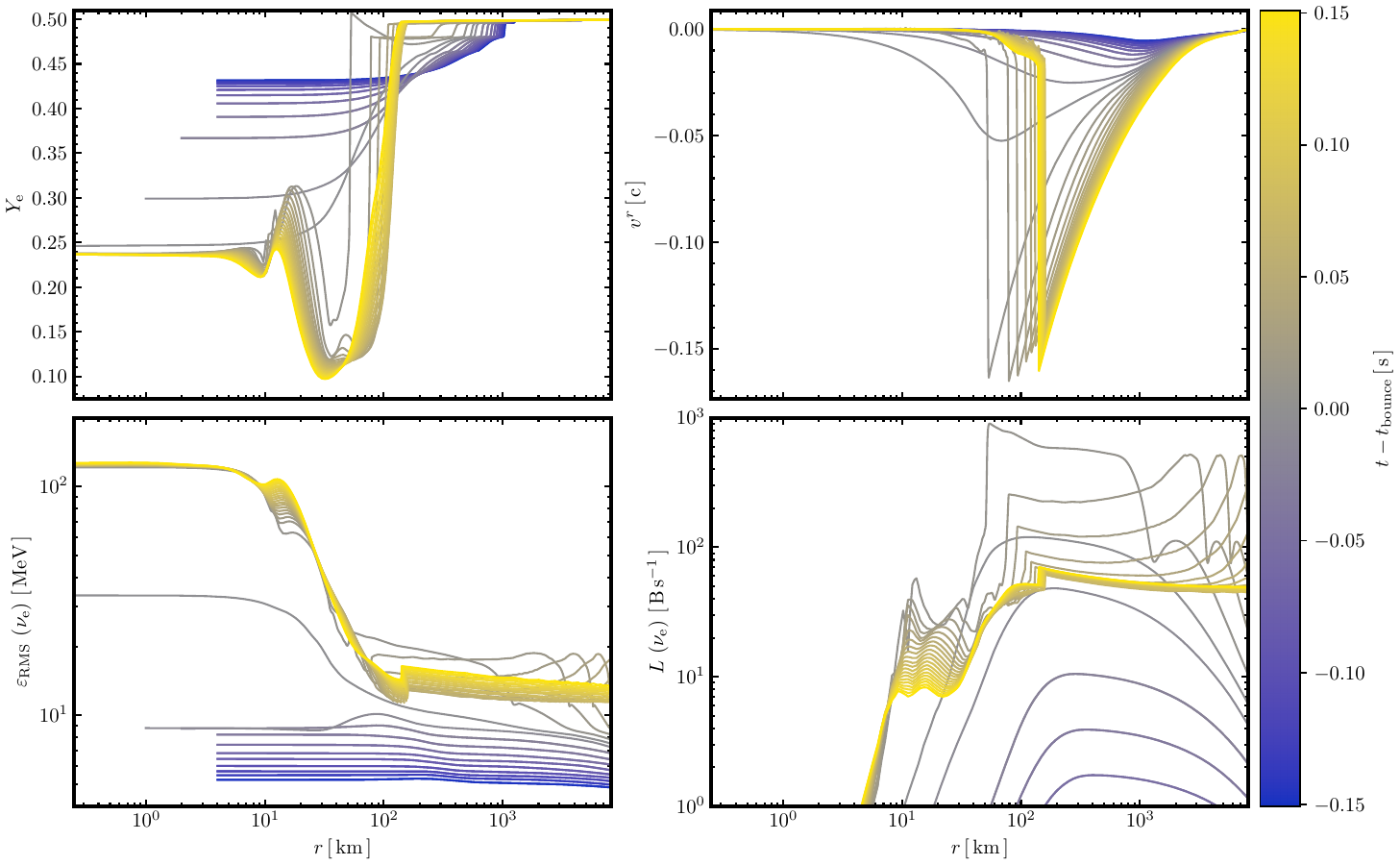}\label{fig:profile_timeseries}
    	\caption{Time evolution (indicated by color) of radial profiles of the electron fraction ($\ye$; top left), radial velocity ($v^r$; top right), electron-neutrino RMS energy ($\varepsilon_{\rm RMS}({\nu_e})$; bottom left), and electron-neutrino luminosity ($L({\nu_e})$; bottom right).   The figure illustrates the collapse and post-bounce evolution over a time window of $\pm 150$~ms around core bounce.  Each curve corresponds to a snapshot separated by $7.5$~ms.}
    \end{center}
\end{figure}

Figure~\ref{fig:profile_timeseries} presents an overview of a neutrino-radiation hydrodynamics simulation performed with \thornado+\flashx.  
The model exhibits characteristic features of spherically symmetric core-collapse simulations reported in the literature \citep[e.g.,][]{liebendorfer_etal_2005,oconnor_etal_2018}.  
During collapse, the electron fraction in the inner core decreases from about $0.43$ to $0.24$.  
The mesh adaptivity is evident by the leftward shift in the starting point of the plotted curves, reflecting the increased spatial resolution as collapse proceeds.  
The central electron-neutrino RMS energy increases from $\sim5$~MeV to over $100$~MeV before bounce, and the luminosity rises steadily as electron-capture neutrinos escape and delepetonize the core.  
For this model, core bounce occurs at $t_{\rm bounce}\approx260$~ms.  
Unless otherwise stated, all subsequent times are given relative to $t_{\rm bounce}$.  
The shock launched from the inner core at bounce appears as a discontinuity in the velocity profiles, first seen located near $r=50$~km.  
The shock radius reaches a maximum of about $160$~km after roughly $75$~ms, after which it recedes slowly, as an accretion shock (see Figure~\ref{fig:shock_and_central_quantities} below for more details).  
The passage of the shock is accompanied by a sharp increase in the electron-neutrino luminosity as the shock propagates into lower-density material that is transparent to neutrinos.  
The outward-moving electron-neutrino burst, clearly seen in both the luminosity and RMS energy profiles, reaches the outer boundary of the computational domain within $50$~ms.  
In the wake of the bounce shock and the neutrino burst, a trough develops in the $\ye$ profiles between $r\approx15$~km and the shock, caused by neutrino transport.  
After the initial neutrino burst, the RMS energy and luminosity profiles ahead of the shock settle into a relatively quiescent state, both exhibiting a trend of gradual increase.  
In the inner core after bounce, where neutrinos remain trapped, the $\ye$ profiles remain essentially unchanged over the plotted time interval.  
Further details on this model, along with results from corresponding simulations using the Levermore and Kershaw two-moment closures, are provided in the following section, which focuses on a code-to-code comparison with \chimera.

\begin{rem}
	For this initial verification study, the \chimera\ comparison is restricted to spherical symmetry.
	CCSNe become convectively unstable within a few milliseconds after bounce, making it nearly impossible to cleanly isolate differences in the neutrino transport sector, which may seed convection differently.  
	\chimera\ employs the ray-by-ray approximation, whereas \thornado+\flashx\ solves the multidimensional two-moment system, leading to potentially different behavior once nonradial flows develop \citep{skinner_etal_2016}.
	The codes also use different coordinate systems in multiple dimensions---spherical-polar in \chimera\ versus cylindrical or Cartesian in \flashx; therefore, truncation errors seed fluid instabilities differently at finite resolution, producing potentially divergent nonlinear evolution unrelated to the transport scheme itself.  
	For multidimensional code comparison studies, see, e.g., \citet{cabezon_etal_2018,glas_etal_2019,varma_etal_2021}.
\end{rem}

\subsection{Comparison with \chimera\ in Spherical Symmetry}

The \chimera\ simulations used for the code-to-code comparison are based on the ``G-series'' version of the code.
To the extent possible, \chimera\ is configured to match \thornado+\flashx\ in terms of physics and phase-space resolution, though some notable differences remain that we highlight here.
\chimera\ employs a moving radial mesh rather than AMR to resolve key regions in the CCSN problem \citep[see][Section~4.5]{bruenn_etal_2020}.
Consequently, the spatial discretizations are fundamentally different between the codes.  
The \chimera\ simulations described here use 596 radial cells covering the same computational domain as \thornado+\flashx.
The \chimera\ grid is roughly geometrically spaced, starting from an inner cell width of 1~km, but is adjusted to maintain approximately 40 cells in the inner core ($\rho>10^{14}$~g~cm$^{-3}$) and 100 cells in the neutrino-heating region ($10^{10}$~g~cm$^{-3}\lesssim\rho\lesssim10^{14}$~g~cm$^{-3}$).  
The resulting spatial resolution around the neutrino spheres and PNS surface is comparable to that of the \flashx\ grid (about 0.5~km).  

A finite-volume discretization is used for the spectral domain, and the energy grid (16~bins) is configured to exactly match that of \thornado's elements.  
\chimera's hydrodynamics is evolved using an explicit Lagrangian-plus-remap scheme with piecewise parabolic reconstruction \citep{CoWo84}, and the neutrino transport uses the fully-implicit multi-group, flux-limited diffusion method of \citet{bruenn_1985}.
The equations for neutrino transport are solved locally by one MPI process in the ray-by-ray approximation using Newton's method.  
Jacobian entries for the zeroth neutrino moment and electron number density are calculated for every iteration using opacities and derivatives that are numerically evaluated from linear interpolation of tabulated kernels.
The resulting linear system for each Newton-Raphson iteration is solved using the biconjugate gradient stabilized (BiCGSTAB) Krylov subspace iterative method with an effective alternating direction implicit (ADI)-like preconditioner adapted from AGILE-BOLTZRAN \citep{DaMeMe05}.  
The matter energy density equation is decoupled from the iterative solver and instead solved in an operator-split fashion.
Unlike \thornado, \chimera\ evolves only four neutrino species ($s\in\{\nu_{\rm e},\bar{\nu}_{\rm e},\nu_{\mu\tau},\bar{\nu}_{\mu\tau}\}$).  
We also include results from a modified \chimera\ configuration (G+) in which corrections to the first neutrino moment resulting from keeping all $\mathcal{O}(v/c)$ terms in the closure relation \citep[][Section~6.12]{bruenn_etal_2020} are applied everywhere, rather than only when the velocity exceeds $0.01~c$ and the density is below $10^{12}$~g~cm$^{-3}$, as in the fiducial setup.  

\begin{figure}[h]
	\centering
	\includegraphics[width=0.5\linewidth]{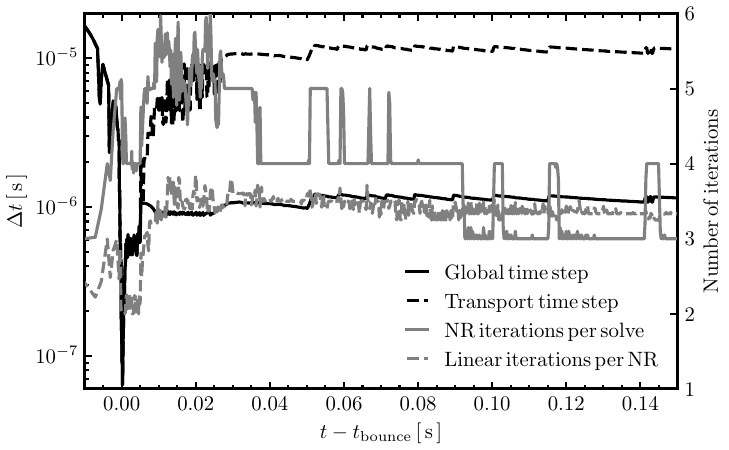}
	\caption{Global (solid black) and transport (dashed black) time step sizes for the fiducial \chimera\ G-series run from shortly before bounce until 150~ms after bounce.
    The average number of Newton-Raphson (NR) iterations per transport solve (solid gray) and number of linear solver iterations per NR iteration (dashed gray) are also shown.}
	\label{fig:chimera_dt}
\end{figure}

The coupled time stepping in \chimera\ is based on Lie--Trotter operator splitting \citep{Trot59}.
At each computational step, all physics operators are advanced relative to a global time step, chosen as the minimum among limits imposed by numerical stability (CFL) and constraints on allowable relative changes in evolved physical quantities.
For the fiducial \chimera\ model, this results in a minimum time step of $\approx 5\times10^{-8}$~s around core bounce, which relaxes to $\approx10^{-6}$~s by $\sim10$~ms after bounce and remains at that level thereafter (solid black line in Figure~\ref{fig:chimera_dt}).  
When the allowed time step for the implicit neutrino transport update exceeds the global hydrodynamics time step, \chimera\ employs a super-cycling strategy, reducing the number of transport solves by as much as a factor of 10 (dashed black line in Figure~\ref{fig:chimera_dt}).  
Consequently, although the \chimera\ simulation executes $\approx1.5\times10^5$ global time steps, only $\approx3.6\times10^{4}$ neutrino transport solutions and $\approx1.4\times10^{5}$ cumulative Newton iterations are performed.  
After bounce, the average number of Newton iterations per transport solve varies from about 6 ($\le30$~ms) to about 4 (varying between 5 and 3 after 40~ms), and the average number of linear iterations per Newton iteration varies between 3 and 4.  
For comparison, the fiducial \thornado+\flashx\ simulation shown in Figure~\ref{fig:profile_timeseries}, whose time step is CFL-limited by the speed of light, requires $\approx2.7\times10^{5}$ time steps for 150~ms of post-bounce evolution.  

Because \chimera's transport is based on flux-limited diffusion, some discrepancies with \thornado's two-moment method are expected from the inherent differences between the models.  
We have verified that \chimera\ and \flashx\ yield excellent agreement in simulations of adiabatic collapse, bounce, and shock propagation (i.e., when neutrino transport is not included) using the same EoS and progenitor model considered here.  
To provide additional context for interpreting differences in simulations with neutrino transport, we have also performed \thornado+\flashx\ runs using, in addition to the fiducial Minerbo closure, the Levermore and Kershaw closures (see Appendix~\ref{app:closures}).  
In this comparison, we run all simulations out to $150$~ms after bounce.  

\begin{figure}[h]
	\centering
	\includegraphics[width=0.98\textwidth]{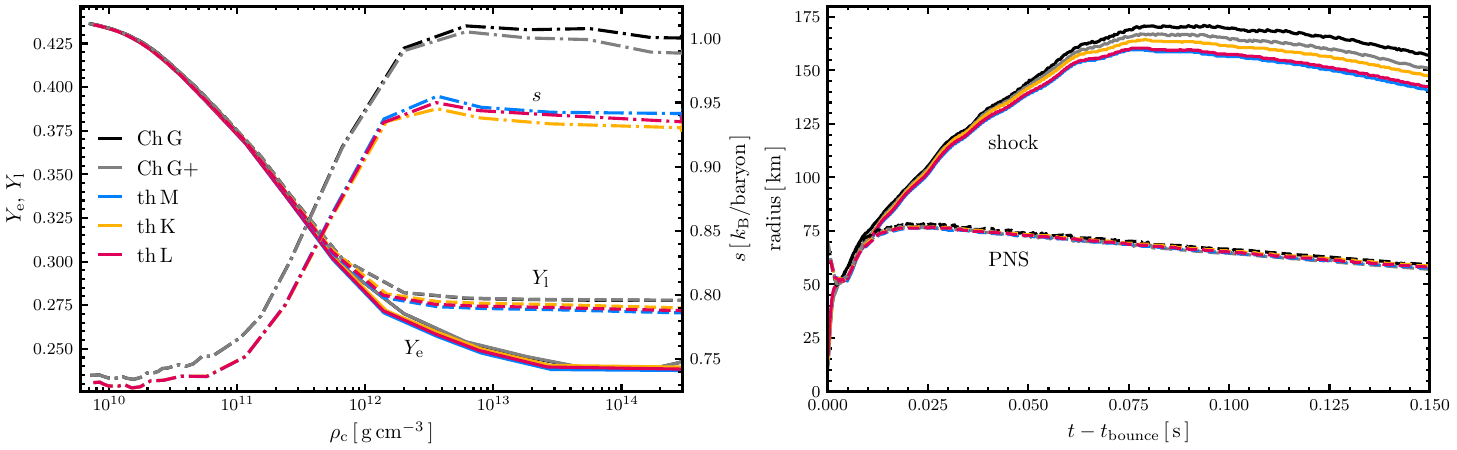}
	\caption{Central electron fraction ($\ye$; solid lines), lepton fraction ($Y_{\rm l}$; dashed lines), and entropy per baryon ($s$; dash-dot lines) versus central density during core-collapse (left panel), and post-bounce evolution of shock (solid lines) and PNS (dashed lines) radii (right panel).  
	The different line colors correspond to a fiducial \chimera\ G-series run (Ch~G; black), a \chimera\ G-series run with modified flux-limiter (Ch~G+; gray), and three \thornado+\flashx\ runs with different closures: Minerbo (th~M; blue), Kershaw (th~K; yellow), and Levermore (th~L; red). 
	In the left panel, due to differences in grid structure between codes, we calculate the central quantities as mass averaged over the innermost $0.01\,M_{\odot}$.}
	\label{fig:shock_and_central_quantities}
\end{figure}

The left panel of Figure~\ref{fig:shock_and_central_quantities} shows selected central quantities as functions of central density ($\rho_{\rm c}$) during collapse.  
The trajectories for the central $\ye$ and total lepton fraction $Y_{\rm l}$ agree well across codes.  
The two quantities are close until $\rho_{\rm c}\approx10^{12}$~g~cm$^{-3}$, when neutrino trapping begins.  
At higher densities, $Y_{\rm l}$ stays approximately constant, whereas $\ye$ continues to decrease as neutrinos and matter equilibrate, until $\rho_{\rm c}\approx3\times10^{13}$~g~cm$^{-3}$.  
For $\ye$ and $Y_{\rm l}$, the two \chimera\ models are indistinguishable, while the \thornado\ models show modest variation among the different closures.  
After trapping, $Y_{\rm l}$ varies by less than one percent across all models, ranging from $0.274$ to $0.276$.  
Larger differences appear in the evolution of the central entropy.  
Although the trajectories have similar shapes, the \chimera\ models reach higher central entropies ($\sim0.98$--$0.99$) than the \thornado\ models ($\sim0.93$--$0.94$).  
Differences of comparable or greater magnitude in this quantity have also been reported in previous code comparison studies \citep[e.g.,][]{oconnor_2015,kuroda_etal_2016}.  

The right panel of Figure~\ref{fig:shock_and_central_quantities} shows the temporal evolution of the PNS radius ($\Rpns$) and the shock radius ($\Rshock$).  
We define $\Rpns$ as the radius at which the mass density equals $10^{11}$~g~cm$^{-3}$.  
The evolution of $\Rpns$ is nearly identical across all models, reaching a maximum of about $77$~km at roughly $20$~ms, and contracting to $55$--$60$~km by the end of the simulations.  
The shock radii also agree closely at early times but begin to separate gradually after $\sim50$~ms, maintaining similar overall trajectories that do not cross during the evolution.  
Among the \chimera\ models, \chimera~G exhibits the largest shock excursion, reaching a maximum of $\sim170$~km and remaining nearly constant between $75$ and $100$~ms.  
\chimera~G+ follows closely, attaining a maximum of $\sim166$~km.  
For the \thornado\ simulations, the Kershaw closure produces the largest shock radius ($\sim165$~km), followed by the Levermove and Minerbo models, which track each other closely and peak at $\sim160$~km, with the Levermore case slightly ahead of Minerbo.  
This ordering among the two-moment closures is consistent with that reported by \citet{wangBurrows_2023}.  

\begin{figure}[h]
	\centering
	\includegraphics[width=0.98\textwidth]{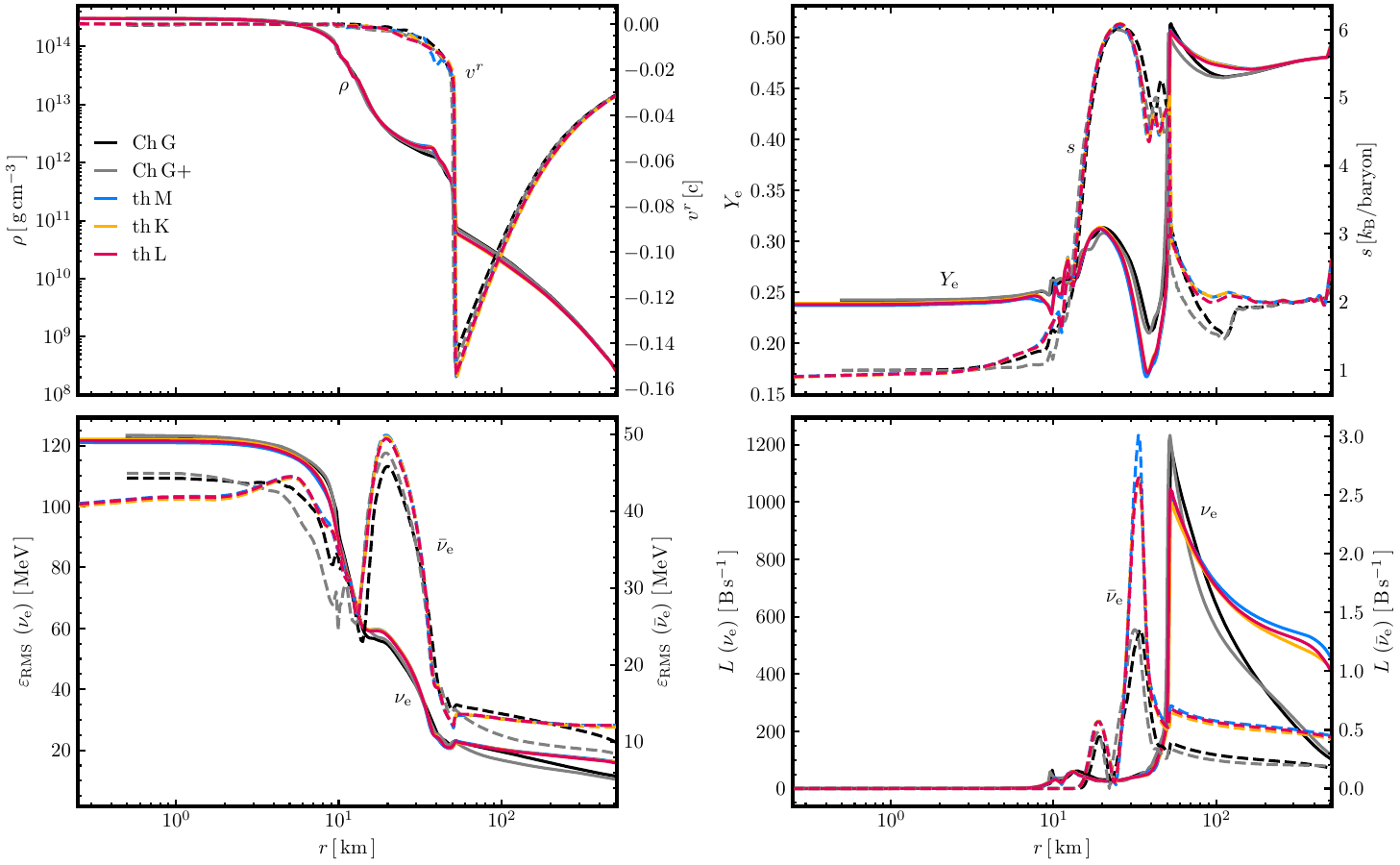}
	\caption{Radial profiles from \chimera\ and \thornado+\flashx\ at 3~ms after core bounce.
	Shown are: (top left) mass density (solid lines) and radial velocity (dashed lines);
	(top right) electron fraction (solid lines) and entropy per baryon (dashed lines);
	(bottom left) electron-neutrino and electron-antineutrino RMS energies (solid and dashed lines, respectively); and
	(bottom right) electron-neutrino and electron-antineutrino luminosities (solid and dashed lines, respectively).}
	\label{fig:radial_profiles_3ms}
\end{figure}

\begin{figure}[h]
	\centering
	\includegraphics[width=0.98\textwidth]{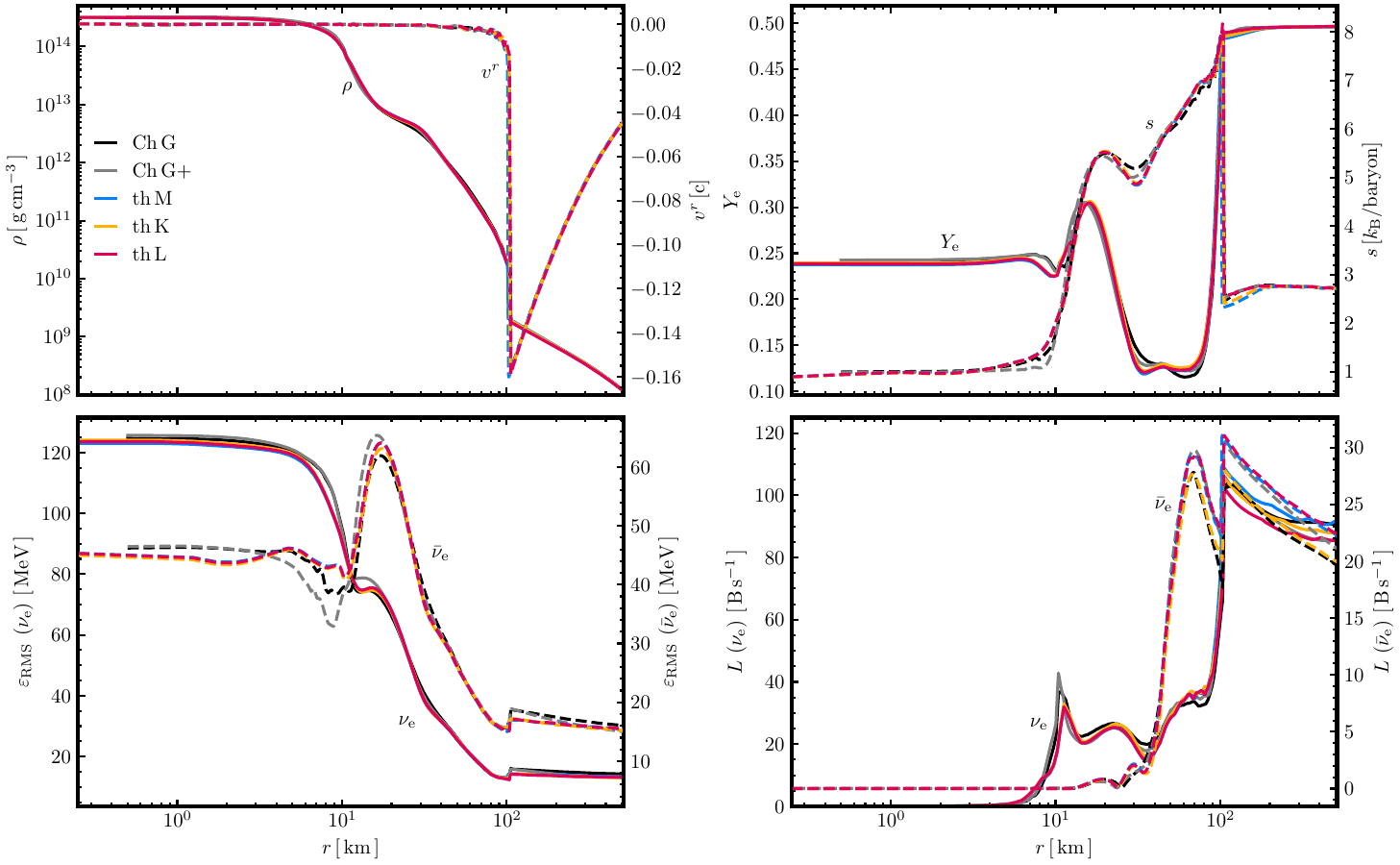}
	\caption{Radial profiles from \chimera\ and \thornado+\flashx\ at 25 ms after core bounce.
	Shown are: (top left) mass density (solid lines) and radial velocity (dashed lines);
	(top right) electron fraction (solid lines) and entropy per baryon (dashed lines);
	(bottom left) electron-neutrino and electron-antineutrino RMS energies (solid and dashed lines, respectively); and
	(bottom right) electron-neutrino and electron-antineutrino luminosities (solid and dashed lines, respectively).}
	\label{fig:radial_profiles_25ms}
\end{figure}

\begin{figure}[h]
	\centering
	\includegraphics[width=0.98\textwidth]{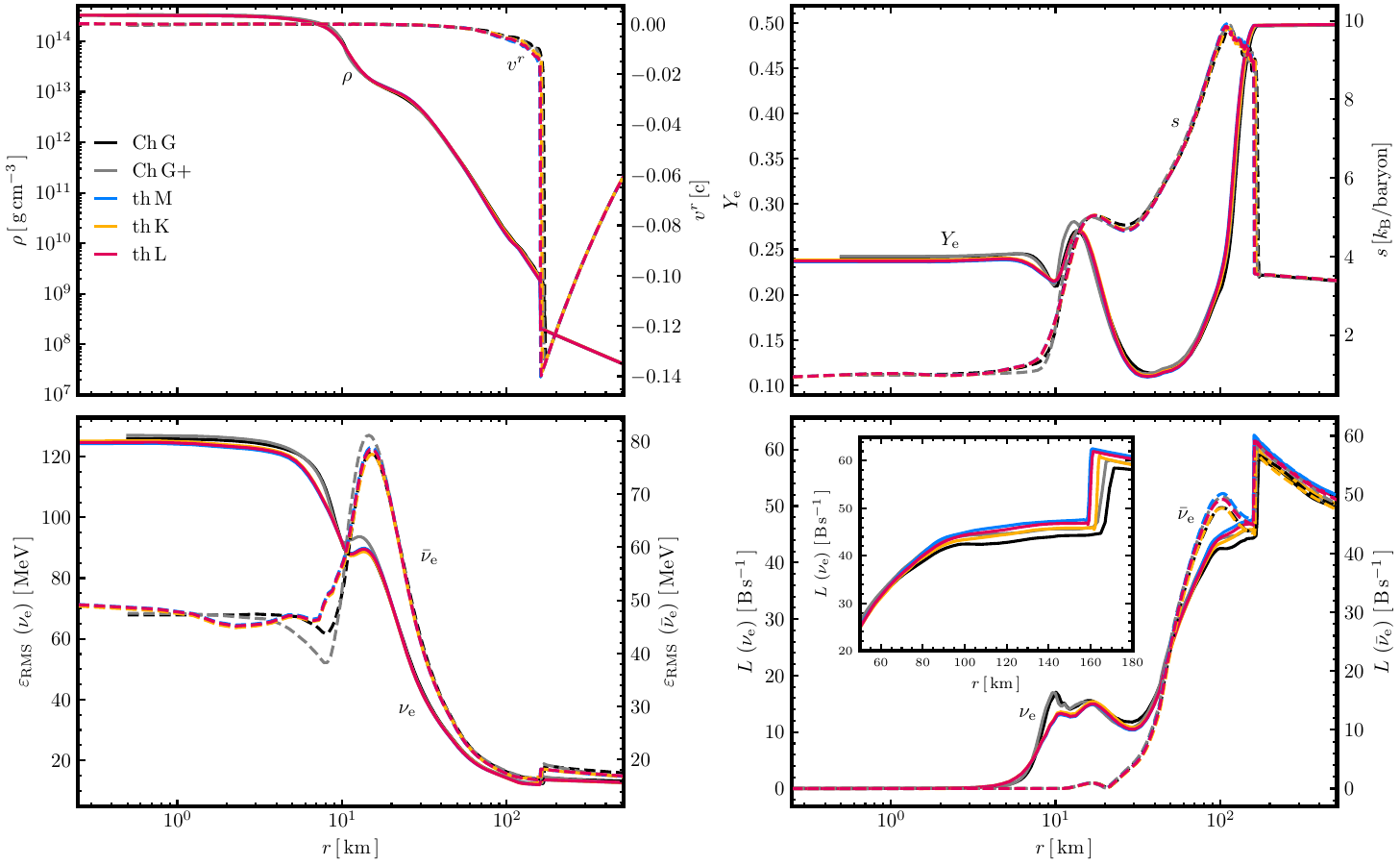}
	\caption{Radial profiles from \chimera\ and \thornado+\flashx\ at 75 ms after core bounce.
	Shown are: (top left) mass density (solid lines) and radial velocity (dashed lines);
	(top right) electron fraction (solid lines) and entropy per baryon (dashed lines);
	(bottom left) electron-neutrino and electron-antineutrino RMS energies (solid and dashed lines, respectively); and
	(bottom right) electron-neutrino and electron-antineutrino luminosities (solid and dashed lines, respectively).  
	The inset shows a zoomed view of the electron-neutrino luminosity below the shock.}
	\label{fig:radial_profiles_75ms}
\end{figure}

Figures~\ref{fig:radial_profiles_3ms}, \ref{fig:radial_profiles_25ms}, and \ref{fig:radial_profiles_75ms} compare selected hydrodynamic and neutrino-radiation quantities as functions of radius at $3$, $25$, and $75$~ms after bounce, respectively.  
At $3$~ms, when the shock is located near $r=50$~km, the velocity and mass density profiles show excellent agreement across all simulations, and this consistency persists at $25$~ms and $75$~ms.  
Within the inner core ($r\lesssim8$~km), the electron fraction in the \chimera\ models is slightly higher than in the \thornado+\flashx\ models.  
Farther out, around $r=40$~km where the $\ye$ trough is developing, the minimum electron fraction is lower in the \thornado+\flashx\ models ($\ye\approx0.17$) than in the \chimera\ models ($\ye\approx0.21$).  
We also note slight discrepancies in the electron fraction and entropy profiles ahead of the shock.  
Below the shock, the electron-neutrino RMS energies and luminosities are in good agreement between the codes, while the electron-antineutrino RMS energies and luminosities exhibit larger differences.  
Across the two-moment closures, the \thornado+\flashx\ results are internally consistent.  
The \thornado+\flashx\ models show slightly higher peak $\bar{\nu}_{\rm e}$ RMS energies around $r=20$~km and roughly a factor of two higher peak luminosity near $r=30$~km.  
Differences in neutrino-radiation quantities ahead of the shock may be attributed in part to differences in how FLD and two-moment transport propagate perturbations associated with the electron-neutrino burst.  
In addition, discrepancies may also be attributed to the transient nature of early post-bounce evolution, which could affect quantities both below and above the shock.  

At $25$~ms and $75$~ms, when the shock is located near $100$~km and $160$-$170$~km, respectively, the electron fraction and entropy profiles have converged further across the codes.  
At $25$~ms, small differences remain near the bottom of the $\ye$ trough, but at $75$~ms these have essentially vanished.  
The neutrino-radiation quantities also show good overall agreement at these times.  
Although the RMS energy and luminosity profiles can be distinguished in some regions, their overall shapes and evolution remain consistent among all models, and the remaining differences likely reflect the continuously evolving post-bounce structure.  
For instance, the peak electron-neutrino RMS energy, located between $10$ and $20$~km, increases from around $62$~MeV at $25$~ms to roughly $80$~MeV at $75$~ms.  
Over the same interval, the local peak in the electron-antineutrino luminosity below the shock, rises from $\sim30$~B~s$^{-1}$ (1~Bethe (B) = $10^{51}$~erg) to $\sim50$~B~s$^{-1}$, while the electron-neutrino luminosity at the shock decreases in the aftermath of the deleptonization burst, from $100$-$120$~B~s$^{-1}$ to about $60$~s$^{-1}$.  
By $75$~ms, the neutrino profiles are settling into overall good agreement; however, a noticeable difference in the $\nu_{\rm e}$ luminosities below the shock has developed (see inset).  
The \chimera\ luminosities are lower than those from \thornado+\flashx, and the ordering of models from low to high luminosity is opposite to the ordering of the corresponding shock radii shown in Figure~\ref{fig:shock_and_central_quantities}.  

\begin{figure}[h]
	\centering
	\includegraphics[width=0.95\textwidth]{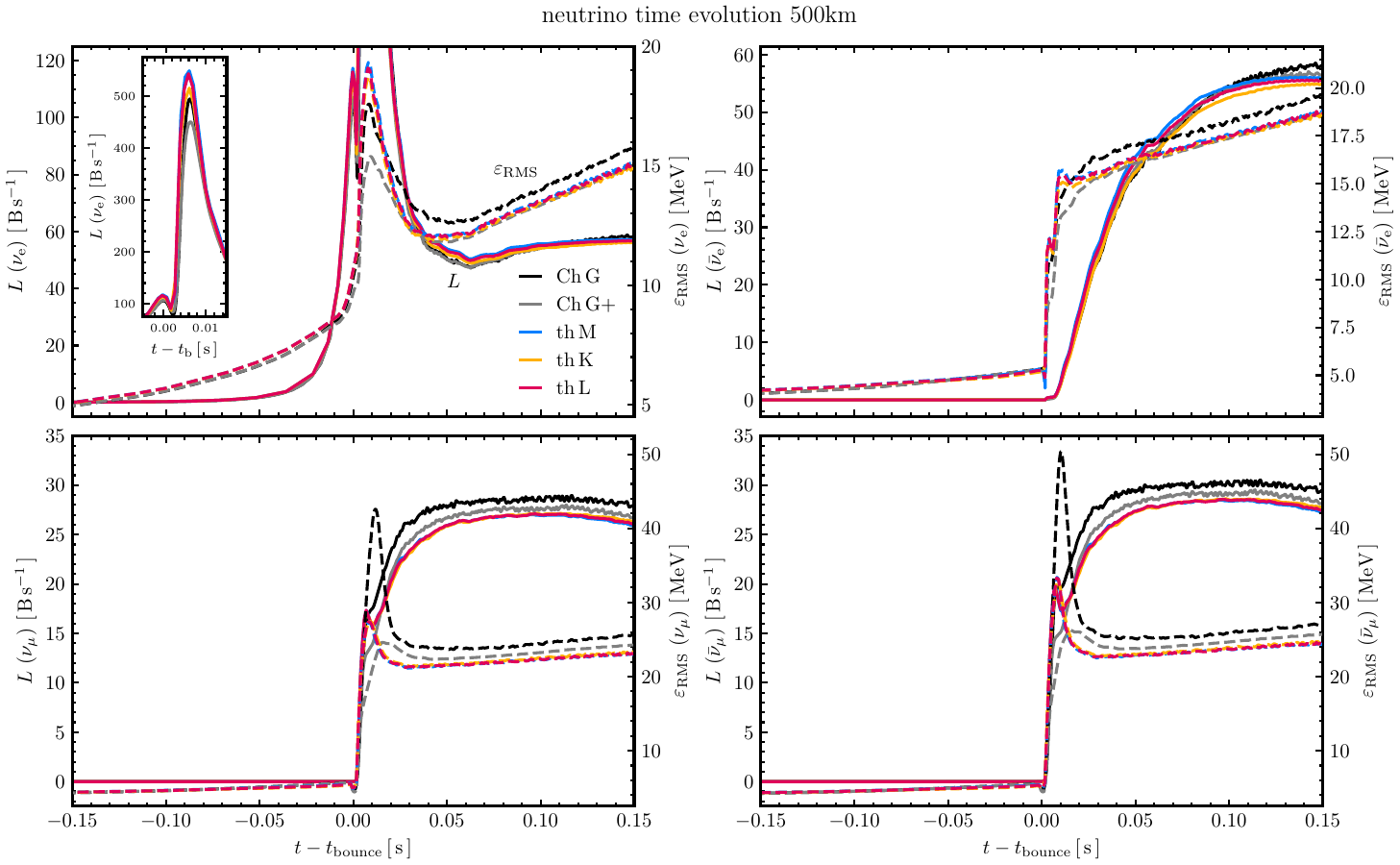}
	\caption{Time evolution of selected neutrino quantities from \chimera\ and \thornado+\flashx, sampled at $r=500$~km.  
	Each panel shows the luminosity (solid lines) and RMS energy (dashed lines) as functions of time after bounce.  
	Results are shown for electron neutrinos ($\nu_{\rm e}$; top left), electron antineutrinos ($\bar{\nu}_{\rm e}$; top right); muon neutrinos ($\nu_{\mu}$; bottom left); and muon antineutrinos ($\bar{\nu}_{\mu}$; bottom right).}
	\label{fig:neutrino_time_evolution}
\end{figure}

Figure~\ref{fig:neutrino_time_evolution} compares the time evolution of neutrino luminosities and RMS energies at $r=500$~km.  
For all neutrino species, these quantities are consistent among the \thornado+\flashx\ models.  
Overall, the electron-neutrino luminosities show very good agreement between \chimera\ and \thornado+\flashx\ models.  
As the burst passes this radius, the Minerbo and Levermore closure models reach peak luminosities of about $550$~B~s$^{-1}$ (see inset in the top left panel).  
The Kershaw and \chimera~G models attain slightly lower maxima of $\sim500$~B~s$^{-1}$, while \chimera~G+ peaks at $\sim450$~B~s$^{-1}$.  
The electron-neutrino RMS energies are consistent among the \thornado+\flashx\ models but differ somewhat from the \chimera\ results.  
During the burst passage, the \chimera\ models produce lower peak RMS energies, while at later times, when the RMS energies gradually increase due to spectral hardening during PNS contraction, the \chimera~G model remains higher by about $1$~MeV compared to the other models.  
Given the good agreement in RMS energies below the shock, these differences may be due to differences in the implementation of observer corrections across the shock in the respective transport schemes.  
For electron antineutrinos, the luminosities remain in good agreement across all models, but the \chimera~G model again exhibits about $1$~MeV higher RMS energy during PNS contraction.  
For muon neutrinos and antineutrinos, both the luminosities and RMS energies in the \chimera\ models are slightly higher than in the \thornado+\flashx\ models during PNS contraction, with \chimera~G slightly above \chimera~G+.  
During the brief passage of the burst (lasting $\sim3$~ms), the peak $\nu_{\mu}$ and $\bar{\nu}_{\mu}$ RMS energies in \chimera~G are notably higher than \thornado+\flashx, while the corresponding peaks in \chimera~G+ are notably lower.  

\subsection{Sensitivity to Solver Parameters}

\begin{figure}[h]
	\centering
	\includegraphics[width=0.95\textwidth]{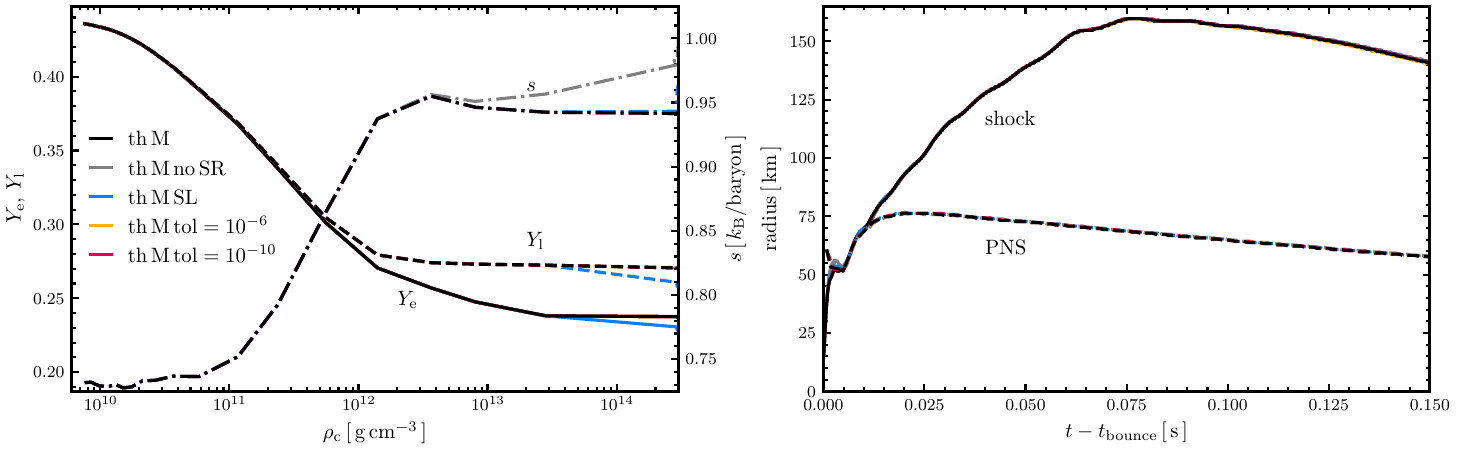}
	\caption{Central electron fraction ($\ye$; solid lines), lepton fraction ($Y_{\rm l}$; dashed lines), and entropy per baryon ($s$; dash-dot lines) versus central density during core collapse (left panel), and post-bounce evolution of the shock (solid lines) and PNS (dashed lines) radii (right panel).  Line colors correspond to the following \thornado+\flashx\ runs: the fiducial Minerbo run with $\texttt{tol}=10^{-8}$ (Th~M; black); a model without the spectral redistribution of \citet{laiu_etal_2025} (Th~M~no~SR; gray); a model using the slope limiter described in Section~\ref{sec:limiters} with $\beta_{\mbox{\tiny TVD}}=1.75$ (th~M~SL; blue); a model with an increased nonlinear solver tolerance of $\texttt{tol}=10^{-6}$ (th~M~tol=$10^{-6}$; yellow); and a model with a decreased tolerance of $\texttt{tol}=10^{-10}$ (th~M~tol=10$^{-10}$; red).}
	\label{fig:central_quantities_shock_radius_solver_params}
\end{figure}

\begin{figure}[h]
	\centering
	\includegraphics[width=0.95\textwidth]{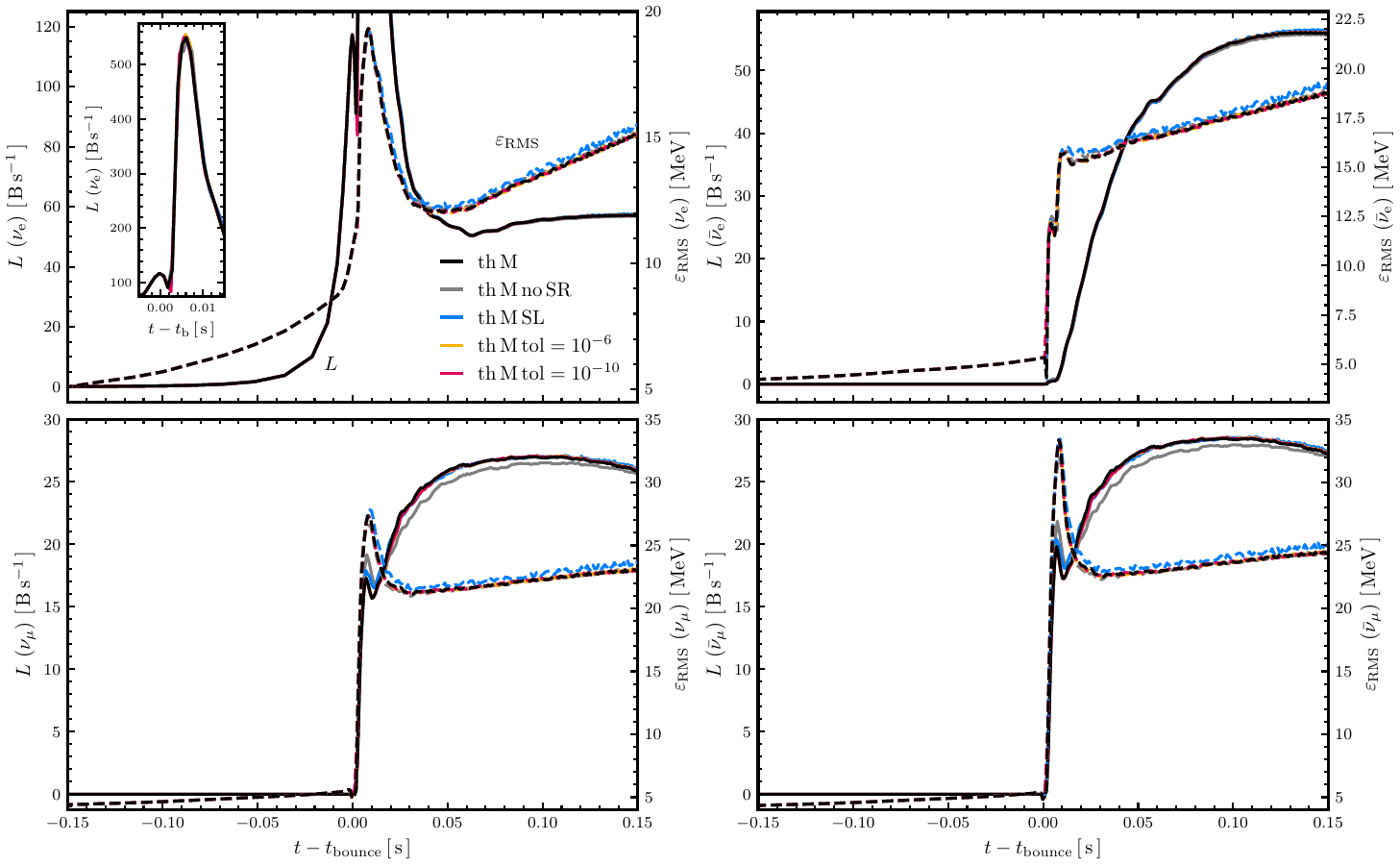}
	\caption{Time evolution of selected neutrino quantities sampled at $r=500$~km, from the same \thornado+\flashx\ models shown in Figure~\ref{fig:central_quantities_shock_radius_solver_params}.  
	Each panel plots the luminosity (solid lines) and RMS energy (dashed lines) as functions of time after bounce.  
	Results are shown for electron neutrinos ($\nu_{\rm e}$; top left), electron antineutrinos ($\bar{\nu}_{\rm e}$; top right), muon neutrinos ($\nu_{\mu}$; bottom left), and muon antineutrinos ($\bar{\nu}_{\mu}$; bottom right).}
	\label{fig:L_RMS_500km_solver_params}
\end{figure}

Next, within the context of spherically-symmetric core-collapse simulations, we examine the sensitivity of results to components and parameters of the neutrino transport method implemented in \thornado, using the fiducial model with the Minerbo closure from the previous section as a reference.  
Figures~\ref{fig:central_quantities_shock_radius_solver_params} and \ref{fig:L_RMS_500km_solver_params}, which display the same quantities as Figures~\ref{fig:shock_and_central_quantities} and \ref{fig:neutrino_time_evolution}, respectively, demonstrate that varying the nonlinear solver tolerance over four orders of magnitude, from $10^{-10}$ to $10^{-6}$, has negligible impact on the hydrodynamic evolution.  
Specifically, the central electron and lepton fractions and entropy during collapse, as well as the PNS and shock radii after bounce, remain effectively unchanged.  
Similarly, the luminosities and RMS energies for all neutrino species show no discernible sensitivity to the solver tolerance in this range.  
The most notable effect of disabling the spectral redistribution algorithm, which restores energy conservation following application of the realizability-enforcing limiter, is a gradual increase in the central entropy per baryon during neutrino trapping and equilibration.  
In this case, the central entropy rises by about three percent, from $\sim0.95$~$\kB$~baryon$^{-1}$ at $\rho_{\rm c}=6\times10^{12}$~g~cm$^{-3}$ to $\sim0.98$~$\kB$~baryon$^{-1}$ at $\rho_{\rm c}=3\times10^{14}$~g~cm$^{-3}$.  
We also observe a slight reduction in the muon neutrino and antineutrino luminosities during the post-bounce evolution.  
We attribute the central entropy behavior to repeated applications of the realizability-enforcing limiter needed to maintain positive neutrino number densities as the electron-neutrino Fermi surface moves to higher energies during this phase \citep[e.g.,][]{mezzacappaBruenn_1993a}.  
In the absence of spectral redistribution, these limiter operations effectively add energy to the neutrino field, which is subsequently shared with the matter through equilibration.  
Regarding the muon neutrino luminosities, it is conceivable that the realizability-enforcing limiter is triggered more frequently in regions affecting the heavy-lepton neutrinos, given that these species decouple at higher densities.  
The model with the slope limiter activated exhibits a reduction of about three percent in the central electron and lepton fractions immediately before bounce, and a slight ($\sim$1-2\%) increase in the RMS energies at $r=500$~km for all species.  
However, the evolution of $\Rshock$ and $\Rpns$ remains unaffected, showing no discernible differences from the fiducial model.  

\begin{figure}[h]
    \centering
    \includegraphics[width=0.95\textwidth]{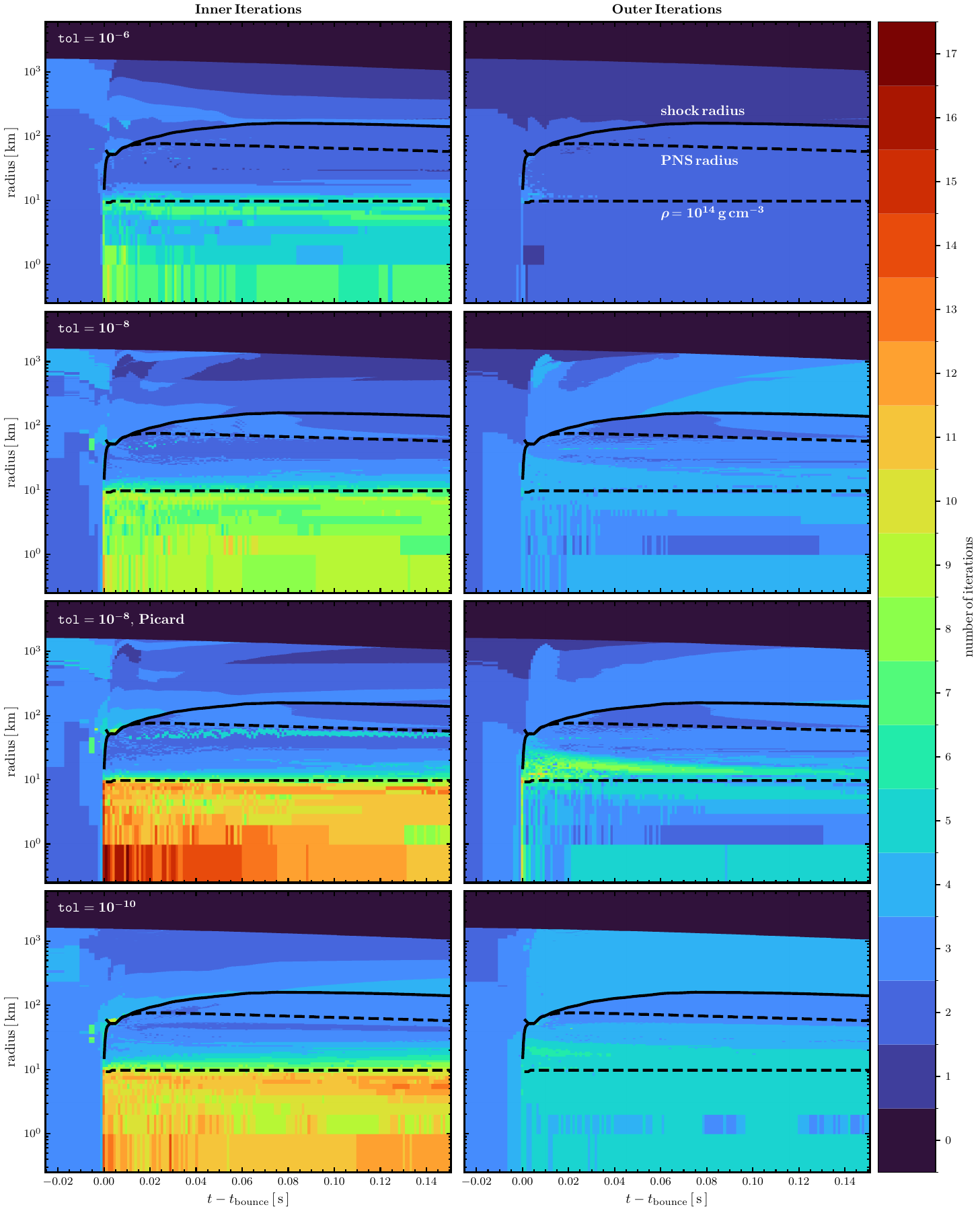}
    \caption{Heat maps of the number of inner (left column) and outer (right column) iterations as functions of radius and time after bounce, from spherically-symmetric core-collapse simulations.  
    Rows correspond to different nonlinear solver tolerances and iteration strategies: 
    the top row shows results for $\texttt{tol}=10^{-6}$ with Anderson acceleration ($\texttt{M{\_}outer}=\texttt{M{\_}inner}=2$); 
    the second row shows $\texttt{tol}=10^{-8}$ with Anderson acceleration ($\texttt{M{\_}outer}=\texttt{M{\_}inner}=2$); 
    the third row shows $\texttt{tol}=10^{-8}$ with Picard iteration ($\texttt{M{\_}outer}=\texttt{M{\_}inner}=1$); 
    and the bottom row shows $\texttt{tol}=10^{-10}$ with Anderson acceleration ($\texttt{M{\_}outer}=\texttt{M{\_}inner}=2$).  
    In each panel, the shock radius (solid black line), the PNS radius (dashed black line), and the $\rho=10^{14}$~g~cm$^{-3}$ contour (dashed black line) are indicated.}
    \label{fig:iteration_count_heatmaps}
\end{figure}

Figure~\ref{fig:iteration_count_heatmaps} illustrates the influence of the nonlinear solver tolerance on the number of inner and outer iterations.  
The data corresponds to the second stage of the PDARS IMEX scheme for selected time steps, with the number of inner iterations averaged over the outer iterations, following the same procedure as in Figure~\ref{fig:deleptonization}.  
Regions with mass density below $10^{7}$~g~cm$^{-3}$, where neutrino--matter coupling is disabled, show zero iterations because the nonlinear solve is skipped.
In the fiducial simulation (second row), the number of inner iterations remains between two and four throughout the evolution, while before bounce the outer iteration count varies within a similar range.  
At bounce, the number of outer iterations increases sharply to about seven or eight for densities above $10^{14}$~g~cm$^{-3}$ ($r\lesssim10$~km) and remains relatively stable below ten thereafter.  
For densities below $10^{14}$~g~cm$^{-3}$, the outer iteration count decreases gradually with decreasing density, from roughly five to two or three iterations just below the shock.  
As the shock expands towards its maximum extent, the iteration counts at a given radius remain relatively constant in time.  
When the solver tolerance is varied, the overall iteration behavior remains qualitatively unchanged.  
Relaxing the tolerance by two orders of magnitude to $\texttt{tol}=10^{-6}$ yields a modest reduction in iterations: inner iterations typically vary between one and two, while outer iterations range from one to three before bounce.  
After bounce, outer iterations increase to between 4 and 8 for densities above $10^{14}$~g~cm$^{-3}$ and decrease with density to between 2 and 3 just below the shock.  
Conversely, tightening the tolerance by two orders of magnitude to $\texttt{tol}=10^{-10}$ results in a modest increase in iterations.  
Here, inner iterations are typically between 3 and 5, while outer iterations are mostly between 3 and 4 before bounce.  
After bounce, outer iterations rise to between 10 and 13 for $\rho\gtrsim10^{14}$~g~cm$^{-3}$, while remaining below $\sim$5 at lower densities below the shock.  
The Anderson-accelerated solver converges consistently faster than the Picard iteration scheme at the same tolerance, particularly in regions of strong neutrino--matter coupling, demonstrating the improved efficiency of the accelerated method.  
Overall, the iteration heat maps confirm that the nonlinear solver converges robustly across a wide range of tolerances, with convergence behavior largely insensitive to parameter variations.  

Because the nonlinear neutrino--matter coupling constitutes the most computationally expensive part of the algorithm, changes in iteration counts are clearly reflected in the overall runtimes.  
Normalizing total runtimes, measured from the onset of collapse through 150~ms of post-bounce evolution, by that of the fiducial Minerbo model with $\texttt{tol}=10^{-8}$, we obtain normalized runtimes of $0.81$, $1.29$, and $1.27$, for the $\texttt{tol}=10^{-6}$, Picard with $\texttt{tol}=10^{-8}$, and $\texttt{tol}=10^{-10}$ simulations, respectively.  
For the same tolerance, Anderson acceleration yields an overall runtime reduction of approximately 22\% compared with the unaccelerated Picard iteration.  

\subsection{Demonstration of an Axially Symmetric Simulation}

We conclude this section with a demonstration of \thornado+\flashx\ applied to a multidimensional CCSN simulation, evolved from the onset of collapse to 150~ms past bounce.  
The fiducial Minerbo-closure model introduced in the beginning of the section is here evolved in axial symmetry using cylindrical coordinates, with a spatial domain $R\in[0,5120]$~km and $z\in[-5120,5120]$~km.
The spatial resolution in this model is coarser than in the spherical \chimera\ comparison (see Appendix~\ref{app:resolution} for a discussion on resolution dependence in 1D).  
We employ a fixed mesh-block size of $16\times16$ FV cells, and a base grid of 5 blocks in $R$ and 10 blocks in $z$, yielding a base resolution of $\Delta R = \Delta z = 64$~km.  
AMR is applied up to a maximum refinement level of $\ell_{\max}=6$, giving a finest-level resolution of $\Delta R = \Delta z = 1$~km.  
For the DG discretization, we use linear basis functions (polynomial degree $k=1$), such that each DG element spans four FV cells (two in each spatial dimension).  
When maximally refined, the time step is identical to that of the 1D models ($\sim5.6\times10^{-7}$~s) because the factor of two increase in cell width is cancelled by the $d_{\bx}=2$ factor in Equation~\eqref{eq:cflConditionTransport}.  
The maximum refinement level increases progressively from $\ell_{\max}=2$ to $\ell_{\max}=6$ during collapse, with additional levels activated when the central density satisfies $\log_{10}(\rho_{\rm c}/[{\rm g}~{\rm cm}^{-3}])=11$, $11.5$, $12$, and $12.5$.  
When $\ell_{\max}\ge3$, the region inside a radius of $\sim600$~km is refined to at least $\ell=3$.  
This constitutes the minimum refinement level within $600$~km, and higher refinement is possible when triggered by refinement criteria.  
In order to enforce a grid structure that attempts to maintain roughly constant angular resolution ($\Delta r/r \sim {\rm const}$), we follow~\cite{oconnorCouch_2018} and do not mark a block for refinement if the angular resolution of the block center is less than $\sim 1.4^{\circ}$, and actively mark a block for de-refinement if the angular resolution of the block center is less than $\sim 0.7^{\circ}$.  

\begin{figure}[h]
    \centering
    \includegraphics[width=0.95\textwidth]{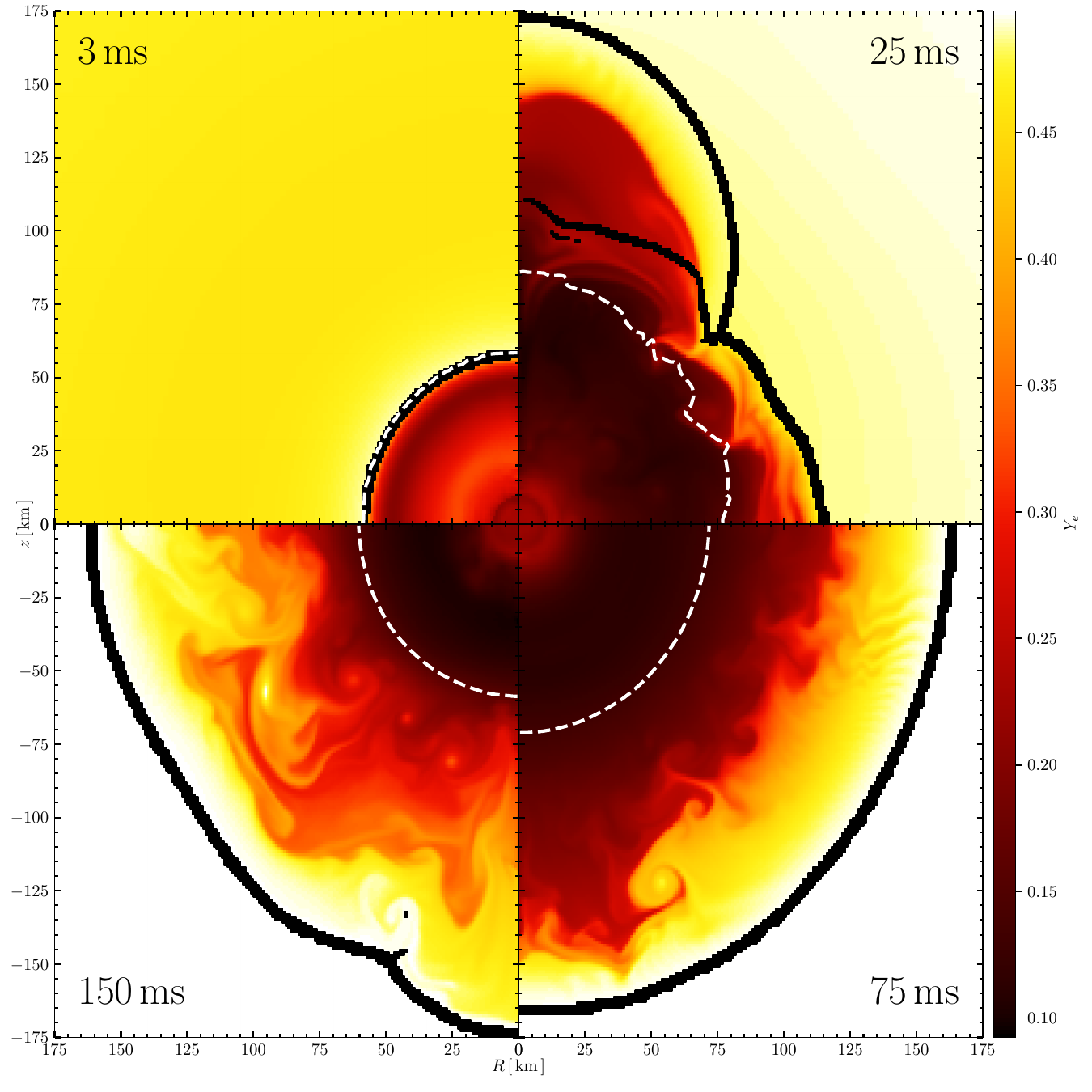}
    \caption{Color maps of the electron fraction at selected post-bounce times from the axisymmetric \thornado+\flashx\ simulation.  
    Clockwise from the upper-left panel: 3~ms, 25~ms, 75~ms, and 150~ms.  
    In each panel, cells where the \flashx\ variable {\sc shok}, 
    which flags finite volume cells containing a shock, equals unity, are plotted in black, 
    and the dashed white contour indicates $\Rpns$, defined by $\rho=10^{11}$~g~cm$^{-3}$.}
    \label{fig:2D_Ye_evolution}
\end{figure}

Figure~\ref{fig:2D_Ye_evolution} shows a sequence of $\ye$ snapshots illustrating the early post-bounce evolution of the axisymmetric model, from 3 to 150~ms after bounce.  
At 3~ms, the shock remains nearly spherical and is located at a radius of about $60$~km.  
By 25~ms, however, it has become noticeably aspherical, assuming a prolate shape with a radius of roughly 175~km along the pole and 115~km in the equatorial region.  
The onset of this deformation, and the associated convective activity, occurs earlier than in the axisymmetric \chimera\ models of \citet{bruenn_etal_2016}, which used spherical-polar coordinates, but on a similar time scale to the cylindrical-coordinate simulation of \citet{zhang_etal_2012}.  
The negative entropy gradient behind the shock, visible in the 1D profiles in Figure~\ref{fig:radial_profiles_3ms}, gives rise to a convectively unstable layer.  
We suspect that the relatively coarse Cartesian resolution amplifies quadrupolar perturbations, contributing to the early and vigorous development of prompt convection.  
At later times (75 and 150~ms), the shock has relaxed toward a more quasi-spherical shape but remains continually perturbed by convective activity in the neutrino-heating region.  
Deeper layers are more spherical, as indicated by the $\rho=10^{11}$~g~cm$^{-3}$ contour (the PNS radius), which contracts noticeably from approximately $70$~km at 75~ms to $60$~km at 150~ms.  

\begin{figure}[h]
    \centering
    \includegraphics[width=0.95\textwidth]{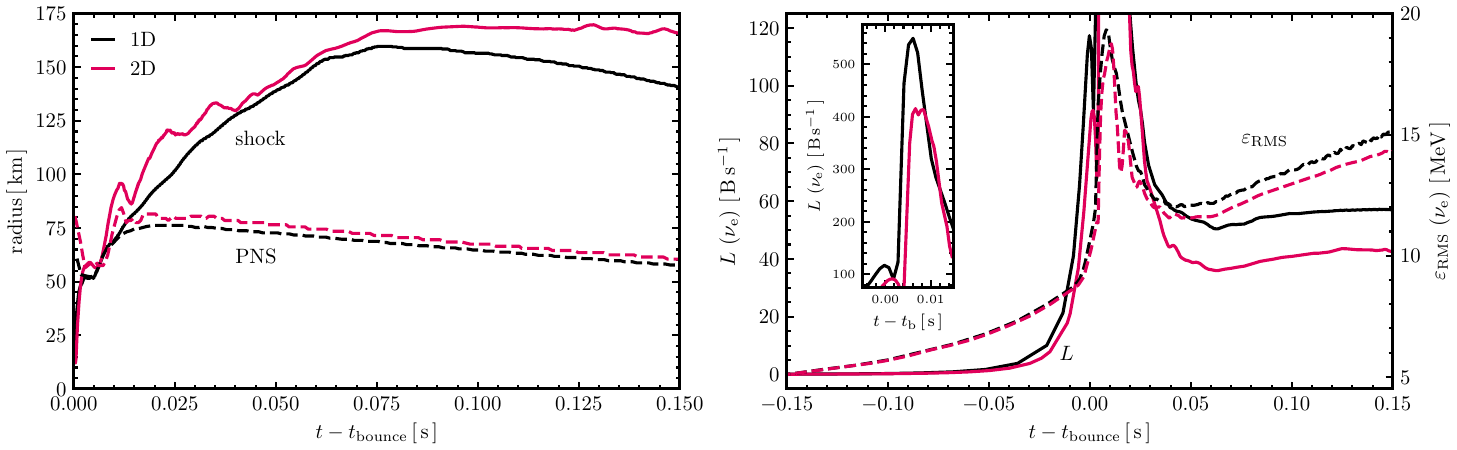}
    \caption{Comparison of spherically symmetric (black) and axisymmetric (red) \thornado+\flashx\ simulations, showing selected quantities as functions of time.  
    The left panel plots the shock (solid lines) and PNS (dashed lines) radii, and the right panel plots the electron-neutrino luminosity (solid lines) and RMS energy (dashed lines) at $r=500$~km.}
    \label{fig:1D_2D_comparison}
\end{figure}

Figure~\ref{fig:1D_2D_comparison} presents quantitative comparisons between the axisymmetric (2D) and corresponding spherically symmetric (1D) models.  
Overall, the two models exhibit qualitatively similar behavior.  
The shock and PNS radii (taken to be the mean radii in 2D) in both cases evolve along comparable trajectories, and the electron-neutrino luminosities and RMS energies follow similar trends over time.  
However, at least two factors contribute to the differences observed between the 1D and 2D results.  
First, the early convective activity in the 2D model influences the evolution of both $\Rshock$ and $\Rpns$, which are systematically larger than in the 1D case.  
Within the first 50~ms, several oscillations appear in the average shock radius and $\Rpns$ of the 2D model that are absent in 1D.  
After about $75$~ms, $\Rshock$ in the 1D model begins to recede, while in the 2D model it stabilizes near $170$~km with smaller oscillations superimposed.  
The peak electron-neutrino luminosity associated with the deleptonization burst (see inset in the right panel) is also significantly lower in 2D, reaching about $400$~B~s$^{-1}$ compared to $550$~B~s$^{-1}$ in 1D.  
At later times, the luminosity and RMS energy in the 2D model remain approximately $25\%$ and $1$~MeV lower, respectively, than in 1D.  
Second, the spatial resolution in the 2D simulation is a factor of two coarser than in 1D.  
As discussed in Appendix~\ref{app:resolution}, reducing the spatial resolution in the 1D model from $0.5$~km to $1$~km produces trends consistent with the differences observed between the 1D and 2D simulations, although the effects are somewhat less pronounced.  

\begin{figure}[h]
        \begin{center}
        \captionsetup[subfigure]{justification=centering}
        {\begin{minipage}{0.47\textwidth}
                        \includegraphics[width=\linewidth]{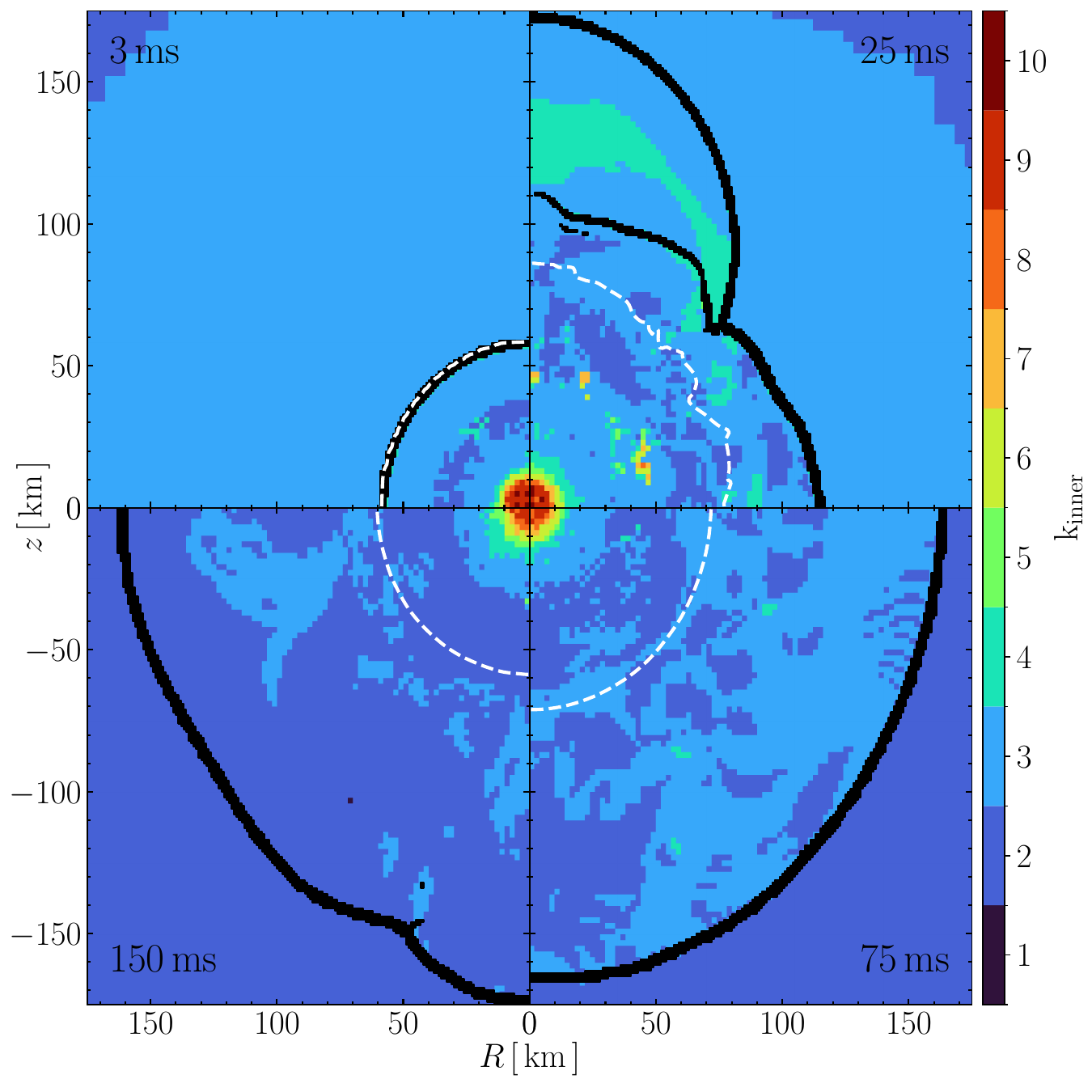}
                        \label{fig:}
                \end{minipage}
        }~~~~~~~
        {\begin{minipage}{0.47\textwidth}
                        \includegraphics[width=\linewidth]{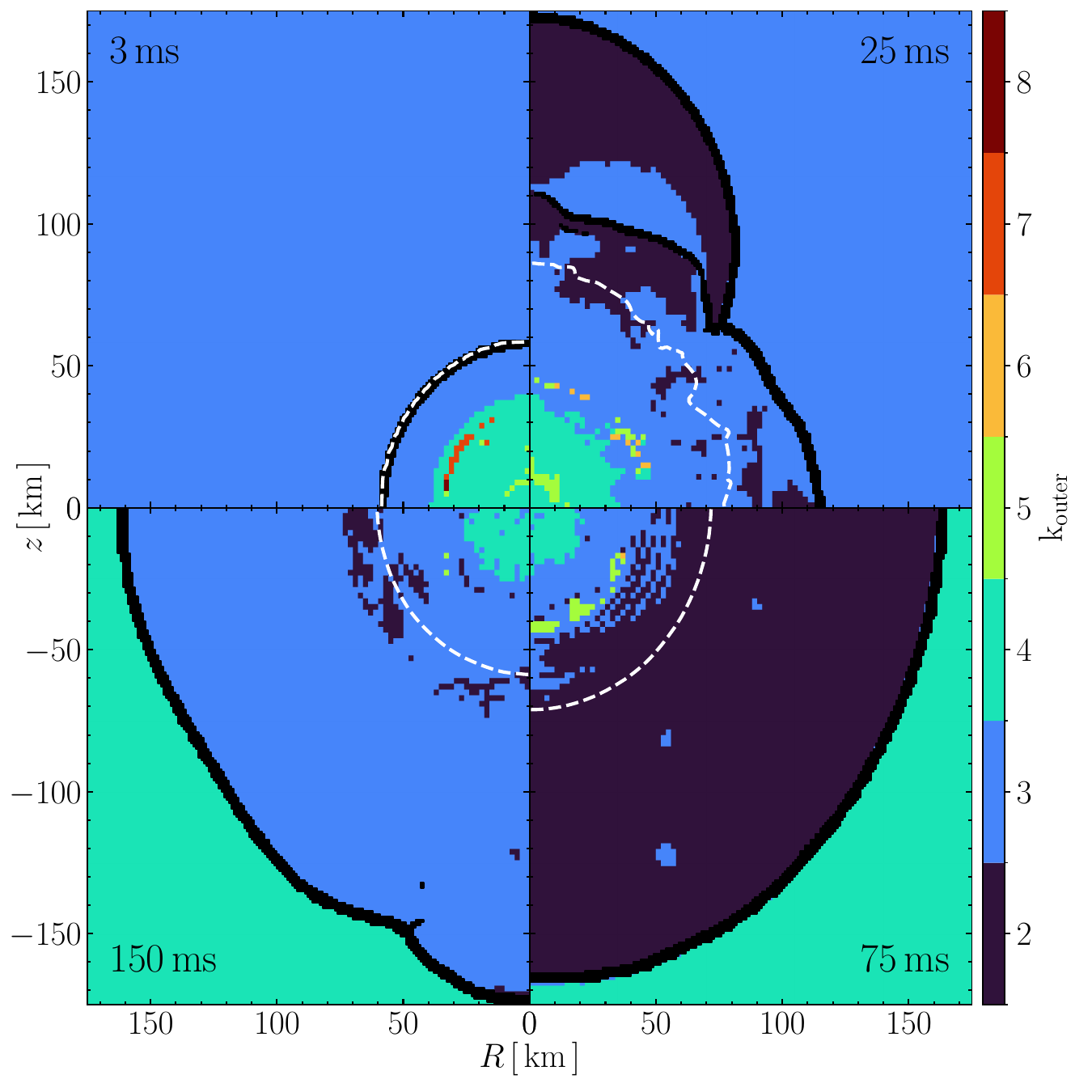}
                        \label{fig:}
                \end{minipage}
        }
        \caption{Heat maps showing the spatial distribution of the number of inner (left panels) and outer (right panels) iterations from the axisymmetric \thornado+\flashx\ simulation at the same post-bounce times in Figure~\ref{fig:2D_Ye_evolution}.
        Clockwise from the upper left: 3~ms, 25~ms, 75~ms, and 150~ms.}
        \label{fig:2D_iteration_overview}
        \end{center}
\end{figure}

Figure~\ref{fig:2D_iteration_overview} shows the spatial distribution of inner and outer iteration counts in the nonlinear neutrino--matter solver, exhibiting behavior consistent with the 1D results in Figure~\ref{fig:iteration_count_heatmaps}.  
The data correspond to the second stage of the PDARS IMEX scheme for the time steps  shown in Figure~\ref{fig:2D_Ye_evolution}, with inner iteration counts averaged over outer iterations.  
The highest iteration counts occur at high densities, well inside $\Rpns$, where localized hotspots intermittently reach eight inner iterations, though values typically lie between two and four.  
Between $\Rpns$ and $\Rshock$, the inner iteration count remains in the range of two to three.  
Outer iterations remain consistently between eight and ten for densities above $\sim10^{14}$, and drop to two--three for lower densities.  
At later times (75 and 150~ms), we observe slightly elevated inner iteration counts ahead of the shock; typically four, compared to two or three just inside the shock.  
Although not investigated in detail, this behavior is likely linked to the velocity-dependent step-size parameter $\lambda$ (Remark~\ref{rem:stepSizeParameter}), which promotes robustness in regions with large flow speeds at the expense of additional iterations.  
Finally, we note that the spatial variation in iteration count is not currently considered with regard to domain decomposition for parallel execution.  
Given the relative cost of the neutrino--matter coupling, accounting for this variation may improve load balance and overall efficiency in future simulations.  

\section{Summary, Conclusions, and Outlook}
\label{sec:conclusions}

We have presented a detailed description of the numerical methods for spectral two-moment neutrino transport in \thornado, including $\cO(v/c)$ velocity-dependent observer corrections, tabulated microphysics, and six neutrino species.  
Our approach combines DG phase-space discretization with IMEX time stepping, and introduces an extension of the nested fixed-point nonlinear solver of \citet{laiu_etal_2021} to enable fully implicit integration of the collision terms within the IMEX framework.  
Following the strategy developed in \cite{harris_etal_2022,thavappiragasam_etal_2024,laiu_etal_2025}, the neutrino transport solver has also been ported to GPUs using either OpenMP offloading or OpenACC, and all the supernova applications presented in this work employ this GPU-enabled implementation.  
To enable large-scale simulations, the solver is importable into \flashx, which provides finite-volume, self-gravitating hydrodynamics, distributed-memory parallelism, and AMR, and we have described the strategy used to transform between DG and finite-volume representations.  
The implementation of neutrino transport in \thornado, as well as the coupled \thornado+\flashx\ framework, has been verified on a hierarchy of test problems and supernova applications.  

Results from the basic verification tests with idealized opacities in Section~\ref{sec:results}, using spherical-polar, cylindrical, and Cartesian spatial coordinates, demonstrate accuracy in the diffusion limit on meshes that are coarse relative to the scattering mean free path, the correct behavior of velocity-dependent observer corrections for time-independent, spatially varying velocity fields, and the ability to advect free-streaming radiation cleanly across an adaptive mesh.  
	
Relaxation and deleptonization problems in Section~\ref{sec:application_idealized} introduce tabulated microphysics, six neutrino species, and an extended set of opacities, including neutrino--electron scattering and thermal pair processes.  
The relaxation problem, designed to assess nonlinear solver performance using backward Euler time stepping, begins from out-of-equilibrium radiation moments and demonstrates convergence to the correct equilibrium Fermi--Dirac distribution, with total lepton number and energy conserved to within the nonlinear solver tolerance.  
By varying the time step over three orders of magnitude, from $10^{-3}$~ms to $1$~ms, we find that the nonlinear solver maintains robust convergence without a corresponding increase in iteration counts, and that Anderson acceleration consistently improves convergence relative to Picard iteration.  
The deleptonization problem, which models transport through a static matter background over 100~ms using IMEX time stepping, exhibits qualitatively correct multi-species transport behavior and conserves total lepton number and energy within limits set by the nonlinear solver tolerance.  
Here, too, the solver shows robust convergence: the largest iteration counts occur in the inner core ($\rho\gtrsim10^{13}$~g~cm$^{-3}$), requiring roughly 3-6 outer iterations and up to 10-15 inner iterations per outer iteration, with the iteration count decreasing as the mass density decreases.  

Results from the CCSN applications in Section~\ref{sec:application_ccsn} further confirm the robustness of the proposed methods under fully dynamic conditions.  
In spherically-symmetric collapse and post-bounce evolution, despite crucial model (flux-limited diffusion (FLD) versus two-moment) and method differences, the \thornado+\flashx\ results exhibit remarkably close qualitative agreement with the \chimera\ reference models in many key quantities, including the temporal evolution of shock and PNS radii, neutrino luminosities, and RMS energies, and in radial profiles at selected post-bounce times.  
The shock trajectories of the individual models diverge gradually over time but remain highly synchronized throughout the 150 ms post-bounce comparison.
Although the precise origin of the separation is difficult to isolate, our results point to details in the transport treatment, for example, the choice of closure, as the primary contributor.  
The \chimera\ G+ model---which applies velocity-dependent observer corrections consistently below the shock---and the \thornado+\flashx\ model with the Kershaw closure track each other to within $\lesssim2\%$.
This close agreement is notable because FLD and Kershaw closures are conceptually similar: both interpolate between diffusive and free-streaming limits, with FLD determining the flux factor and Kershaw determining the Eddington factor.  
The nonlinear neutrino--matter coupling solver remains convergent throughout collapse, bounce, and early post-bounce evolution, with iteration counts following trends consistent with those observed in the idealized tests.  
The axisymmetric simulation demonstrates, \emph{for the first time using DG neutrino transport}, the capability of the coupled neutrino-radiation hydrodynamics framework to model multidimensional CCSN dynamics.  
The model develops prompt convection and shock deformation on physically reasonable time scales, and the evolution of neutrino luminosities, RMS energies, and global shock behavior remains broadly consistent with that of the corresponding 1D model, taking into account differences attributable to dimensionality and spatial resolution.  
Iteration counts in 2D are likewise consistent with the behavior observed in 1D and again indicate robust solver performance across the computational domain.  
Together, the CCSN applications provide an important validation of the neutrino transport implementation, the nonlinear solver strategy under realistic conditions, and the hybrid DG--finite-volume coupling.  
	
The successful implementation of neutrino transport in \thornado, and its integration with \flashx, provides a strong foundation for extending the framework's physics and numerical capabilities, as well as for improving performance in preparation for future applications in nuclear and multimessenger astrophysics.  

On the physics side, the incorporation of general relativistic effects is crucial for realistic modeling \citep[e.g.,][]{bruenn_etal_2001,lentz_etal_2012,muller_etal_2012,oconnorCouch_2018}.  
In the near term, we will adopt an effective-potential approach \citep[e.g.,][]{marek_etal_2006} to augment the Newtonian potential, and include gravitational redshift terms in \thornado's neutrino transport.  
Longer-term efforts will move toward fully relativistic models, building on recent progress toward relativistic neutrino transport with DG methods in \thornado\ \citep{hunter_etal_2025}.  
Additional physics capabilities---such as inelastic neutrino--nucleon scattering \citep[e.g.,][]{muller_etal_2012}, muonic effects \citep[e.g.,][]{bollig_etal_2017,fischer_etal_2020}, and coupling to nuclear reaction networks in \flashx\ (enabling computational domains that cover $\sim10^{5}$~km to follow explosion development)---represent natural extensions of the current framework.  

On the numerical side, the development of a robust fully implicit solver for nonlinear neutrino--matter coupling establishes a baseline for advancing multiphysics time-stepping strategies in CCSN models.  
In the current implementation, \thornado's IMEX schemes advance both explicit and implicit components using a single time step set by explicit stability constraints.  
This can be suboptimal for the collision term, whose dynamics become stiff but slow in near-equilibrium regions.  
Multi-rate methods \cite[e.g.,][]{Sandu_2019,rujekoReynolds_2021}, which evolve different operators with distinct time steps and may be designed to evaluate expensive implicit collision updates less frequently than cheaper explicit transport terms, present a promising direction made feasible by the solver infrastructure developed in this work.  

The GPU-enabled neutrino transport solver opens opportunities for large-scale simulations with improved fidelity.  
With \thornado\ now coupled to the \flashx\ framework and verified against \chimera, we have sufficient confidence in the implementation to begin a systematic, data-driven performance assessment to guide targeted optimizations of the transport solver---for example, consolidating kernels to reduce launch overhead, improving initial guesses for iterative solvers, and selective computation with reduced precision.  
Such refinements have the potential to lower the overall cost of neutrino transport, allowing improved physical fidelity in other model components.  
In particular, applications that require evolution of large nuclear reaction networks, such as electron-capture supernova models, stand to benefit from improved performance of the GPU-enabled transport solver.  

\begin{acknowledgements}
	We thank Margot Fitz Axen for a careful reading of the manuscript and helpful comments.  
	This research was partially supported by the Exascale Computing Project (17-SC-20-SC), a collaborative effort of the U.S. Department of Energy Office of Science and the National Nuclear Security Administration.
	The work at Oak Ridge National Laboratory is supported under contract DE-AC05-00OR22725 with the U.S. Department of Energy (DOE).
	The work at Lawrence Berkeley National Laboratory was supported by the U.S. Department of Energy under contract No. DE-AC02-05CH11231.
	This work was supported in part by the U.S. Department of Energy, Office of Science, Office of Advanced Scientific Computing Research and Office of Nuclear Physics, Scientific Discovery through Advanced Computing (SciDAC) program.
	E.~Endeve and A.~Mezzacappa acknowledge support from the National Science Foundation's (NSF's) Gravitational Physics program under Grant No.~2409148 and NSF's Cyberinfrastructure for Sustained Scientific Innovation program under Grant No.~2535874.  
	An award of computer time was provided by the INCITE program.  
	This research also used resources of the Oak Ridge Leadership Computing Facility, which is a DOE Office of Science User Facility supported under Contract DE-AC05-00OR22725.
	This research used resources of the National Energy Research Scientific Computing Center (NERSC), a Department of Energy User Facility using NERSC award NP-ERCAP0032339.
\end{acknowledgements}

\appendix

\section{Levermore and Kershaw Closures}
\label{app:closures}

In terms of the flux factor, $h$, the Eddington and heat-flux factor of the Kershaw-type closure implemented in \thornado\ are given by \citep{kershaw_1976,schneider_2016,banachLarecki_2017a}
\begin{equation}
	\psi_s^{\rm Ke}(h)
	= \frac{1}{3} \left( 1 + 2 h^2 \right)
	\qquad\text{and}\qquad
	\zeta_s^{\rm Ke}(h)
	= \frac{h}{9} \left( 4 h^2 + 5 \right), 
\end{equation}
respectively.  
These expressions have been derived for non-negative distribution functions without an upper bound.  

For the Levermore closure, the Eddington factor is given by \citep{levermore_1984}
\begin{equation}
	\psi_s^{\rm Le}(h) \equiv \frac{1}{3} \left( 5 - 2 \sqrt{ 4 - 3 h^2 } \right).
\end{equation}
The heat-flux factor of the Levermore closure was provided by \citet{vaytet_etal_2011}.  
In \thornado, we use an equivalent form provided by \citet{wangBurrows_2023}
\begin{align}
	\zeta(h)
	&= \left( \vphantom{{\rm arctanh}(\frac{-2 + a}{h} ) }4 h^3 (286 - 89 a) + 576 h (-2 + a) + 3 h^5 (-80 + 9 a) \right. \\ \nonumber
	&\hspace{32pt}
	- 48 \left[ h^6 + h^2 (42 - 15 a) + 3 h^4 (-5 + a) + 16 (-2 + a) \vphantom{h^6 + h^2 (42 - 15 a) + 3 h^4 (-5 + a)} \right] \left.  {\rm arctanh}\left[ \frac{-2 + a}{h} \right] \right) / (-2 + a)^5,
\end{align}
where $a=\sqrt{4 - 3\,h^2}$. 
While $\zeta(h)$ possesses well-defined limits as $h\to 0$ and $h\to1$ (yielding 0 and 1, respectively), it cannot be numerically evaluated directly at these endpoints, even with arbitrary precision floating-point arithmetic.
In double-precision floating-point arithmetic (which we use in \thornado), $\zeta(h)$ becomes highly oscillatory and numerically divergent for small $h$ that are typically attained in simulations. 
To avoid numerical instability related to these regimes, we perform a sixth-order Taylor expansion of $\zeta(h)$ around $h=0$, and use the resulting polynomial for $h\leq0.2$, as well as a safety factor for evaluation when $h$ is close to 1, to arrive at the following expression for the Levermore heat-flux factor
\begin{equation}
	\zeta_s^{\rm Le}(h)
	=
  	\begin{cases}
		\frac{3 h \left( 63 h^4 + 144 h^2 + 448\right)}{2240} & h \leq 0.2, \\
		\zeta (h) & 0.2 < h < 1-10^{-12}, \\
		1 & 1-10^{-12} \leq h.
	\end{cases}
\end{equation}

\section{Collision Terms}
\label{app:collisionTerms}

For completeness, in this appendix we provide explicit expressions for the individual processes on the right-hand side of Equation~\eqref{eq:collisionTerm_AllProcesses}.  

The term for emission and absorption can be written as
\begin{equation}
  \mathcal{C}_{s}^{\EmAb}(f_{s}) = \eta_{s}^{\EmAb} - \chi_{s}^{\EmAb}\,f_{s},
\end{equation}
where $\chi_{s}^{\EmAb}(\varepsilon)$ is the absorption opacity (corrected for stimulated absorption), which is independent of angle $\omega$, and the emissivity is obtained through the detailed balance relation $\eta_{s}^{\EmAb}=\chi_{s}^{\EmAb}\,f_{s}^{\Eq}$, where the equilibrium Fermi-Dirac distribution is
\begin{equation}
  f_{s}^{\Eq} = \f{1}{e^{(\varepsilon-\mu_{s})/T} + 1},
  \label{eq:FermiDiracDistribution}
\end{equation}
where $\mu_{s}$ is the chemical potential of neutrino species $s$.  

Scattering on nucleons and nuclei is treated as isoenergetic.  
The collision term for these processes is written as \citep{bruenn_1985}
\begin{align}
  \mathcal{C}_{s}^{\Iso}(f_{s}) 
  &= \f{4\pi\varepsilon^{2}}{h^{3}}
  \Big[\,
  	\f{1}{4\pi}\int_{\bbS^{2}}R_{s}^{\Iso}(\varepsilon,\cos\alpha)f_{s}(\omega',\varepsilon)d\omega' 
	- f_{s}(\omega,\varepsilon)\f{1}{4\pi}\int_{\bbS^{2}}R_{s}^{\Iso}(\varepsilon,\cos\alpha)d\omega
  \,\Big] \label{eq:collisionTerm_Iso} \\
  &=\f{4\pi\varepsilon^{2}}{h^{3}}
  \Big[\,
  	\Phi_{s,0}^{\Iso}(\varepsilon)\,\big(\,\mathcal{D}_{s}(\varepsilon)-f_{s}(\omega,\varepsilon)\,\big)
	+\Phi_{s,1}^{\Iso}(\varepsilon)\,\ell^{i}(\omega)\,{\mathcal{I}_{s}}_{i}
  \,\Big], \label{eq:collisionTermKernelExpanded_Iso}
\end{align}
where in Equation~\eqref{eq:collisionTermKernelExpanded_Iso} the angular dependence of the scattering kernel $R_{s}^{\Iso}$ is approximated by a truncated (to linear order) Legendre expansion in the scattering angle cosine, $\cos\alpha=\ell^{i}(\omega)\ell_{i}(\omega')$,
\begin{equation}
	R_{s}^{\Iso}(\varepsilon,\cos\alpha)
	:=\Phi_{s,0}^{\Iso}(\varepsilon)+\Phi_{s,1}^{\Iso}(\varepsilon)\,\cos\alpha.
	\label{eq:kernelExpansion_Iso}
\end{equation}

Neutrino--electron scattering (NES) is treated with energy exchange between neutrinos and electrons.  
The NES collision term is written as \citep{bruenn_1985,mezzacappaBruenn_1993c}
\begin{align}
  \mathcal{C}_{s}^{\NES}(f_{s})
  &=
  (1-f_{s}(\omega,\varepsilon))\int_{\bbR^{+}}\f{1}{4\pi}\int_{\bbS^{2}}R_{s}^{\In}(\varepsilon,\varepsilon',\cos\alpha)\,f_{s}(\varepsilon',\omega')\,d\omega'dV_{\varepsilon'} \nonumber \\
  &\hspace{12pt}
  -f_{s}(\omega,\varepsilon)\int_{\bbR^{+}}\f{1}{4\pi}\int_{\bbS^{2}}R_{s}^{\Out}(\varepsilon,\varepsilon',\cos\alpha)\,(1-f_{s}(\omega',\varepsilon'))\,d\omega'\,dV_{\varepsilon'} \label{eq:collisionTerm_NES} \\
  &=
  (1-f_{s}(\omega,\varepsilon))\,\int_{\bbR^{+}}\Phi_{s,0}^{\In}(\varepsilon,\varepsilon')\,\mathcal{D}_{s}(\varepsilon')\,dV_{\varepsilon'}
  -f_{s}(\omega,\varepsilon)\,\int_{\bbR^{+}}\Phi_{s,0}^{\Out}(\varepsilon,\varepsilon')\,(1-\mathcal{D}_{s}(\varepsilon'))\,dV_{\varepsilon'}
  \nonumber \\
  &\hspace{12pt}
  +(1-f_{s}(\omega,\varepsilon))\,\ell^{i}(\omega)\int_{\bbR^{+}}\Phi_{s,1}^{\In}(\varepsilon,\varepsilon')\,{\mathcal{I}_{s}}_{i}(\varepsilon')\,dV_{\varepsilon'}
  +f_{s}(\omega,\varepsilon)\,\ell^{i}(\omega)\int_{\bbR^{+}}\Phi_{s,1}^{\Out}(\varepsilon,\varepsilon')\,{\mathcal{I}_{s}}_{i}(\varepsilon')\,dV_{\varepsilon'}, \label{eq:collisionTermKernelExpanded_NES}
\end{align}
where in Equation~\eqref{eq:collisionTermKernelExpanded_NES} we have used the linear-order Legendre expansion of the scattering kernels $R_{s}^{\In}$ and $R_{s}^{\Out}$,
\begin{equation}
	R_{s}^{\In/\Out}(\varepsilon,\varepsilon',\cos\alpha)
	:=\Phi_{s,0}^{\In/\Out}(\varepsilon,\varepsilon')+\Phi_{s,1}^{\In/\Out}(\varepsilon,\varepsilon')\,\cos\alpha.  
	\label{eq:kernelExpansion_NES}
\end{equation}

Finally, the collision term due to pair processes is written as \citep{bruenn_1985}
\begin{align}
  \mathcal{C}_{s}^{\Pair}(f_{s},\bar{f}_{s})
  &=
  (1-f_{s}(\omega,\varepsilon))\int_{\bbR^{+}}\f{1}{4\pi}\int_{\bbS^{2}}R_{s}^{\Pro}(\varepsilon,\varepsilon',\cos\alpha)\,(1-\bar{f}_{s}(\omega',\varepsilon'))\,d\omega'\,dV_{\varepsilon'} \nonumber \\
  &\hspace{12pt}
  -f_{s}(\omega,\varepsilon)\int_{\bbR^{+}}\f{1}{4\pi}\int_{\bbS^{2}}R_{s}^{\Ann}(\varepsilon,\varepsilon',\cos\alpha)\,\bar{f}_{s}(\omega',\varepsilon')\,d\omega'\,dV_{\varepsilon'} \label{eq:collisionTerm_Pair} \\
  &=
  (1-f_{s}(\omega,\varepsilon))\,\int_{\bbR^{+}}\Phi_{s,0}^{\Pro}(\varepsilon,\varepsilon')\,(1-\bar{\mathcal{D}}_{s}(\varepsilon'))\,dV_{\varepsilon'}
  -f_{s}(\omega,\varepsilon)\,\int_{\bbR^{+}}\Phi_{s,0}^{\Ann}(\varepsilon,\varepsilon')\,\bar{\mathcal{D}}_{s}(\varepsilon')\,dV_{\varepsilon'}
  \nonumber \\
  &\hspace{12pt}
  -(1-f_{s}(\omega,\varepsilon))\,\ell^{i}(\omega)\int_{\bbR^{+}}\Phi_{s,1}^{\Pro}(\varepsilon,\varepsilon')\,{\bar{\mathcal{I}_{s}}}_{i}(\varepsilon')\,dV_{\varepsilon'}
  -f_{s}(\omega,\varepsilon)\,\ell^{i}(\omega)\int_{\bbR^{+}}\Phi_{s,1}^{\Ann}(\varepsilon,\varepsilon')\,{\bar{\mathcal{I}_{s}}}_{i}(\varepsilon')\,dV_{\varepsilon'}, \label{eq:collisionTermKernelExpanded_Pair}
\end{align}
where in Equation~\eqref{eq:collisionTermKernelExpanded_Pair} we have used a Legendre expansion similar to Equation~\eqref{eq:kernelExpansion_NES} to approximate the angular dependence of the production and annihilation kernels, $R_{s}^{\Pro}$ and $R_{s}^{\Ann}$, respectively.  

\section{Resolution Study of Core-Collapse Simulations in Spherical Symmetry}
\label{app:resolution}

In this Appendix, we present results from resolution studies of the fiducial, Minerbo-closure CCSN model in spherical symmetry from Section~\ref{sec:application_ccsn}, varying spatial and spectral resolution independently.
The studies are performed independently to isolate their respective effects: varying the AMR structure probes sensitivity to hydrodynamic and transport discretization in position space, whereas varying the spectral grid isolates the influence of energy-space resolution on neutrino heating and cooling rates, as well as on the discretization of velocity-dependent observer corrections; specifically Doppler shifts.  

For the spatial-resolution study with AMR, keeping the spectral resolution fixed at Grid~C from Table~\ref{tab:energyGrids}, we hold the coarsest-level resolution fixed at $(\Delta r)^{\ell=0} = 64$~km and vary the maximum refinement level from $\ell_{\max}=5$ to $\ell_{\max}=9$.
This corresponds to finest-level resolutions ranging from $(\Delta r)^{\ell_{\rm max}}=2$~km down to $0.125$~km.  
We apply the same refinement strategy during collapse as described in Section~\ref{sec:application_ccsn}, increasing $\ell_{\rm max}$ progressively from $\ell_{\max}-4$ up to $\ell_{\max}$ when the central density satisfies $\log_{10}(\rho_{\rm c}/[{\rm g}~{\rm cm}^{-3}])=11$, $11.5$, $12$, and $12.5$.  
As before, the innermost $600$~km is always refined to the maximum level.  
With this setup, the initial grid at the onset of collapse differs across models because refinement is initially restricted to $\ell_{\max}-4$.  
The DG discretization is unchanged: we use linear elements, so that each DG element spans two FV cells.  

\begin{figure}[h]
    \centering
    \includegraphics[width=0.95\textwidth]{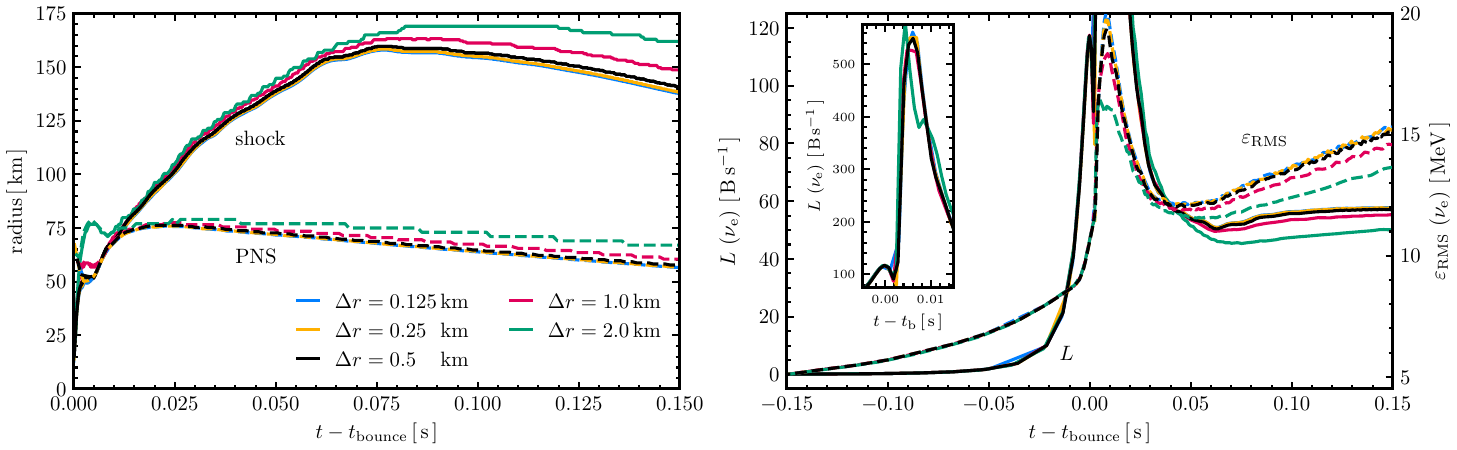}
    \caption{Results from the spatial resolution study, showing selected quantities versus time from simulations in which the minimum finite-volume cell width was varied, while the spectral resolution corresponds to Grid~C in Table~\ref{tab:energyGrids}.  
    The left panel shows the shock (solid lines) and PNS (dashed lines) radii as functions of time, and the right panel shows the electron-neutrino luminosity (solid lines) and RMS energy (dashed lines) at $r=500$~km as functions of time.  
    Line colors indicate the minimum finite-volume radial cell width at the highest AMR level: 0.125~km (blue), 0.25~km (yellow), 0.5~km (black), 1.0~km (red), and 2.0~km (green).}
    \label{fig:ccsn_spatial_resolution}
\end{figure}

Figure~\ref{fig:ccsn_spatial_resolution} shows that key global quantities are effectively converged once the finest-level spatial resolution reaches $0.5$~km.  
At this resolution, the major features of the shock and PNS radii closely match those of higher-resolution simulations.  
Coarser meshes, however, introduce progressively larger deviations: the early shock excursion ($\lesssim15$~ms after bounce) differs noticeably, the later-time shock recession proceeds more slowly, and PNS contraction is reduced.  
Between $50$ and $150$~ms, both the electron-neutrino luminosity and RMS energy also decrease as the spatial resolution is degraded.  

\begin{figure}[h]
    \centering
    \includegraphics[width=0.95\textwidth]{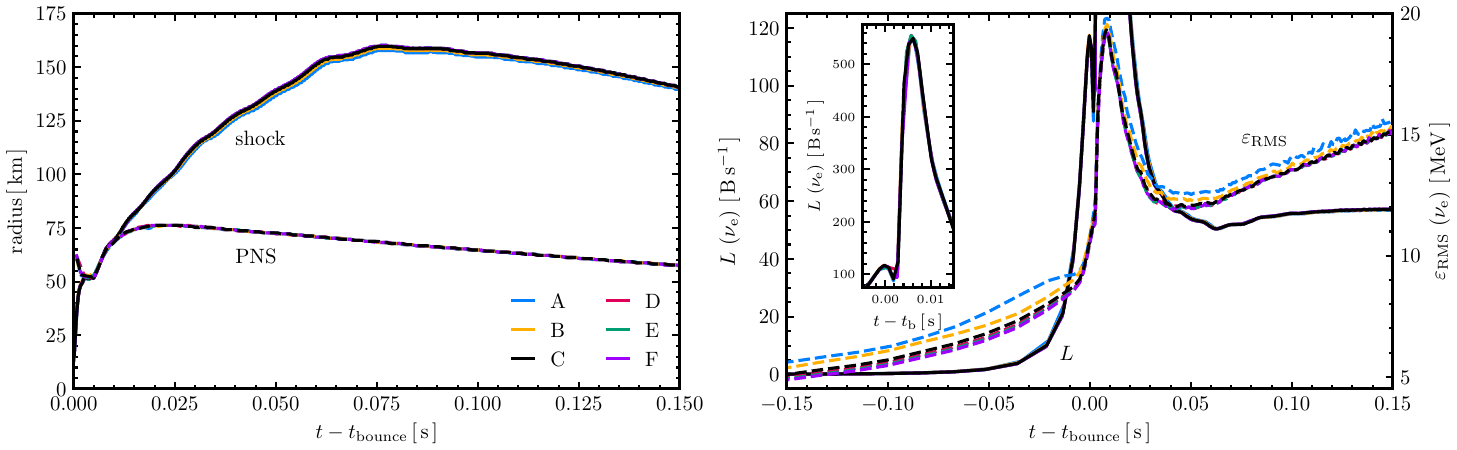}
    \caption{Results from the spectral resolution study, showing selected quantities versus time from simulations in which the parameters of the spectral grid were varied, while the spatial resolution corresponds to a minimum finite-volume radial cell width of $0.5$~km at the highest AMR level.  
    The left panel shows the shock (solid lines) and PNS (dashed lines) radii as functions of time, and the right panel shows the electron-neutrino luminosity (solid lines) and RMS energy (dashed lines) at $r=500$~km as functions of time.  
    Line colors indicate the spectral grids listed in Table~\ref{tab:energyGrids}: Grid~A (blue), Grid~B (yellow), Grid~C (black), Grid~D (red), Grid~E (green), and Grid~F (purple).}
    \label{fig:ccsn_spectral_resolution}
\end{figure}

For the spectral-resolution study, we use the fiducial model from Section~\ref{sec:application_ccsn} with $(\Delta r)^{\ell_{\rm max}} = 0.5$~km, and vary the spectral grid using all options in Table~\ref{tab:energyGrids}.
Figure~\ref{fig:ccsn_spectral_resolution} shows that the shock and PNS radii exhibit only minor variations across the different spectral resolutions.  
The peak deleptonization-burst luminosity is similarly insensitive to the choice of grid.  
The electron-neutrino RMS energy at $r=500$~km displays some pre-bounce sensitivity, whereas post-bounce only the coarsest resolution (Grid~A) deviates appreciably from the others.  
Overall, the solutions are effectively converged with spectral Grid~C or finer.  

\software{Matplotlib~\citep{Hunter2007_matplotlib},~NumPy~\citep{harris2020_numpy},~SciPy~\citep{2020SciPy},~\weaklib~\footnote{\href{https://github.com/starkiller-astro/weaklib}{github.com/starkiller-astro/weaklib}},~\thornado~\footnote{\href{https://github.com/endeve/thornado}{github.com/endeve/thornado}},~\flashx~\footnote{\href{https://flash-x.org}{flash-x.org}}~\citep{dubey_etal_2022},~\chimera~\citep{bruenn_etal_2020}}

\clearpage

\bibliography{references}

\end{document}